\documentclass{article}

\usepackage[authoryear]{natbib}
\usepackage{graphicx}
\usepackage{subfigure}
\usepackage{float}
\usepackage{amsmath}
\usepackage{color}
\usepackage{multirow}
\usepackage{amssymb}
\usepackage[top=1in, bottom=1in, left=1.0in, right=1.0in, letterpaper]{geometry}

\begin{document}

\title{An Efficient Coarse Grid Projection Method for Quasigeostrophic Models of Large-Scale Ocean Circulation}


\author{Omer San and Anne E. Staples \\ Department of Engineering Science and Mechanics \\ Virginia Tech, Blacksburg, VA 24061, USA}





\date{}


\maketitle

\section*{Abstract}
This paper puts forth a coarse grid projection (CGP) multiscale method to accelerate computations of quasigeostrophic (QG) models for large scale ocean circulation. These models require solving an elliptic sub-problem at each time step, which takes the bulk of the computational time. The method we propose here is a modular approach that facilitates data transfer with simple interpolations and uses black-box solvers for solving the elliptic sub-problem and potential vorticity equations in the QG flow solvers. After solving the elliptic sub-problem on a coarsened grid, an interpolation scheme is used to obtain the fine data for subsequent time stepping on the full grid. The potential vorticity field is then updated on the fine grid with savings in computational time due to the reduced number of grid points for the elliptic solver. The method is applied to both single layer barotropic and two-layer stratified QG ocean models for mid-latitude oceanic basins in the beta plane, which are standard prototypes of more realistic ocean dynamics. The method is found to accelerate these computations while retaining the same level of accuracy in the fine-resolution field. A linear acceleration rate is obtained for all the cases we consider due to the efficient linear-cost fast Fourier transform based elliptic solver used. We expect the speed-up of the CGP method to increase dramatically for versions of the method that use other, suboptimal, elliptic solvers, which are generally quadratic cost. It is also demonstrated that numerical oscillations due to lower grid resolutions, in which the Munk scales are not resolved adequately, are effectively eliminated with CGP method.\\

\emph{\textbf{Keywords:}}Coarse grid projection, multigrid, forced-dissipative quasigeostropic ocean models, large-scale ocean circulation, single-layer barotropic model, two-layer stratified model, climate modeling.

\section{Introduction}
\label{sec:intro}

Oceanic and atmospheric flows display an enormous range of spatial and temporal scales, from seconds to decades and from centimeters to thousands of kilometers. Scale interactions, both spatial and temporal, are the dominant feature of all aspects of general circulation models in geophysical fluid dynamics \citep{klein2010scale,hurrell2009unified} and bridging the scales in geophysical systems is of paramount importance in numerous critical areas and industries. Atmospheric and oceanic flows have intrinsic complex multiscale interactions. The accurate and efficient numerical simulation of these geophysical flows is of great importance in weather and climate models and could perhaps benefit from a dedicated investigation of the application of classes of modern \emph{multiscale modeling and simulation (MMS)} methods \citep{brandt2002multiscale,barth2002multiscale,fish2009multiscale,weinan2011principles}, which exploit vastly different temporal and spatial scales in problems in order to speed up their computations. Similar to the parallel developments in applied mathematics and engineering science, interest in the development and testing of the MMS methods in geophysical flow settings has increased in recent years \citep{iskandarani2002multi,majda2003systematic,khairoutdinov2008evaluation,alam2011towards,campin2011super,ringler2011exploring}, leading to several successful applications and research efforts at numerous weather and climate centers. The MMS framework could provide a significant increase in the accuracy and computational efficiency of numerous interconnected physical oceanic and atmospheric flow models for advanced numerical weather prediction, data assimilation and climate modeling strategies.

The ocean is a forced-dissipative fluid system, with forcing largely at the boundaries and dissipation at the molecular scale.
The investigation of characteristics of the forced-dissipative ocean circulation models is of primary importance in developing our understanding of the complex large-scale nonlinear motions of large-scale oceanic flows that move in great circular sweeps (called ``gyres"). As one of the main circulation sources, most of the surface currents in the ocean are shaped by wind, which drives the general circulation associated with the subtropical and subpolar gyres, which can be identified with the strong, persistent, sub-tropical and sub-polar western boundary currents in the North Atlantic Ocean (the Gulf Stream and the Labrador Current) and North Pacific Ocean (the Kuroshio and the Oyashio Currents) and sub-tropical counterparts in the southern hemisphere \citep{stommel1972gulf,kelly2010western}. One of the major similarities between the various ocean basins is the asymmetry of the gyres: strong western boundary currents and weaker flow in the interior; weak and shallow eastern boundary currents. The most obvious motivation for being interested in forced-dissipative wind-driven ocean circulation is the connection between ocean currents and climate dynamics \citep{ghil2008climate,lynch2008origins,stocker2011introduction}.

Ocean eddies have a horizontal scale of typically several hundred kilometers and are often much stronger than the mean flow, leading to a highly turbulent, chaotic flow. The mean pattern of currents only emerges after averaging over many years.
The wind-driven circulation in an enclosed, midlatitude rectangular or square oceanic basin is a classical problem in physical oceanography, studied extensively by modelers \citep{allen1980models,holland1980example,griffa1989wind,vallis2006atmospheric,miller2007numerical}. To decrease the computational cost required for an accurate representation of large-scale oceanic flows, several class of simplified models are derived from the full-fledged equations of geophysical flows, Boussinesq equations (BEs) or the primitive equations (PEs), to guide the theoretical studies on boundary currents, alternating zonal flows, or jet formations, as well as to identify some key issues related to the robustness of the model dynamics to the changes of parameters that is closely linked to a dynamical system point of view \citep{speich1995successive,meacham2000low,chang2001transition,nauw2004regimes,dijkstra2005nonlinear,dijkstra2005low}. The quasigeostropic (QG) model is a simplification of the primitive equation model that retains many of the essential features of geophysical fluid flows. Details of the mathematical and physical approximations may be found in standard textbooks on geophysical fluid dynamics, such as \cite{pedlosky1987geophysical}, \cite{vallis2006atmospheric}, and \cite{mcwilliams2006fundamentals}. The main assumptions that go into the QG models are: the hydrostatic balance, the $\beta$-plane approximation (i.e., the variation of the Coriolis parameter with latitude), the geostrophic balance, and the eddy viscosity parameterization. Despite the fact that the QG models are a simplified version of the full-fledged equations of geophysical flows, their numerical simulation is still computationally challenging when long-time integration is required, as is the case in climate modeling.

The {\em barotropic vorticity equation (BVE)} represents one of the most commonly used mathematical models for this type of geostrophic flows with various dissipative and forcing terms \citep{majda2006non}. In reality, the ocean is a stratified fluid on a rotating Earth driven from its upper surface by patterns of momentum and buoyancy fluxes \citep{marshall1997hydrostatic}. While the barotropic model is not stratified, it exhibits many of the features that are observed in the stratified case. To explore some of the effects of the stratification, the one-layer barotropic equation can be extended to the 1.5-layer model, also called the reduced gravity QG model \citep{özgökmen2001connection}. There are two layers in this model, but the second layer is infinitely deep and at rest (passive), and the dynamics are effectively barotropic. The two-layer model takes the next step in increasing the complexity of stratification by adding a second dynamically active layer \citep{holland1978role,özgökmen1998emergence,berloff1999large,dibattista2001equilibrium,berloff2009mechanism}. The dynamics in this model include the first baroclinic modes. The complexity of the models could be increased by adding more active layers, resulting in the N-layer models \citep{siegel2001eddies}, which, in turn, yield the three dimensional primitive equations when N goes to infinity \citep{mcwilliams2006fundamentals}. In this study, we use both the {\it one-layer quasigeostropic (QG1)} and {\it two-layer quasigeostropic (QG2)} models.

Although large-scale ocean dynamics are well represented by QG ocean models, primitive equation models have been used in state of art weather
prediction softwares. Several reasons have been addressed for the recent lack of popularity in of these quasigeostropic models \citep{miller2007numerical}. One is the ready availability of full-fledged primitive equation codes on the web, but that is not the only reason. There is a need for computational strategies that can significantly decrease the computational cost of the geophysical models without compromising their physical accuracy. Large eddy simulation (LES) approaches appear as a natural choice to accelerate the simulation on coarser grid in which the subgrid scale terms are modeled to capture the under-resolved flow, i.e., the flow in the regions where the grid size becomes greater than the specified Munk scale. \cite{san2011approximate} proposed an approximate deconvolution large eddy simulation technique for one-layer barotropic quasiqeostropic large-scale ocean model. It was shown that the approximate deconvolution model provides an accurate approximation for under-resolved subfilter-scale effects.

Here, we approach to the problem from a different point of view. Instead of filtering the governing equations and modeling underresolved quantities on a coarser grid, we separate the problem to two parts based on the nature of the QG models: (i) the elliptic sub-problem and (ii) the potential vorticity evolution. Most of the demand on computational resources by QG models comes in the solution of the elliptic inversion sub-problem which states the relationship between the potential vorticity and stream function. The natural advantages of QG models may well lead to increased application when efficient methods available for elliptic sub-problem. One straightforward way to accelerate QG simulations is to reduce the number of grid points for the most time consuming part of the problem, the elliptic sub-problem. Previous studies have demonstrated that scale dependant computational slowness can be overcome by multiscale and multiresolution algorithms \citep{brandt2005multiscale}. Along this direction, the coarse grid projection (CGP) framework was proposed by \cite{lentine2010anovel} and successfully applied to three-dimensional incompressible flow simulations by coarsening the number of grid points for the Poisson equation. A systematic error analysis on the coarse grid projection method has been recently performed by \cite{san2012a} for Navier-Stokes equations. The main goal of this report is to extend the CGP approach to the QG large scale ocean circulation models. The cost of the flow computations is reduced by coarsening the resolution of the numerical grid on which the elliptic sub-problem is solved by factors of two in each direction according to $M = 2^{-\ell}N$, where $N$ is the fine resolution of the numerical grid on which the potential vorticity equation is solved, and $M$ is the coarse resolution for the solution of the elliptic sub-problem. When $\ell=0$ no coarsening is applied and the CGP method reduces to the underlying standard QG solver. In our numerical investigations, we serendipitously discovered that the CGP procedure can actually predict the fine level simulation details without loss of accuracy for a reduced computational cost.

The coarse grid projection methodology is a modular approach that facilitates data transfer with simple restriction and prolongation interpolations and uses black-box solvers for the advection-diffusion and elliptic parts of the QG models. A particular version of the method is applied here using a third-order Runge-Kutta method for advection-diffusion part and a fast Fourier transform based direct solver for the elliptic part. The full weighting operation for mapping from the fine to coarse grids is used to obtain the data for the elliptic sub problem. After solving the elliptic part on a coarsened grid, bilinear interpolation is used to obtain the fine data for consequent time stepping on the full grid. Similar mapping operators have been used in multigrid algorithms \citep{brandt1977multi,hackbusch1985multi,briggs2000multigrid,trottenberg2001multigrid}. The efficiency of the interpolation based methods highly depends on the interpolation scheme \citep{fish2004discrete}, and has been substantially investigated in the multigrid literature. Since we are using a second-order spatial discretization scheme, however, the full weighting operator for the restriction and bilinear interpolation for the prolongation are consistent and efficient for our study. The mapping procedures used in this study could potentially be improved by introducing higher-order spline formulas, or more advanced methods for deriving interpolations \citep{lee1997scattered,hollig2002multigrid,brezina2005adaptive,brandt2010principles}. The coarse grid projection approach proposed here can be classified within the systematic upscaling methodologies that have been used in the context of the multigrid/multiresolution branch of multiscale methods.

The rest of the paper is organized as follows:
Both barotropic one-layer BVE and two-layer stratified QG models, the mathematical models used in this report, are presented in Section \ref{sec:mm}.
Section \ref{sec:cgp} presents the coarse grid projection methodology for these QG models. The numerical methods used in our simulations are briefly discussed in Section \ref{sec:numerics}. The results for the new CGP method for the large-scale ocean models are presented in Section \ref{sec:results}.
Finally, the conclusions are summarized in Section \ref{sec:summary}.

\section{Mathematical Models}
\label{sec:mm}
In this section, we present the quasigeostropic models used in the numerical investigation of the proposed coarse grid projection method for large scale ocean circulation. We first present the BVE which is one of the most used mathematical models for geostrophic flows with various dissipative
and forcing terms \citep{majda2006non}. Next, we present the two-layer QG model which is the simple extension of single-layer barotropic model to the stratified ocean by including another active layer.

\subsection{One-layer barotropic model}

In this section, we present the BVE, one of the most used mathematical models for forced-dissipative large scale ocean circulation problem. Studies of wind-driven circulation using an idealized double-gyre wind forcing have played an important
role in understanding various aspects of ocean dynamics, including the role of mesoscale eddies and
their effect on mean circulation.
Following \cite{greatbatch2000four}, we briefly describe the BVE.
For more details on the physical mechanism and various formulations utilized, the reader is referred to
\cite{greatbatch2000four,munk1982observing,cummins1992inertial,nadiga2001dispersive,fox2005reevaluating}.

The barotropic vorticity equation for one-layer quasigeostropic forced-dissipative ocean model can be written as
\begin{equation}
\frac{\partial q}{\partial t} + J(\psi,q) = D + F .
\label{eq:ge}
\end{equation}
where $D$ and $F$ represent the dissipation and forcing terms, respectively. In Eq.~(\ref{eq:ge}), $q$ is the potential vorticity, defined as
\begin{equation}
q = \omega + \beta y,
\label{eq:pv}
\end{equation}
where $\beta$ is the gradient of the Coriolis parameter at the basin center ($y=0$). Here, $\omega$ is the local vorticity, the curl of the velocity field, and the kinematic relationship between the vorticity and the stream function yields the following definition
\begin{equation}
\omega = \nabla^2 \psi = \frac{\partial^2 \psi}{\partial x^2} + \frac{\partial^2 \psi}{\partial y^2} .
\label{eq:ke}
\end{equation}
in which $\nabla^2$ is the Laplacian operator, and $\psi$ symbolizes the velocity stream function. Flow velocity components can be found
from the stream function:
\begin{equation}
u = -\frac{\partial \psi}{\partial y}; \quad v = \frac{\partial \psi}{\partial x}
\label{eq:vell}
\end{equation}
The nonlinear convection term in Eq.~(\ref{eq:ge}), called the Jacobian, is defined as
\begin{equation}
J(\psi,q) = \frac{\partial \psi}{\partial x}\frac{\partial q}{\partial y} - \frac{\partial \psi}{\partial y}\frac{\partial q}{\partial x}.
\label{eq:jac}
\end{equation}
The viscous dissipation in Eq.~(\ref{eq:ge}) has the form
\begin{equation}
D
= \nu
\left(\frac{\partial^2 \omega}{\partial x^2} + \frac{\partial^2 \omega}{\partial y^2}\right) ,
\label{eq:dis}
\end{equation}
where $\nu$ is the uniform eddy viscosity coefficient. Viscous term can be written in terms of stream function in the following form
\begin{equation}
D=\nu \nabla^4 \psi
\label{eq:dis}
\end{equation}
The double-gyre wind forcing is given by
\begin{equation}
F = \frac{\pi \tau_0}{\rho H} \sin(\pi y) ,
\label{eq:forc}
\end{equation}
where $\tau_0$ is the maximum amplitude of double-gyre wind stress, $\rho$ is the mean density, and $H$ is the mean depth of the ocean basin. In order to nondimensionalize the BVE equation we use the following definitions
\begin{equation}
x = \frac{\tilde{x}}{L}, \ y = \frac{\tilde{y}}{L}, \ t = \frac{\tilde{t}}{L/V}, \ q = \frac{\tilde{q}}{\beta L}, \ \psi= \frac{\tilde{\psi}}{V L},
\label{eq:sverdrup}
\end{equation}
where the tilde denotes the corresponding dimensional variables. In the nondimensionalization, $L$ represents the characteristic horizontal length scale (i.e., in our study $L$ is the basin dimension in $x$ direction), and $V$ represents the characteristic velocity scale. The Sverdrup velocity scale used for nondimensionalization can be written in the follwing form:
\begin{equation}
V = \frac{\pi \ \tau_0}{\rho H \beta L}.
\label{eq:vscale}
\end{equation}
Finally, the governing equations for two-dimensional incompressible barotropic flows can be written in
dimensionless form of the potential vorticity formulation in the beta plane as the BVE:
\begin{equation}
\frac{\partial \tilde{q}}{\partial \tilde{t}} + J(\tilde{\psi},\tilde{q}) = \left(\frac{\delta_{M}}{L}\right)^{3} \tilde{\nabla}^4 \tilde{\psi} + sin(\pi \tilde{y}) .
\label{eq:ge}
\end{equation}
where $\delta_{M}$ is the Munk scale. The elliptic sub-problem which relates the potential vorticity and stream function becomes
\begin{equation}
\tilde{q} = \left(\frac{\delta_{I}}{L}\right)^{2} \, \tilde{\nabla}^2 \tilde{\psi} + \tilde{y} ,
\label{eq:pv}
\end{equation}
where $\delta_{I}$ is defined as Rhines scale. In dimensionless form, there are only two physical parameters, the Rhines scale and the Munk scale, which are related to the physical parameters in the following way:
\begin{equation}
\frac{\delta_{I}}{L} = \left(\frac{V}{\beta  L^2}\right)^{1/2}; \quad \quad \frac{\delta_{M}}{L} = \left(\frac{\nu}{\beta  L^3}\right)^{1/3}.
\label{eq:relation}
\end{equation}
The physical parameters in the BVE (\ref{eq:ge}), the Rhines scale
$\delta_{I}$
and the Munk scale $\delta_{M}$,
are related to the Reynolds and Rossby numbers through the following formulas:
\begin{equation}
\frac{\delta_{I}}{L}= (\mbox{Ro})^{1/2} ,
\label{eq:Rhines}
\end{equation}
\begin{equation}
\frac{\delta_{M}}{L}= (\mbox{Re}^{-1}\mbox{Ro})^{1/3} ,
\label{eq:Munk}
\end{equation}
where $\mbox{Ro}$ is the Rossby number and $\mbox{Re}$ is the Reynolds number based on the basin dimension, $L$.
We note that some authors use a boundary layer Reynolds number, which is written as
\begin{equation}
\mbox{Re}_B=\mbox{Re} \frac{\delta_{I}}{L} = \frac{\delta_{I}^{3}}{\delta_{M}^{3}} ,
\label{eq:Reb}
\end{equation}
where $\mbox{Re}_B \sim O(10)-O(10^3)$ for oceanic flows \cite{fox2005reevaluating}.
Finally, in order to completely specify the mathematical model, boundary and initial conditions need to be prescribed. In many theoretical studies of large scale ocean circulation, slip or no-slip boundary conditions are used.  Following these studies \cite{greatbatch2000four,nadiga2001dispersive,holm2003modeling,cummins1992inertial,munk1950wind,bryan1963numerical}, we use slip boundary conditions for the velocity, which translate into homogenous Dirichlet boundary conditions for the vorticity:
$\omega |_{\Omega} = 0$.
The impermeability boundary condition is imposed as $\psi |_{\Omega} = 0$.
For the initial condition, we start our computations from a quiescent state and integrate Eq.~(\ref{eq:ge}) until a statistically steady state
is obtained in which the wind forcing, dissipation, and Jacobian balance each other.

\subsection{The two-layer quasigeostrophic equations}
The two-layer quasigeostrophic model used in this study is one of the simplified forced-dissipative oceanic models that considers baroclinic effects. The stratified ocean is partitioned into two isopycnal layers, each of constant depth, density and temperature. The governing quasigeostrophic potential vorticity equations for the two dynamically active layers are \citep{pedlosky1987geophysical,salmon1998lectures,mcwilliams2006fundamentals}
\begin{eqnarray}
\frac{\partial q_1}{\partial t} + J(\psi_1,q_1) &=& D_1 + F_1, \\
\frac{\partial q_2}{\partial t} + J(\psi_2,q_2) &=& D_2 + F_2,
\label{eq:ge}
\end{eqnarray}
where the layer index starts from top, $q_i$ represents potential vorticities, and $\psi_i$ denotes for streamfunctions.
The dissipation and forcing (Ekman pumping) terms are represented by $D_i$, and $F_i$, respectively. The potential vorticities for each layer are related to the velocity streamfunctions through the following elliptic coupled system of equations:
\begin{eqnarray}
q_1 &=& \nabla^2\psi_1 + \beta y + \frac{f_{0}^{2}}{g'H_1}(\psi_2-\psi_1), \\
q_2 &=& \nabla^2\psi_2 + \beta y + \frac{f_{0}^{2}}{g'H_2}(\psi_1-\psi_2).
\label{eq:pv}
\end{eqnarray}
The isopycnal flow velocity components can be found from the velocity streamfunctions:
\begin{equation}
u_i = -\frac{\partial \psi_i}{\partial y}; \quad v_i = \frac{\partial \psi_i}{\partial x}.
\label{eq:vel}
\end{equation}
The two symbols $\beta$ and $f_0$ are parts of the linearized $\beta$-plane approximation to the Coriolis parameter $f=f_0 + \beta y$. Here $f_0=2 \,  \Omega \, \mbox{sin}(\phi_0)$ is the local rotation rate at $y=0$, where $\Omega$ is the rotational speed of the earth and $\phi_0$ is the latitude at $y=0$. This is equivalent to approximating the spherical Earth with a tangent plane at $y=0$. Stratification is represented by two stacked isopycnal layers with thicknesses $H_1$ and $H_2$, starting from the top, and $g'=g \frac{\Delta \rho}{\rho_1}$ is reduced gravity associated with the density jump between the two layers in which $\Delta \rho$ is the density difference between the two layers, $\rho_1$ is the reference (upper layer) density, and $g$ is the gravitational acceleration. The inertial radius of deformation between layers, a measure of stratification strength, is defined as the Rossby deformation radius $R_d=\sqrt{\frac{g' H_1H_2}{f_{0}^2 H}}$, where $H=H_1 + H_2$.
In this study, the top and bottom layers of the ocean are forced by an Ekman pumping of the form
\begin{eqnarray}
F_1 &=& \frac{1}{\rho_1 H_1} \hat{k} \cdot \nabla \times \vec{\tau} ,  \\
F_2 &=& -\gamma \nabla^2\psi_2 ,
\label{eq:force}
\end{eqnarray}
where $\vec{\tau}=(\tau^{(x)},\tau^{(y)})$ is the stress vector for surface wind forcing, and $\hat{k}$ is unit vector in vertical direction. In the present model, we use a double-gyre wind forcing only for zonal direction: $\tau^{(x)}=\tau_0 \,  \mbox{cos}\bigg(\frac{2\pi}{L} y\bigg)$, where $L$ is the meridional length of the ocean basin centered at $y=0$, and $\tau_0$ is the maximum amplitude of the wind stress. This form of wind stress represents the meridional profile of easterly trade winds, mid-latitude westerlies, and polar easterlies from South to North. The bottom Ekman layer is parameterized by a linear bottom friction with coefficient $\gamma$.  In the equations above, $\nabla$ and $\nabla^2$ are the gradient and Laplacian operators, respectively.
For the dissipation terms, the following EV parameterizations are used:
\begin{eqnarray}
D_1 &=& \nu \nabla^4\psi_1 , \\
D_2 &=& \nu \nabla^4\psi_2 ,
\label{eq:dis}
\end{eqnarray}
where $\nu$ is eddy viscosity coefficient. Similar to the previous analysis, the governing equations can be written in dimensionless form by using the Sverdrup balance to set the velocity scale of the form
\begin{equation}
V = \frac{2 \pi \tau_0}{\rho_1 H_1 \beta L }.
\label{eq:sverdrup}
\end{equation}
Then the two-layer quasigeostrophic equations in dimensionless form become
\begin{eqnarray}
\frac{\partial \tilde{q}_1}{\partial \tilde{t}} + J(\tilde{\psi}_1,\tilde{q}_1) &=&A \tilde{\nabla}^4\tilde{\psi}_1 + \mbox{sin}(2\pi \tilde{y}) , \\
\frac{\partial \tilde{q}_2}{\partial \tilde{t}} + J(\tilde{\psi}_2,\tilde{q}_2) &=&A \tilde{\nabla}^4\tilde{\psi}_2 - \sigma \tilde{\nabla}^2\tilde{\psi}_2 .
\label{eq:ge-non}
\end{eqnarray}
In dimensionless form, the kinematic relationships between potential vorticities and streamfunctions yield the following elliptic sub-problem:
\begin{eqnarray}
\tilde{q}_1 &=& \mbox{Ro} \tilde{\nabla}^2\tilde{\psi}_1 + \tilde{y} + \frac{\mbox{Fr}}{\delta}(\tilde{\psi}_2-\tilde{\psi}_1) , \\
\tilde{q}_2 &=& \mbox{Ro} \tilde{\nabla}^2\tilde{\psi}_2 + \tilde{y} + \frac{\mbox{Fr}}{1-\delta}(\tilde{\psi}_1-\tilde{\psi}_2) .
\label{eq:pv-non}
\end{eqnarray}
For clarity of exposition, in the remainder of the paper we will drop the tilde symbol used for the dimensionless variables.
In the two-layer QG model, $\delta=\frac{H_1}{H}$ is the aspect ratio of vertical layer thicknesses, $\mbox{Ro}$ is the Rossby number, $\mbox{Fr}$ is the Froude number, $A$ is the lateral eddy viscosity coefficient, and $\sigma$ is the Ekman bottom later friction coefficient. The definitions of these dimensionless parameters are:
\begin{equation}
\mbox{Ro}=\frac{V}{\beta L^2}; \  \mbox{Fr}=\frac{f_{0}^{2}V}{g'\beta H}; \ A=\frac{\nu}{\beta L^3}; \ \sigma =\frac{\gamma}{\beta L}.
\label{eq:ndim}
\end{equation}

The following three length scales are useful for setting the two-layer problem parameters: (i) the Munk scale, $\delta_M=\bigg(\frac{\nu}{\beta}\bigg)^{1/3}$, for the viscous boundary layer; this is related to the smaller scale dissipation; (ii) the Stommel scale, $\delta_S=\frac{\gamma}{\beta}$, for the bottom boundary layer thickness; this is accounting for larger scale damping; and (iii) the Rhines scale, $\delta_I=\bigg(\frac{V}{\beta}\bigg)^{1/2}$, for the inertial boundary layer; this is measuring the strength of the nonlinearity.

In order to complete the mathematical model, boundary and initial conditions should be prescribed. In many theoretical studies of ocean circulation, the modelers either use free-slip boundary conditions or no-slip boundary conditions. Following \cite{cummins1992inertial,özgökmen1998emergence}, we use free-slip boundary conditions
for the velocity for both isopycnal layers, which translates into homogenous Dirichlet boundary conditions for the vorticity (Laplacian of streamfunction):
$\nabla^2 \psi |_{\Omega} = 0$. The impermeability boundary condition is imposed as $\psi |_{\Omega} = 0$. Similar to the one-layer problem, we start from a rest state,  integrate the model until a statistically steady state is obtained, and continue for several decades to compute time-averaged results.

\section{Coarse Grid Projection}
\label{sec:cgp}
\begin{figure}[!b]\vspace{-5pt}
\centering
\includegraphics[width=0.6\textwidth]{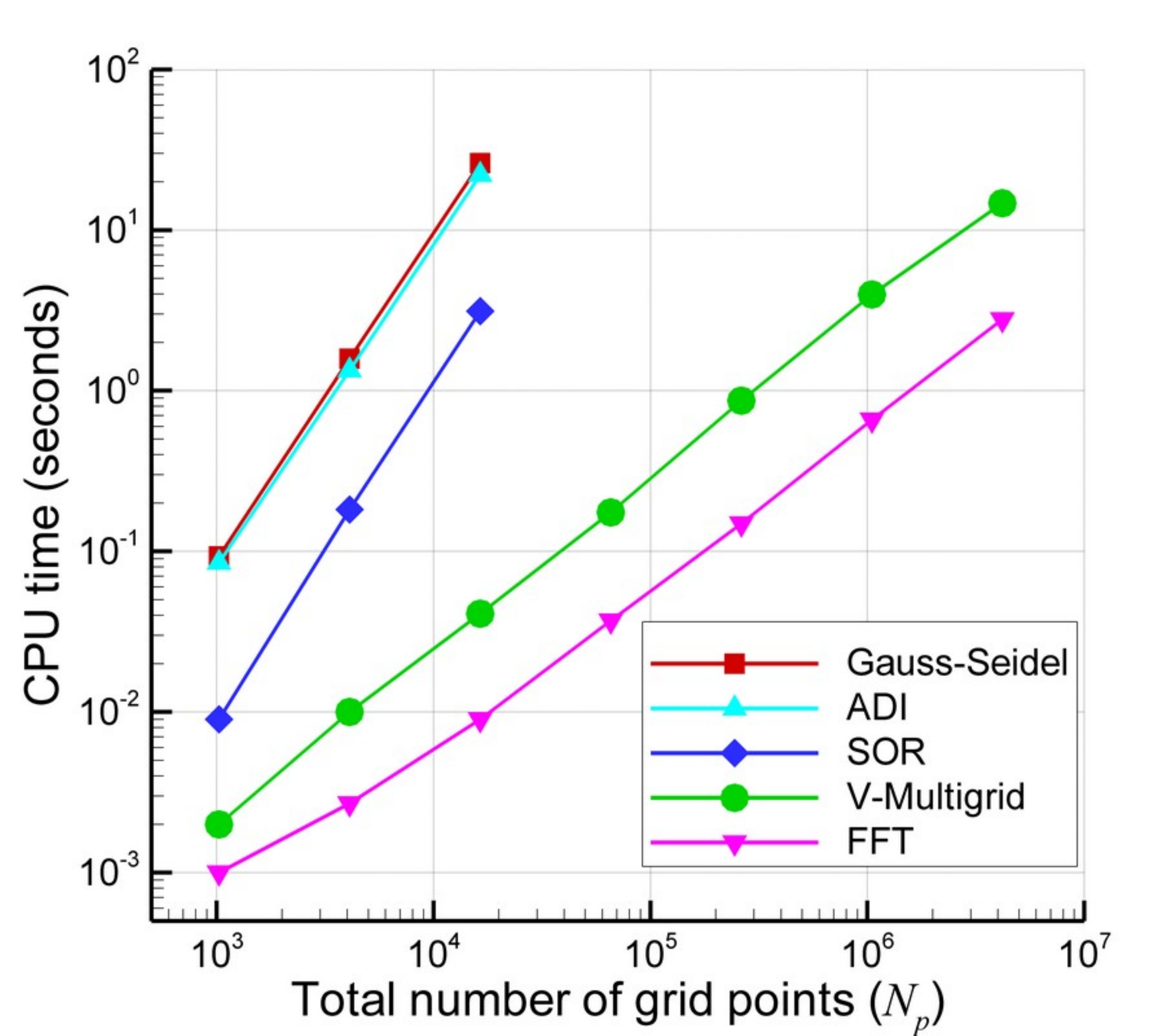}
\caption{Computational efficiency of Poisson solvers. ADI, GS, and SOR are classical iterative solvers that scale as $N_{p}^{2}$. FFT is the fast Fourier transform based direct solver, and V-Multigrid is the V-cycle iterative multigrid solver, which both scale as $N_p$.}
\label{fig:poisson}
\end{figure}

In QG models solving the elliptic sub-problem takes considerably more computational time than solving the time dependent part (i.e., potential vorticity evolution equation) of the problem. Within each time step, solving the time dependant part of the problem is usually of $O(N_p)$ where ``$N_p$" is the number of degrees of freedom (total grid points, $N_x \times N_y$ for one-layer setting and $2 N_x \times 2 N_y$ for two-layer setting) of the problem. In general, the alternating direction implicit (ADI), Gauss-Seidel (GS) or  successive over relaxation (SOR) types of iterative algorithms for solving the elliptic equation are of $O(N_{p}^{2})$ \citep{saad2003iterative}. The practical consequence is that it is not feasible to use these types of iterative elliptic solvers for high resolution (and therefore high Reynolds number) computations along with long time integration. In order to accelerate these solvers, very successful multigrid algorithms have been developed that reduce the computational effort to close to $O(C_{MG}N_p)$ where $C_{MG}$ is a proportionality constant \citep{wesseling1995introduction,gupta1997comparison,zhang1998fast}. On the other hand, for certain ideal problems on equally-spaced grids, fast Fourier transform (FFT) based fast Poisson solvers can be used that are $O(C_{FFT}N_p log(N_p))$, and are presently the fastest algorithms ($C_{FFT}log(N_p) < C_{MG}$ in the relevant resolutions) for solving Poisson equations \citep{moin2001fundamentals}. The computational efficiencies of different elliptic solvers are shown in Fig.~(\ref{fig:poisson}) for solving a standard Poisson equation on a square domain with equidistant grid spacing. This preliminary comparison shows that FFT based elliptic solver is the most efficient solver for our sub-problems in the QG models. The computational gains in the CGP method are directly related to selection of the elliptic solver. We have intentionally selected most efficient linear-cost elliptic solver to show the concept. We highlight that any computational gain obtained with this optimal linear-cost FFT based elliptic solver would be more immense if we used some other sub-optimal iterative type elliptic solver.

The basic idea behind coarse grid projection (CGP) is to use a smaller number of grid points for solving the elliptic sub-problem. The CGP approach we propose here is modular and independent of the elliptic solver that is used. As shown by our preliminary analysis for solving Poisson equation, usually FFT based elliptic solvers are optimal for rectangular domain problems because they are fast and have minimal storage requirements. They are preferred because of their efficiency when using orthogonal coordinate systems in which there are no mixed derivatives. The general domains treated with body-fitted coordinates are out of bounds for FFT solvers due to the presence of mixed derivatives in the transformed generalized curvilinear coordinates. In that case, the V-cycle multigrid solver becomes the optimal elliptic solver to use. Since we used Drichlet type boundary conditions (i.e., free-slip boundary contions) for our oceanic basins, a fast sine transform are utilized which we will briefly address on the following section where we present the numerical methods.

Even though we use the fastest elliptic solver available for our computations, the most time consuming part of the computations is still solving the elliptic sub-problem. With this in mind, we propose a new multiresolution approach that reduces the number of degrees of freedom for the elliptic solver part of the problem to accelerate the computation. The procedure is as follows: first, we solve the potential vorticity part of the problem using a fine resolution, $N$ (the resolution in one direction). Next, we restrict these data to a coarser grid with resolution $M = 2^{-\ell}N$, where $\ell$ is an integer that determines the level of coarsening ($\ell=0$ corresponds to no grid coarsening). After solving the Poisson equation on this coarser grid, we then perform a prolongation of the coarse data to the fine-resolution grid for subsequent time stepping. The procedure can be summarized in the following form:
\begin{enumerate}
\item[(i)]
Compute potential vorticity equation on fine grid
\item[(ii)]
Map potential vorticity data from fine grid to coarse grid to provide source term for elliptic sub-problem
\item[(iii)]
Solve elliptic sub-problem for stream function on coarse grid
\item[(iv)]
Remap stream function data from coarse grid to fine grid, continue to (i) for subsequent time step
\end{enumerate}
We use prolongation and restriction operators between the data for the potential vorticity part and the elliptic sub-problem part of the problem. This seems to be effectively a low-pass filter on the solution to the elliptic sub-problem, similar to the large eddy simulation (LES) methods. However, the concept in CGP is fundamentally different than LES approaches. The philosophy of the LES concept is based on decomposition of the flow variables into resolved and unresolved scales by applying a filter to Navier-Stokes equations \citep{sagaut2006large,berselli2006mathematics}. The idea of spatial filtering is central in LES: the large, spatially filtered flow variables are approximated, whereas the effect of the small scales is modeled. This allows for much coarser spatial meshes and thus a computational cost that is significantly lower than that of a direct numerical simulation (DNS). To achieve the same order of physical accuracy as fine resolved DNS, however, LES needs to correctly treat the closure problem: the effect of the small scales on the large ones needs to be modeled. Since the inception of the LES for numerical solution of turbulent flows, substantial effort has been devoted to the turbulence models. These closure models have been suggested to relate the effects of subfilter scales into resolved scales.

On the other hand, the CGP method requires the reduction of degree of freedom just for the most time consuming part of the problem. Although the CGP method can be classified in multigrid methods, the underpinning idea behind CGP is different than the classical multigrid methods. In CGP method we use a smaller number of grid points for solving the most time consuming part of the mathematical model (i.e., elliptic sub-problems in QG models). We demonstrate that it is possible to accelerate simulations without a loss of accuracy by utilizing simple averaging and interpolations between the time dependant part of the solver and the elliptic part. We do this by reducing the size of the problem (via a restriction operation) before we solve the elliptic sub-problem, and solving a smaller sized elliptic system (using any Poisson solver), and then prolongating the results for the consequent time step. We add the costs of the restriction and prolongation operations to the solver, but we save much more computational time in solving the elliptic system because the Poisson equation is solved on a coarser grid. In the classical multigrid approach, the elliptic equation is solved iteratively in such a way that the error in the residual is linearly smoothed. Therefore, it is an optimal linear-cost iterative solution technique for elliptic equations. It doesn't require changing anything in the existing black-box elliptic solver. The same is true for the algorithm for the time dependant part of mathematical model. Therefore, the multigrid type of elliptic solver can be used in CGP framework. As shown in our preliminary computations for a simple elliptic Possion equation on Cartesian domain, FFT based direct elliptic solver is the most efficient. Since we had chosen only standard oceanic basins on regular Cartesian grids, the FFT-based direct solver became the optimal one for our study. If, instead of using the FFT-based direct solver, we used a multigrid solver, we would see the same speed-up ratios but longer CPU times in the reported results. The speed-up would become larger if we would use a sub-optimal classical solver for the elliptic sub-problem.

\subsection{Mapping operators}
\label{sec:operators}
The only modifications to the standard flow solver computational procedures are the mapping procedures from fine to coarse data and vice versa. In the current study, the full weighting averaging operation is used for data coarsening (restriction), which is given for a two-dimensional equally spaced data array as \citep{moin2001fundamentals}:
\begin{eqnarray}
\bar{\phi}_{i,j} &=&\frac{1}{16}[4\phi_{2i,2j} +2(\phi_{2i,2j-1}+\phi_{2i,2j+1}+\phi_{2i-1,2j}+\phi_{2i+1,2j}) \nonumber\\ &+&\phi_{2i+1,2j+1}+\phi_{2i+1,2j-1}+\phi_{2i-1,2j+1}+\phi_{2i-1,2j-1} ]
\label{eq:res}
\end{eqnarray}
where $\bar{\phi}_{i,j}$ and $\phi_{i,j}$ are the coarse and fine data arrays, and $i,j$ are the coarse grid indices. For almost all multiscale computations, the mapping from the coarse data to the fine data is a critical issue \citep{weinan2007heterogeneous,kevrekidis2009equation,fish2009multiscale,wagner2010equation}. The bilinear interpolation procedure that we use is given for two-dimensional equally spaced grid as:
\begin{eqnarray}
\phi_{2i,2j}   &=& \bar{\phi}_{i,j} \nonumber \\
\phi_{2i+1,2j} &=& \frac{1}{2}(\bar{\phi}_{i,j}+\bar{\phi}_{i+1,j}) \nonumber \\
\phi_{2i,2j+1} &=& \frac{1}{2}(\bar{\phi}_{i,j}+\bar{\phi}_{i,j+1}) \nonumber \\
\phi_{2i+1,2j+1} &=& \frac{1}{4}(\bar{\phi}_{i,j}+\bar{\phi}_{i+1,j}+\bar{\phi}_{i,j+1}+\bar{\phi}_{i+1,j+1}).
\label{eq:pro}
\end{eqnarray}

The half mapping procedures given by Eq.~\ref{eq:res} for restriction and Eq.~\ref{eq:pro} for prolongation can be performed multiple times to obtain different levels of coarsening. The mapping procedure does not take significant computational time compared to the elliptic solver. The mapping procedures used in this study could potentially be improved by introducing higher-order spline formulas. Since we are using a second-order spatial discretization scheme, however, bilinear interpolation is suitable for this study.

\subsection{CGP algorithm}
\label{sec:vs-alg}
The coarse grid projection (CGP) method is independent from the time integration method used. It can be implemented using any time stepping algorithm, for example, the backward difference method, or a method from the Adams-Bashforth or Adams-Moulton families \citep{schafer2006computational}. Here, the CGP method is used in conjunction with the third-order Runge-Kutta method for the QG models. The joint CGPRK3 algorithm for one-layer barotropic model is presented below. Starting with the value of the potential vorticity, $q^{n}$, at the current time step, the CGPRK3 algorithm for computing the potential vorticity at the next time step, $q^{n+1}$, consists of the following steps:
\begin{eqnarray}
q^{n} &\Rightarrow& \bar{q}^{n} \\
\frac{\partial^2 \bar{\psi}^{n}}{\partial x^2} + \frac{\partial^2 \bar{\psi}^{n}}{\partial y^2} &=& \left(\frac{\delta_{I}}{L}\right)^{-2} (\bar{q}^{n}-y) \\
\psi^{n} &\Leftarrow& \bar{\psi}^{n} \\
q^{(1)} &=& q^{n} + \Delta t G^{n} \\
q^{(1)} &\Rightarrow& \bar{q}^{(1)} \\
\frac{\partial^2 \bar{\psi}^{(1)}}{\partial x^2} + \frac{\partial^2 \bar{\psi}^{(1)}}{\partial y^2} &=& \left(\frac{\delta_{I}}{L}\right)^{-2} (\bar{q}^{(1)}-y) \\
\psi^{(1)} &\Leftarrow& \bar{\psi}^{(1)} \\
q^{(2)} &=& \frac{3}{4}  q^{n} + \frac{1}{4} q^{(1)} + \frac{1}{4}\Delta t G^{(1)} \\
q^{(2)} &\Rightarrow& \bar{q}^{(2)} \\
\frac{\partial^2 \bar{\psi}^{(2)}}{\partial x^2} + \frac{\partial^2 \bar{\psi}^{(2)}}{\partial y^2} &=&\left(\frac{\delta_{I}}{L}\right)^{-2} (\bar{q}^{(2)}-y)  \\
\psi^{(2)} &\Leftarrow& \bar{\psi}^{(2)} \\
q^{n+1} &=& \frac{1}{3}  q^{n} + \frac{2}{3} q^{(2)} + \frac{2}{3}\Delta t G^{(2)}
\label{eq:TVDRK}
\end{eqnarray}
where
\begin{equation}
G = - J(\psi,q)+D+F.
\label{eq:jac}
\end{equation}
The arrows in the algorithm represent the mapping operators. The procedure for the two-layer QG model for stratified flow computations is similar to the one-layer algorithm except that two potential vorticity equations for each isopycnal layer along with a coupled system for elliptic sub-problem are be solved. We will address the solution procedure for the elliptic sub-problem in the following section.

\section{Numerical Methods}
\label{sec:numerics}
In many physically relevant situations, where the Munk and Rhines scales being close to each other, the solutions to oceanic models, such as the QG models, do not converge to a steady state as time goes to infinity \citep{medjo2000numerical}. Rather they remain time dependent by producing statistically steady state with one or multiple equilibria. Therefore, numerical schemes designed for numerical integration of such phenomena should be suited for such behavior of the solutions and for the long-time integration. In this study, the governing equations are solved by a fully conservative finite difference scheme along with a third-order Runge-Kutta adaptive time stepping algorithm. An efficient, linear-cost, fast sine transform method is utilized for solving the elliptic subproblems.

\subsection{Arakawa scheme for the Jacobian}
\cite{arakawa1966computational} suggested that the conservation of energy,
enstrophy, and skew-symmetry is sufficient to avoid computational instabilities stemming
from nonlinear interactions.
The second-order Arakawa scheme for the Jacobian (i.e., the nonlinear term in the governing equations) is
\begin{equation}
J(\psi,q) =\frac{1}{3}\bigl(J_{1}(\psi,q)+J_{2}(\psi,q)+J_{3}(\psi,q)\bigr) ,
\label{eq:ja1}
\end{equation}
where the discrete Jacobians have the following forms:
\begin{eqnarray}
J_{1}(\psi,q) &=& \frac{1}{4 \, \Delta_x \, \Delta_y} \bigl[-(q_{i+1,j}-q_{i-1,j})(\psi_{i,j+1}-\psi_{i,j-1}) \nonumber \\
& &   +(q_{i,j+1}-q_{i,j-1})(\psi_{i+1,j}-\psi_{i-1,j}) \bigr] ,
\label{eq:j1} \\
J_{2}(\psi,q) &=& \frac{1}{4 \, \Delta_x \, \Delta_y} \bigl[-q_{i+1,j}(\psi_{i+1,j+1}-\psi_{i+1,j-1})
+q_{i-1,j}(\psi_{i-1,j+1}-\psi_{i-1,j-1}) \nonumber \\
& &  +q_{i,j+1}(\psi_{i+1,j+1}-\psi_{i-1,j+1})
-q_{i,j-1}(\psi_{i+1,j-1}-\psi_{i-1,j-1}) \bigr] ,
\label{eq:j2} \\
J_{3}(\psi,q) &=& \frac{1}{4 \, \Delta_x \, \Delta_y} \bigl[-q_{i+1,j+1}(\psi_{i,j+1}-\psi_{i+1,j})
+q_{i-1,j-1}(\psi_{i-1,j}-\psi_{i,j-1}) \nonumber \\
& & +q_{i-1,j+1}(\psi_{i,j+1}-\psi_{i-1,j})
-q_{i+1,j-1}(\psi_{i+1,j}-\psi_{i,j-1}) \bigr] .
\label{eq:j3}
\end{eqnarray} 
Note that $J_{1}$, which corresponds to the central second-order difference scheme,
is not sufficient for the conservation of energy, enstrophy, and skew-symmetry by the
numerical discretization.
\cite{arakawa1966computational} showed that the judicious combination of $J_{1}, J_{2}$, and $J_{3}$ in
Eq.~\ref{eq:ja1} achieves the above discrete conservation properties.

\subsection{Time integration scheme}
\label{sec:time}
For the time discretization, as illustrated in the CGP algorithm, we employ an optimal third-order total variation diminishing Runge-Kutta (TVDRK3) scheme \citep{gottlieb1998total}. For clarity of notation, we rewrite the governing equations in the following form:
\begin{equation}
\frac{dq_i}{dt} = R_i ,
\label{eq:rk}
\end{equation}
where subscript $i$ represents the layer index and $R_i$ denotes the discrete spatial derivative operator, including the nonlinear Jacobian of the convective term, the linear biharmonic diffusive term, and the forcing term. For each layer, the TVDRK3 scheme then becomes:
\begin{eqnarray}
q_{i}^{(1)} &=& q^{n} + \Delta t R_{i}^{(n)} , \nonumber \\
q_{i}^{(2)} &=& \frac{3}{4}  q_{i}^{n} + \frac{1}{4} q_{i}^{(1)} + \frac{1}{4}\Delta t R_{i}^{(1)} , \\
q_{i}^{n+1} &=& \frac{1}{3}  q_{i}^{n} + \frac{2}{3} q_{i}^{(2)} + \frac{2}{3}\Delta t R_{i}^{(2)} , \nonumber
\label{eq:TVDRK}
\end{eqnarray}
where $\Delta t$ is the adaptive time step size, which can be computed at the end of each time step by:
\begin{equation}
\Delta t = c \, \frac{\mbox{min}(\Delta_x,\Delta_y)}{\mbox{max}\left\{ |\frac{\partial \psi_i}{\partial x}|,|\frac{\partial \psi_i}{\partial y}| \right\} } ,
\label{eq:dt}
\end{equation}
where $c$ is known as Courant-Friedrichs-Lewy (CFL) number which is restricted to $c\leq1$ due to numerical stability of the TVDRK3 scheme.

\subsection{Elliptic sub-problems}
\label{sec:subproblem}
Most of the demand on computing resources posed by QG models comes in the solution of the elliptic inversion subproblem \citep{miller2007numerical}. This is also true for our study.  However, we take advantage of the simple square shape of our domain and utilize one of the fastest available techniques, which is the FFT based direct inversion to solve the elliptic sub-problems. For example, the linear coupled elliptic sub-problem for the two-layer QG model can be written in the following form:
\begin{eqnarray}
Q_1 &=& \mbox{Ro} \nabla^2\psi_1  + \frac{\mbox{Fr}}{\delta}(\psi_2-\psi_1) , \\
Q_2 &=& \mbox{Ro} \nabla^2\psi_2  + \frac{\mbox{Fr}}{1-\delta}(\psi_1-\psi_2) ,
\label{eq:sub-pr}
\end{eqnarray}
where $Q_1=q_1-y$ and $Q_2=q_2-y$. The impermeability boundary condition imposed as $\psi |_{\Omega} = 0$ suggests the use of a fast sine transform (an inverse transform) for each layer:
\begin{equation}
\hat{Q}_{1_{k,l}}=\frac{2}{N_x}\frac{2}{N_y} \sum_{i=1}^{N_x-1} \sum_{j=1}^{N_y-1} Q_{1_{i,j}} \sin \left(\frac{\pi k i}{N_x}\right) \sin \left(\frac{\pi l j}{N_y}\right) ,
\label{eq:ifft-1}
\end{equation}
\begin{equation}
\hat{Q}_{2_{k,l}}=\frac{2}{N_x}\frac{2}{N_y} \sum_{i=1}^{N_x-1} \sum_{j=1}^{N_y-1} Q_{2_{i,j}} \sin \left(\frac{\pi k i}{N_x}\right) \sin \left(\frac{\pi l j}{N_y}\right) ,
\label{eq:ifft-2}
\end{equation}
where $N_x$ and $N_y$ are the total number of grid points in $x$ and $y$ directions. Here the symbol hat is used to represent the corresponding Fourier coefficient of the physical grid data with a subscript pair $i,j$, where $i=0,1, ... N_x$ and $j=0,1, ... N_y$. As a second step, we directly solve the subproblem in Fourier space:
\begin{equation}
\hat{\psi}_{1_{k,l}}= \frac{\alpha_{k,l} \hat{Q}_{1_{k,l}} - \frac{\mbox{Fr}}{1-\delta}\hat{Q}_{1_{k,l}} - \frac{\mbox{Fr}}{\delta}\hat{Q}_{2_{k,l}}}{\alpha_{k,l} \left(\alpha_{k,l} -\frac{\mbox{Fr}}{\delta} - \frac{\mbox{Fr}}{1-\delta} \right)} ,
\label{eq:isub-1}
\end{equation}
\begin{equation}
\hat{\psi}_{2_{k,l}}= \frac{ \alpha_{k,l} \hat{Q}_{2_{k,l}} - \frac{\mbox{Fr}}{1-\delta}\hat{Q}_{1_{k,l}} - \frac{\mbox{Fr}}{\delta}\hat{Q}_{2_{k,l}}}{\alpha_{k,l} \left(\alpha_{k,l} -\frac{\mbox{Fr}}{\delta} - \frac{\mbox{Fr}}{1-\delta} \right)} ,
\label{eq:isub-2}
\end{equation}
where
\begin{equation}
\alpha_{k,l} = \frac{\mbox{Ro}}{\Delta_x^2}\left[2 \cos \left(\frac{\pi k }{N_x} \right) - 2  \right] + \frac{\mbox{Ro}}{\Delta_y^2}\left[2 \cos \left(\frac{\pi l }{N_y} \right)- 2  \right] .
\label{eq:ialpha}
\end{equation}
Finally, the streamfunction arrays for each layer are found by performing a forward sine transform:
\begin{equation}
\psi_{1_{i,j}}= \sum_{k=1}^{N_x-1} \sum_{l=1}^{N_y-1} \hat{\psi}_{1_{k,l}} \sin \left(\frac{\pi k i}{N_x}\right) \sin \left(\frac{\pi l j}{N_y}\right) ,
\label{eq:ffft-1}
\end{equation}
\begin{equation}
\psi_{2_{i,j}}= \sum_{k=1}^{N_x-1} \sum_{l=1}^{N_y-1} \hat{\psi}_{2_{k,l}} \sin \left(\frac{\pi k i}{N_x}\right) \sin \left(\frac{\pi l j}{N_y}\right) ,
\label{eq:ffft-2}
\end{equation}
The computational cost of this elliptic solver is $\displaystyle \mathcal{O}\left(N_x \, N_y \, \log(N_x) \, \log(N_y) \right)$. The FFT algorithm given by \cite{press1992numerical} is used for forward and inverse sine transforms. The elliptic sub-problem for the one-layer barotropic model can be easily inverted by using similar approach.

\section{Results}
\label{sec:results}

To investigate the performance of the CGP method we consider two different large-scale QG ocean circulation models, one with one-layer barotropic ocean basin, one with two-layer stratified ocean basin. For each model we present several numerical experiments having different physical setting to evaluate the  characteristics of the CGP multiscale method.

\subsection{CGP experiments for one-layer QG model}
\label{sec:results1}

The main goal of this section is to test the proposed CGP method in the numerical simulation of the wind-driven circulation in a barotropic shallow ocean basin, a standard prototype of more realistic ocean dynamics.
The model employs the BVE driven by a symmetric double-gyre wind forcing, which yields a
four-gyre circulation in the time mean.
This test problem has been used in numerous studies \citep{cummins1992inertial,
greatbatch2000four,nadiga2001dispersive,holm2003modeling,fox2005reevaluating,san2011approximate}.
This problem represents an ideal test for the one-layer QG models.
Indeed, as showed in \cite{greatbatch2000four}, although a double gyre wind forcing is used,
the long time average yields a four gyre pattern, which is challenging to capture on coarse
spatial resolutions. Thus, we will investigate numerically whether our CGP model can reproduce the four gyre time average on a coarse mesh.

The mathematical model used in the four gyre problem is the BVE \ref{eq:ge}. We utilize two different parameter sets, corresponding to two physical oceanic settings: Experiment I with a Rhines scale of $\delta_{I}/L = 0.04$  and a Munk scale of $\delta_{M}/L = 0.02$, which corresponds to a Reynolds number of $Re=200$ (or a boundary layer based Reynolds number of $Re_B=8$) and a Rossby number of $Ro=0.0016$; and Experiment II with a Rhines scale of $\delta_{I}/L = 0.05$ and a Munk scale of $\delta_{M}/L = 0.02$, which corresponds
to a Reynolds number of $Re=312.5$ (or a boundary layer based Reynolds number of $Re_B=15.625$) and a Rossby number of $Ro=0.0025$. Since we set the Munk scale to $\delta_{M}/L = 0.02$ in our study, the uniform eddy viscosity coefficient embedded in the model can be calculated from Eq.~\ref{eq:relation}. For example, if we set the mid-latitude ocean basin length to $L=2000$ km and the gradient of the Coriolis parameter to $\beta=1.75\times10^{-11} \ \mbox{m}^{-1}\mbox{s}^{-1}$, then our model uses $\nu=1120 \ \mbox{m}^{2}\mbox{s}^{-1}$ as its eddy viscosity parametrization. Table~\ref{tab:sets1} summarizes the physical variables associated with the one-layer QG experiments considered in this study.
All numerical experiments conducted here are solved for a maximum dimensionless time of $T_{max}=50$. This value corresponds to the dimensional times of $28.3$ and $18.1$ years for Experiment I and Experiment II, respectively, which are long enough to capture statistically steady states.

\begin{table}[!t]
\centering
\caption{Physical parameter sets used in the numerical experiments for the one-layer QG model. }
\label{tab:sets1}       
\begin{tabular}{lll}
\hline\noalign{\smallskip}
Variable (unit) & Experiment I & Experiment II \\
\noalign{\smallskip}\hline\noalign{\smallskip}
  $L$ ($km$)   & 2000 & 2000 \\
  $\beta$ ($m^{-1}s^{-1}$) & $1.75\times10^{-11}$ & $1.75\times10^{-11}$ \\
  $\nu$ ($m^{2}s^{-1}$)    & 1120    & 1120  \\
  $V$ ($ms^{-1}$)    & 0.112   & 0.175 \\
  $L/V$ ($year$)    & 0.566   & 0.362 \\
  $\delta_{M}/L$    & 0.02    & 0.02 \\
  $\delta_{I}/L$    & 0.04   & 0.05 \\
  Ro   & 0.0016   & 0.0025 \\
  Re   & 200   & 312.5\\
\noalign{\smallskip}\hline
\end{tabular}
\end{table}

\begin{figure*}[!b]
\centering
\mbox{
\subfigure[Experiment I]{\includegraphics[width=0.5\textwidth]{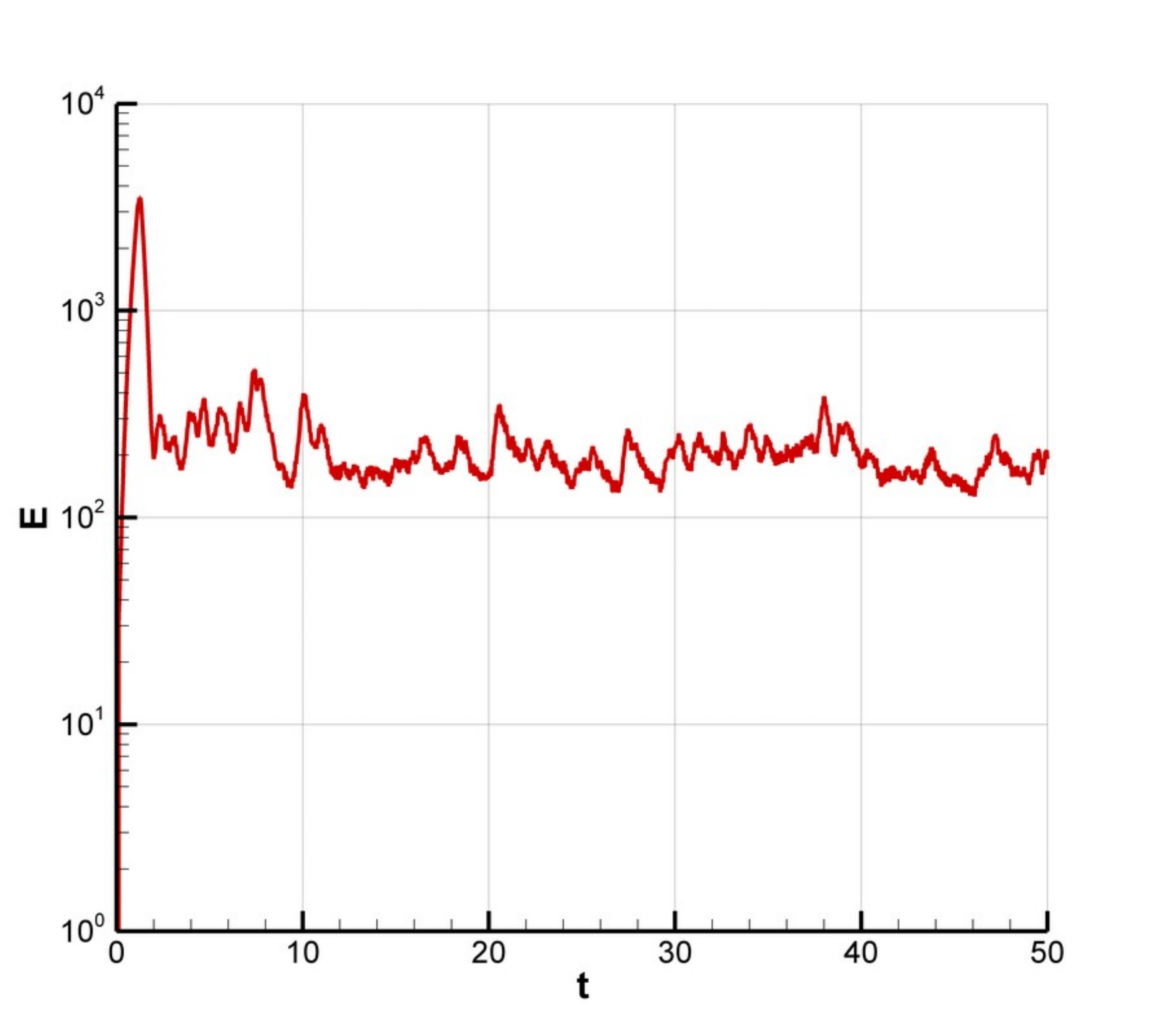}}
\subfigure[Experiment II]{\includegraphics[width=0.5\textwidth]{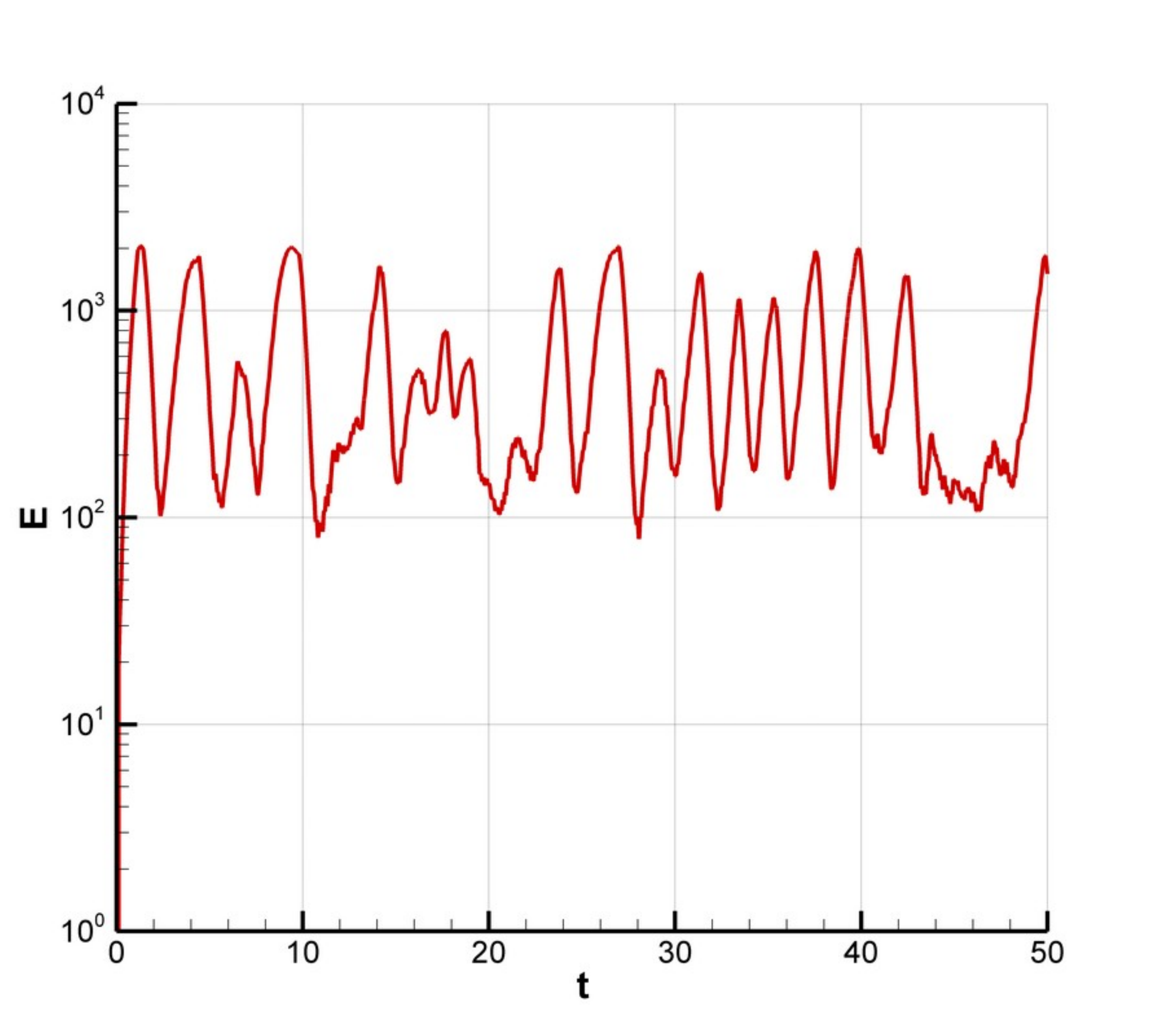}}
}
\caption{
Time histories of basin integrated total kinetic energy.
}
\label{fig:hist-1}
\end{figure*}

\begin{figure*}[!b]
\centering
\mbox{
\subfigure[Experiment I]{\includegraphics[width=0.4\textwidth]{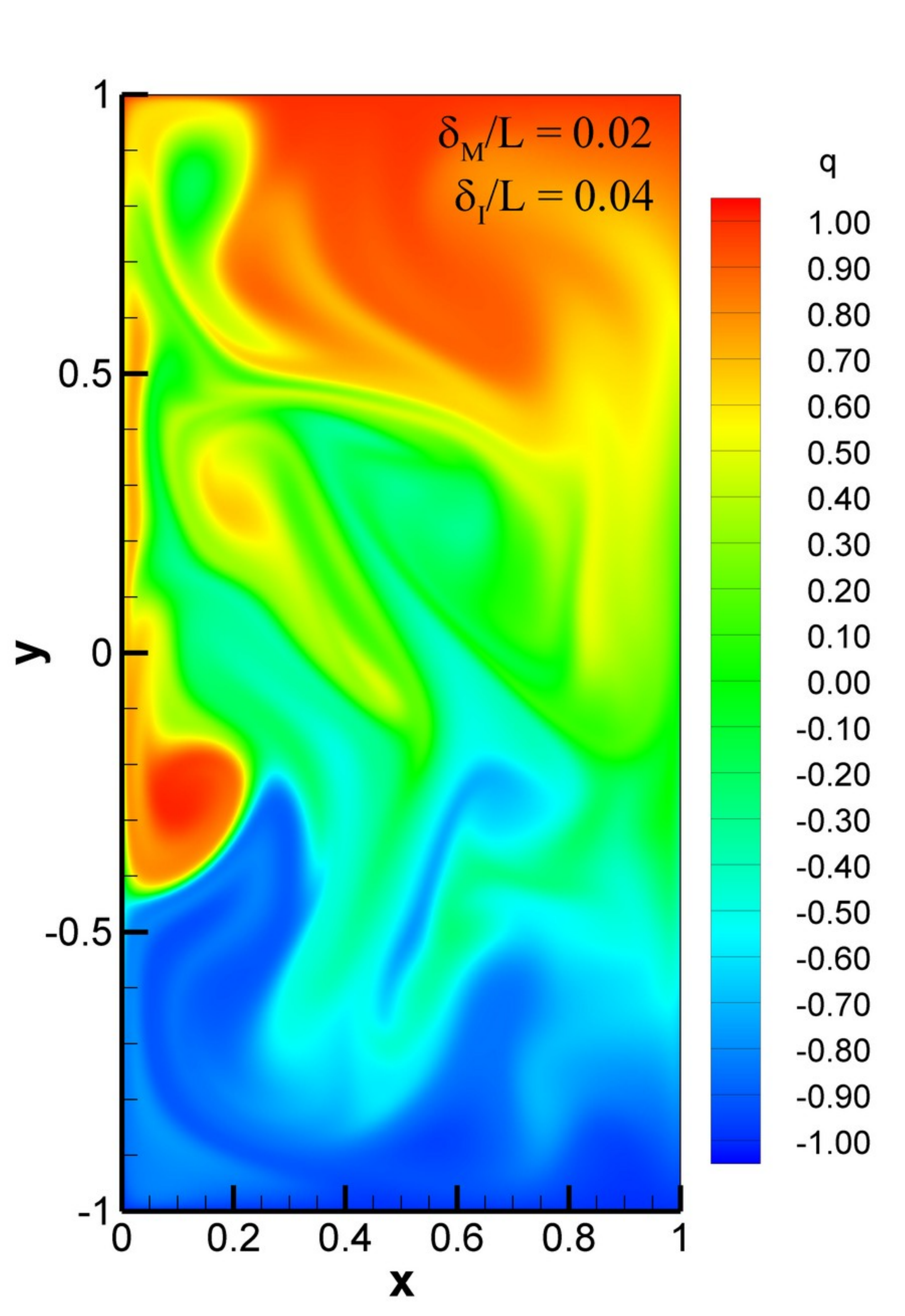}}
\subfigure[Experiment II]{\includegraphics[width=0.4\textwidth]{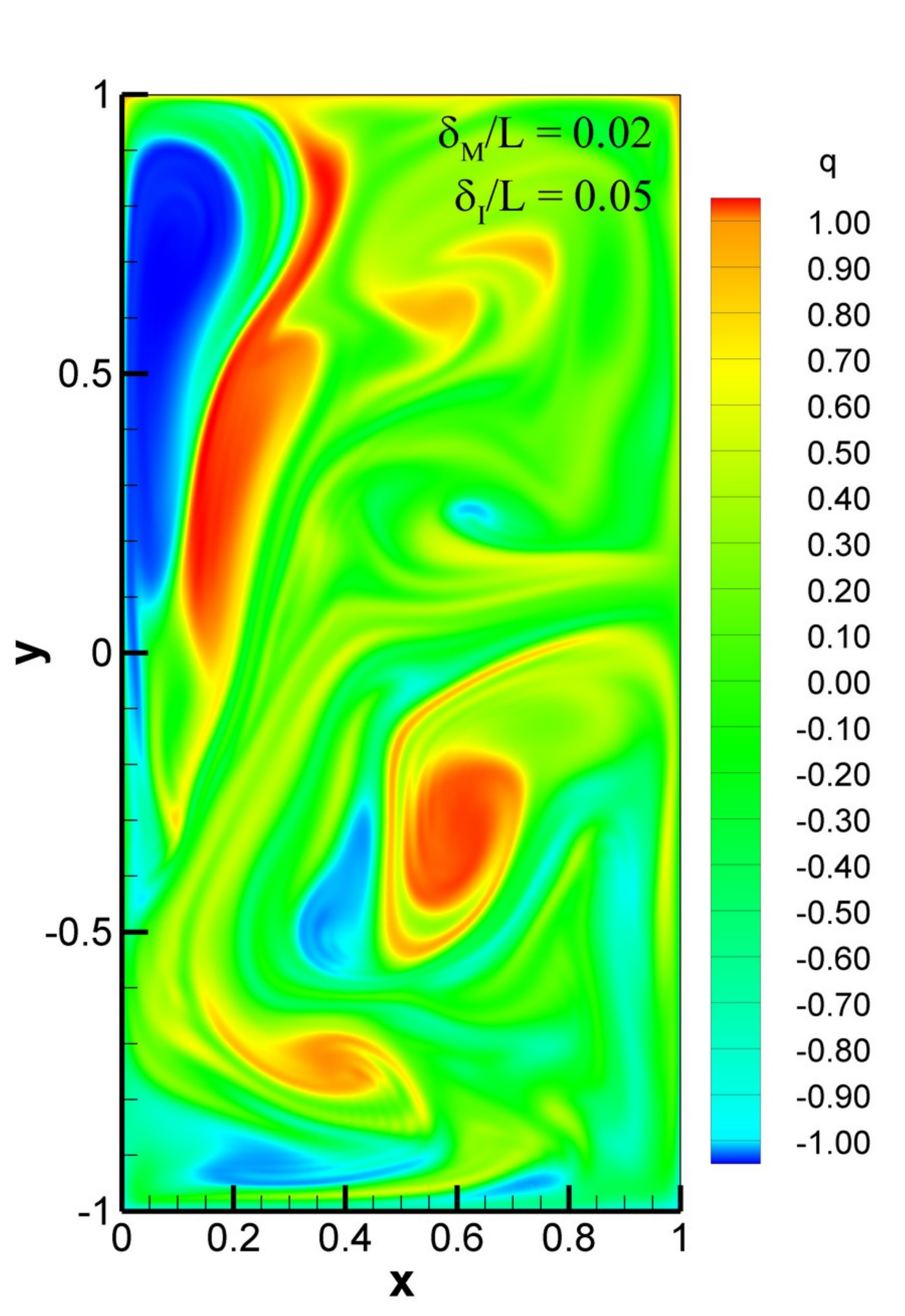}}
}
\caption{
Instantaneous potential vorticity contour plots at time $t=50$.
}
\label{fig:inst-1}
\end{figure*}

\begin{figure*}
\centering
\mbox{
\subfigure{\includegraphics[width=0.22\textwidth]{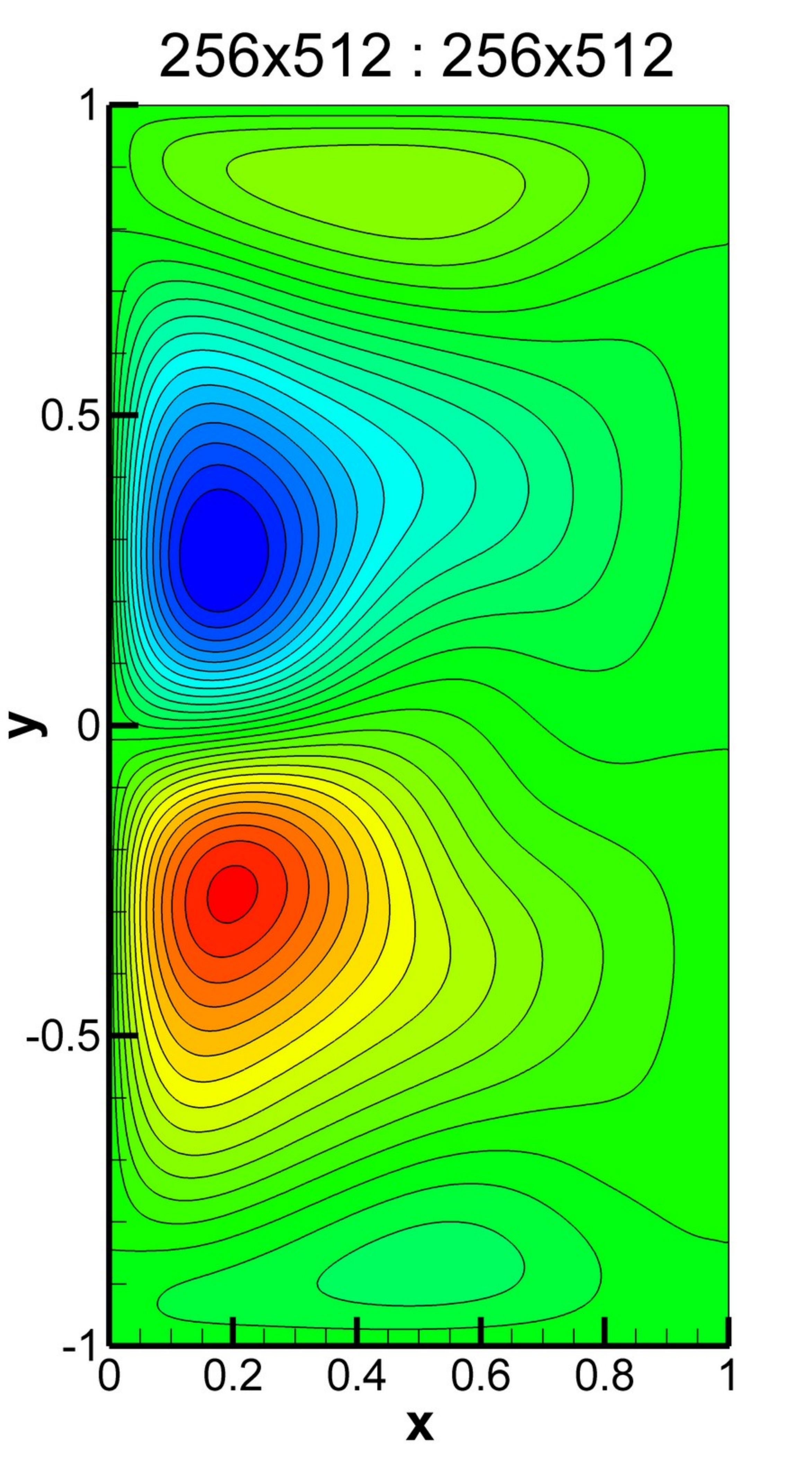}}
\subfigure{\includegraphics[width=0.22\textwidth]{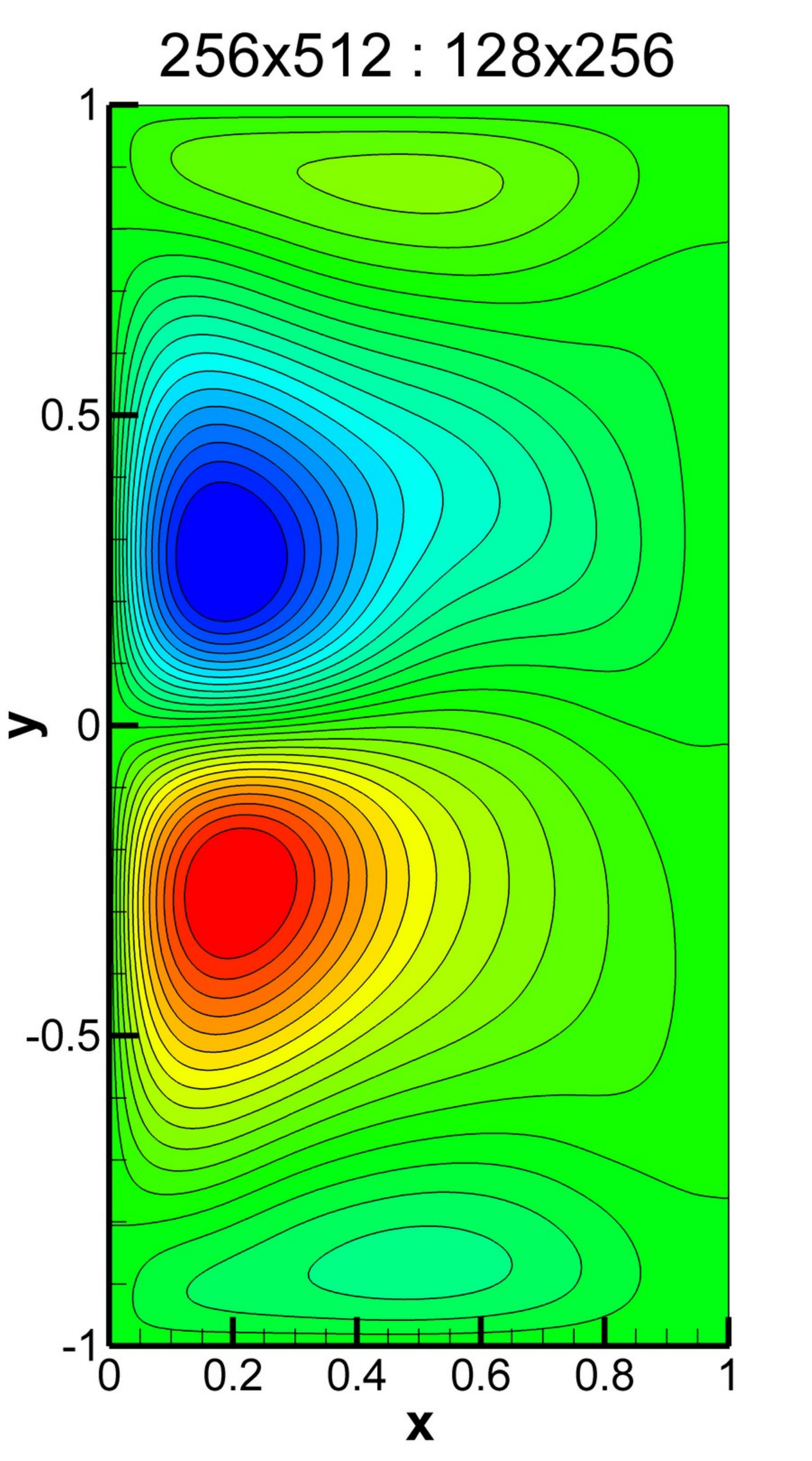}}
\subfigure{\includegraphics[width=0.22\textwidth]{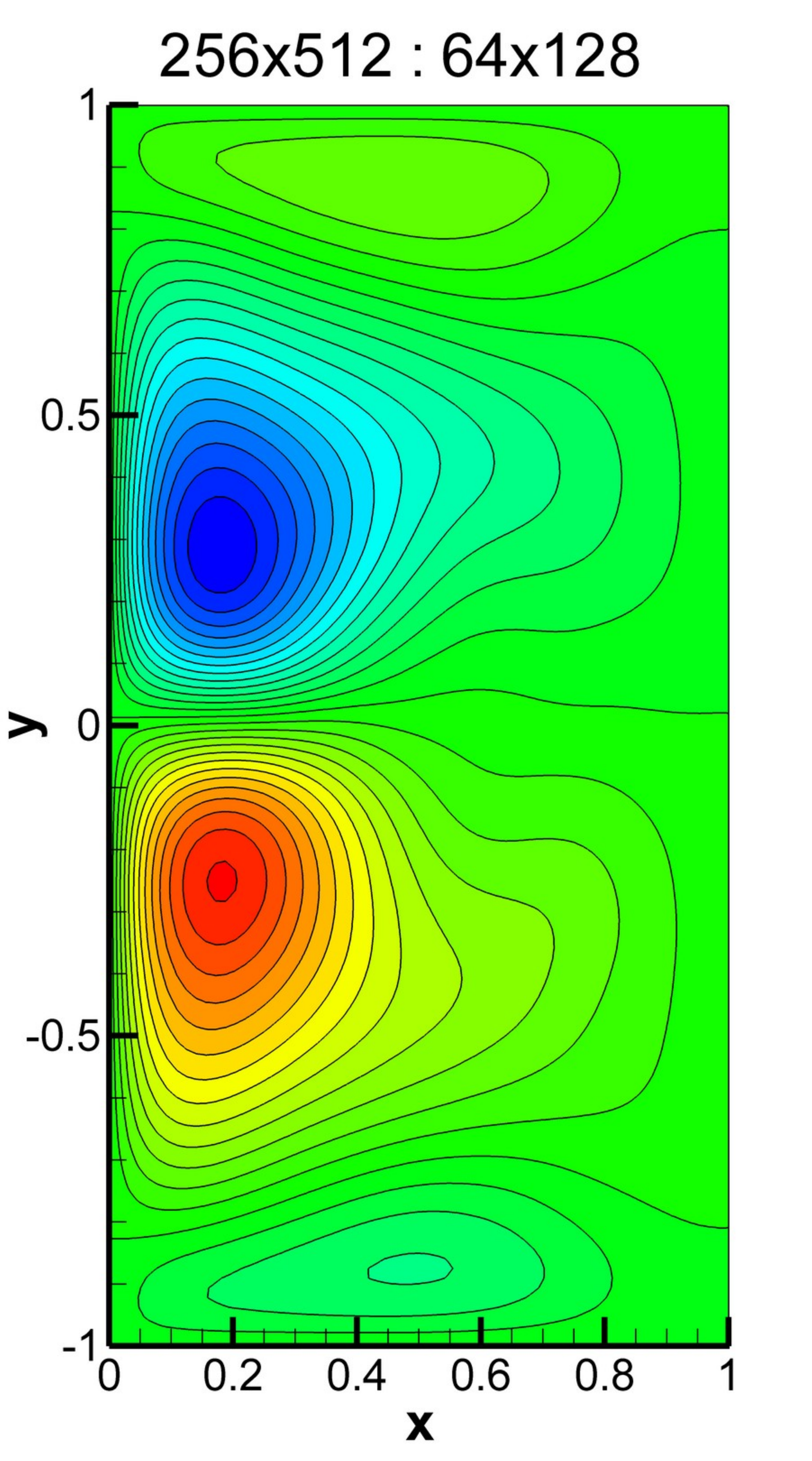}}
\subfigure{\includegraphics[width=0.22\textwidth]{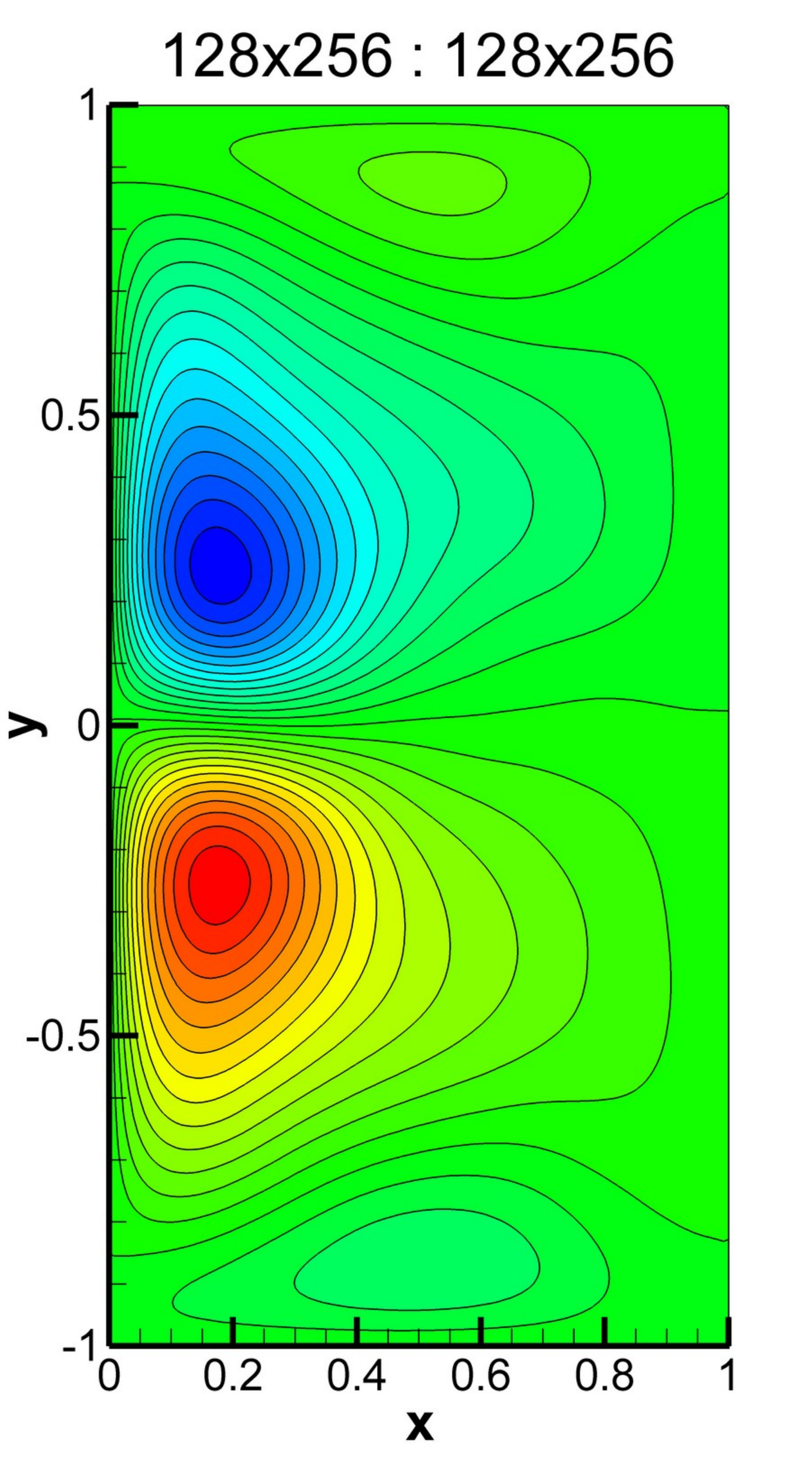}} }
\\
\mbox{
\subfigure{\includegraphics[width=0.22\textwidth]{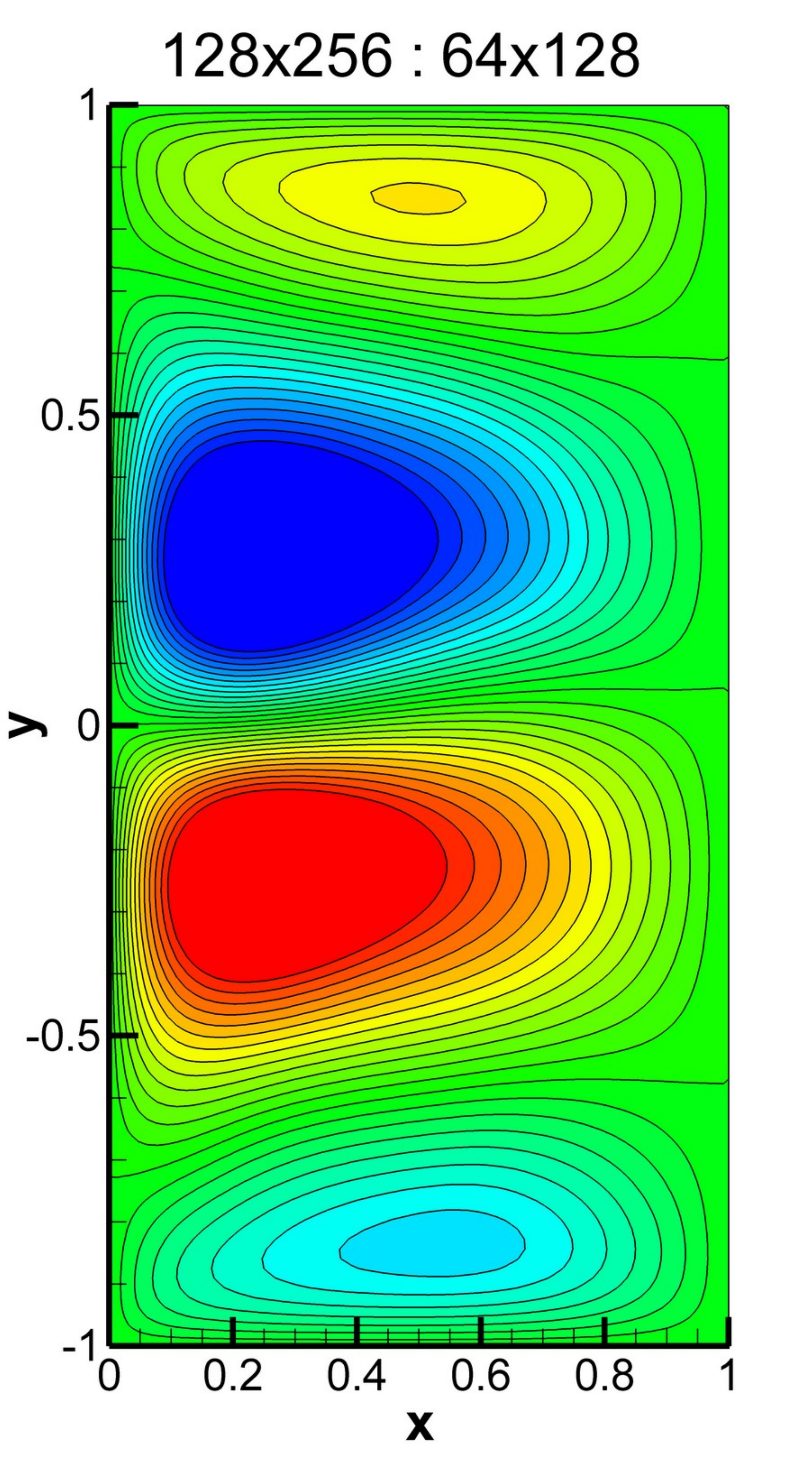}}
\subfigure{\includegraphics[width=0.22\textwidth]{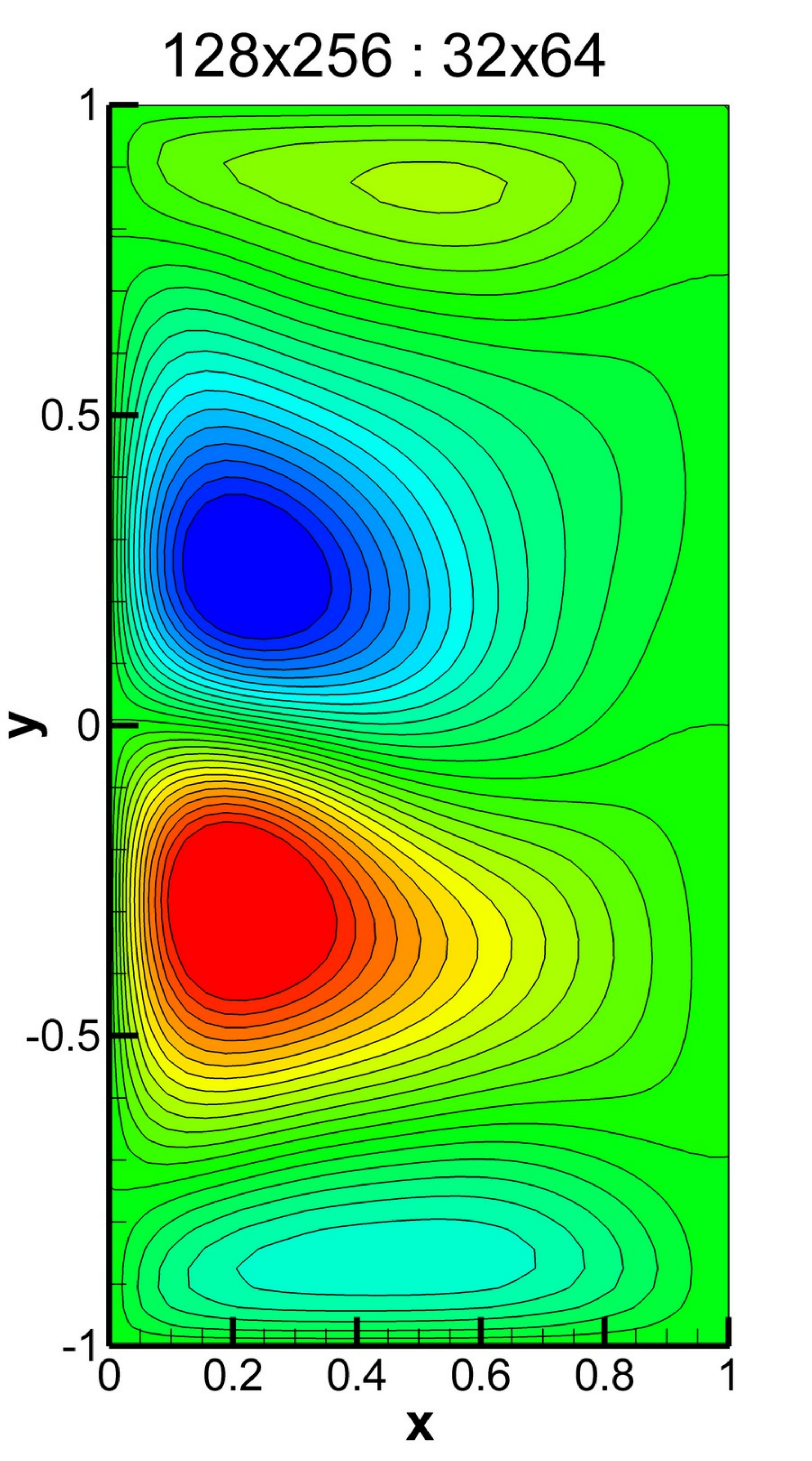}}
\subfigure{\includegraphics[width=0.22\textwidth]{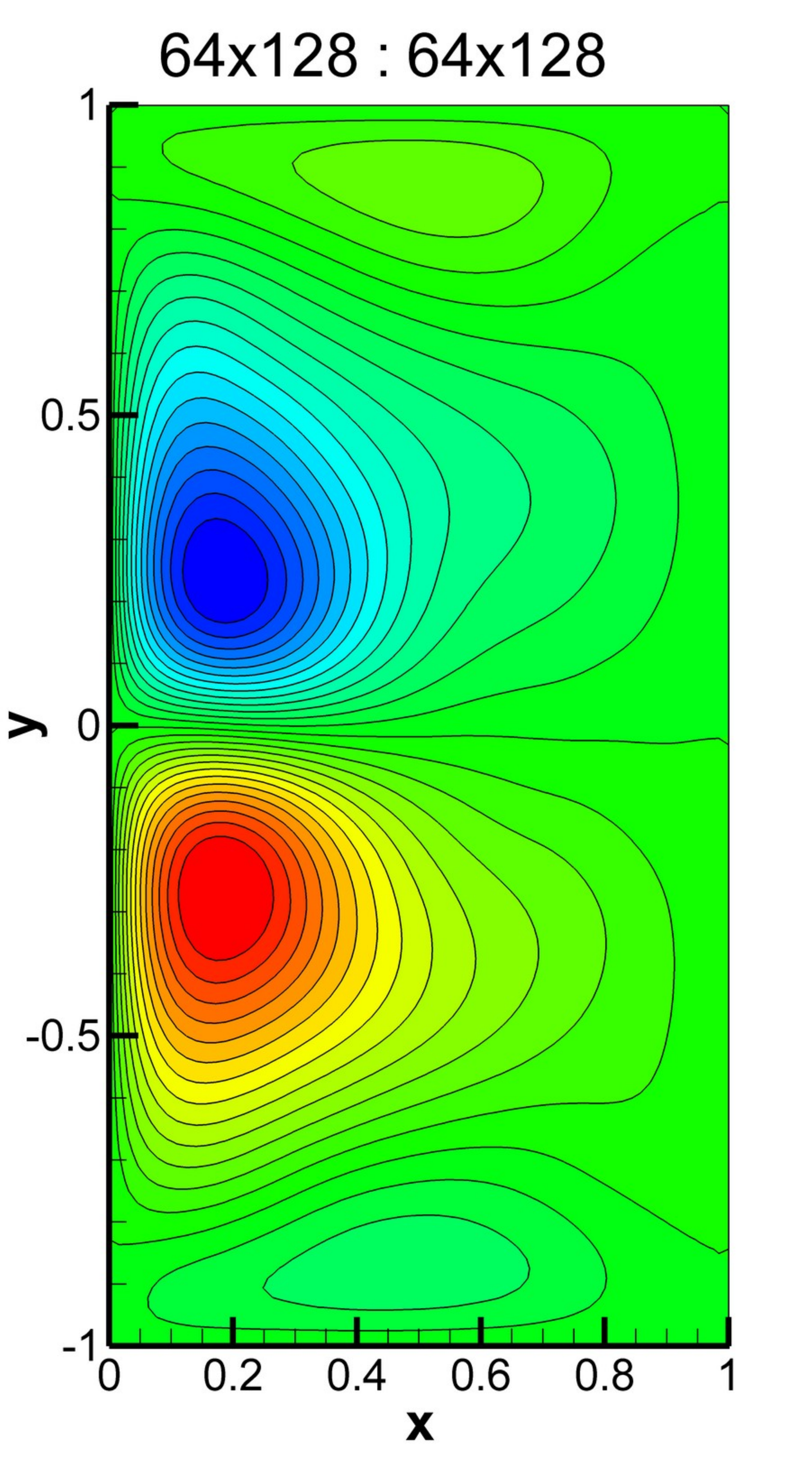}}
\subfigure{\includegraphics[width=0.22\textwidth]{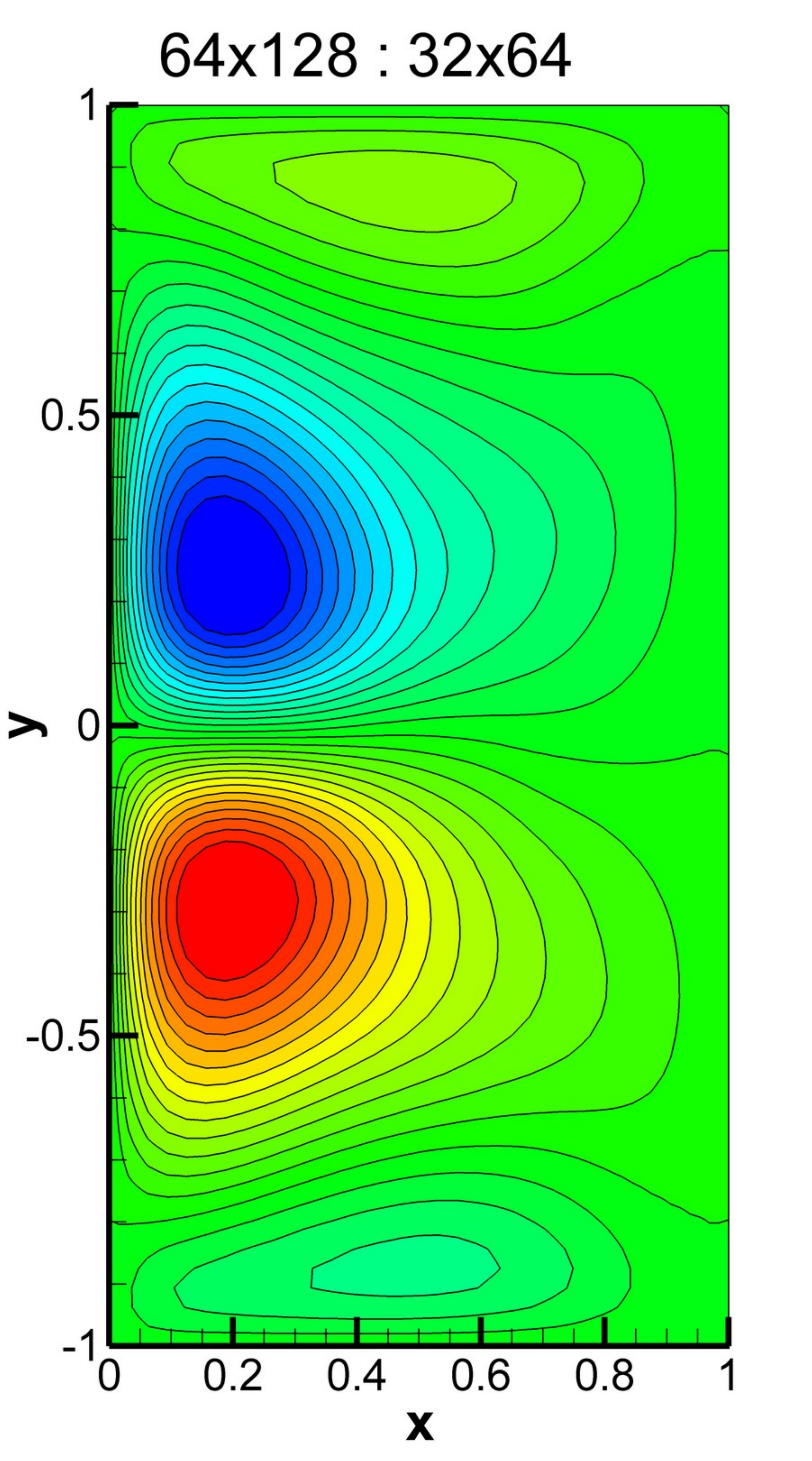}} }
\\
\mbox{
\subfigure{\includegraphics[width=0.22\textwidth]{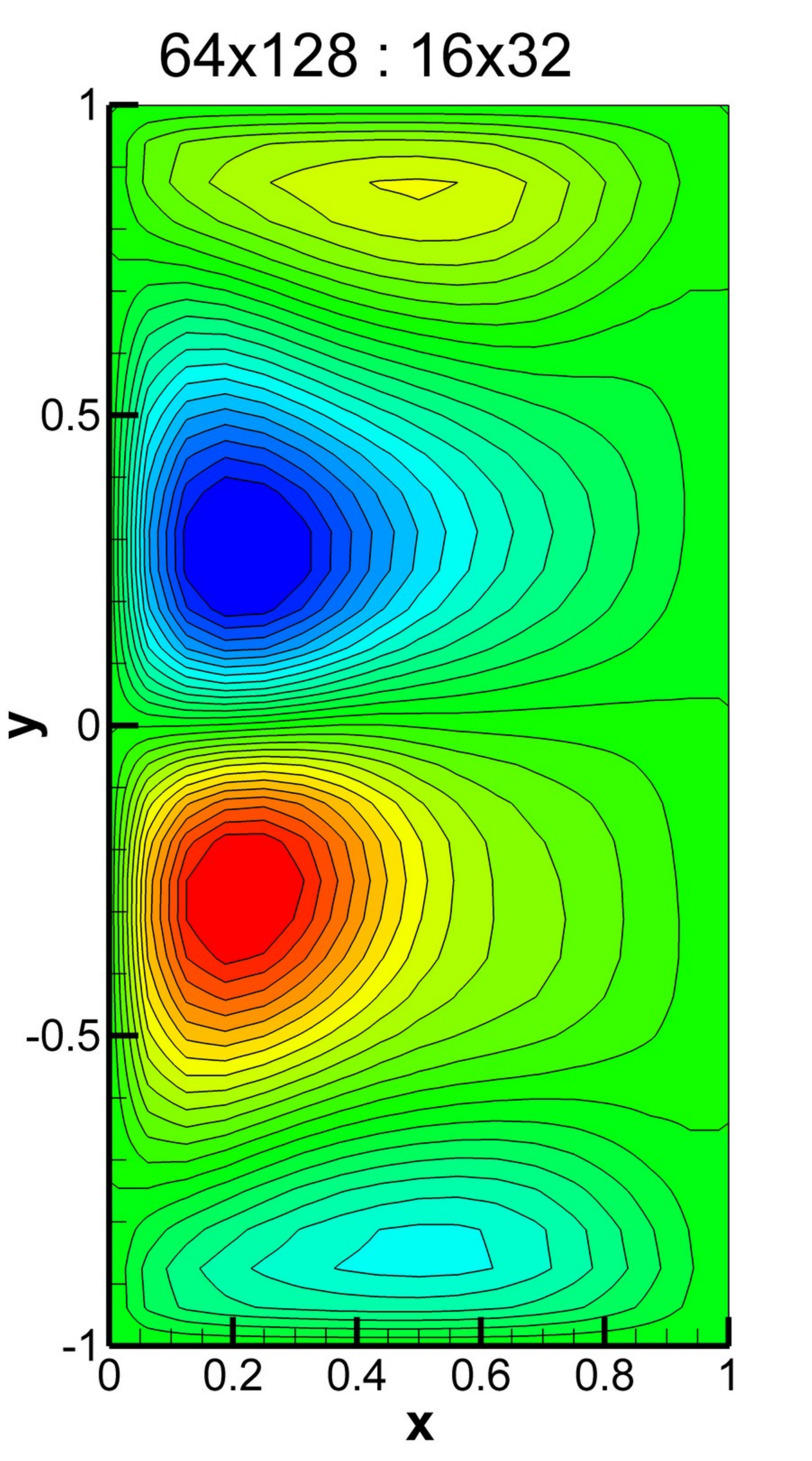}}
\subfigure{\includegraphics[width=0.22\textwidth]{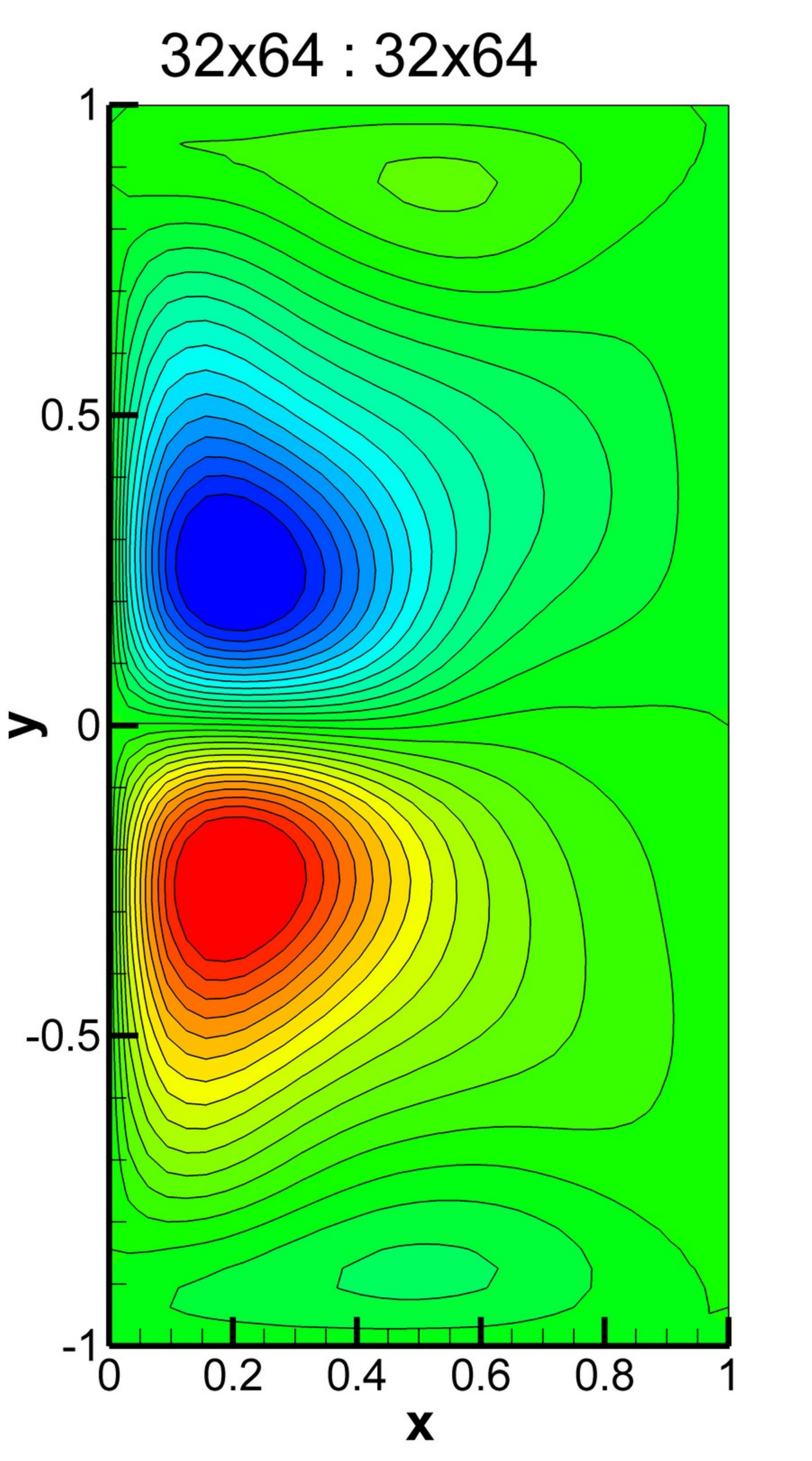}}
\subfigure{\includegraphics[width=0.22\textwidth]{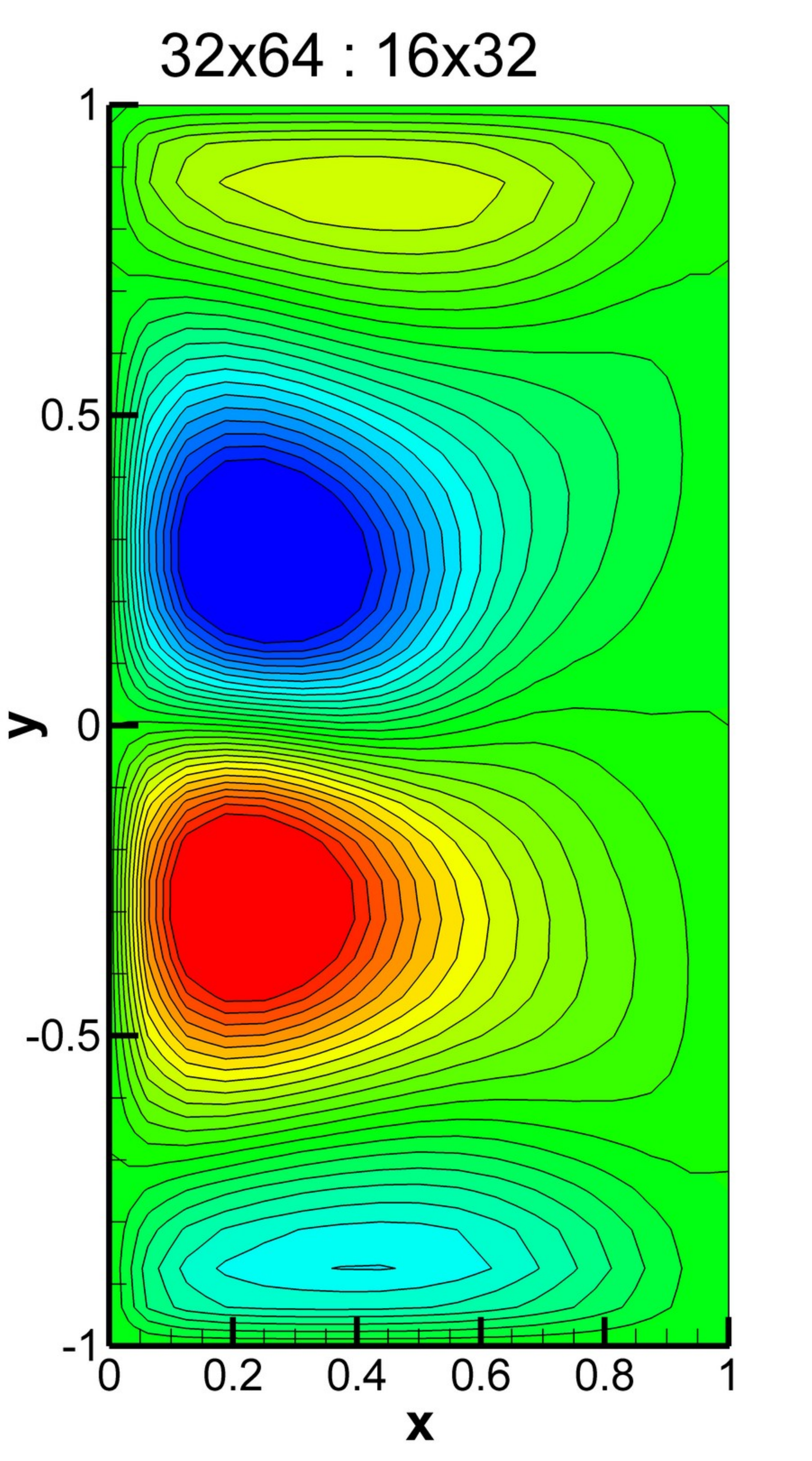}}
\subfigure{\includegraphics[width=0.22\textwidth]{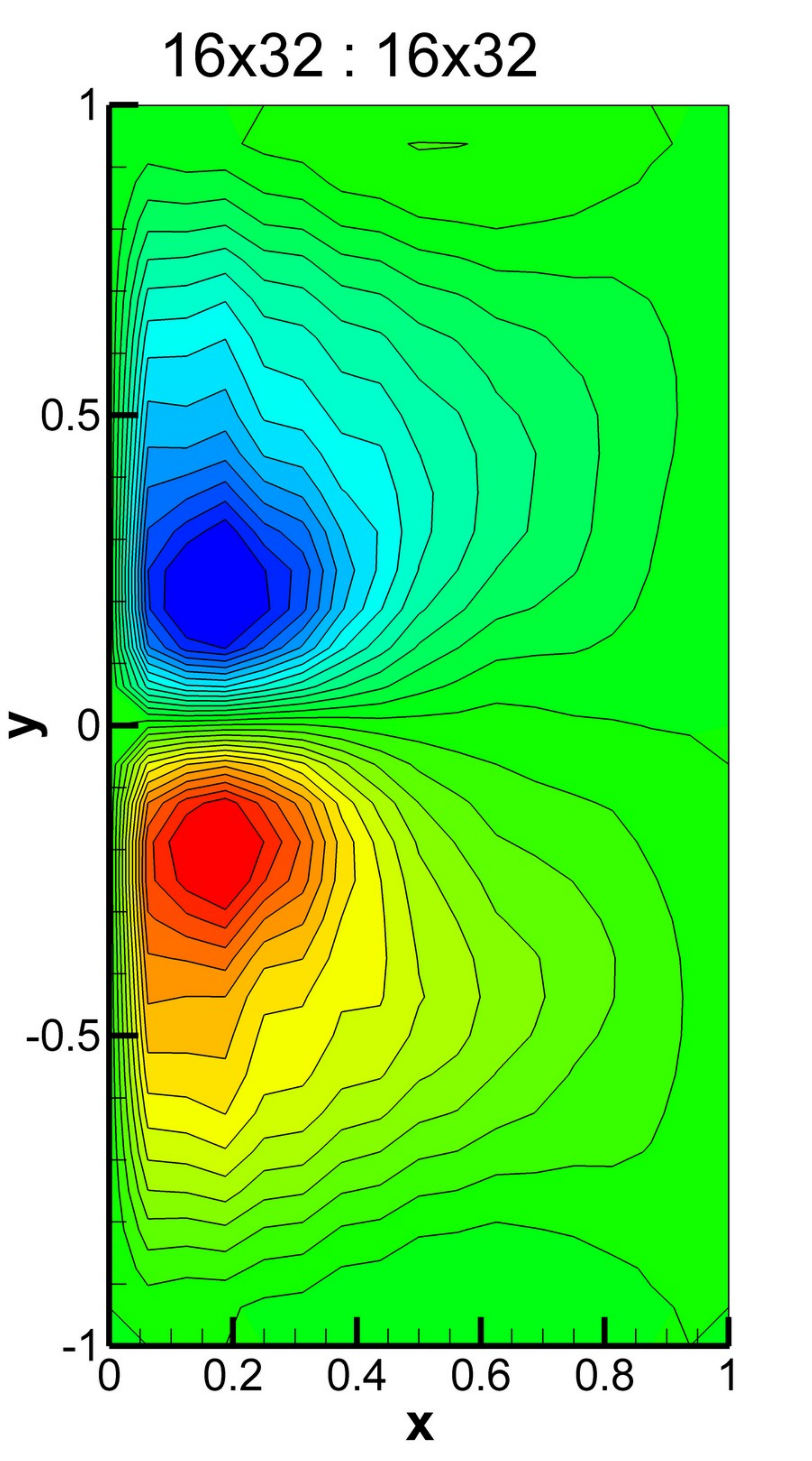}} }
\caption{Experiment I:
Comparison of mean stream functions for $Re=200$ and $Ro=0.0016$ (i.e., $\delta_M/L = 0.02$, and $\delta_I/L = 0.04$). Labels include the resolutions for both parts of the solver in the form $N_x \times N_y : M_x \times M_y$, where $N_x \times N_y$ is the resolution for the barotropic vorticity transport equation, and $M_x \times M_y$ is the resolution for the elliptic sub-problems.
}
\label{fig:sB}
\end{figure*}

\begin{figure*}
\centering
\mbox{
\subfigure{\includegraphics[width=0.22\textwidth]{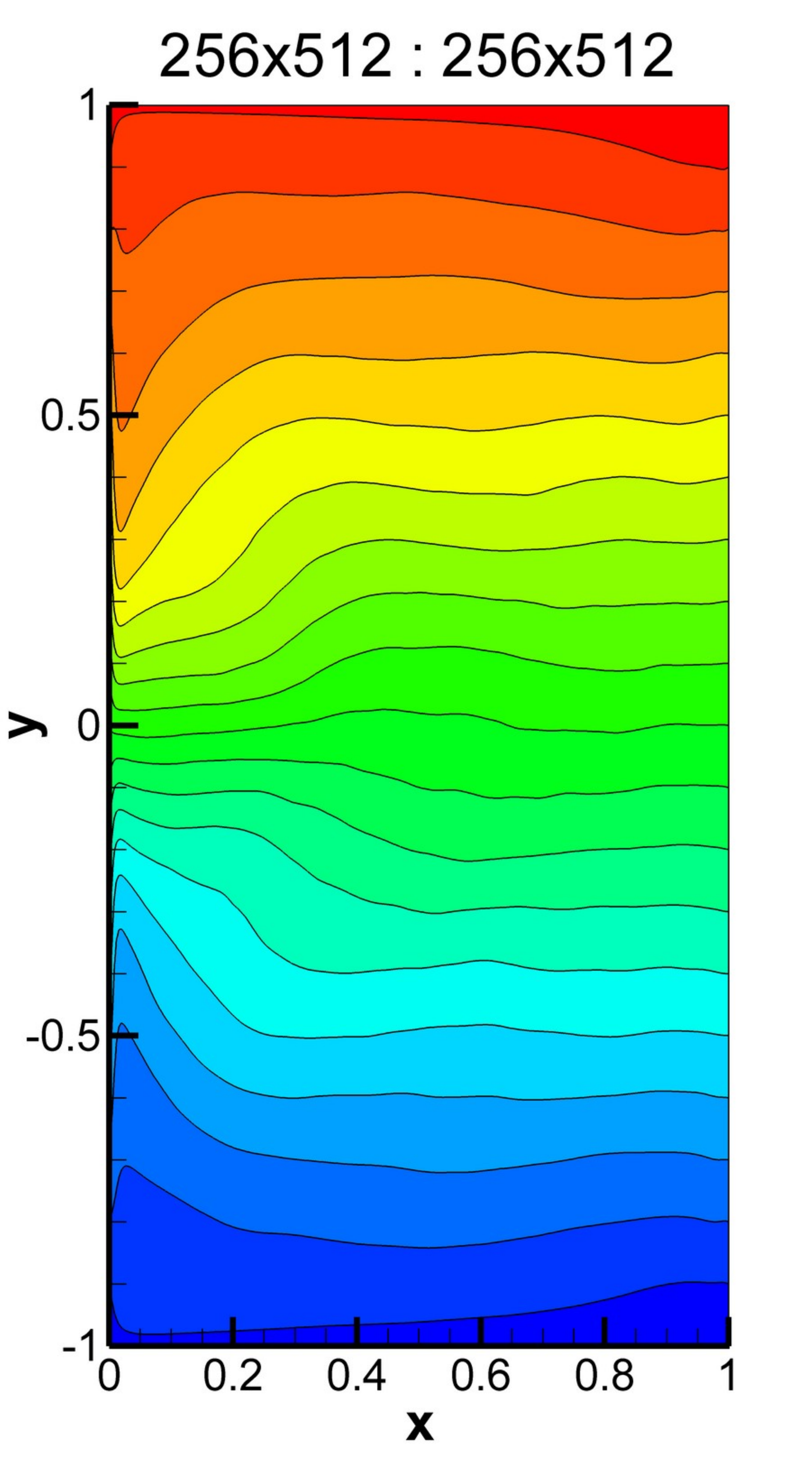}}
\subfigure{\includegraphics[width=0.22\textwidth]{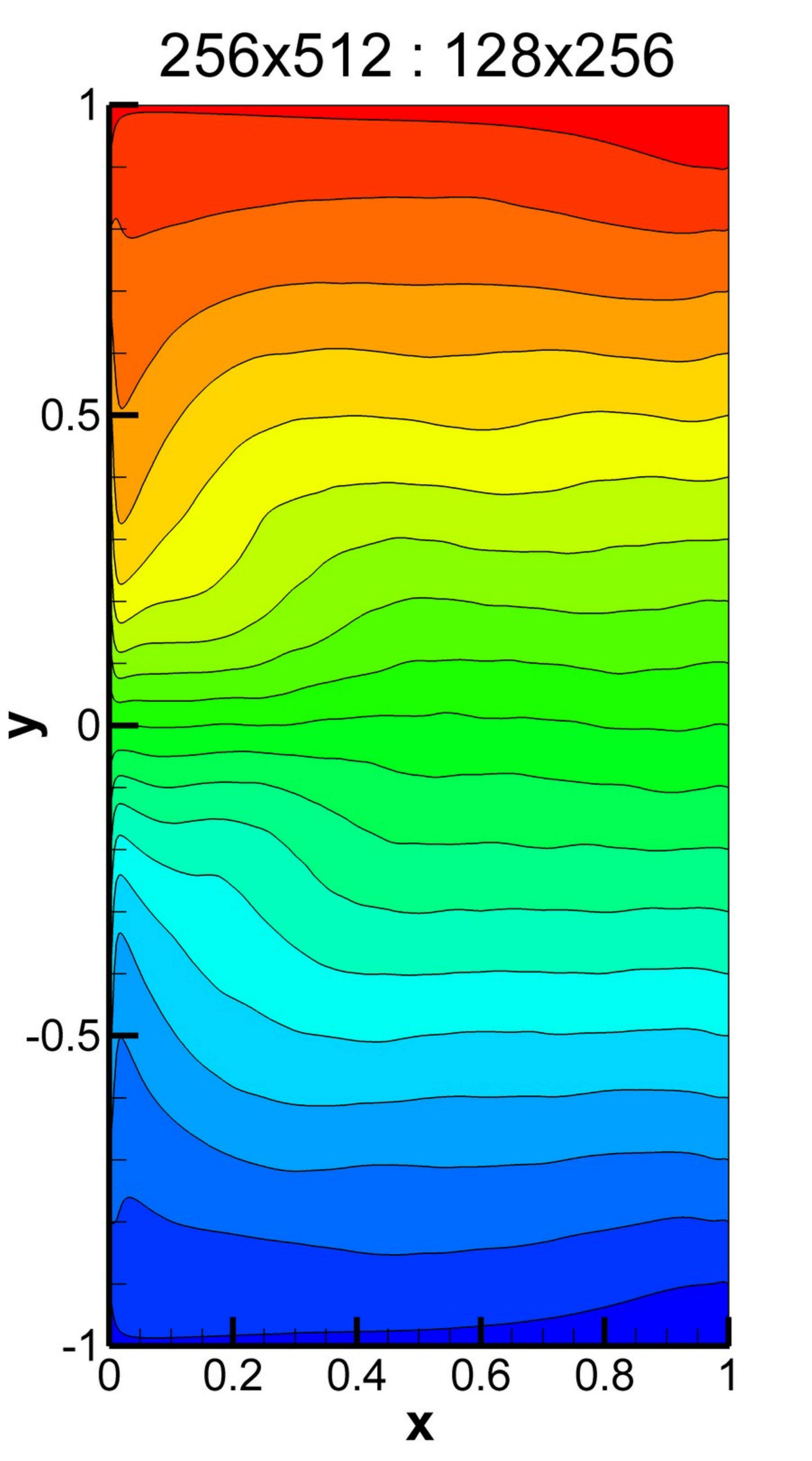}}
\subfigure{\includegraphics[width=0.22\textwidth]{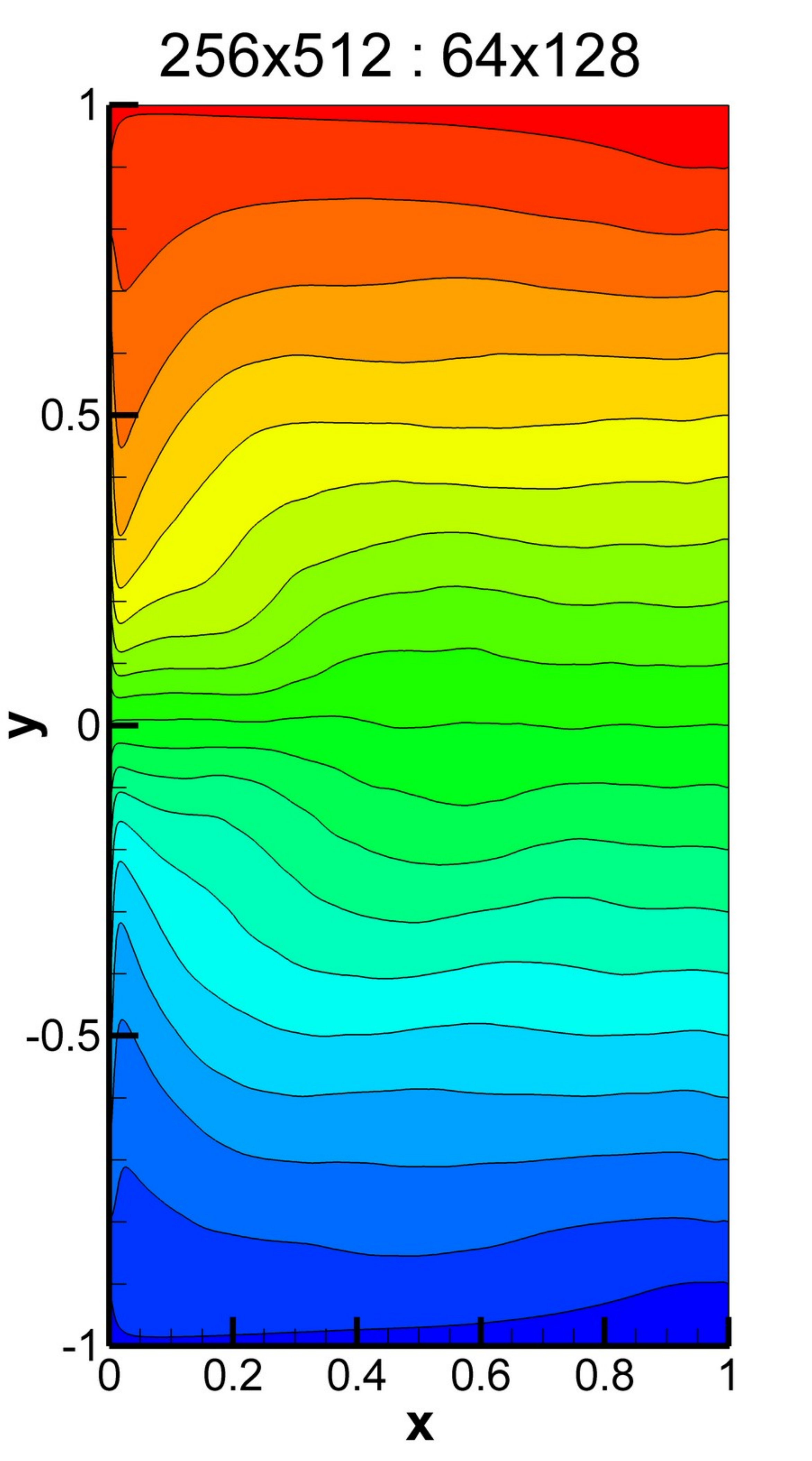}}
\subfigure{\includegraphics[width=0.22\textwidth]{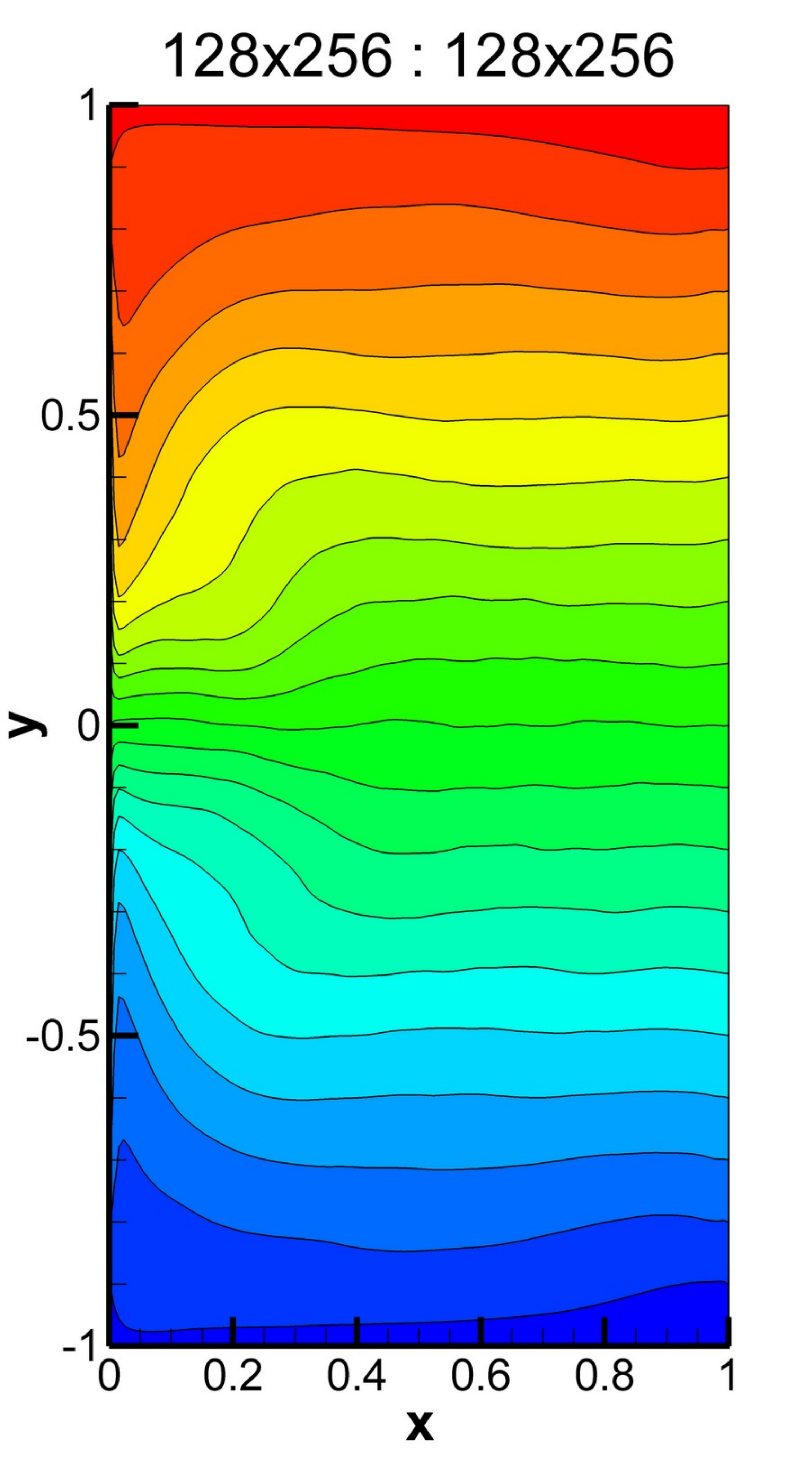}} }
\\
\mbox{
\subfigure{\includegraphics[width=0.22\textwidth]{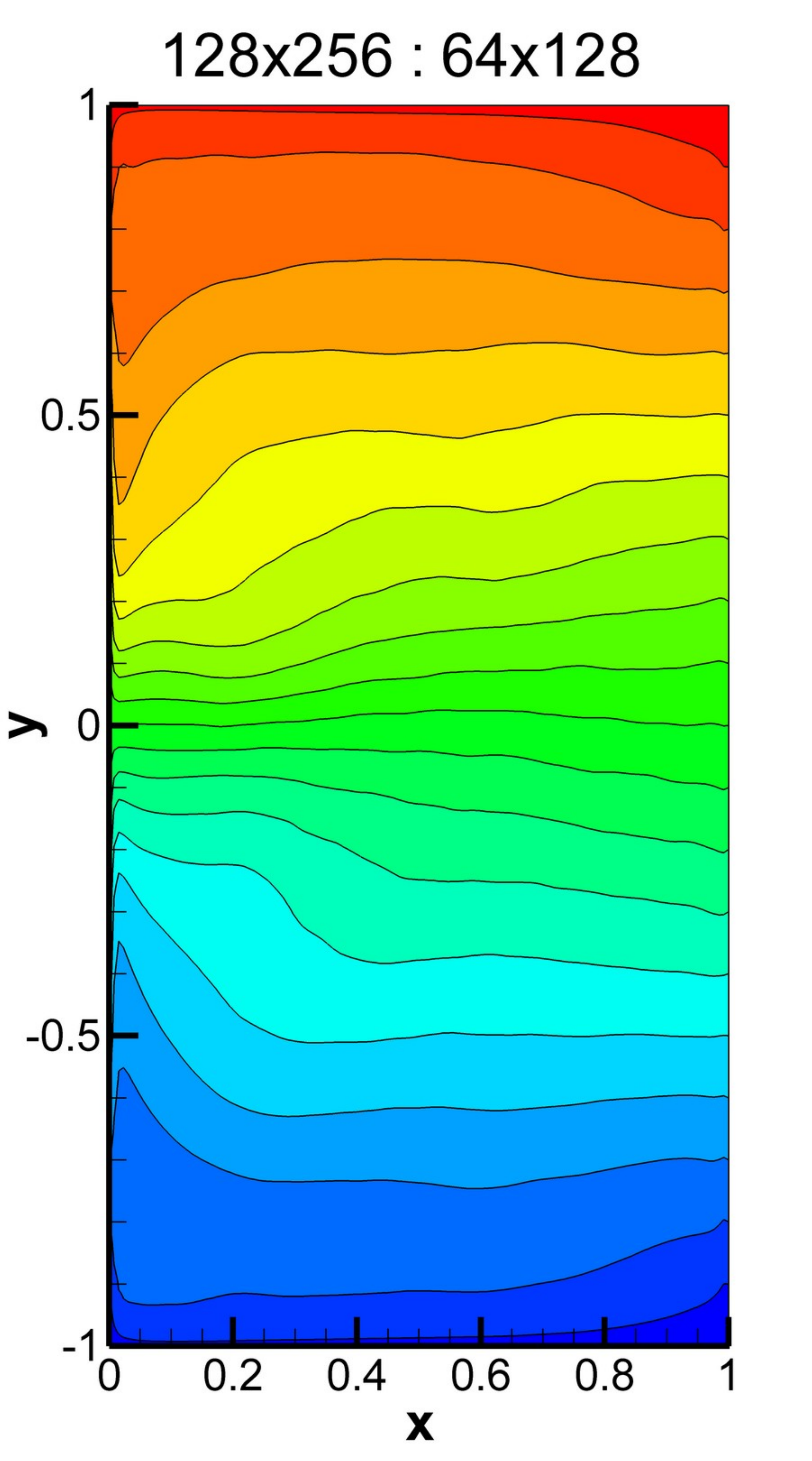}}
\subfigure{\includegraphics[width=0.22\textwidth]{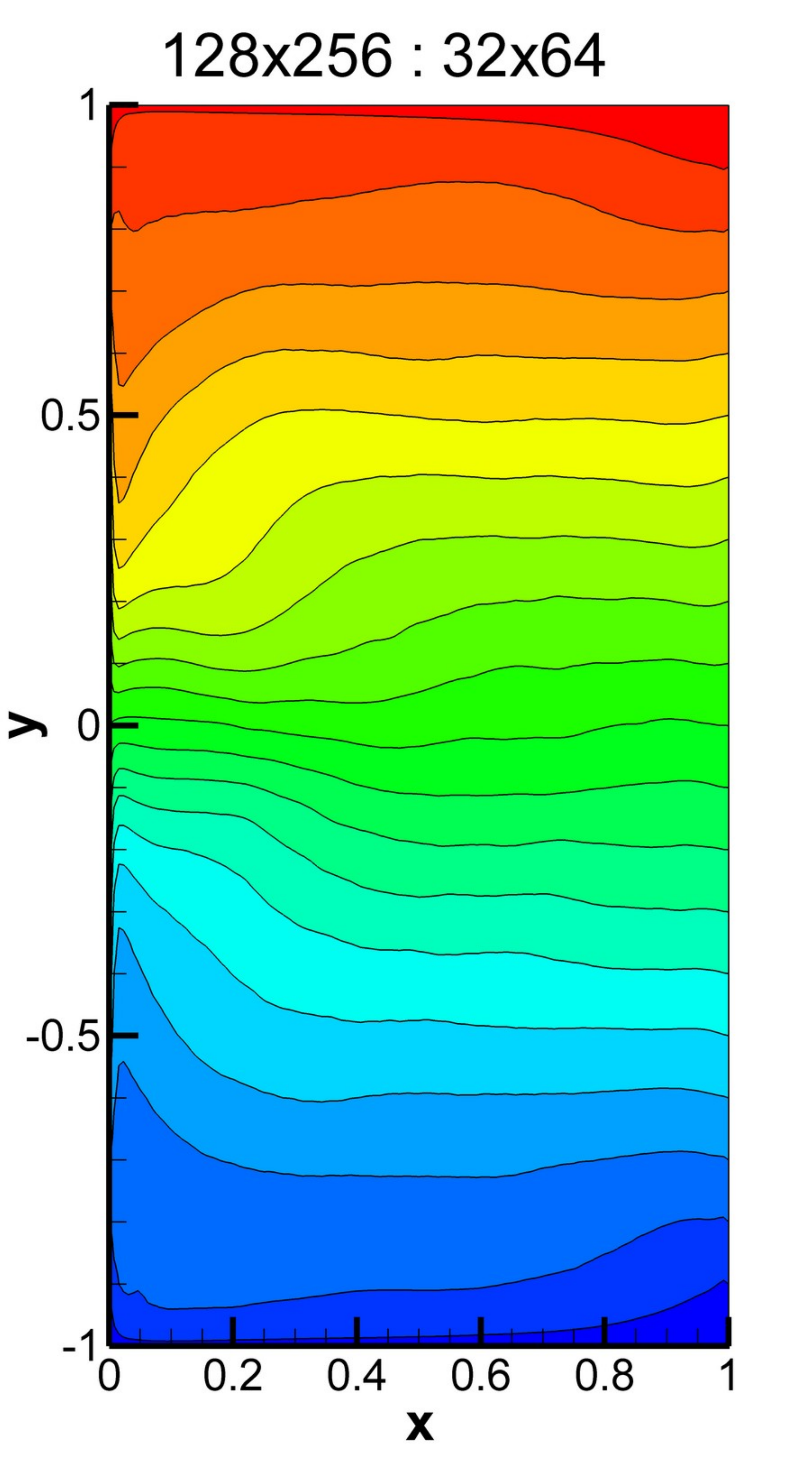}}
\subfigure{\includegraphics[width=0.22\textwidth]{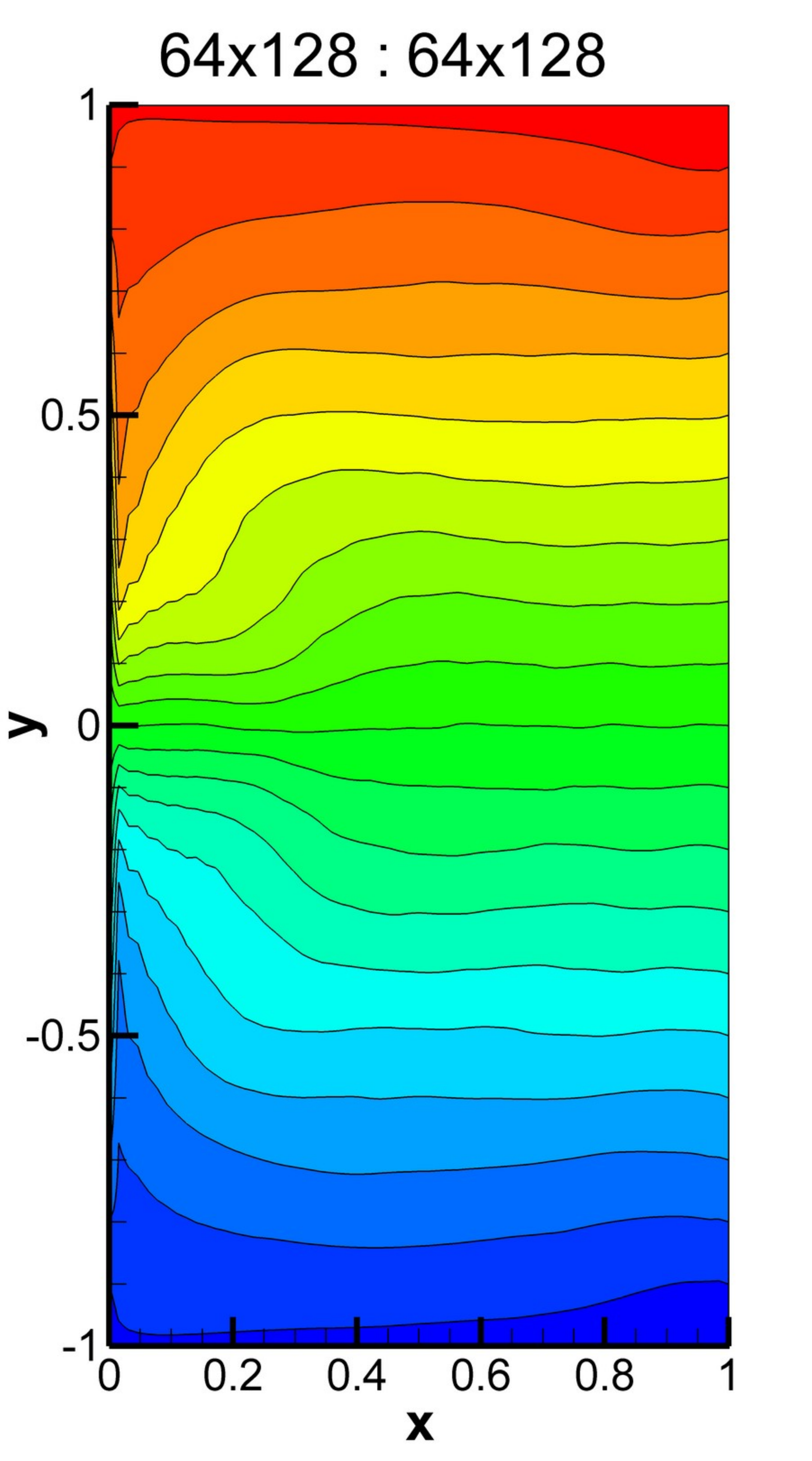}}
\subfigure{\includegraphics[width=0.22\textwidth]{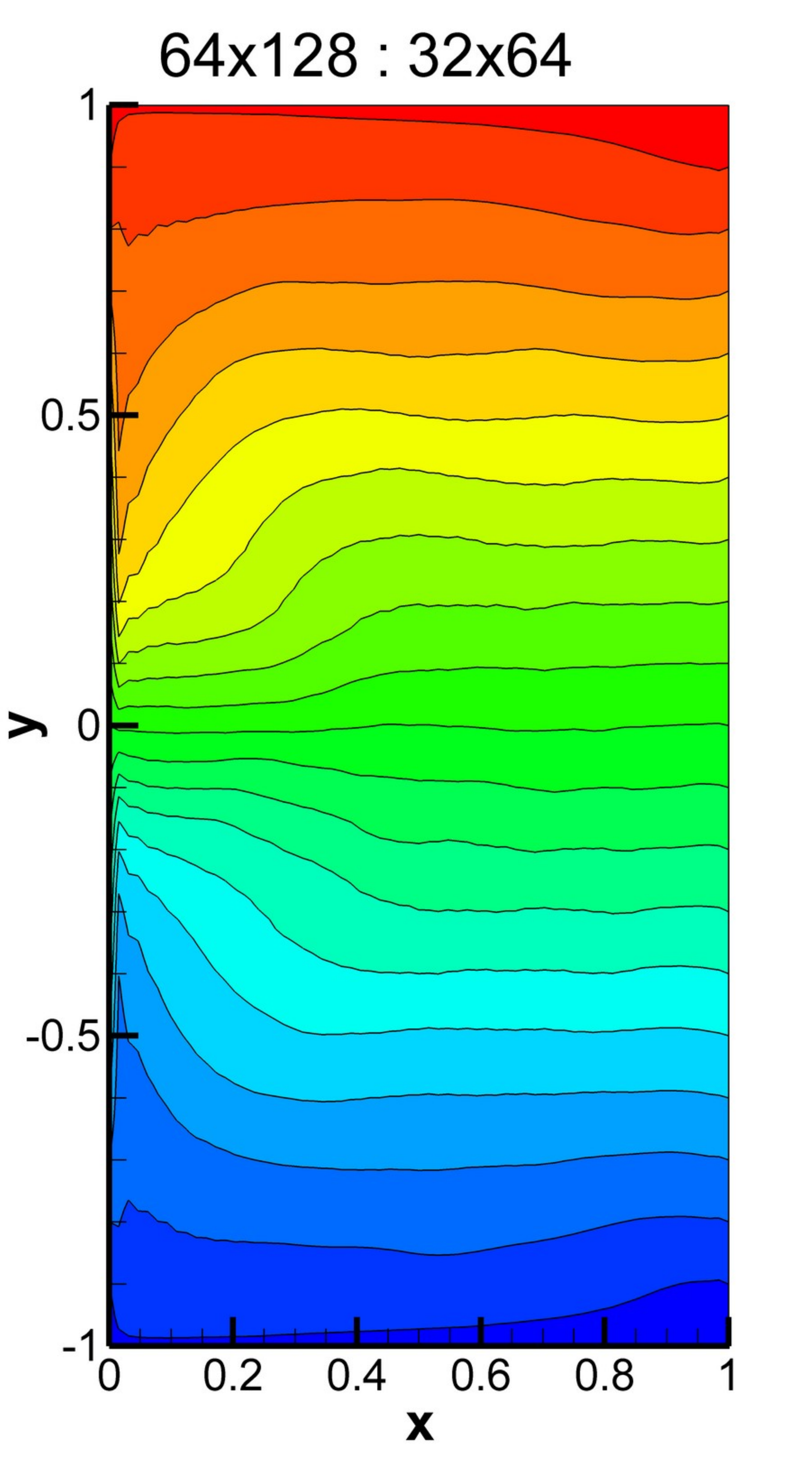}} }
\\
\mbox{
\subfigure{\includegraphics[width=0.22\textwidth]{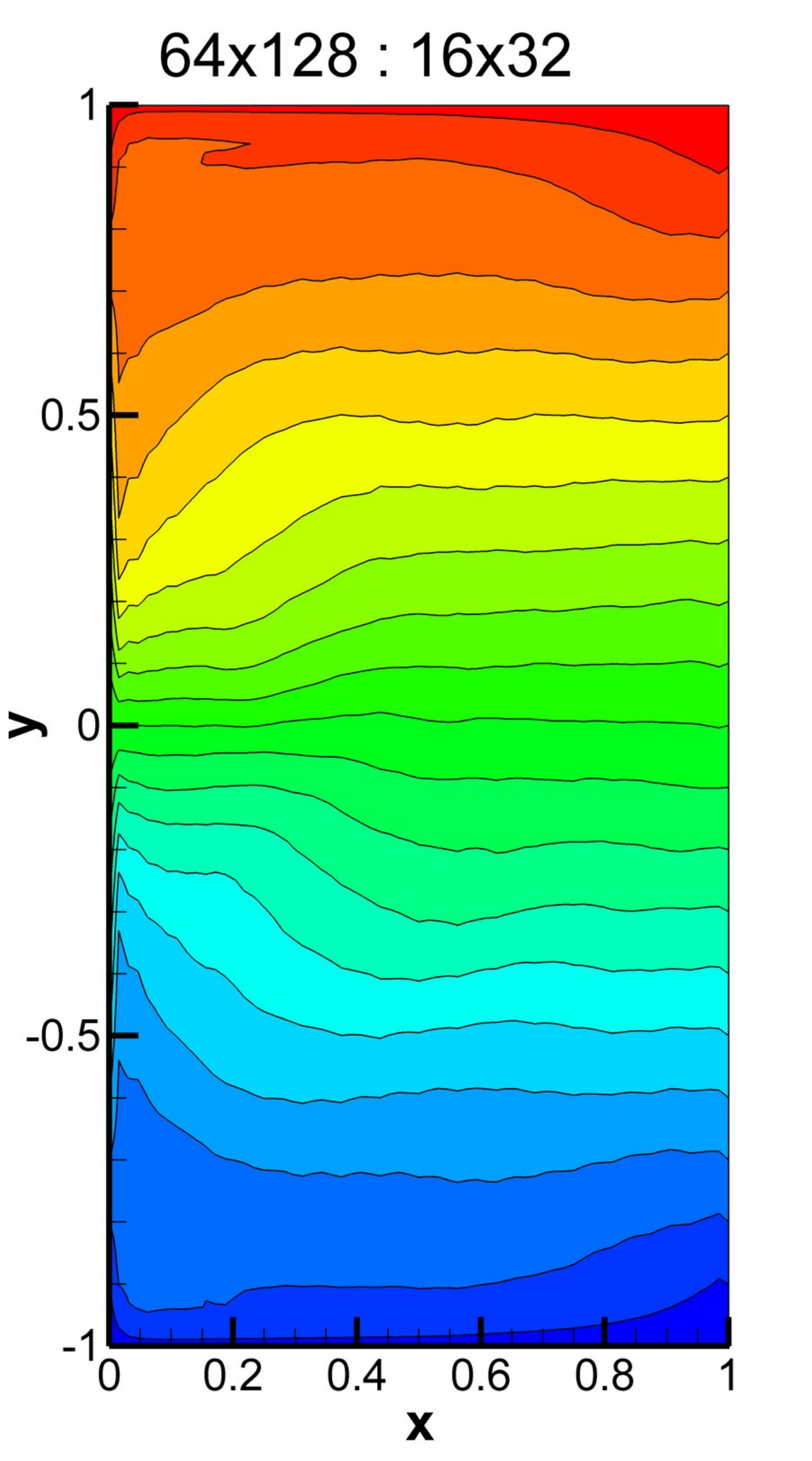}}
\subfigure{\includegraphics[width=0.22\textwidth]{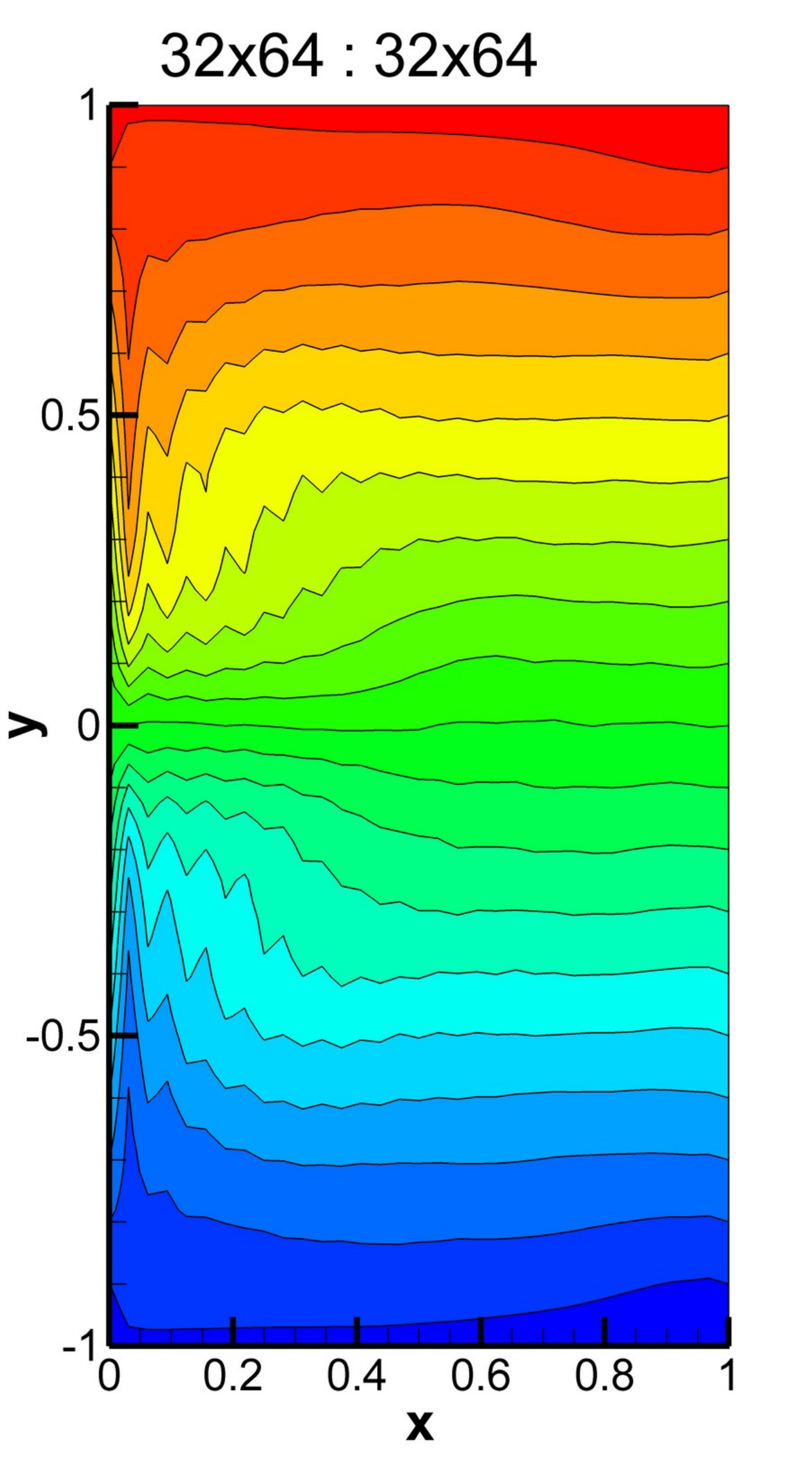}}
\subfigure{\includegraphics[width=0.22\textwidth]{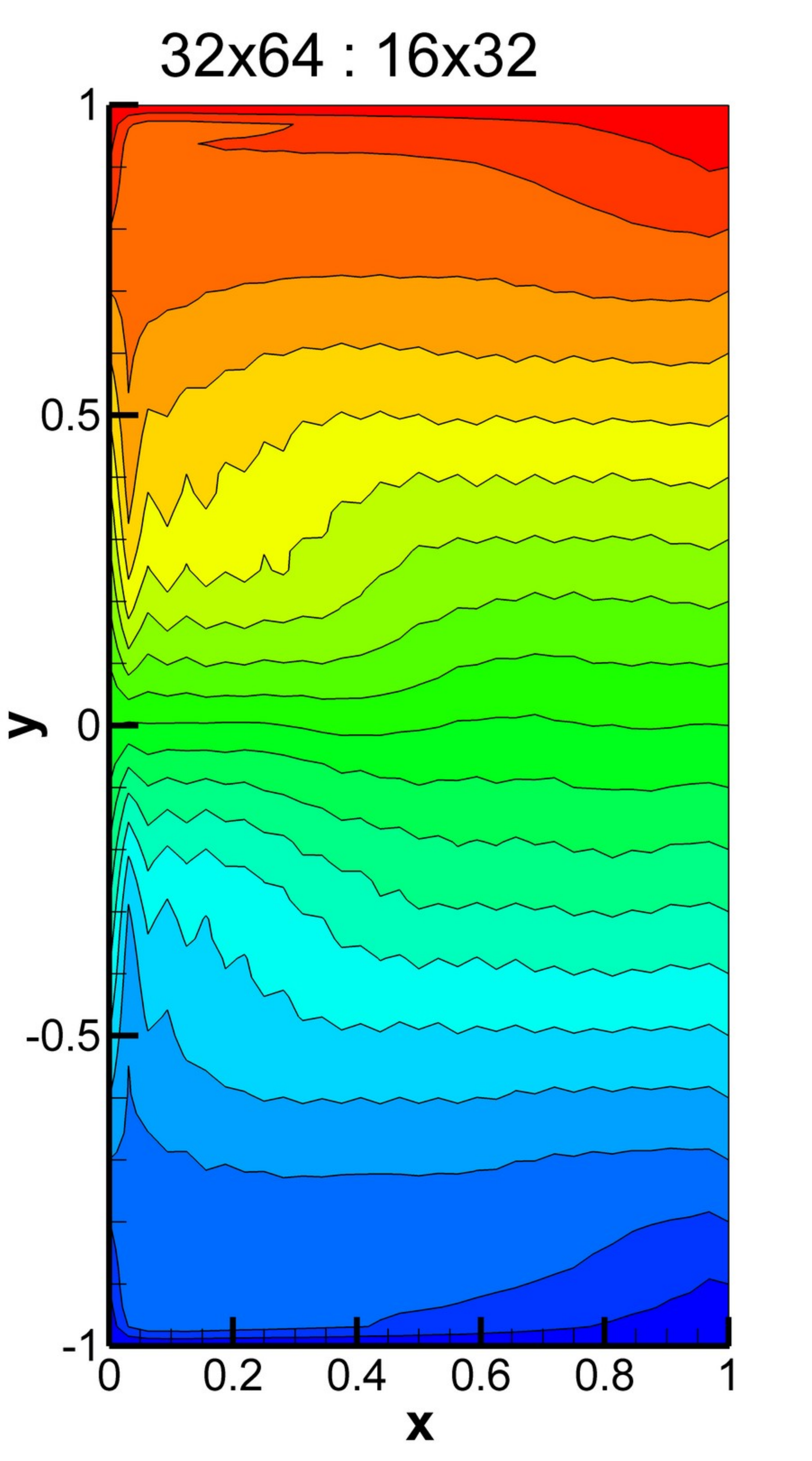}}
\subfigure{\includegraphics[width=0.22\textwidth]{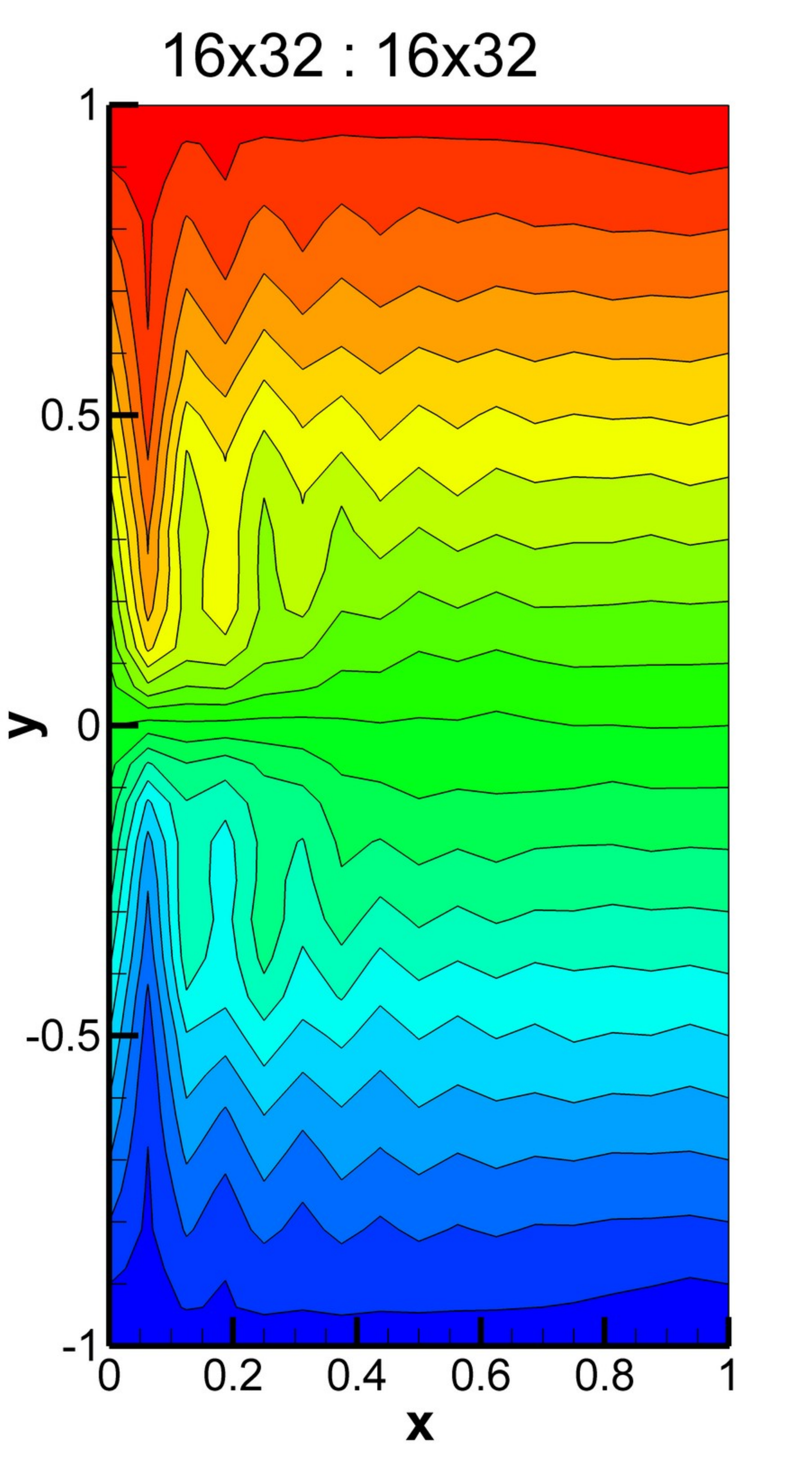}} }
\caption{Experiment I:
Comparison of mean potential vorticity for $Re=200$ and $Ro=0.0016$ (i.e., $\delta_M/L = 0.02$, and $\delta_I/L = 0.04$). Labels include the resolutions for both parts of the solver in the form $N_x \times N_y : M_x \times M_y$, where $N_x \times N_y$ is the resolution for the barotropic vorticity transport equation, and $M_x \times M_y$ is the resolution for the elliptic sub-problems.
}
\label{fig:qB}
\end{figure*}

\begin{figure*}
\centering
\mbox{
\subfigure{\includegraphics[width=0.22\textwidth]{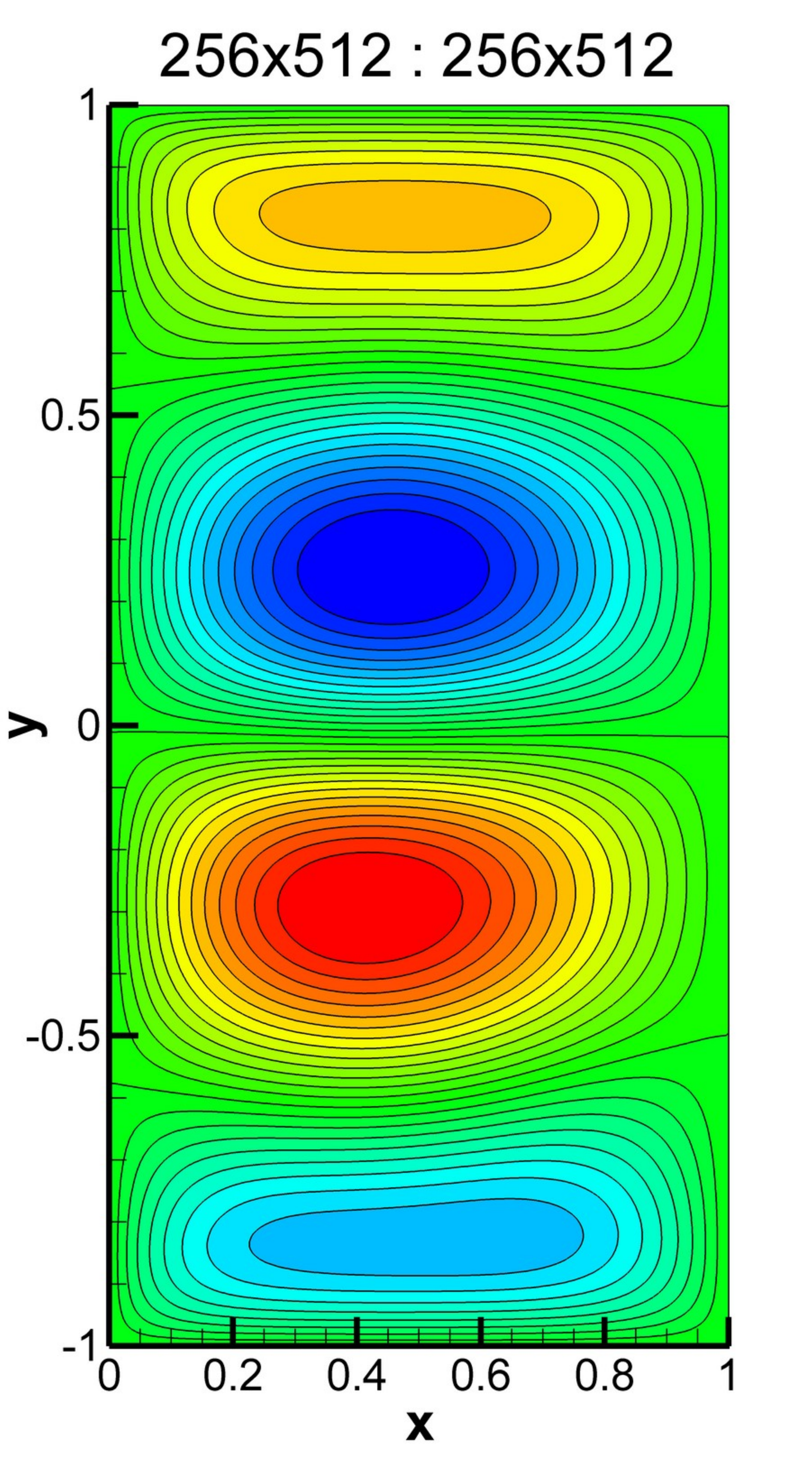}}
\subfigure{\includegraphics[width=0.22\textwidth]{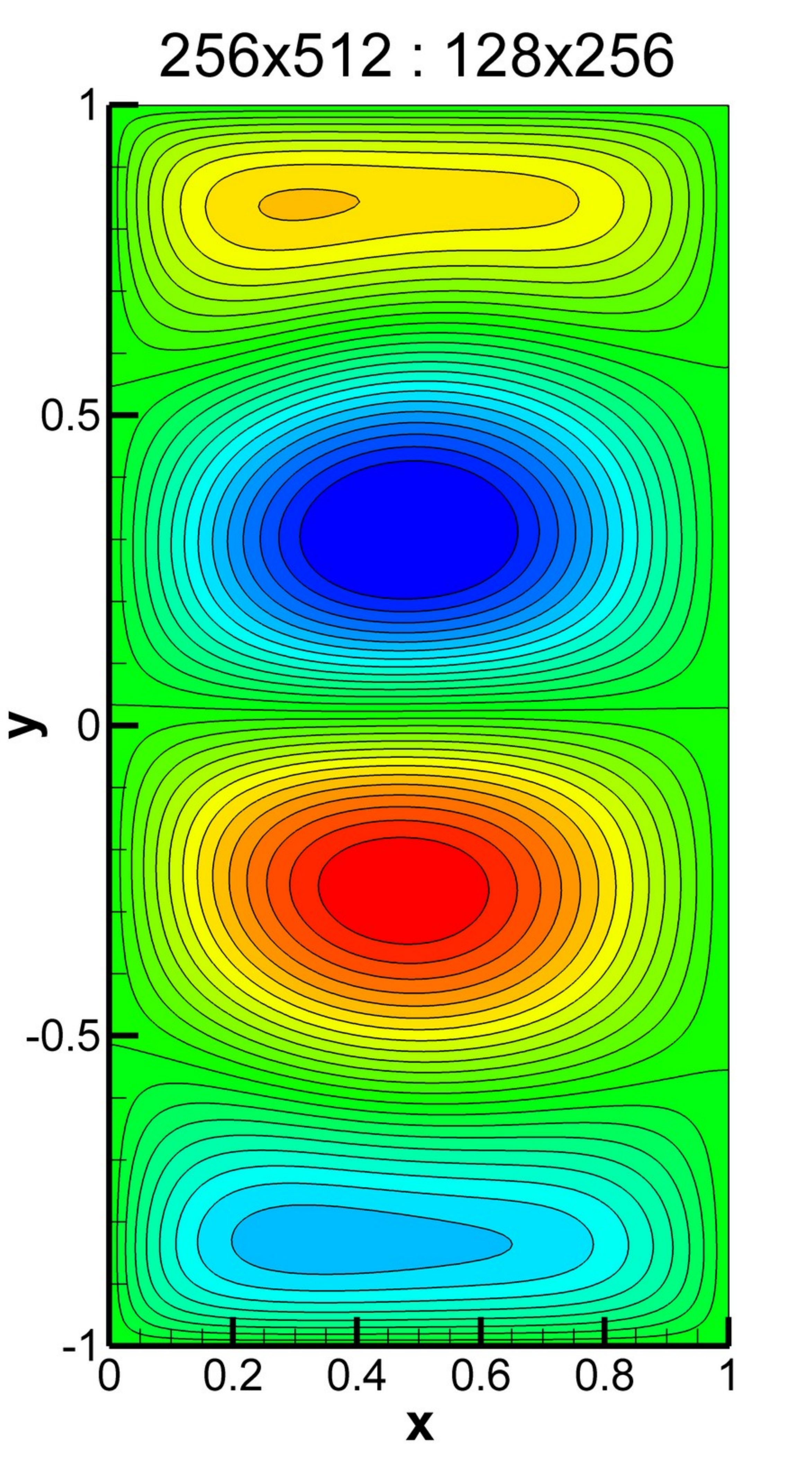}}
\subfigure{\includegraphics[width=0.22\textwidth]{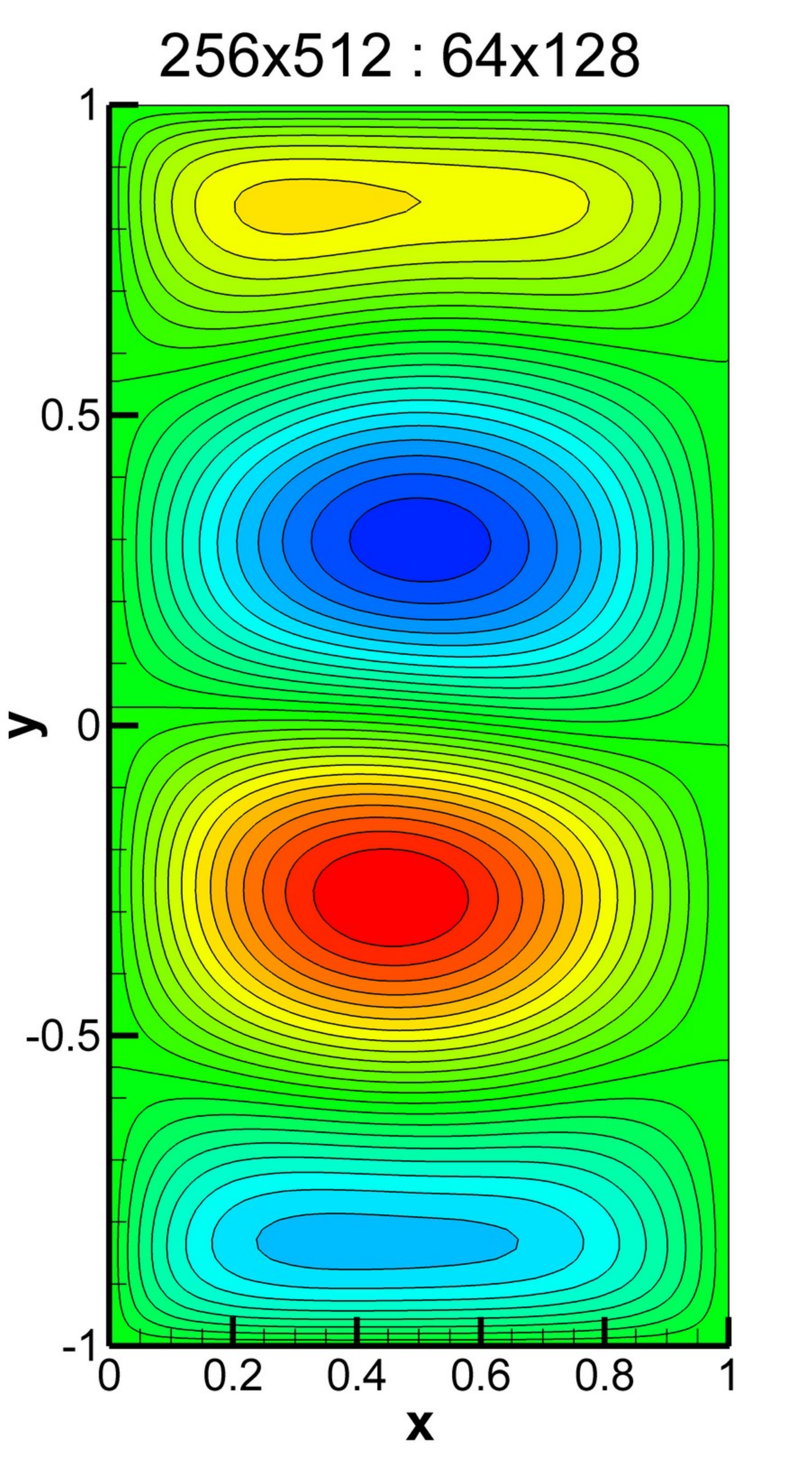}}
\subfigure{\includegraphics[width=0.22\textwidth]{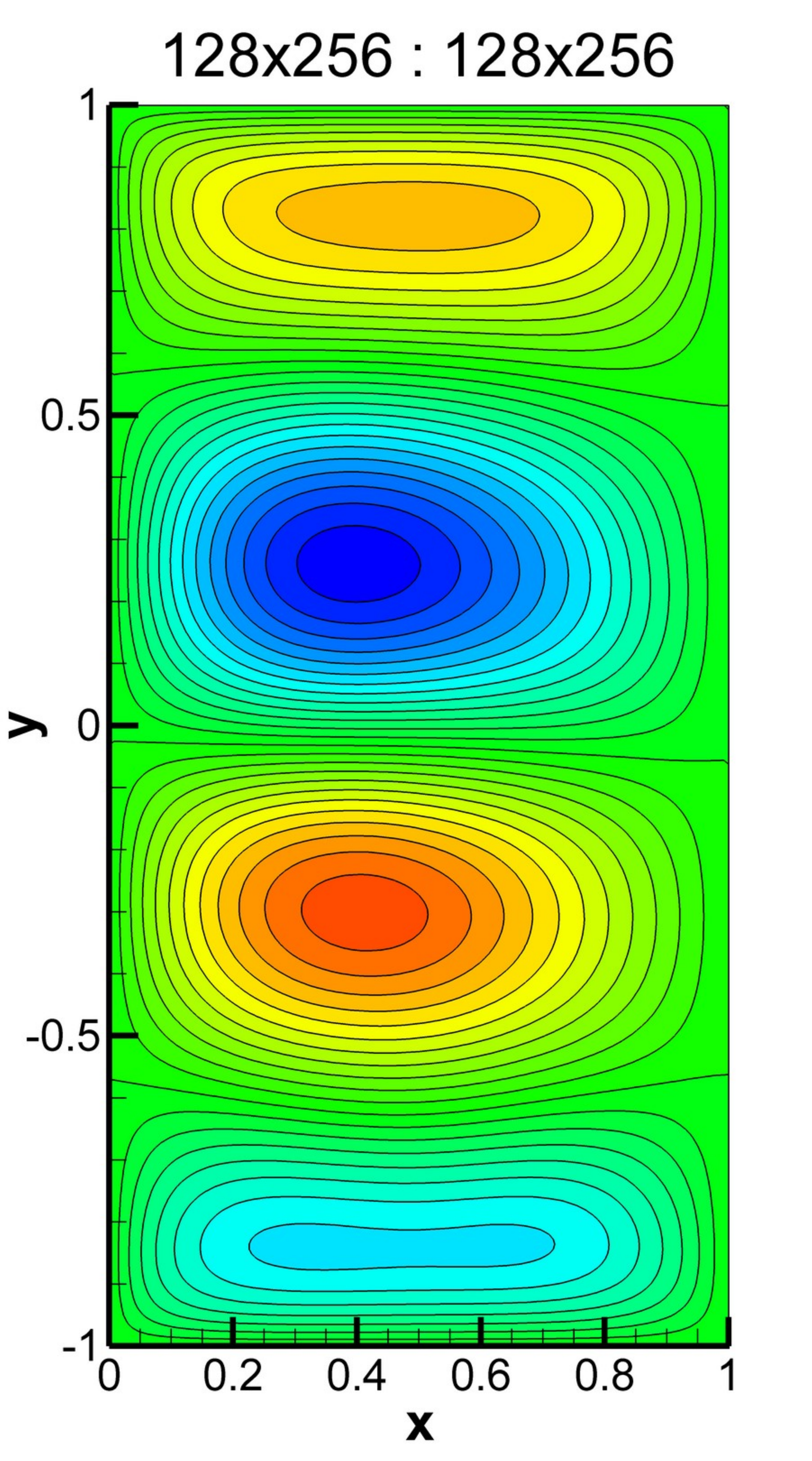}} }
\\
\mbox{
\subfigure{\includegraphics[width=0.22\textwidth]{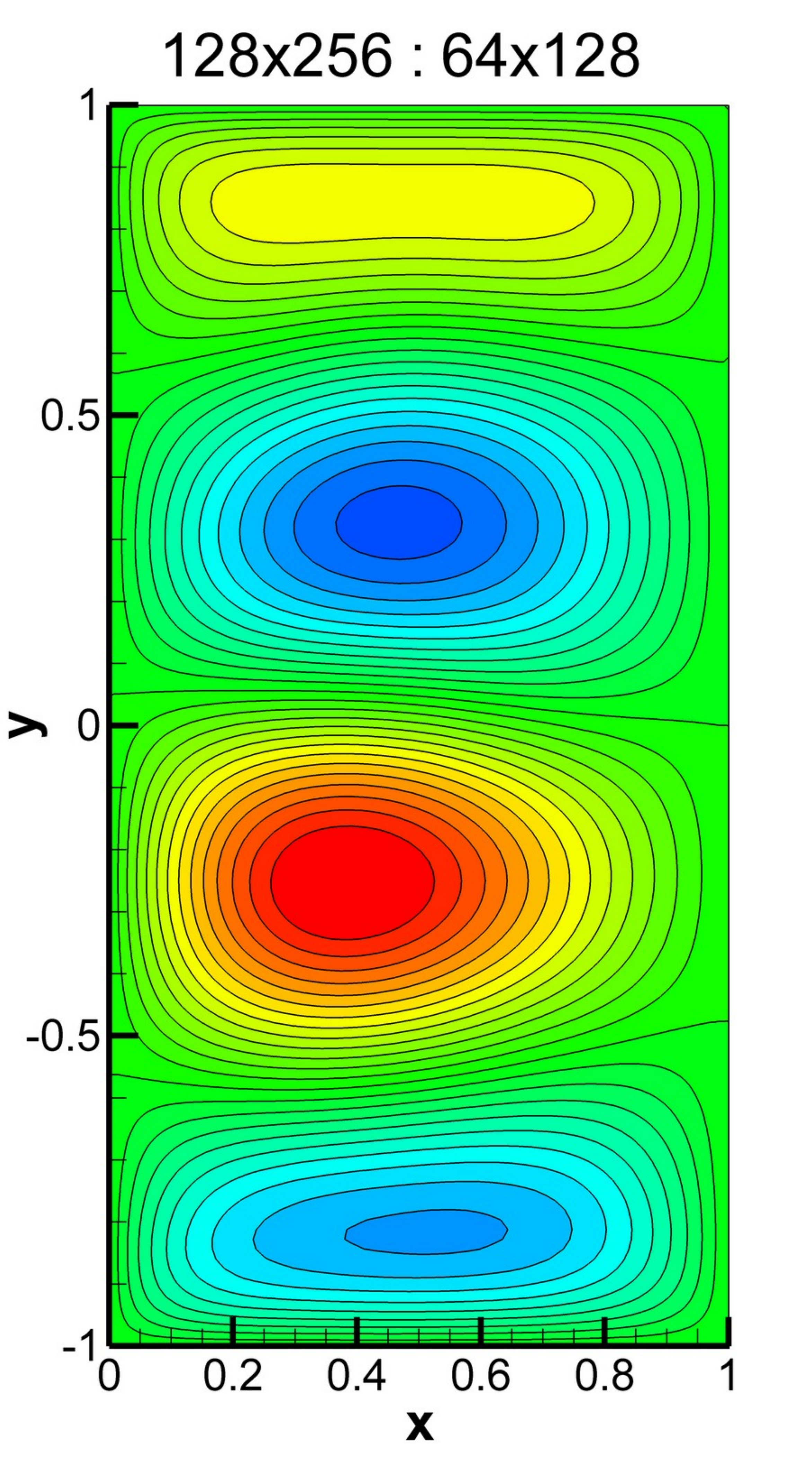}}
\subfigure{\includegraphics[width=0.22\textwidth]{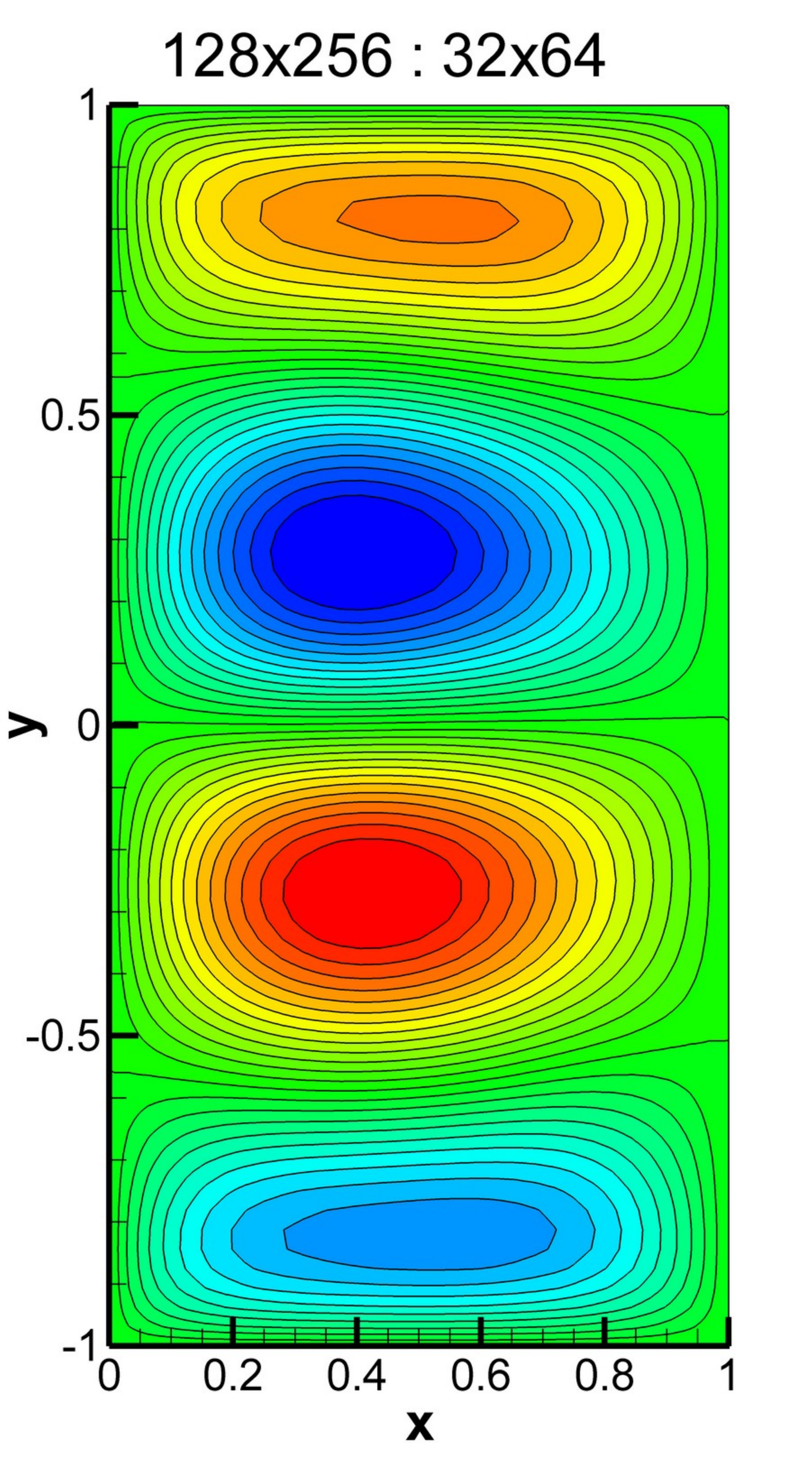}}
\subfigure{\includegraphics[width=0.22\textwidth]{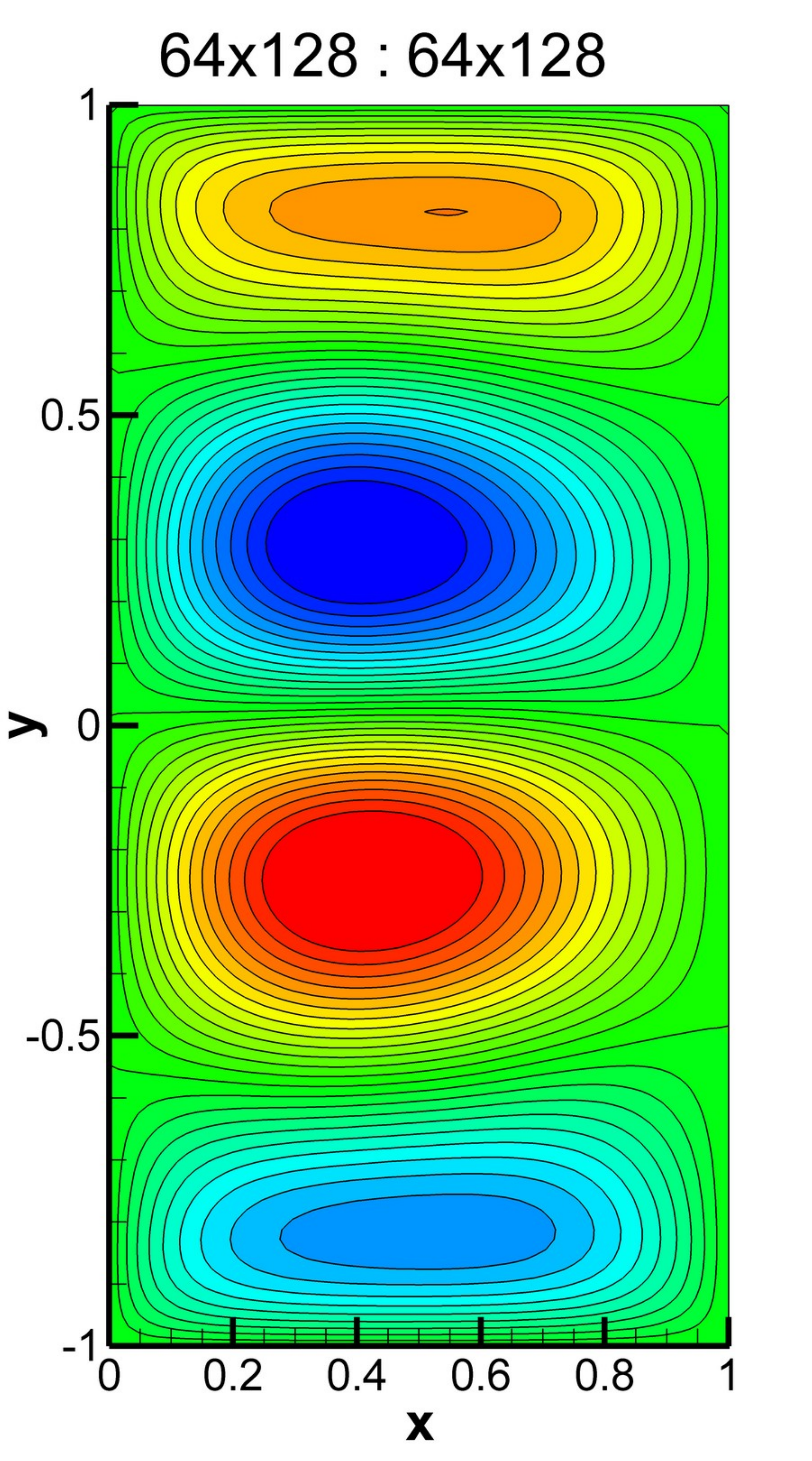}}
\subfigure{\includegraphics[width=0.22\textwidth]{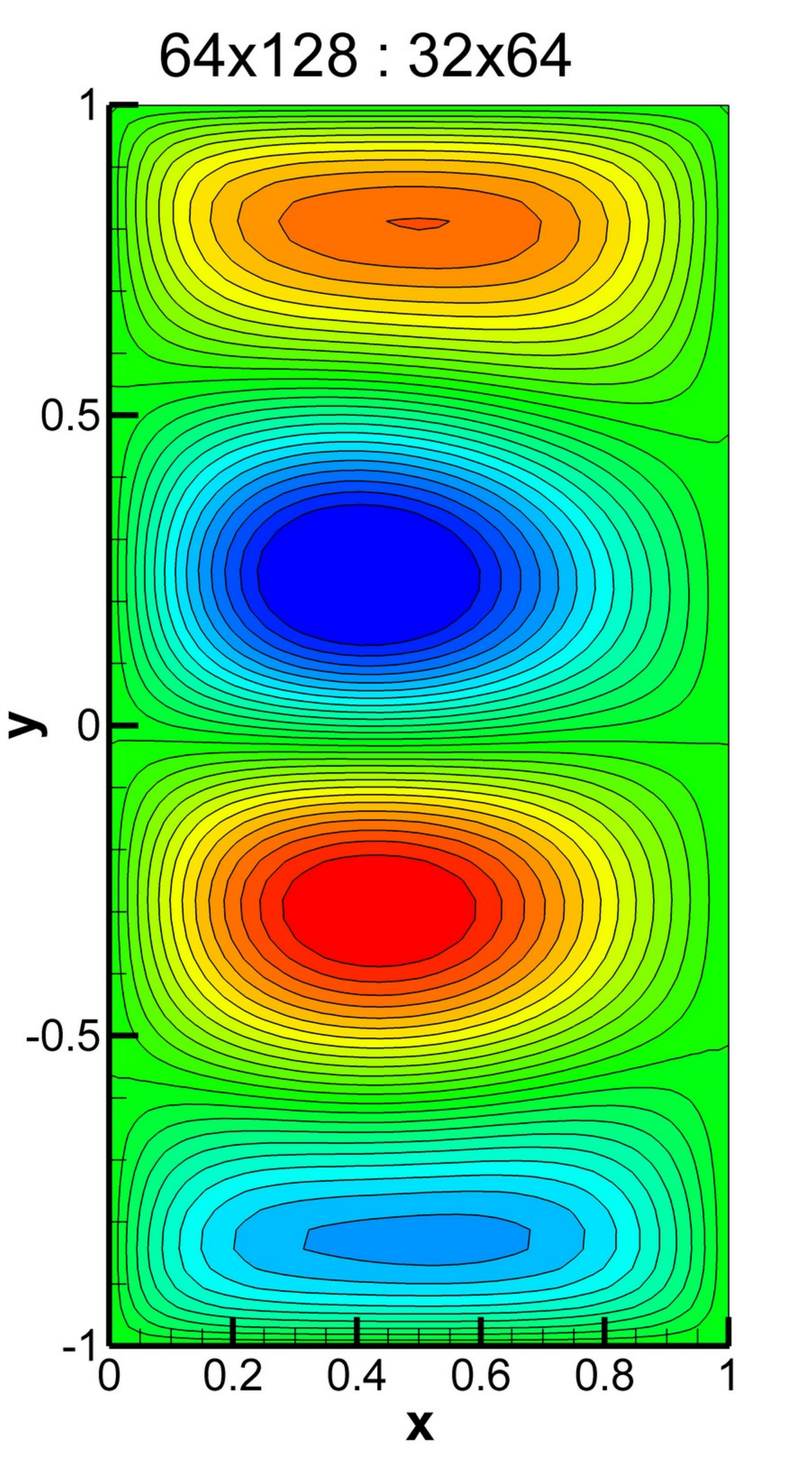}} }
\\
\mbox{
\subfigure{\includegraphics[width=0.22\textwidth]{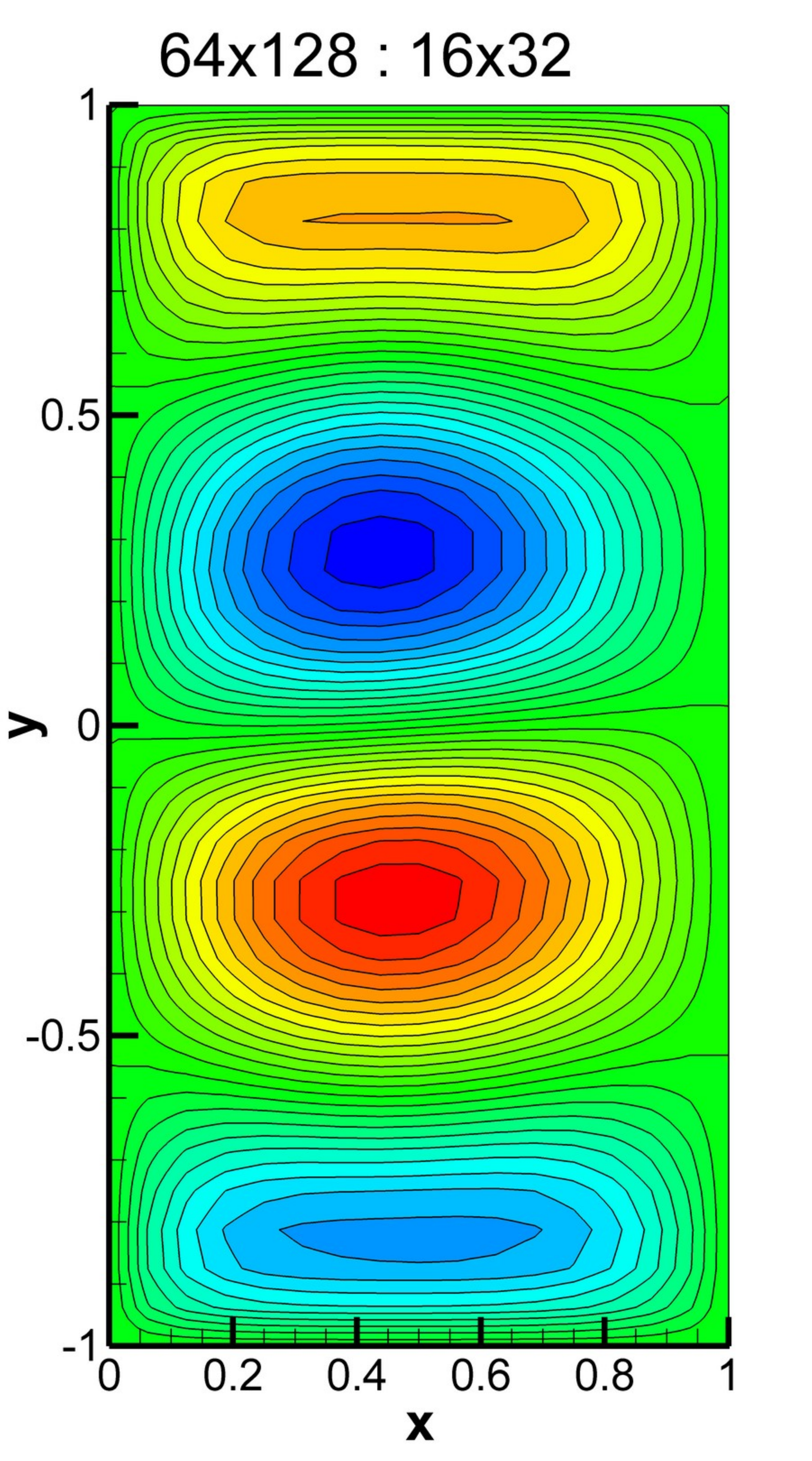}}
\subfigure{\includegraphics[width=0.22\textwidth]{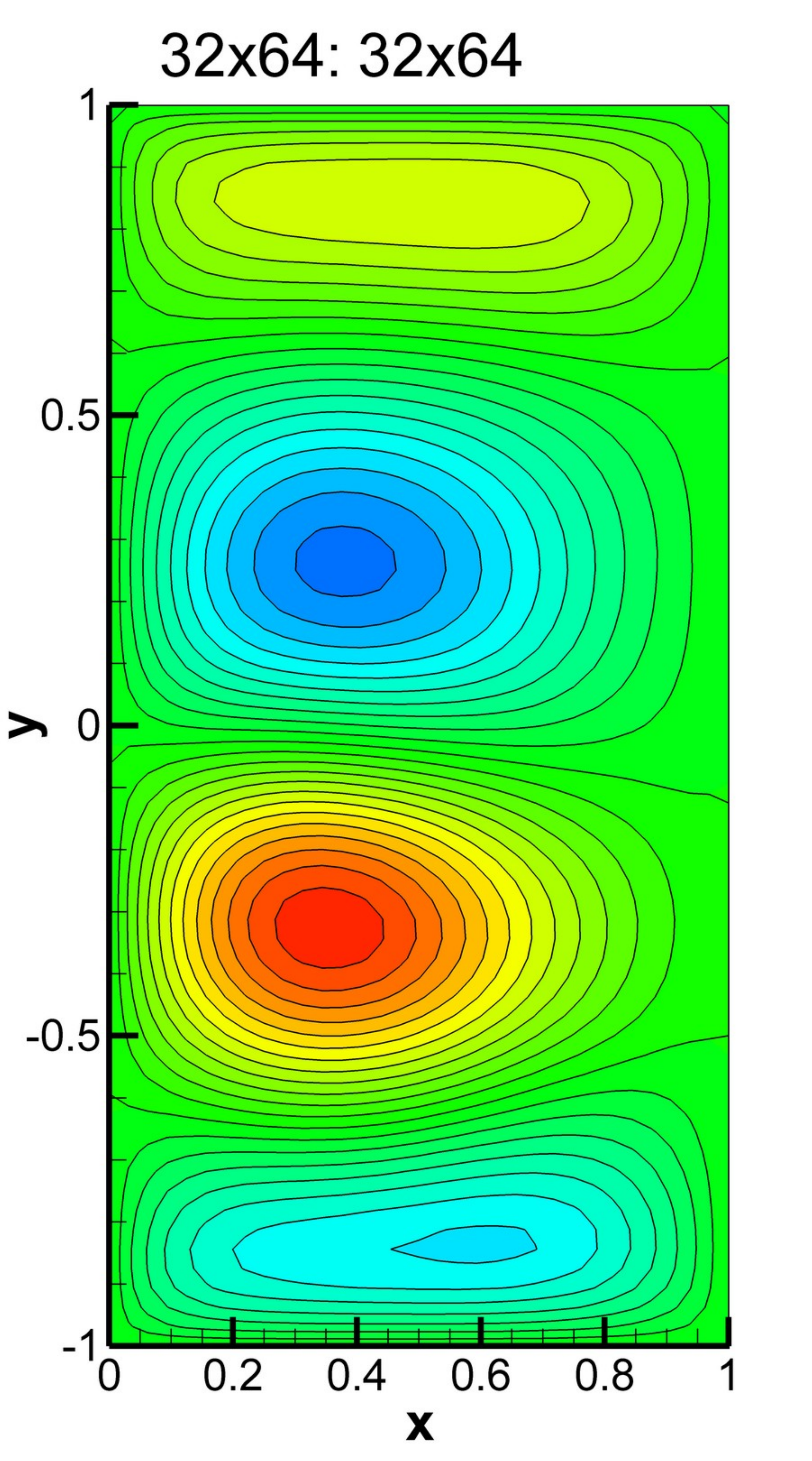}}
\subfigure{\includegraphics[width=0.22\textwidth]{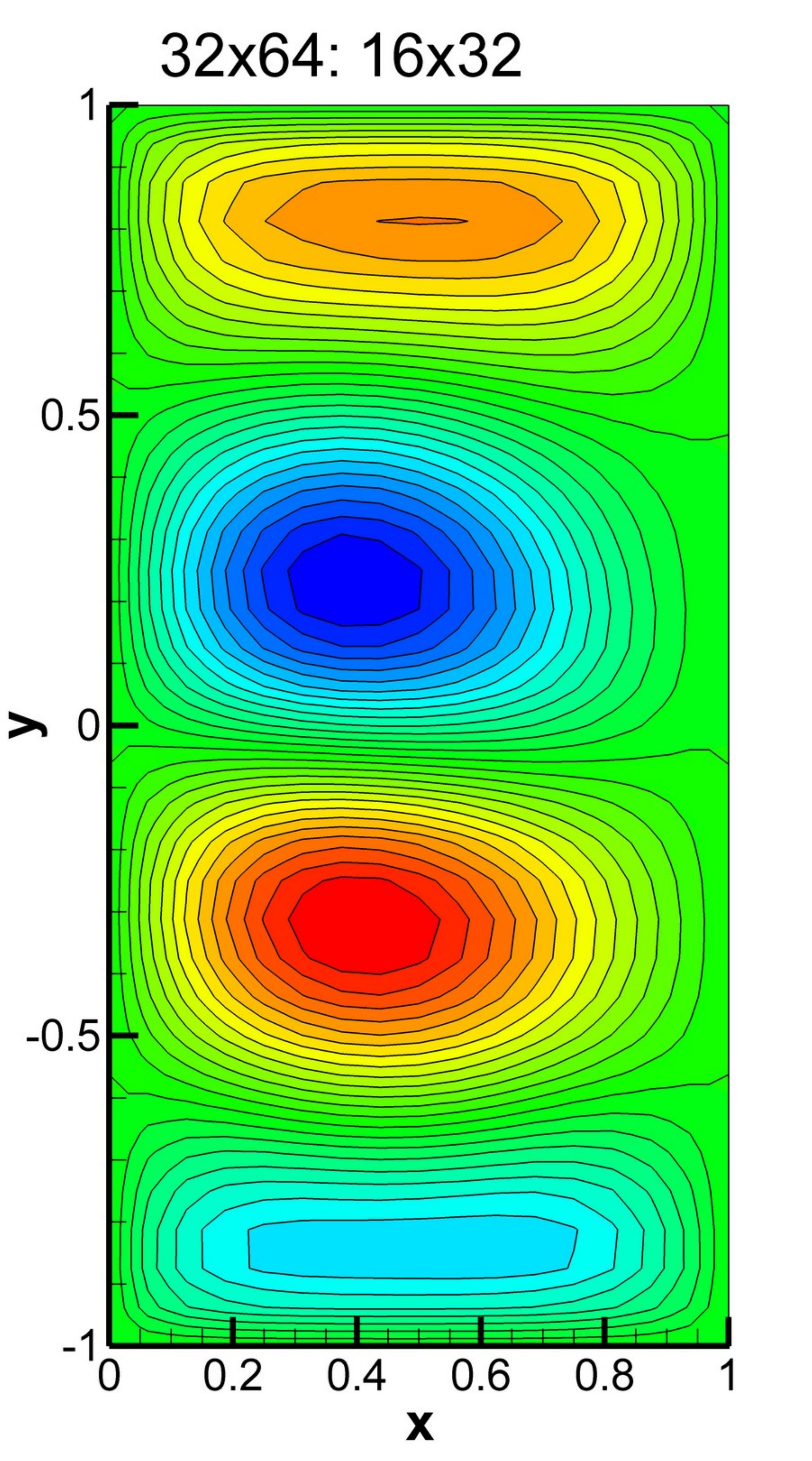}}
\subfigure{\includegraphics[width=0.22\textwidth]{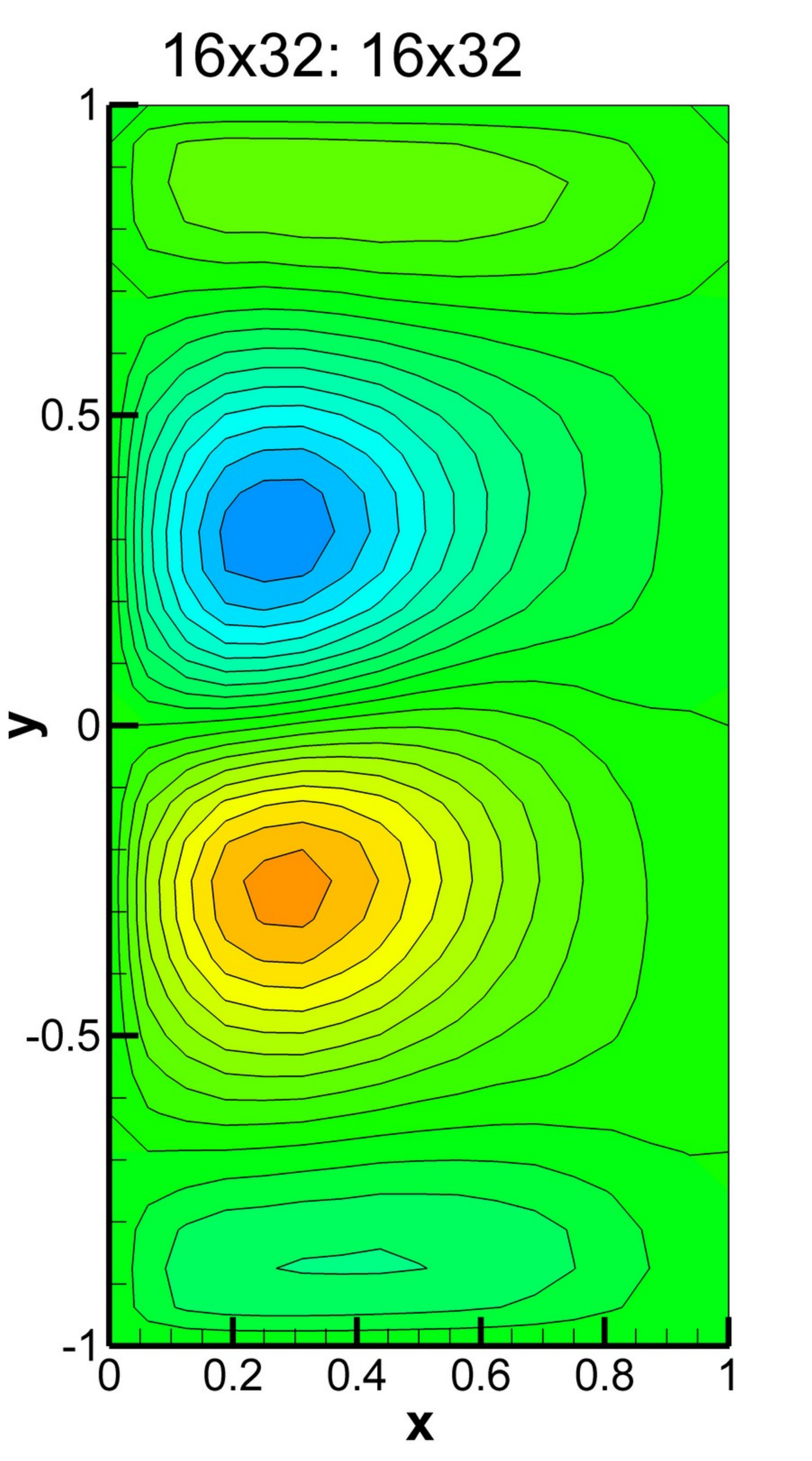}} }
\caption{Experiment II:
Comparison of mean stream functions for $Re=312.5$ and $Ro=0.0025$ (i.e., $\delta_M/L = 0.02$, and $\delta_I/L = 0.05$). Labels include the resolutions for both parts of the solver in the form $N_x \times N_y : M_x \times M_y$, where $N_x \times N_y$ is the resolution for the barotropic vorticity transport equation, and $M_x \times M_y$ is the resolution for the elliptic sub-problems.
}
\label{fig:sC}
\end{figure*}

\begin{figure*}
\centering
\mbox{
\subfigure{\includegraphics[width=0.22\textwidth]{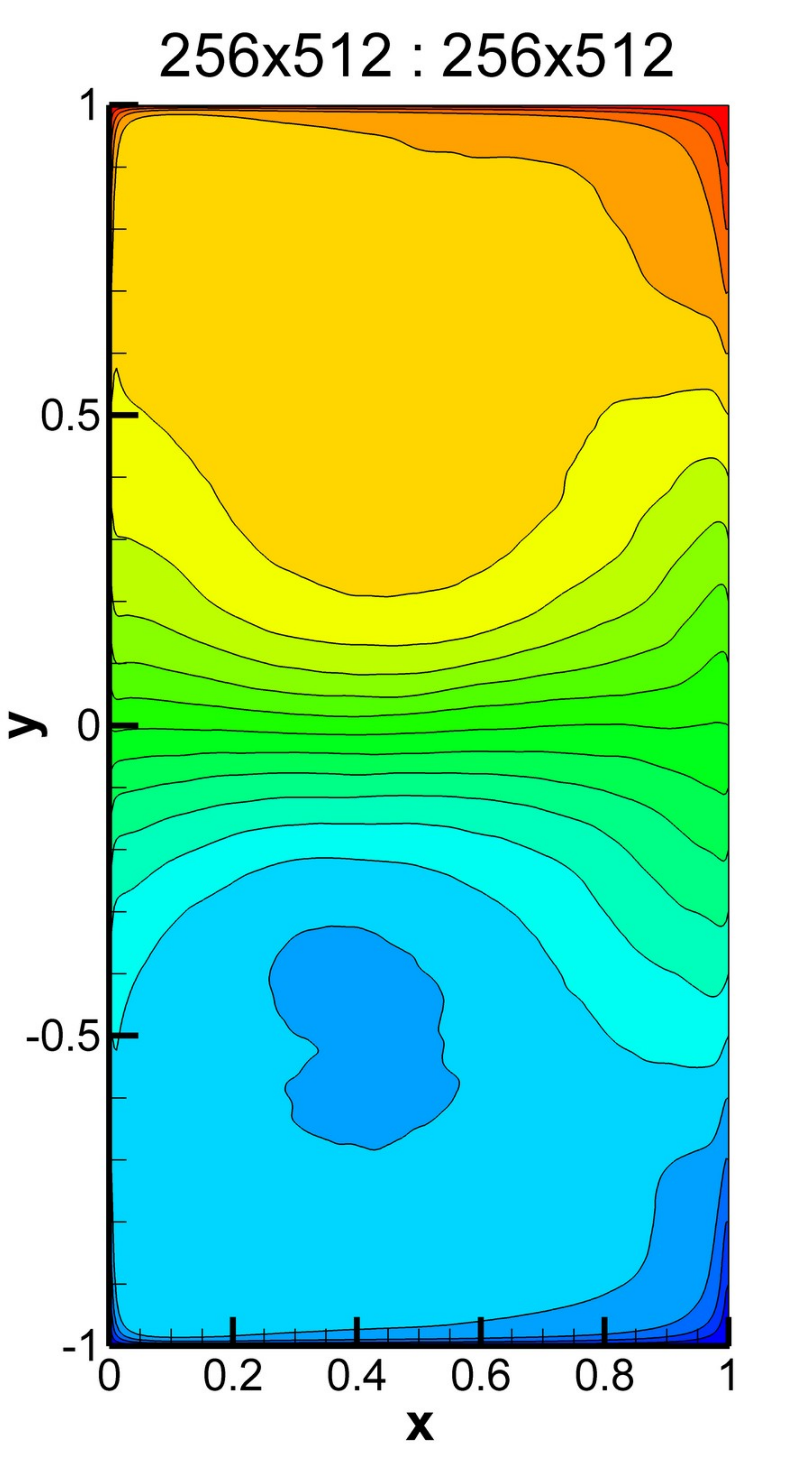}}
\subfigure{\includegraphics[width=0.22\textwidth]{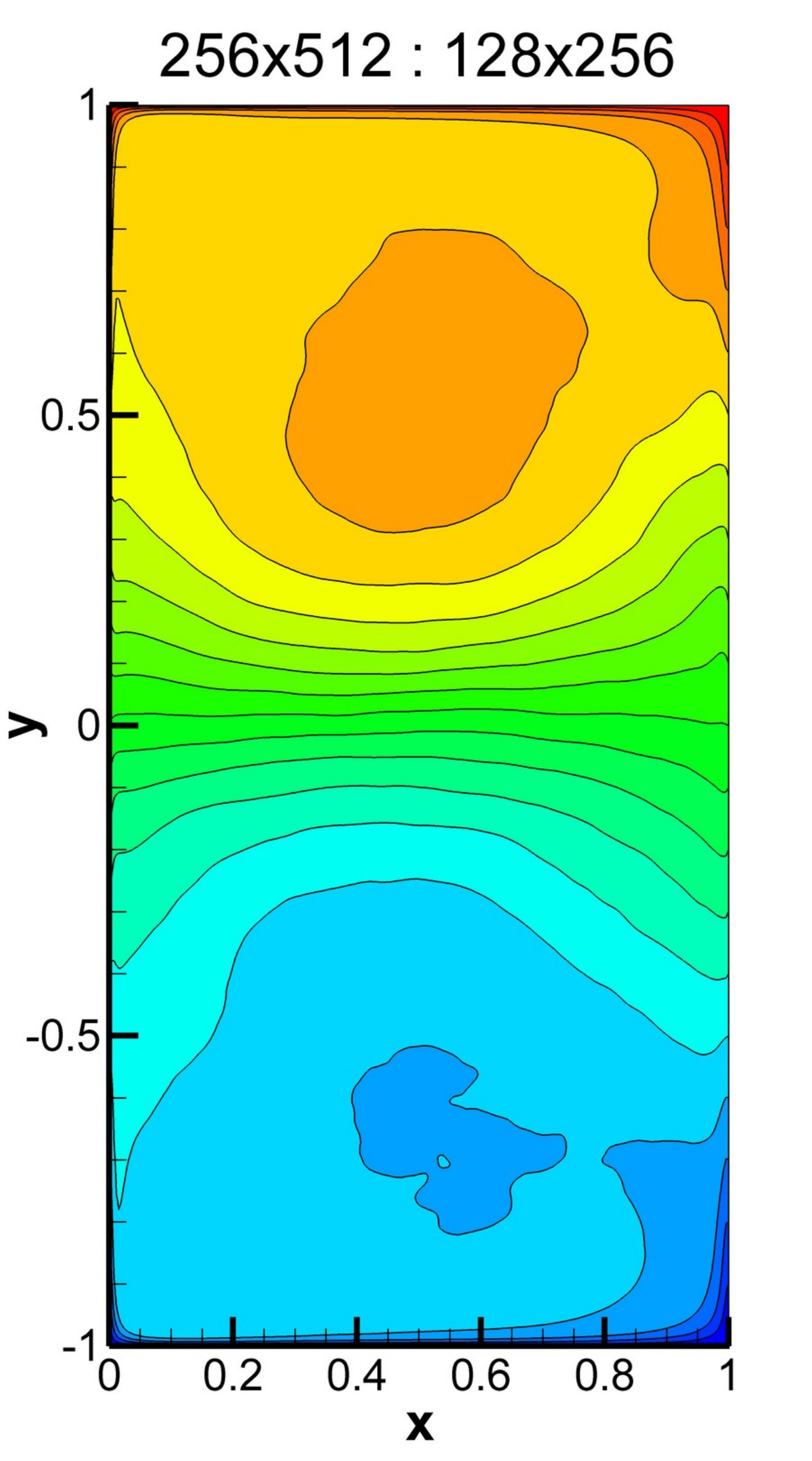}}
\subfigure{\includegraphics[width=0.22\textwidth]{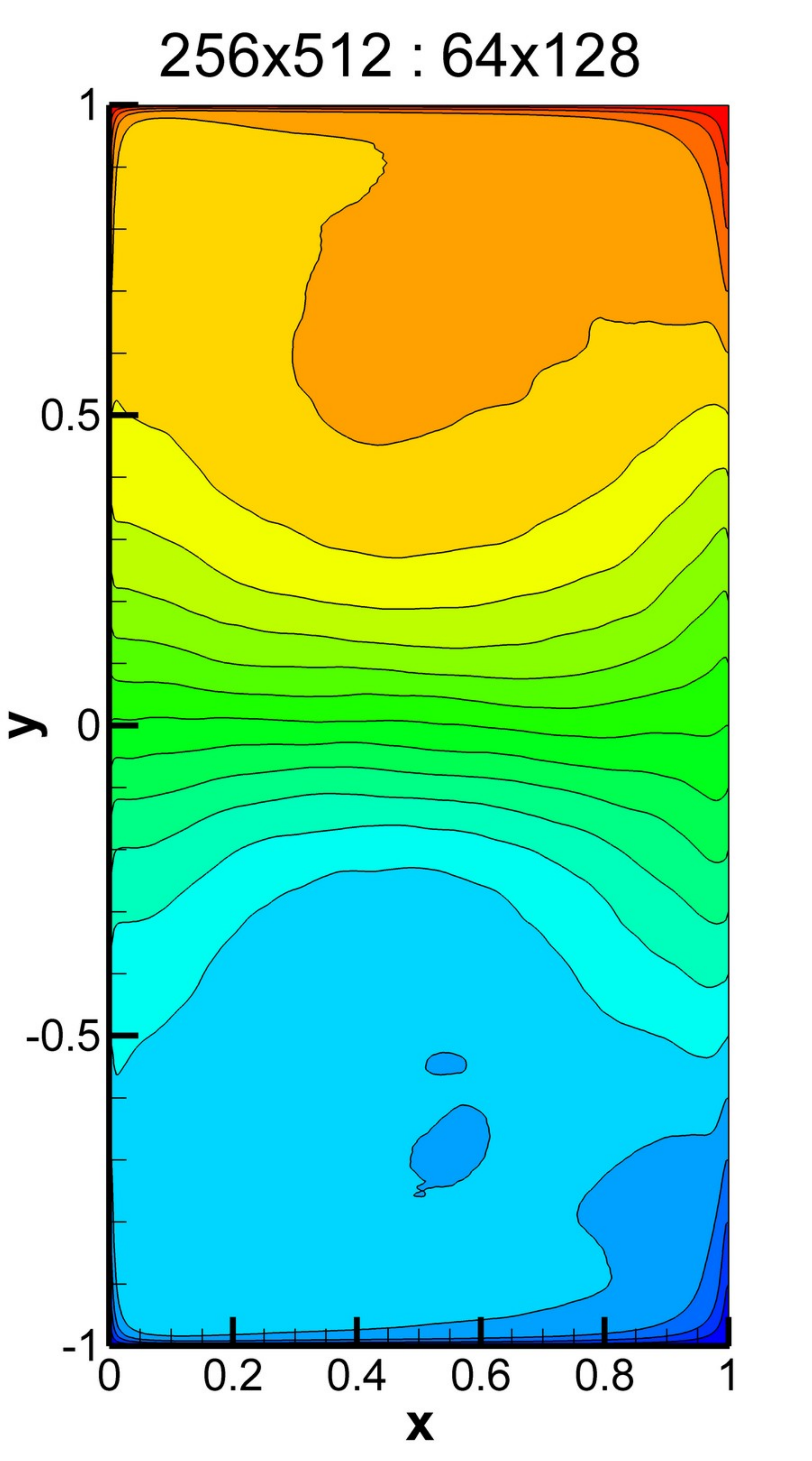}}
\subfigure{\includegraphics[width=0.22\textwidth]{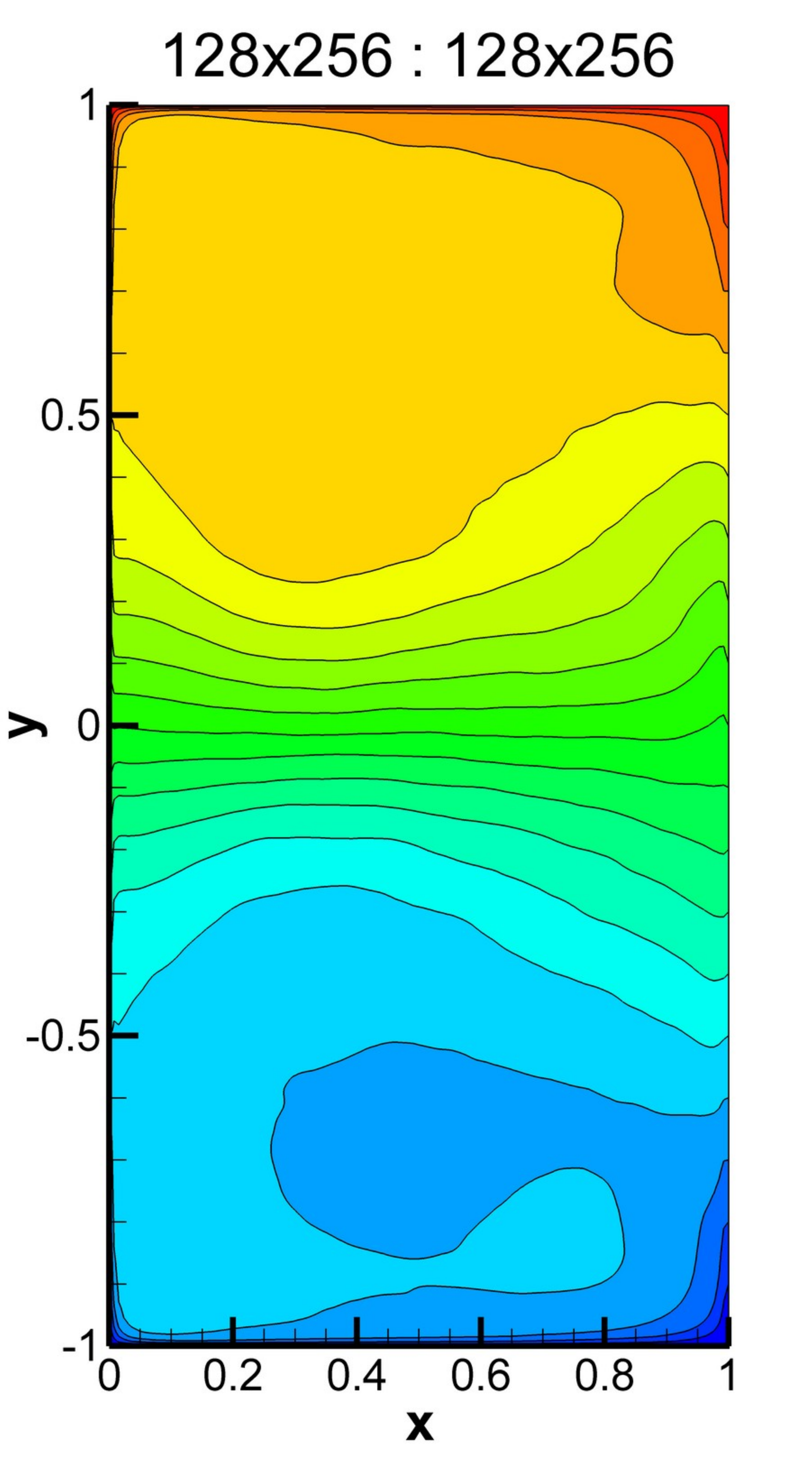}} }
\\
\mbox{
\subfigure{\includegraphics[width=0.22\textwidth]{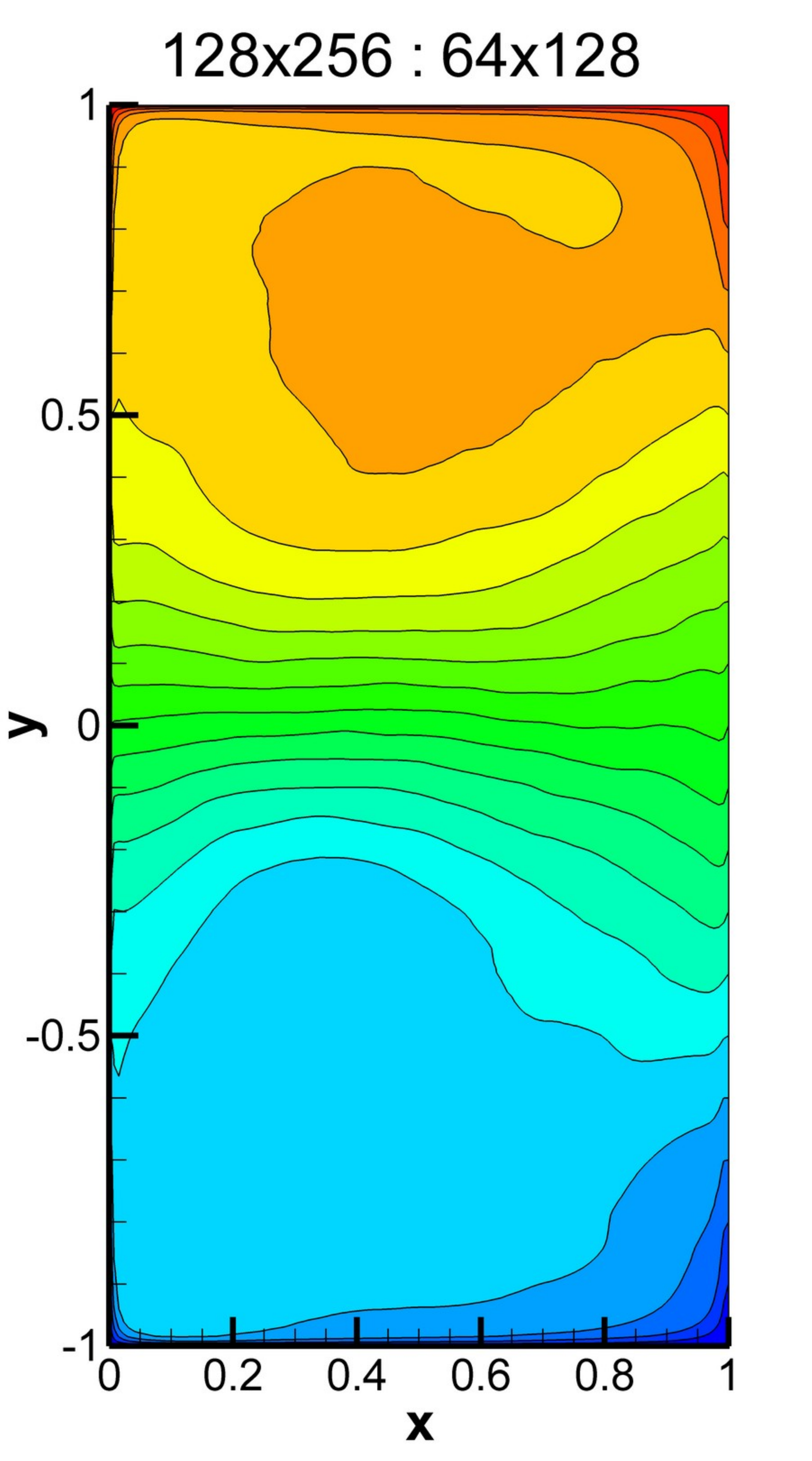}}
\subfigure{\includegraphics[width=0.22\textwidth]{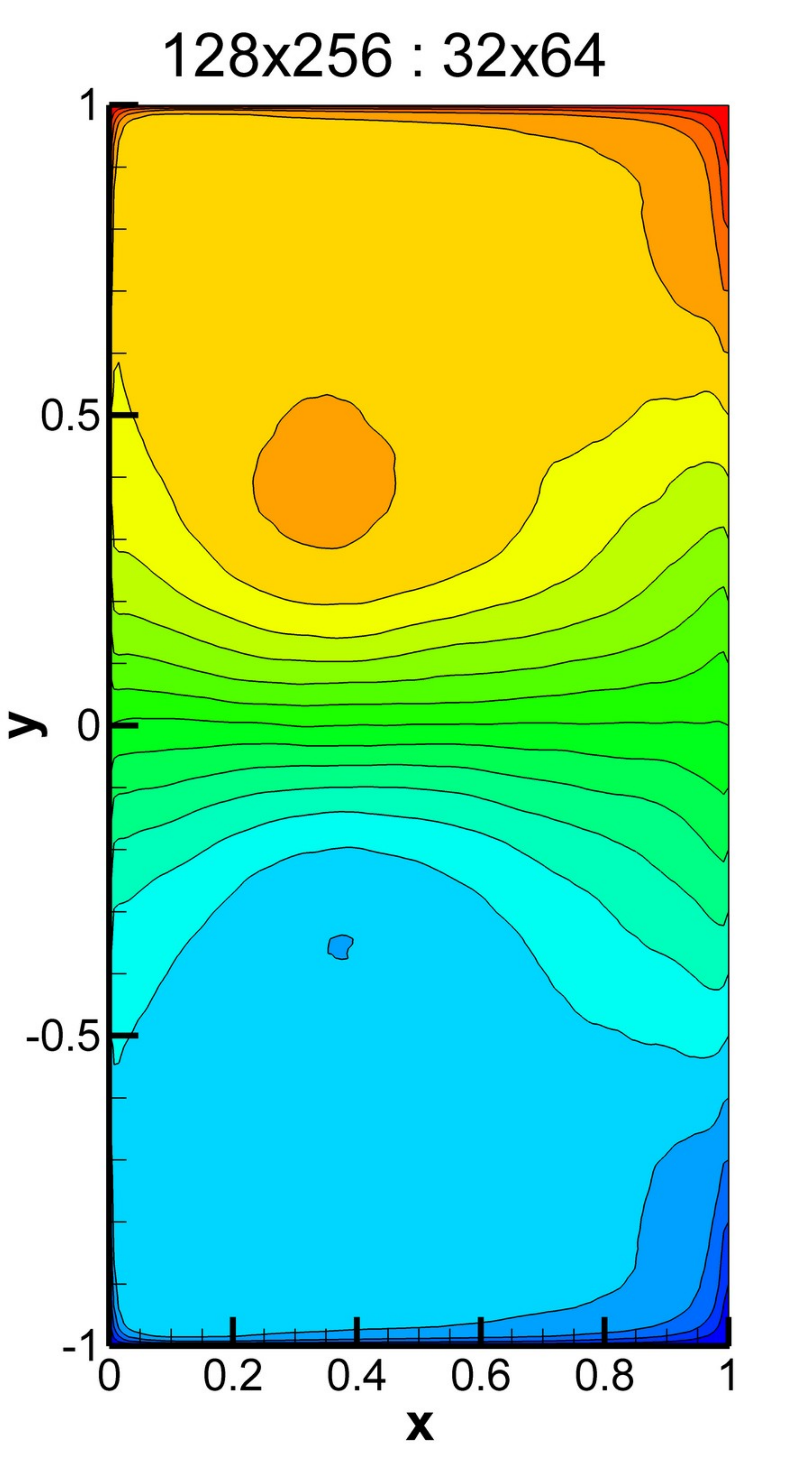}}
\subfigure{\includegraphics[width=0.22\textwidth]{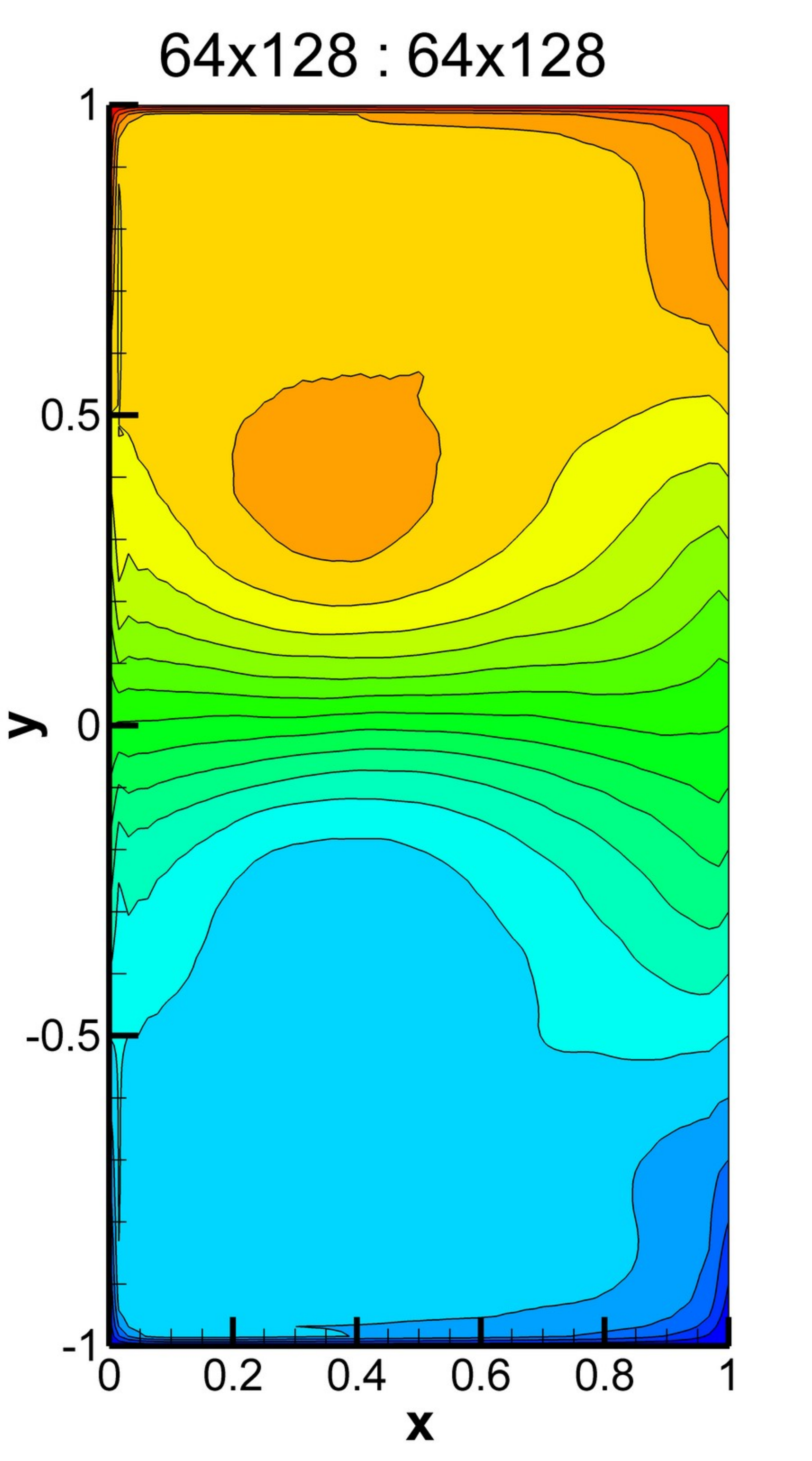}}
\subfigure{\includegraphics[width=0.22\textwidth]{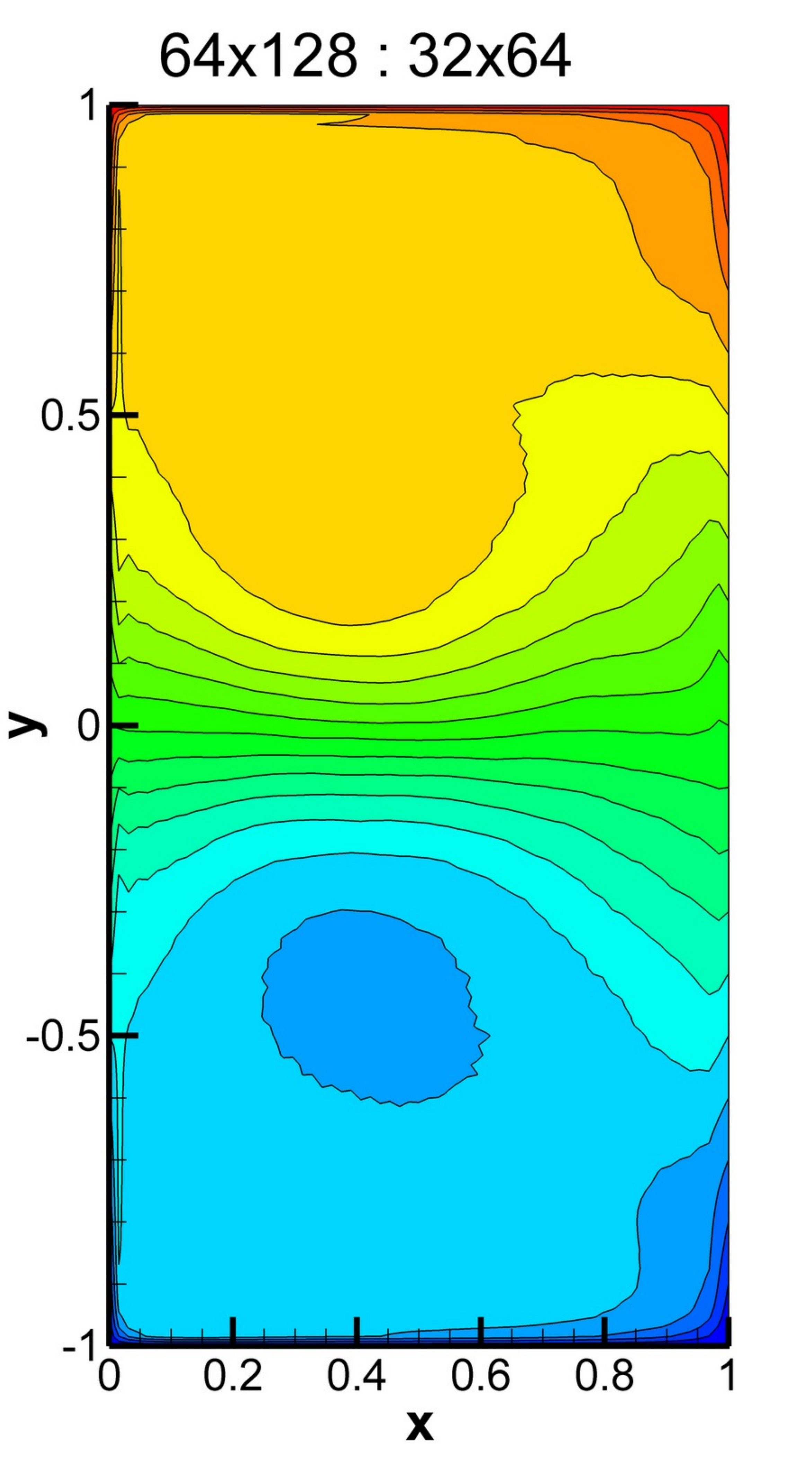}} }
\\
\mbox{
\subfigure{\includegraphics[width=0.22\textwidth]{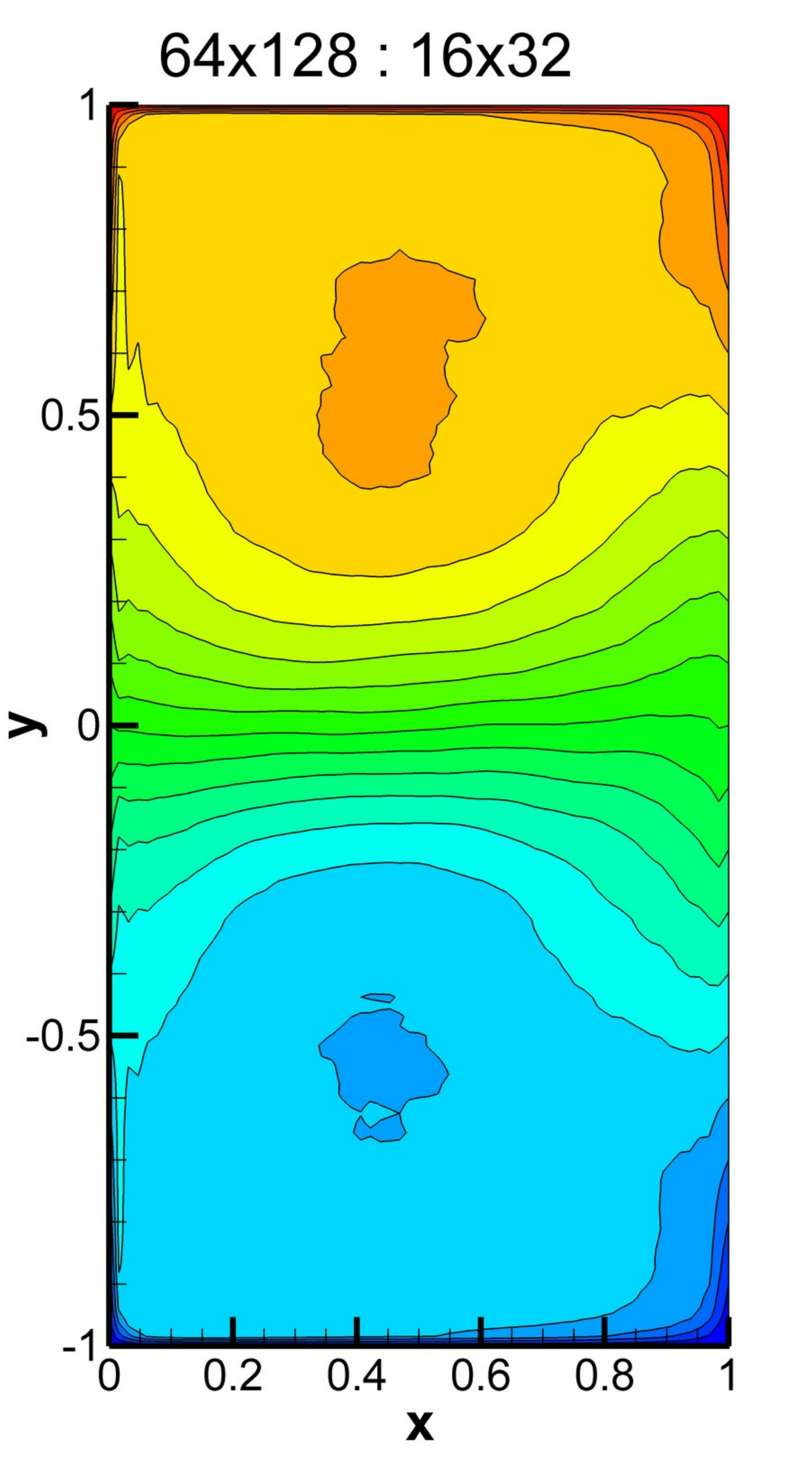}}
\subfigure{\includegraphics[width=0.22\textwidth]{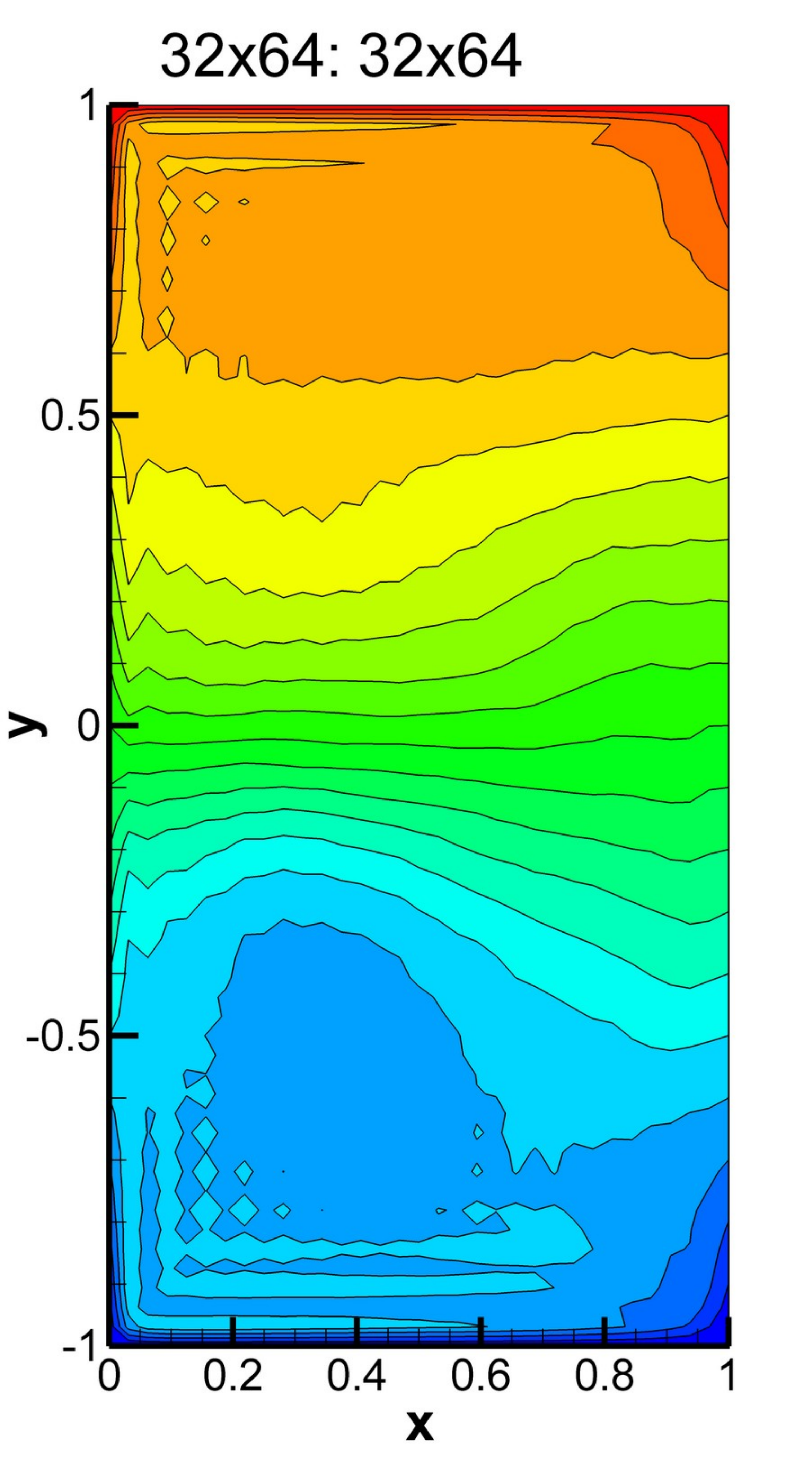}}
\subfigure{\includegraphics[width=0.22\textwidth]{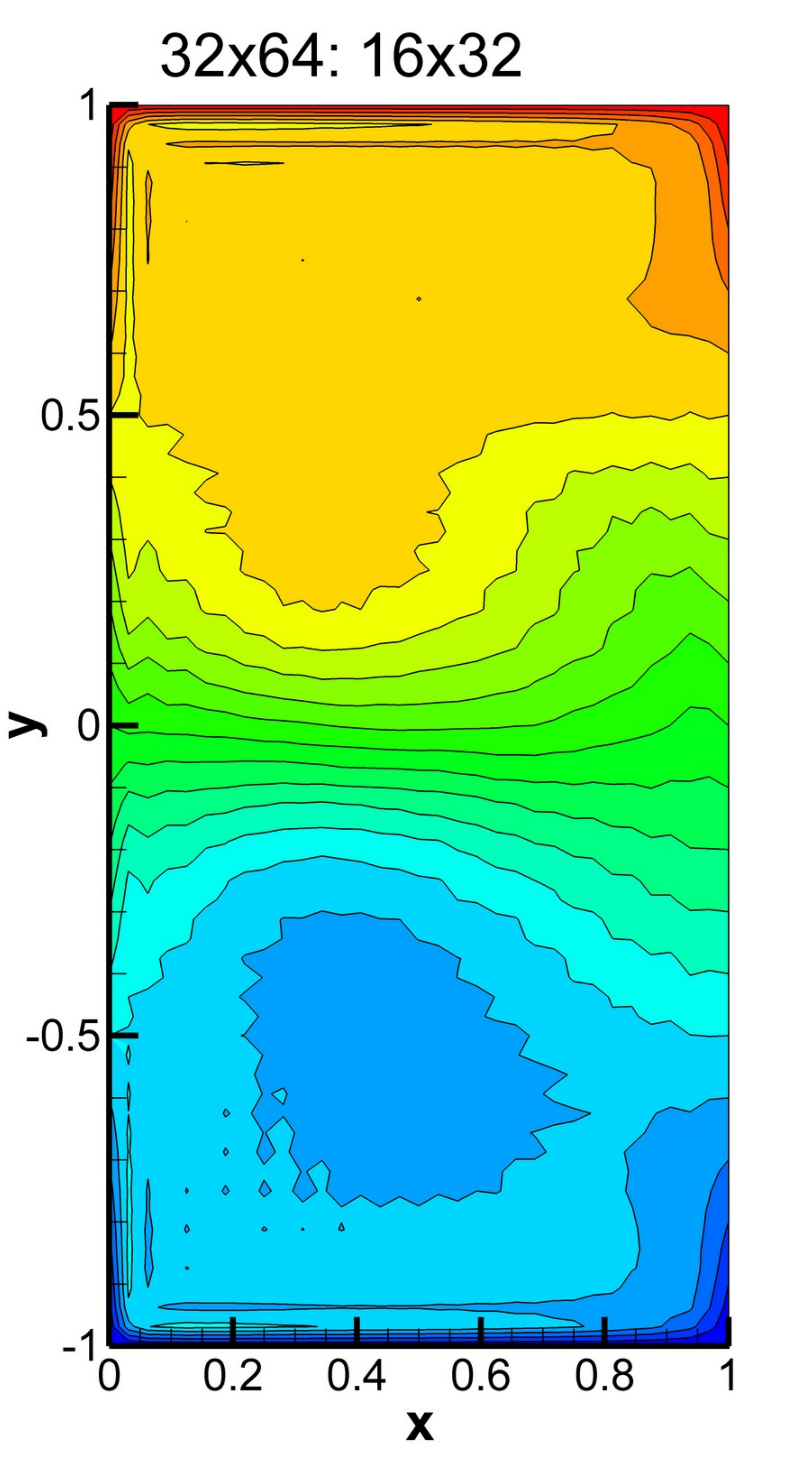}}
\subfigure{\includegraphics[width=0.22\textwidth]{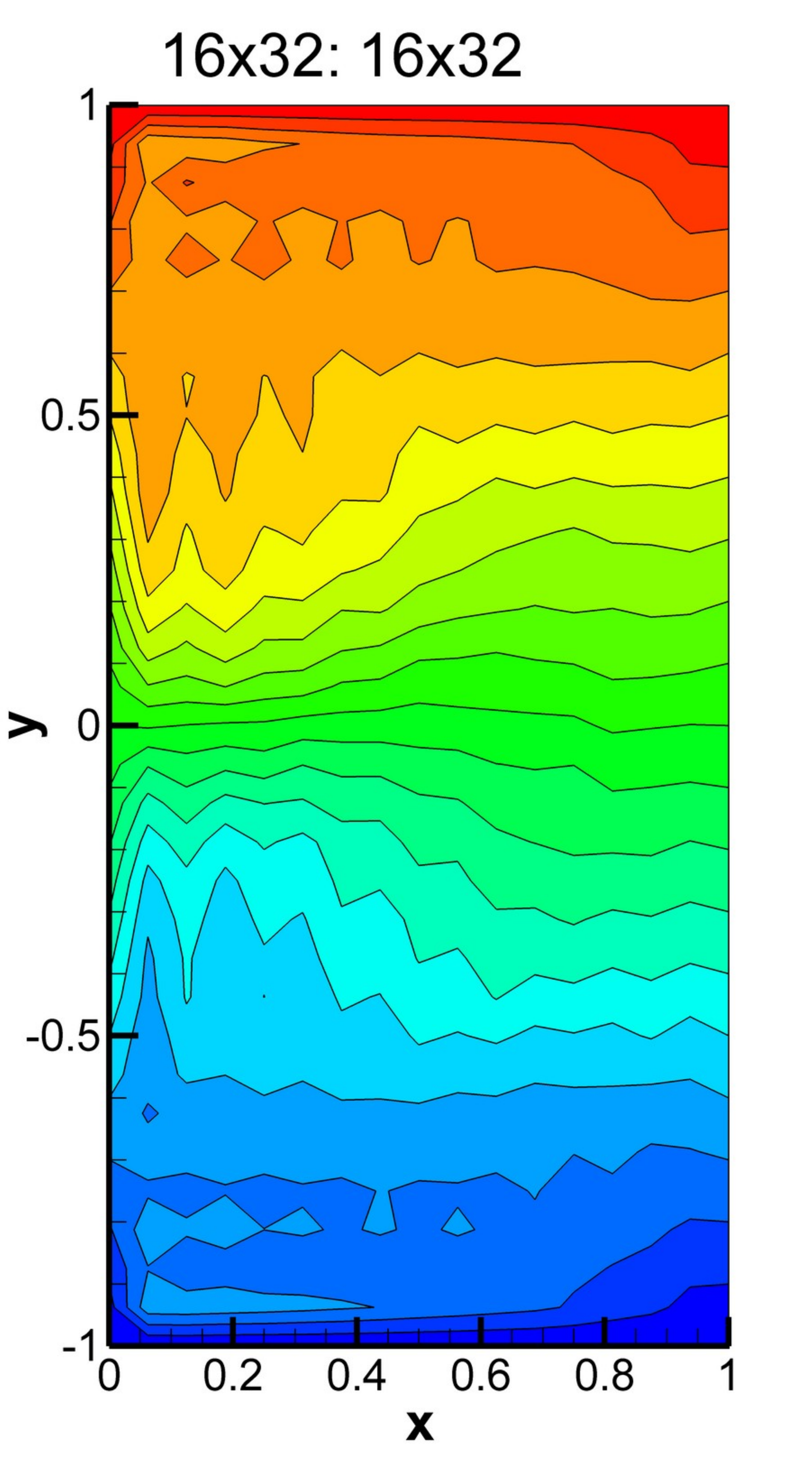}} }
\caption{Experiment II:
Comparison of mean potential vorticity for $Re=312.5$ and $Ro=0.0025$ (i.e., $\delta_M/L = 0.02$, and $\delta_I/L = 0.05$). Labels include the resolutions for both parts of the solver in the form $N_x \times N_y : M_x \times M_y$, where $N_x \times N_y$ is the resolution for the barotropic vorticity transport equation, and $M_x \times M_y$ is the resolution for the elliptic sub-problems.
}
\label{fig:qC}
\end{figure*}

\begin{figure*}
\centering
\mbox{
\subfigure{\includegraphics[width=0.25\textwidth]{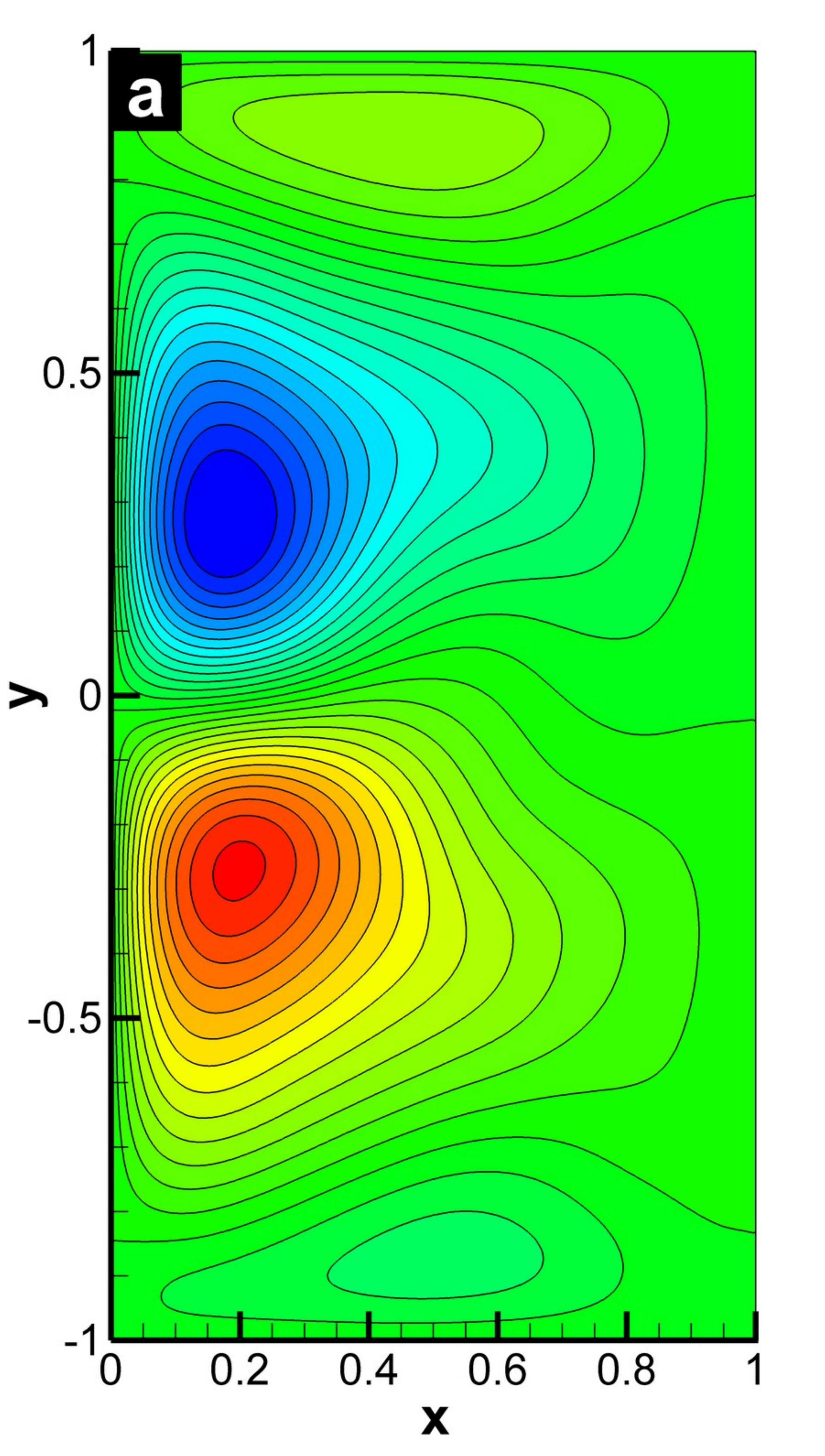}}
\subfigure{\includegraphics[width=0.25\textwidth]{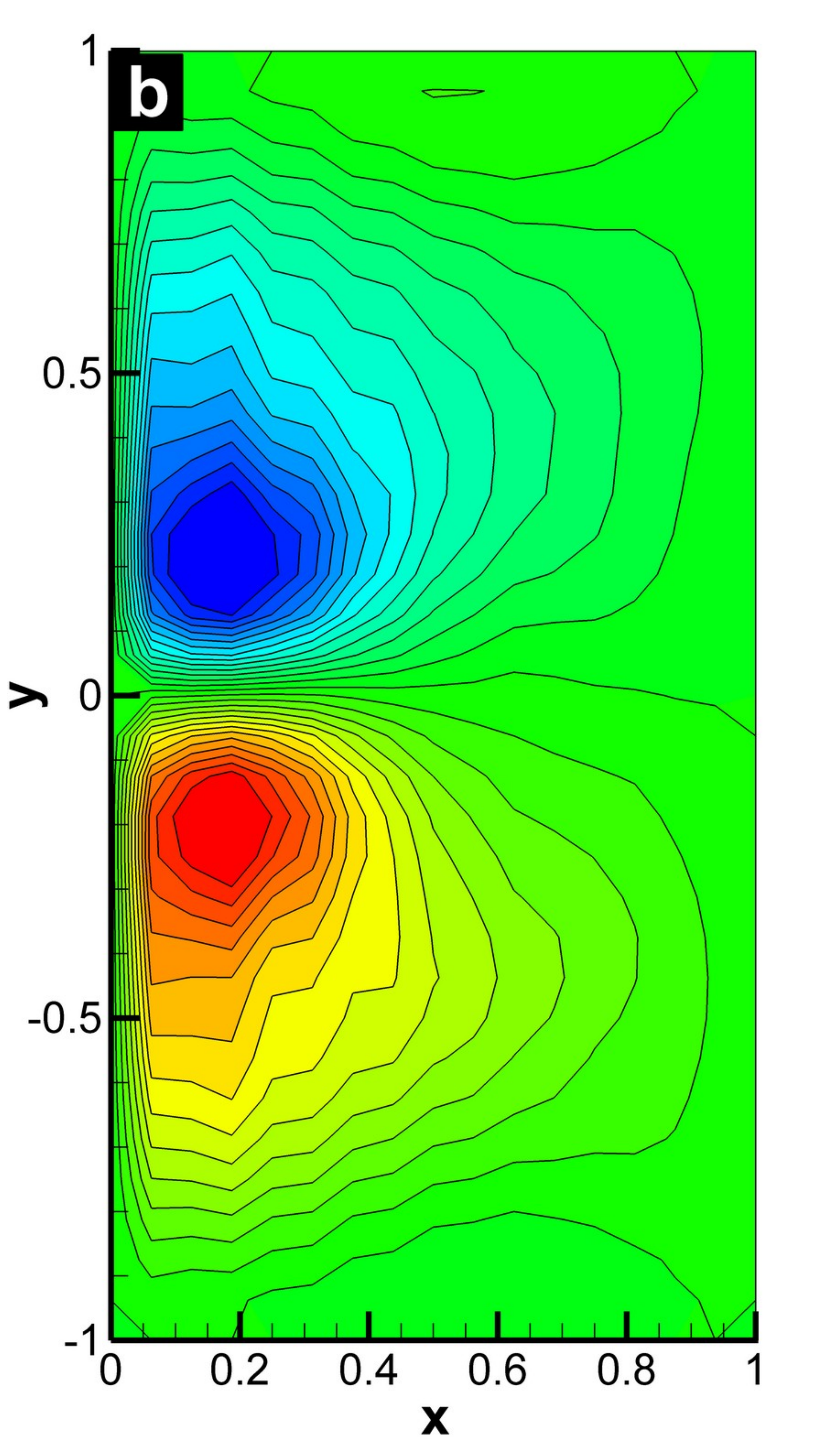}}
\subfigure{\includegraphics[width=0.25\textwidth]{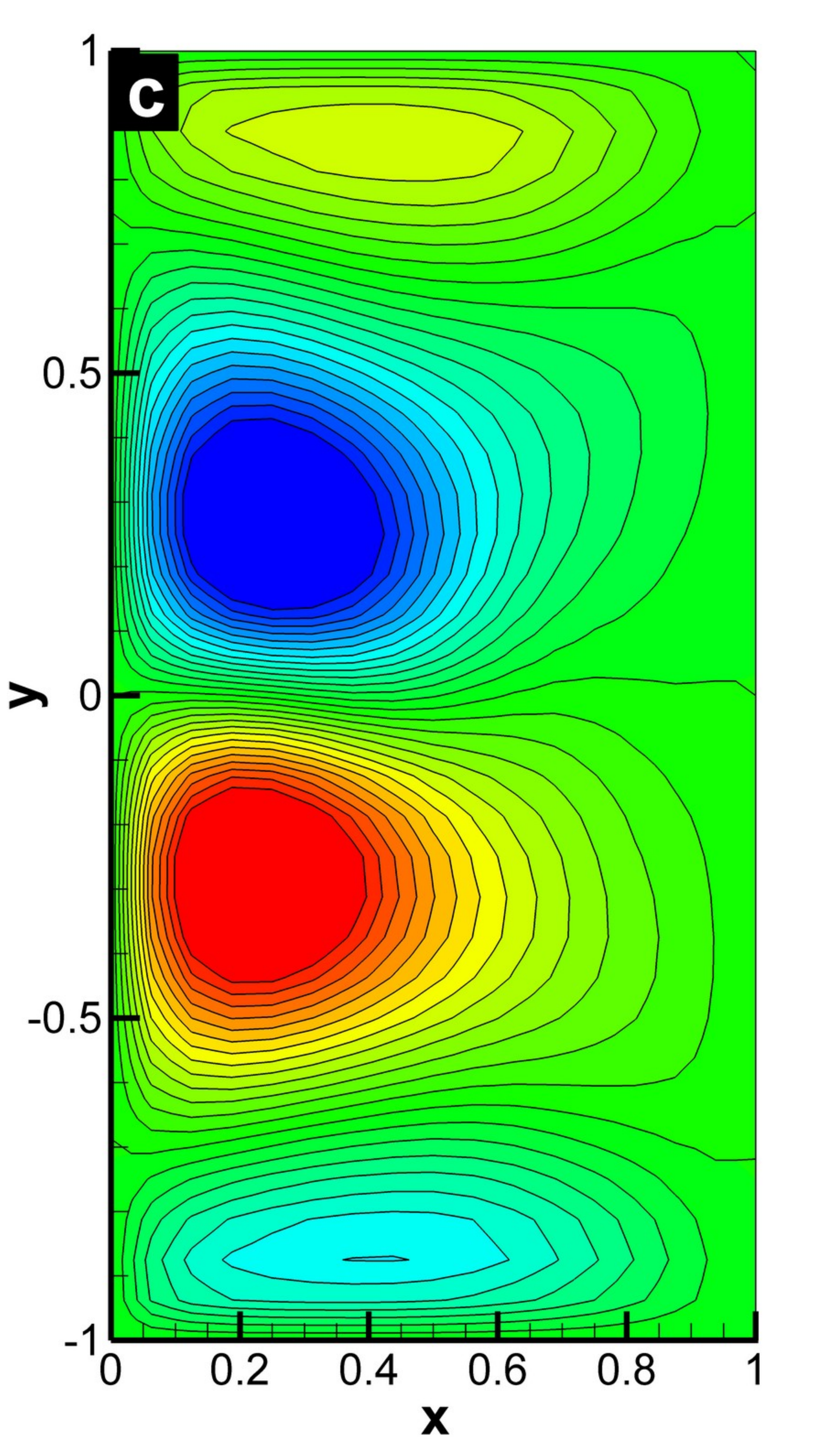}}
\subfigure{\includegraphics[width=0.25\textwidth]{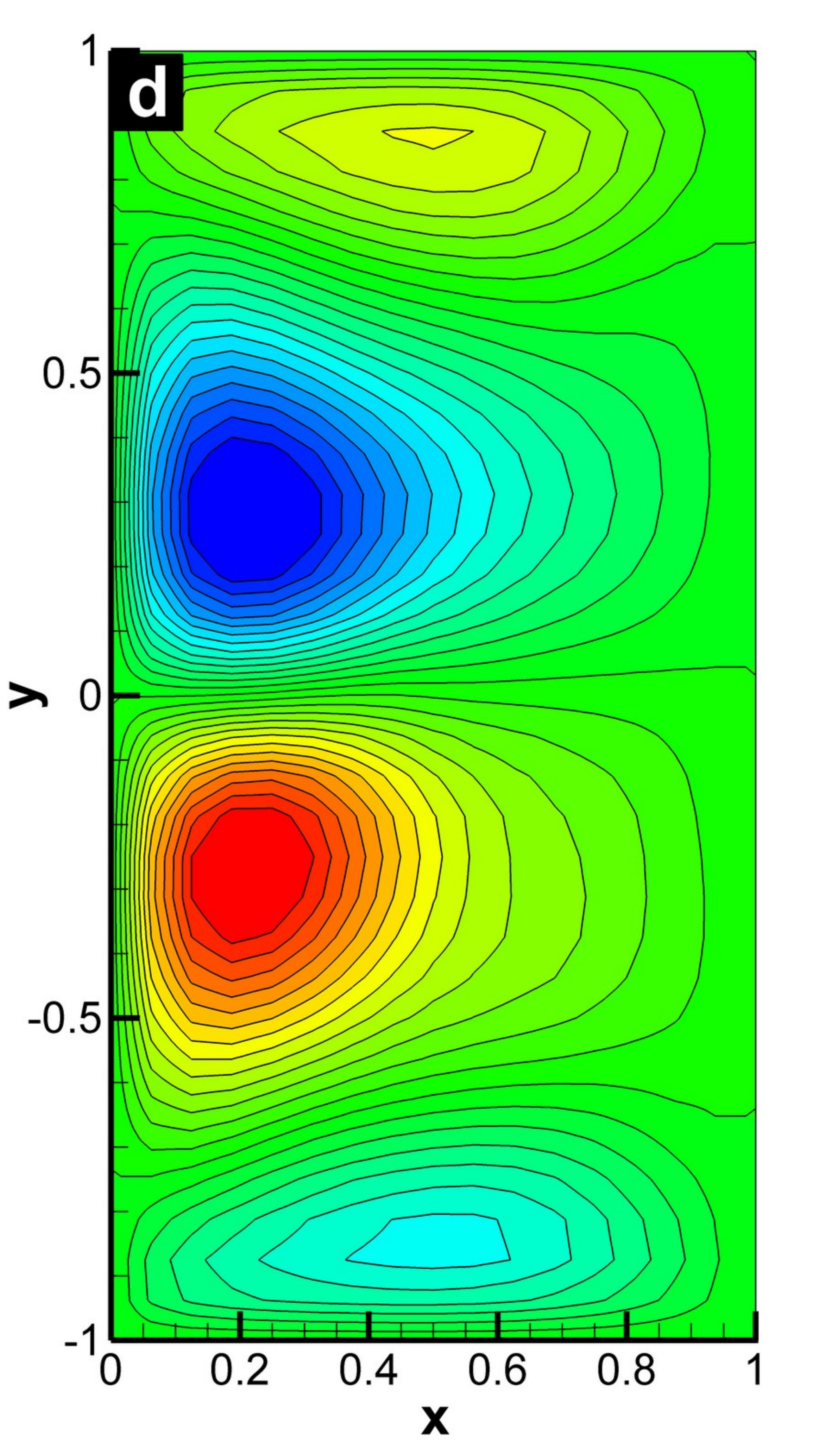}}
}
\caption{Experiment I:
Comparison of mean stream function for $Re=200$ and $Ro=0.0016$ (i.e., $\delta_M/L = 0.02$, and $\delta_I/L = 0.04$). (a) DNS result ($256 \times 512 : 256 \times 512$), (b) standard coarse simulation without CGP ($16 \times 32 : 16 \times 32$), (c) CGP with one-level coarsening ($32 \times 64 : 16 \times 32$), and (d) CGP with two-level coarsening ($64 \times 128 : 16 \times 32$). The contour interval layouts are identical in all cases. Cases (b)-(d) have the same resolutions for the elliptic part of the problem.
}
\label{fig:sB4}
\end{figure*}

\begin{figure*}
\centering
\mbox{
\subfigure{\includegraphics[width=0.25\textwidth]{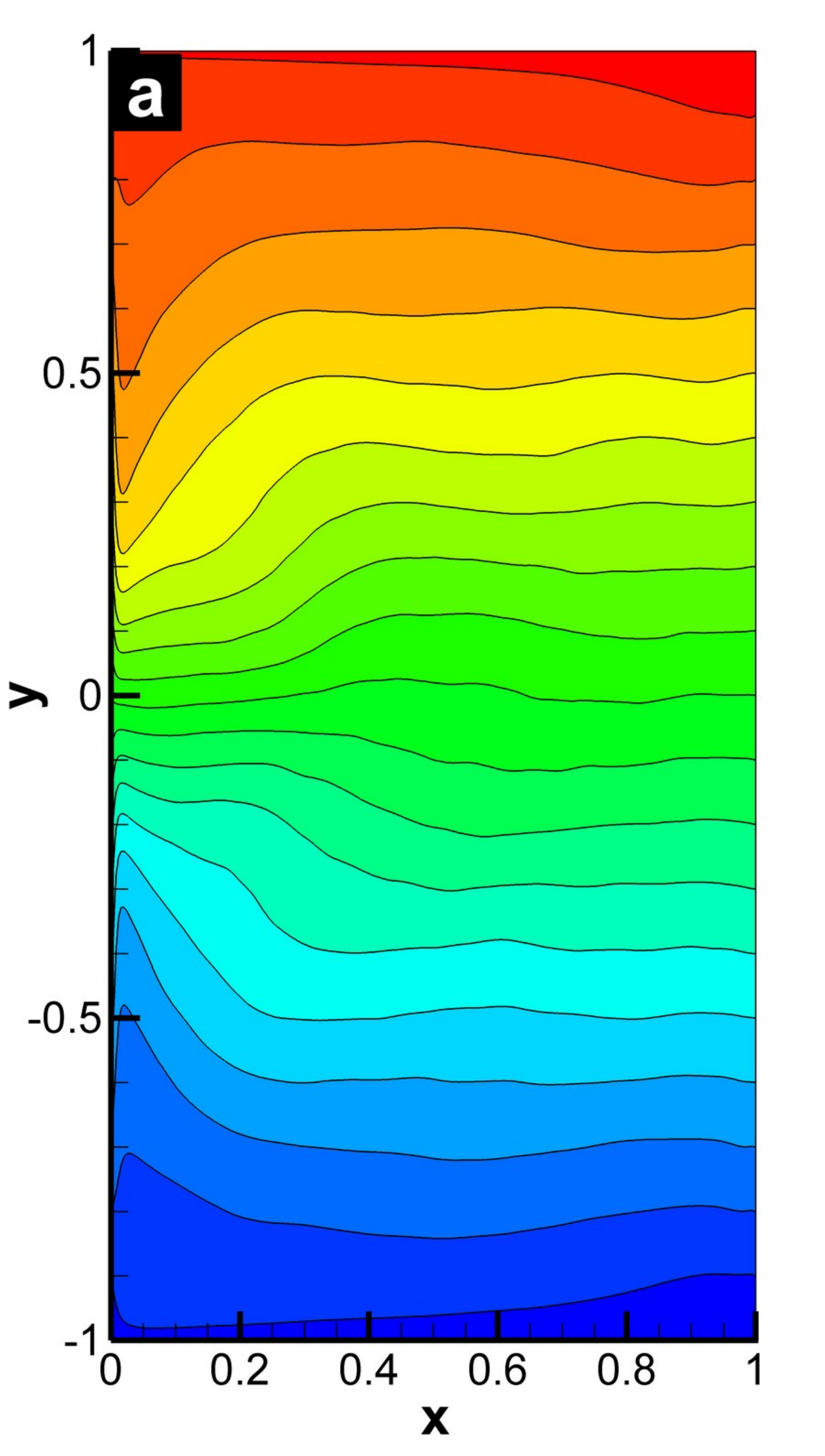}}
\subfigure{\includegraphics[width=0.25\textwidth]{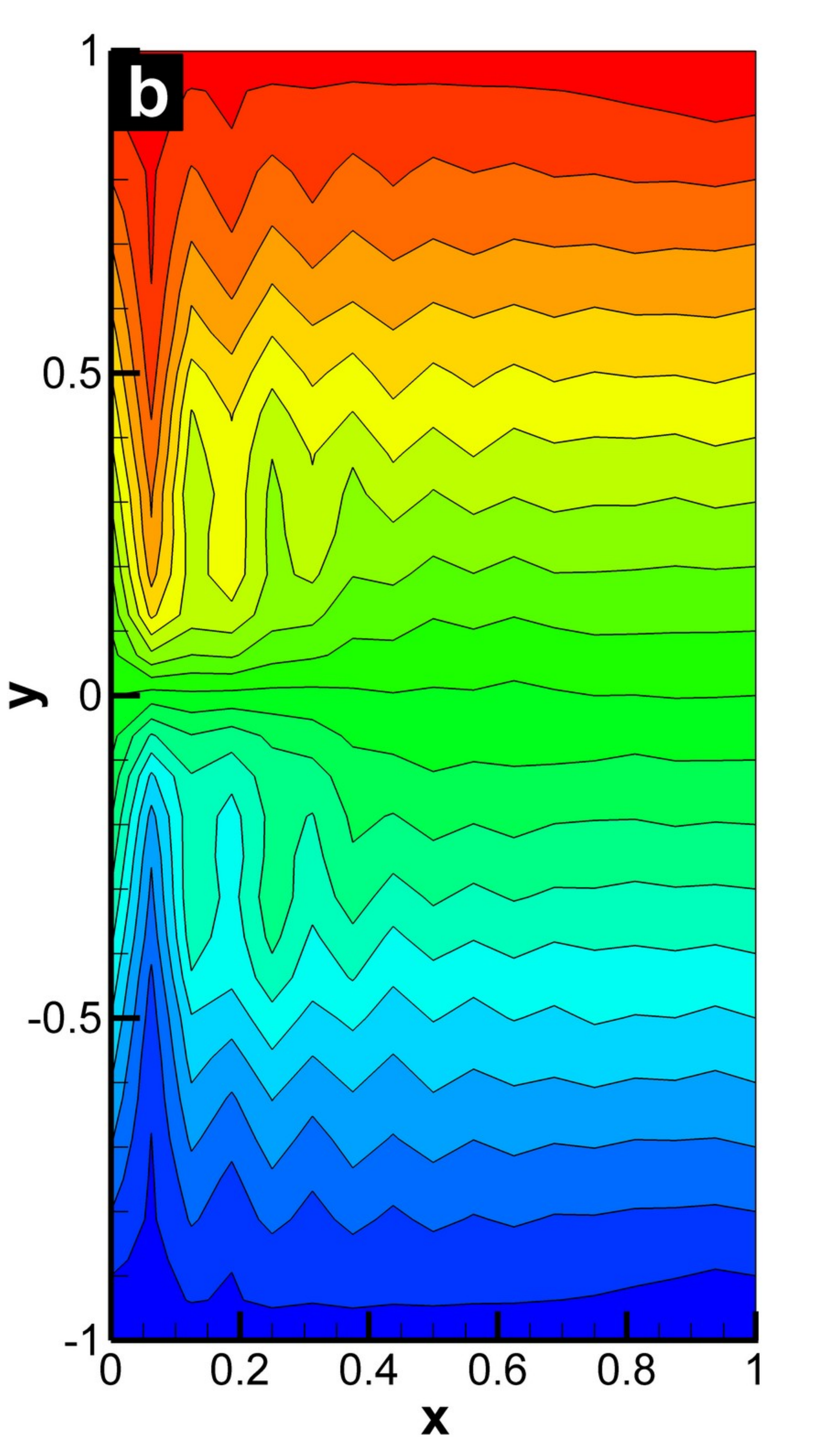}}
\subfigure{\includegraphics[width=0.25\textwidth]{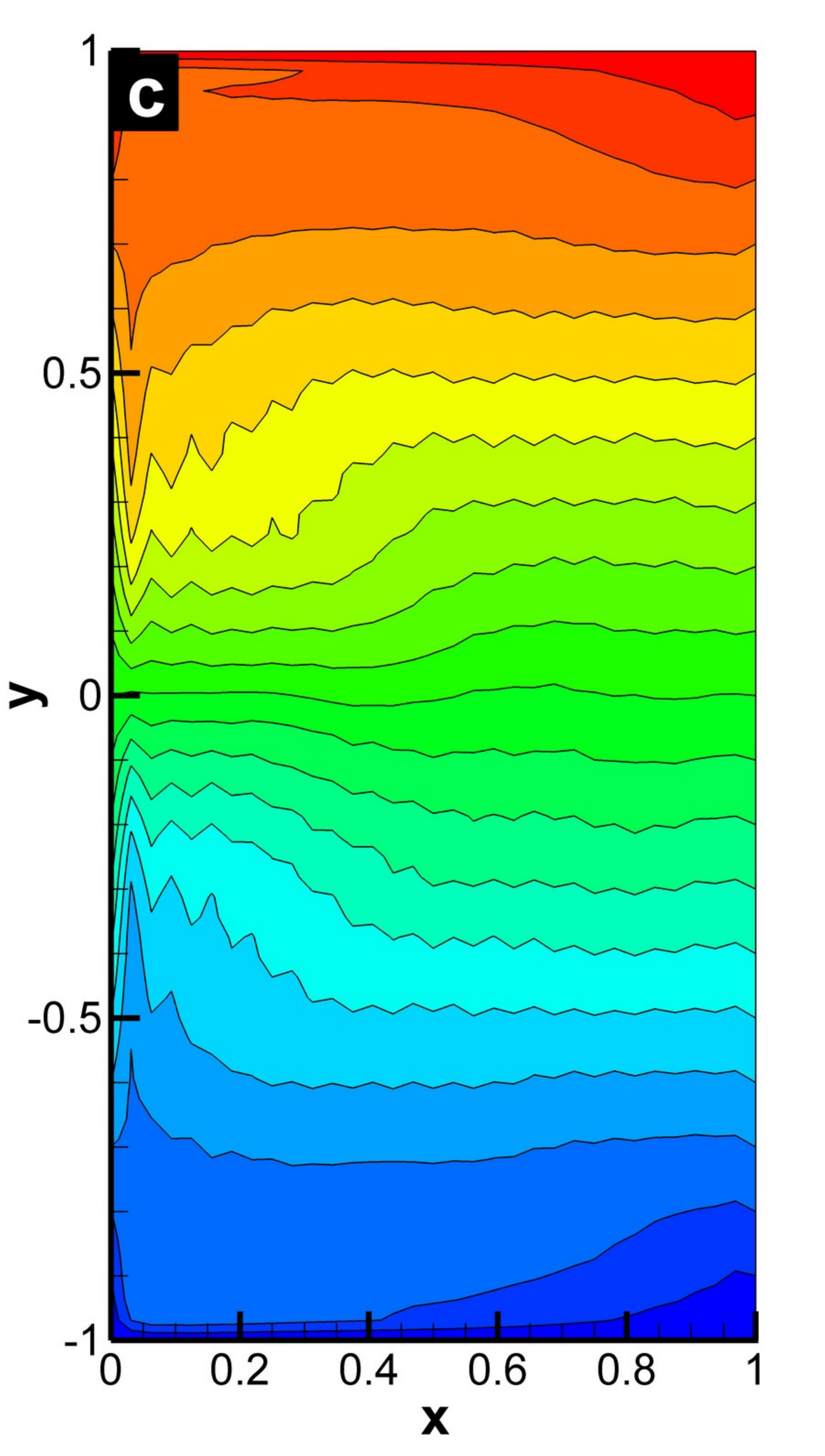}}
\subfigure{\includegraphics[width=0.25\textwidth]{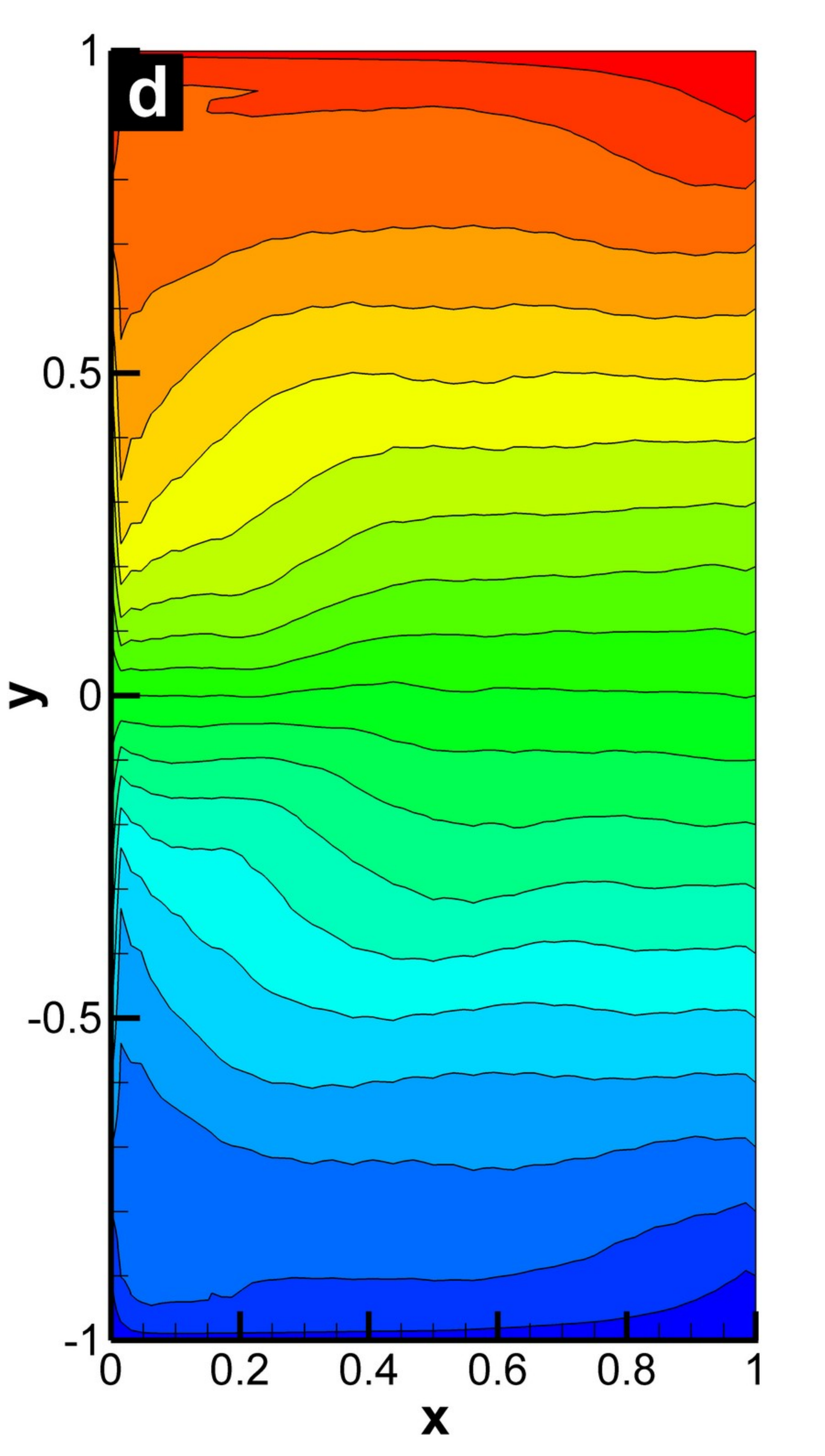}}
}
\caption{Experiment I:
Comparison of mean potential vorticity for $Re=200$ and $Ro=0.0016$ (i.e., $\delta_M/L = 0.02$, and $\delta_I/L = 0.04$). (a) DNS result ($256 \times 512 : 256 \times 512$), (b) standard coarse simulation without CGP ($16 \times 32 : 16 \times 32$), (c) CGP with one-level coarsening ($32 \times 64 : 16 \times 32$), and (d) CGP with two-level coarsening ($64 \times 128 : 16 \times 32$). The contour interval layouts are identical in all cases. Cases (b)-(d) have the same resolutions for the elliptic part of the problem.
}
\label{fig:qB4}
\end{figure*}

\begin{figure*}
\centering
\mbox{
\subfigure{\includegraphics[width=0.25\textwidth]{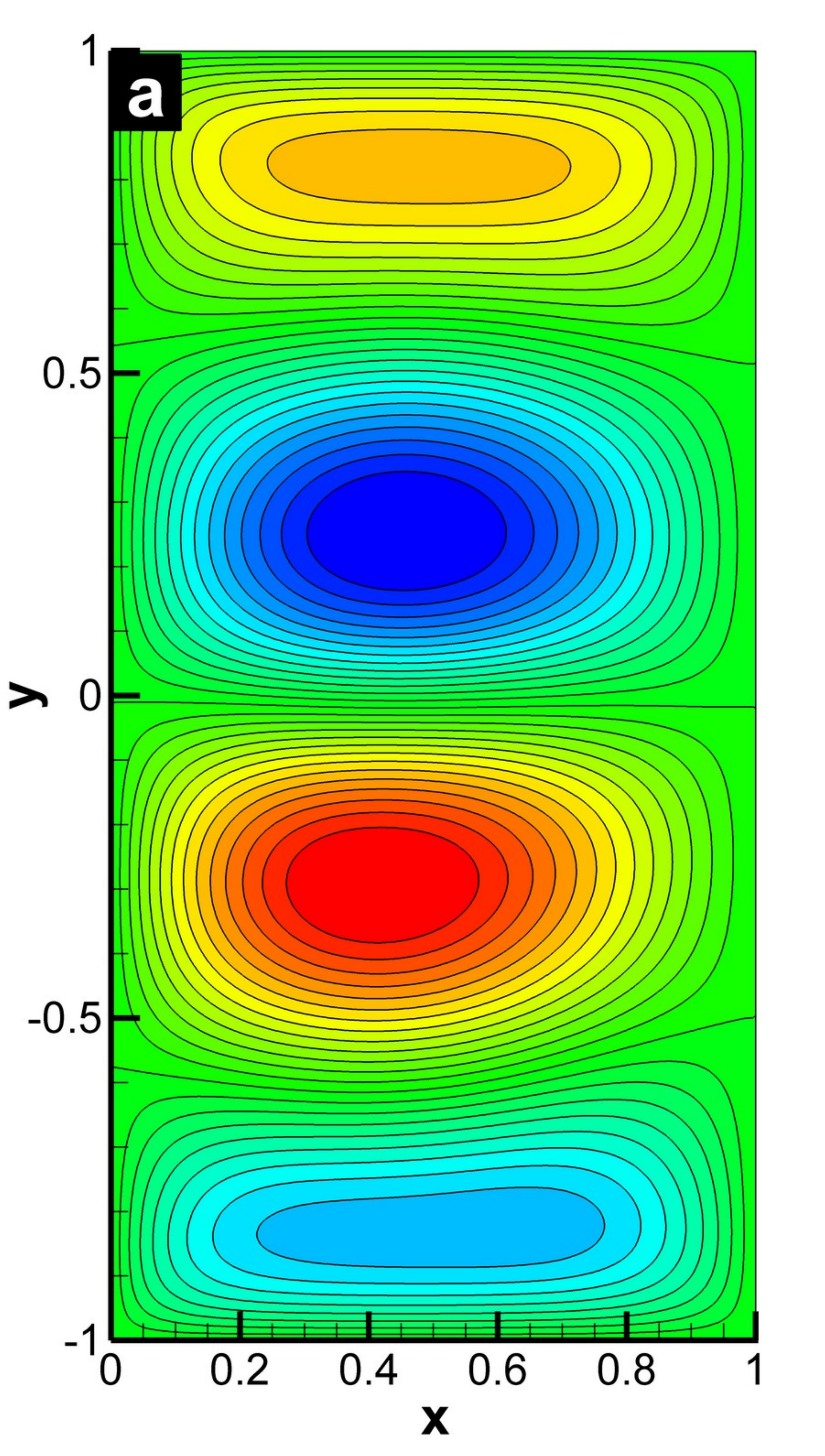}}
\subfigure{\includegraphics[width=0.25\textwidth]{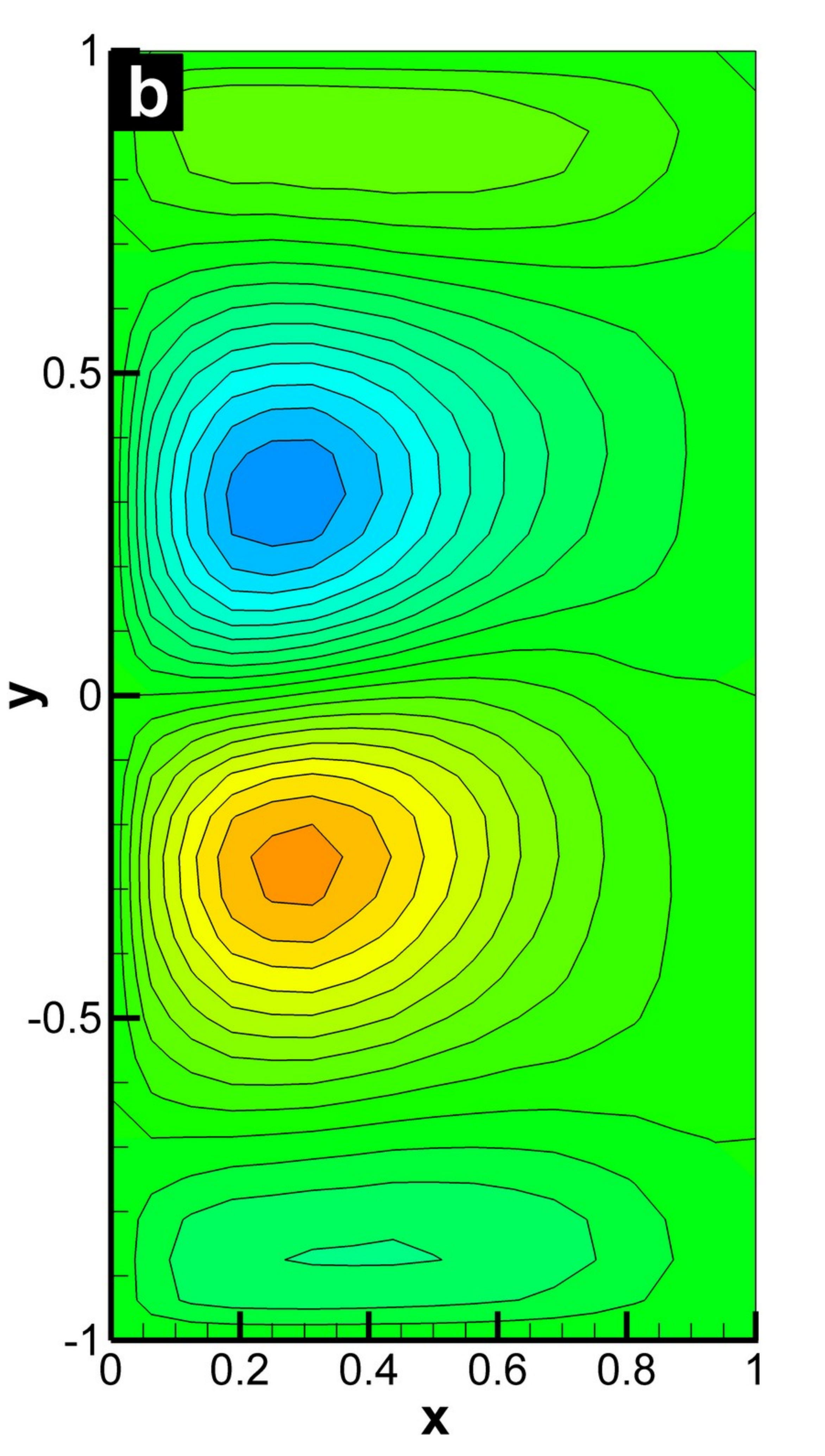}}
\subfigure{\includegraphics[width=0.25\textwidth]{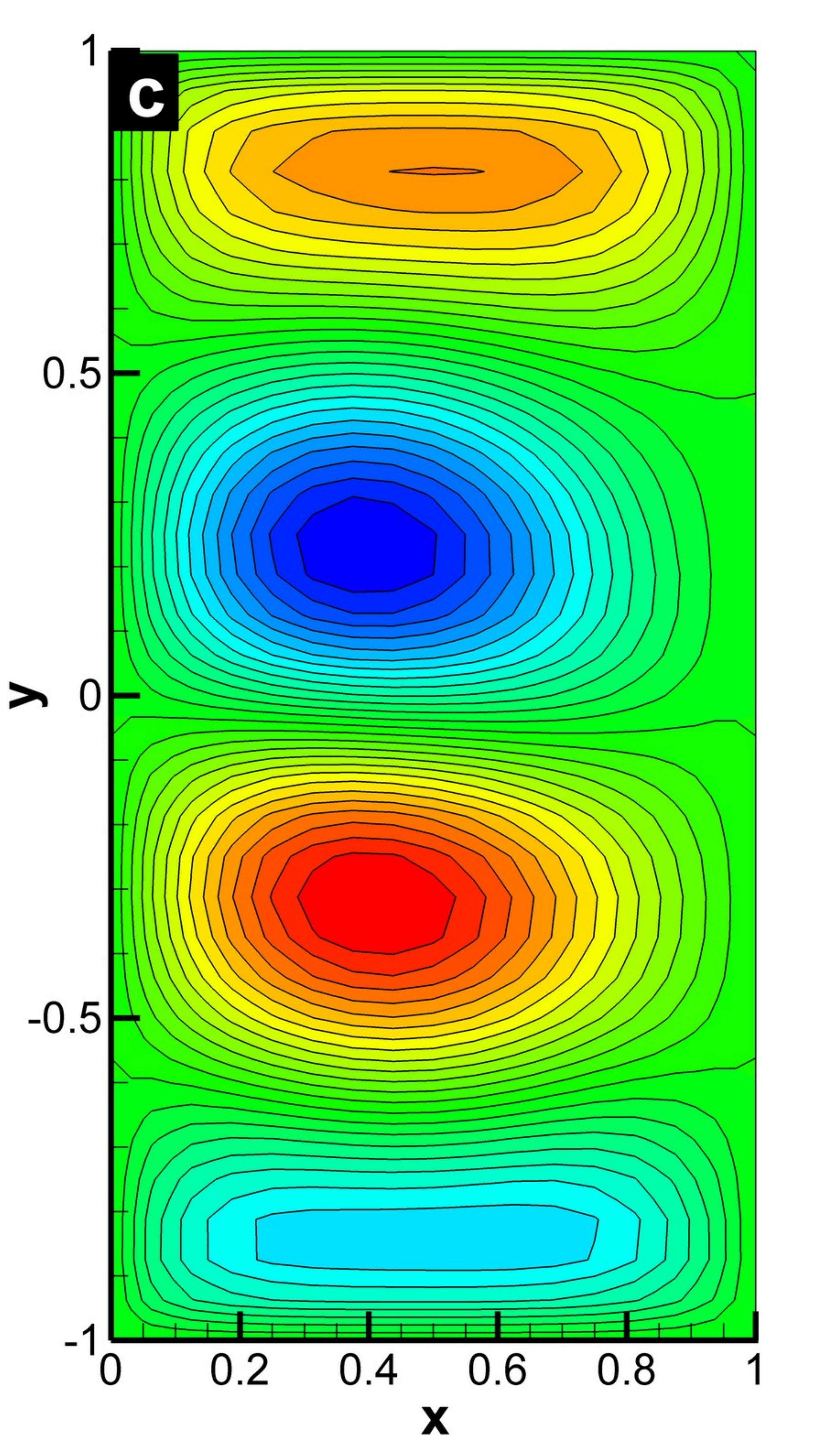}}
\subfigure{\includegraphics[width=0.25\textwidth]{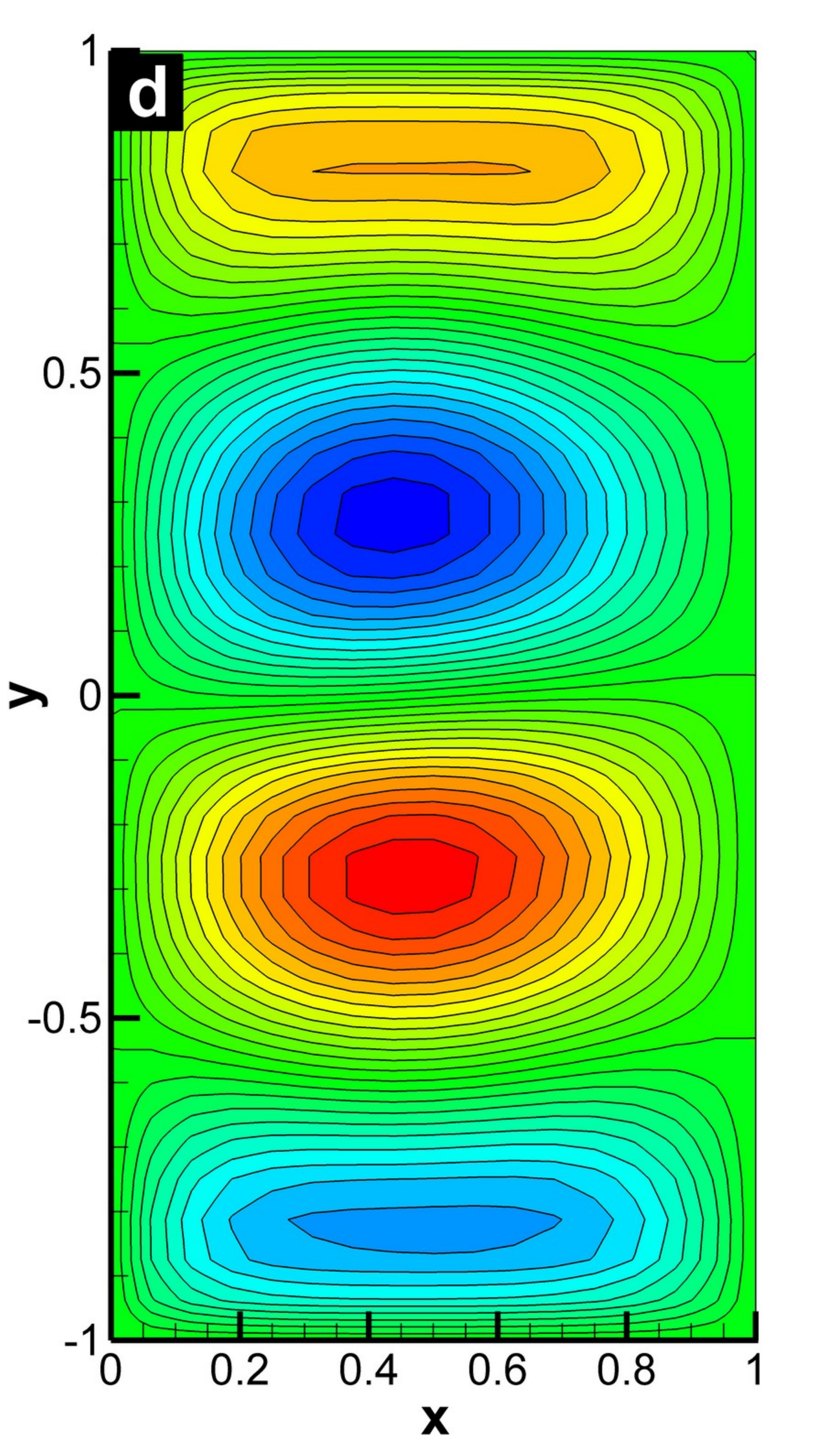}}
}
\caption{Experiment II:
Comparison of mean stream function for $Re=312.5$ and $Ro=0.0025$ (i.e., $\delta_M/L = 0.02$, and $\delta_I/L = 0.05$). (a) DNS result ($256 \times 512 : 256 \times 512$), (b) standard coarse simulation without CGP ($16 \times 32 : 16 \times 32$), (c) CGP with one-level coarsening ($32 \times 64 : 16 \times 32$), and (d) CGP with two-level coarsening ($64 \times 128 : 16 \times 32$). The contour interval layouts are identical in all cases. Cases (b)-(d) have the same resolutions for the elliptic part of the problem.
}
\label{fig:sC4}
\end{figure*}

\begin{figure*}
\centering
\mbox{
\subfigure{\includegraphics[width=0.25\textwidth]{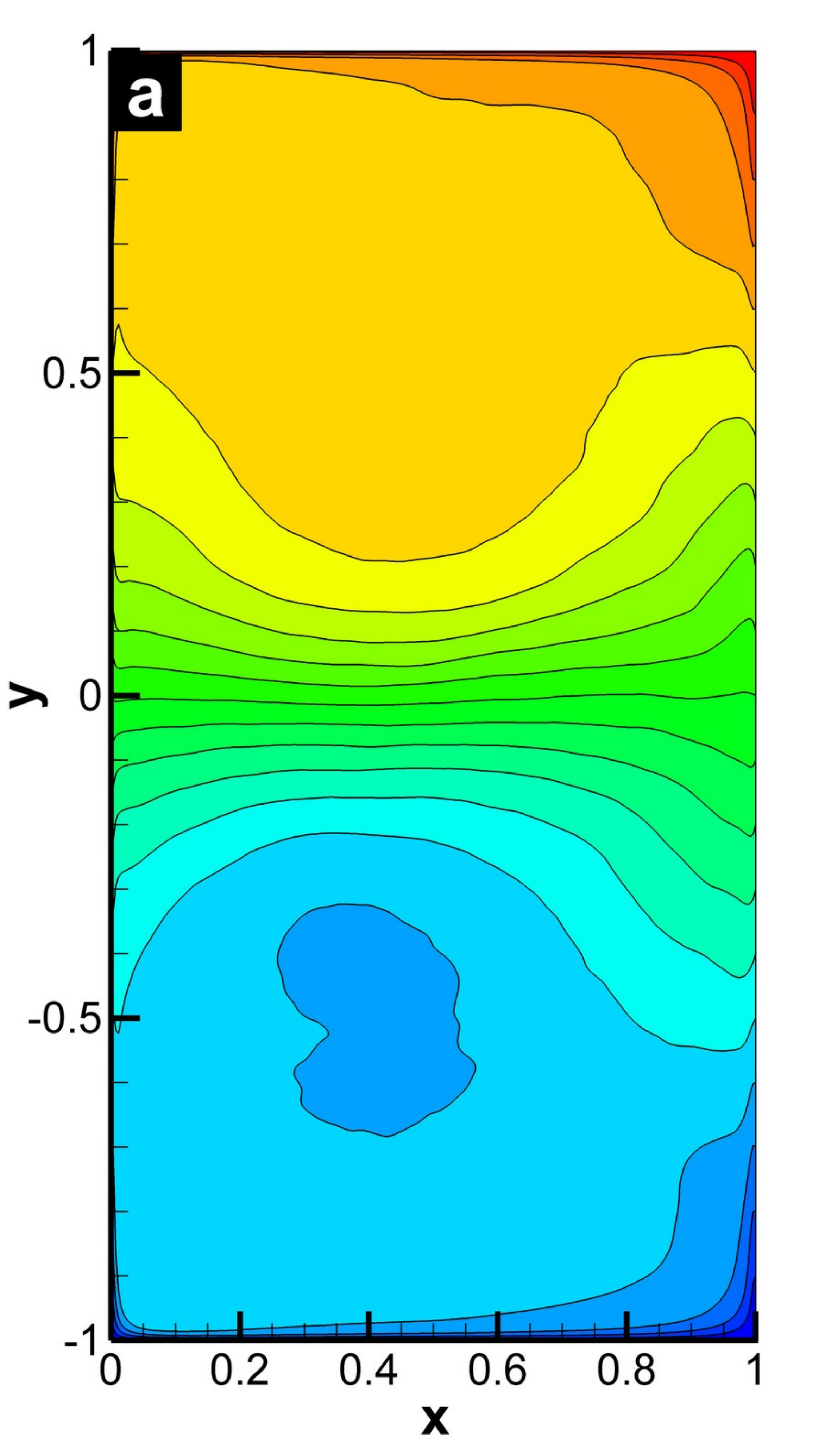}}
\subfigure{\includegraphics[width=0.25\textwidth]{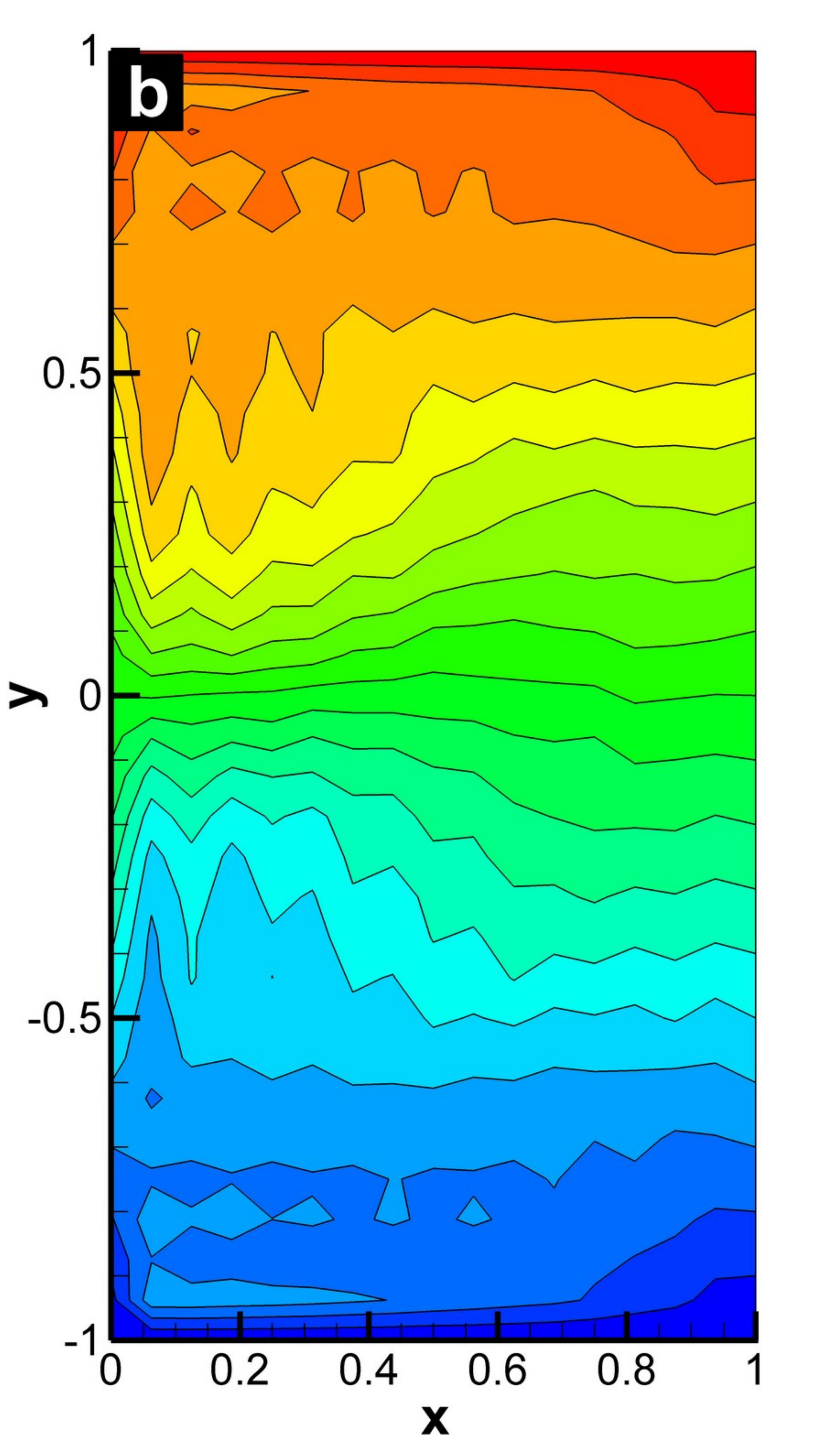}}
\subfigure{\includegraphics[width=0.25\textwidth]{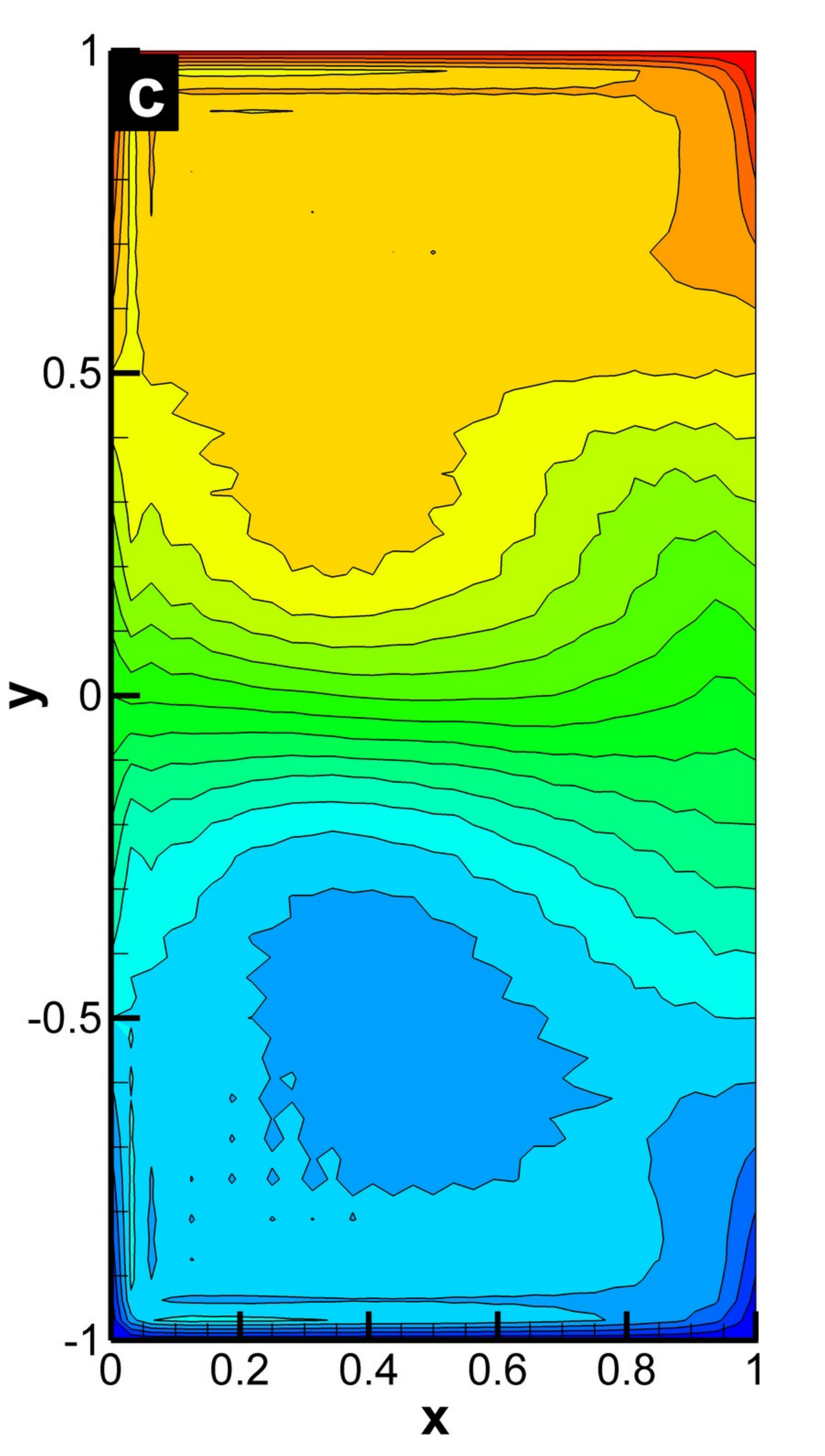}}
\subfigure{\includegraphics[width=0.25\textwidth]{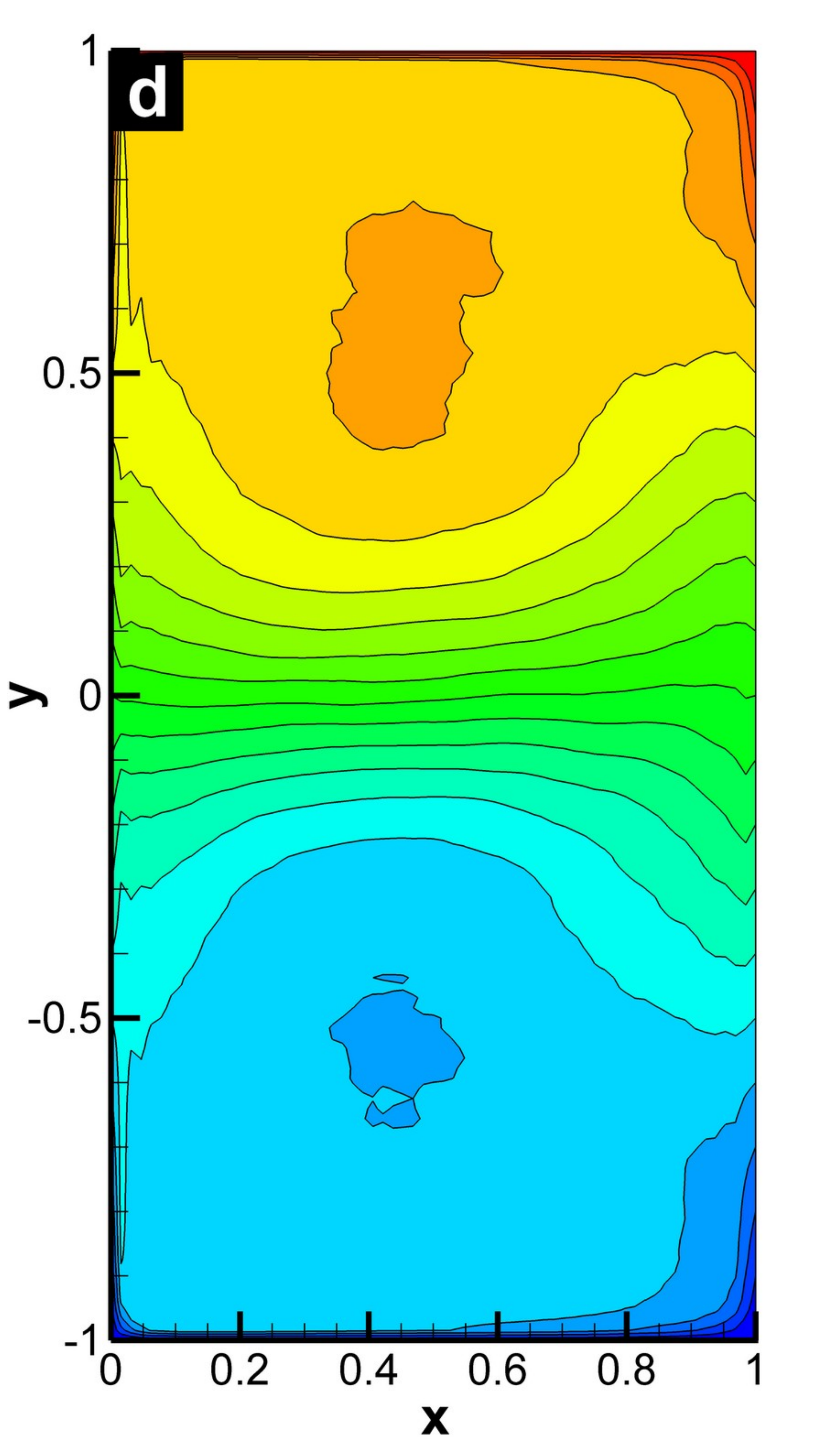}}
}
\caption{Experiment II:
Comparison of mean potential vorticity for $Re=312.5$ and $Ro=0.0025$ (i.e., $\delta_M/L = 0.02$, and $\delta_I/L = 0.05$). (a) DNS result ($256 \times 512 : 256 \times 512$), (b) standard coarse simulation without CGP ($16 \times 32 : 16 \times 32$), (c) CGP with one-level coarsening ($32 \times 64 : 16 \times 32$), and (d) CGP with two-level coarsening ($64 \times 128 : 16 \times 32$). The contour interval layouts are identical in all cases. Cases (b)-(d) have the same resolutions for the elliptic part of the problem.
}
\label{fig:qC4}
\end{figure*}

We start by performing a DNS computation on a fine mesh with a spatial resolution of $256\times512$. As shown in Fig.~(\ref{fig:hist-1}), for both experiments, after a transient  period, a statistically steady state solution is obtained at a time of around $t=5$. However, the behavior of the system in statistically steady state is quite different due to regime transition. In the classification of \cite{berloff1999large}, Experiment I lie in the chaotic regime, whereas in the Experiment II the flow regime showed a quasi-periodic variability. Instantaneous contour plots at the final time $t=50$ for the potential vorticity are shown in Fig.~(\ref{fig:inst-1}) for both experiments. In all the one-layer experiments presented in this study, the time average of the data is taken between time $t=10$ and $t=50$ using $20$ thousands snapshots in order to quantify the statistically steady state. The DNS results are included as a reference value in the following analysis when we present the results with the CGP method.

Three different levels of coarsening: half-coarsening ($\ell=1$, $M=N/2$), $1/4$-coarsening ($\ell=2$, $M=N/4$) and $1/8$-coarsening ($\ell=3$, $M=N/8$), are performed to investigate the behavior of the CGP method. Mean stream function and potential vorticity field contours obtained using the CGP method are plotted in Fig.~(\ref{fig:sB}) and Fig.~(\ref{fig:qB}) along with the regular fine and coarse computations. In these figures, the labels show the resolutions in the form of $N_x\times N_y : M_x \times M_y$, where $N_x \times N_y$ is the resolution for the time dependant potential vorticity equation, and $M_x \times M_y$ is the resolution for the elliptic sub-problem. The computed results with one level of coarsening are very close to those from the fine scale computations. Furthermore, results with two and three levels of coarsening also agree well with the fine scale computations. These results demonstrate that high fidelity numerical simulations can be obtained using the CGP method. For example, if we consider two numerical experiments with $32\times 64 : 32 \times 64$ and $32\times 64 : 16 \times 32$ resolutions, they have almost the same resulting flow field, but the latter results (with a half-coarsened grid for the elliptic sub-problem) were obtained almost 3 times faster than the former results (without CGP), and more importantly, the resulting field obtained using the CGP method with the $32\times 64 : 16 \times 32$ resolution are better than those obtained on the $16\times 32 : 16 \times 32$ grid without the CGP method.

Similar analysis is performed for the physical setting for Experiment II which shows a quasi-periodic flow regime. Comparisons of mean stream function and potential vorticity fields are plotted in Fig.~(\ref{fig:sC}) and Fig.~(\ref{fig:qC}), respectively. Similar to previous analysis, it is clear from these figures that the results using the CGP method agree well with the results of the fine scale computations using the standard method with a considerable reduction in computational cost. We emphasize that, for both parameter sets, {\em four gyres} are clearly visible in the stream function plots. It is more clear in Experiment II due to higher variability in quasi statistically steady state. These results demonstrate that the CGP methodology can provide an accelerated method for solving large scale QG models for ocean circulation problem.

In order to elucidate the fidelity of the CGP method for coarser resolutions, for Experiment I, we plot the time-averaged stream function and potential vorticity contours in Fig.~(\ref{fig:sB4}) and Fig.~(\ref{fig:qB4}), respectively. We include the high-resolution DNS results as well. In these figures, all results except the DNS are obtained with a constant resolution of $16\times 32$ for the elliptic sub-problem. One and two level coarsening methods are compared with non-coarsening standard procedure. We note that the new CGP models with both one-level and two-level coarsening yield improved results by smoothing out the numerical
oscillations present in the results obtained by under-resolved standard computations without CGP method. This improvement is more clearly displayed in the potential vorticity contour plot in Fig.~(\ref{fig:qB4}). Similar observations can be drawn from Fig.~(\ref{fig:sC4}) and Fig.~(\ref{fig:qC4}) for Experiment II. The CGP model yields results that are significantly better than those corresponding to the under-resolved BVE simulations having the same resolution for the elliptic sub-problem. In computational point of view, the price of CGP simulations are close to the price of the coarse grid simulation (i.e., the price of mapping operators and explicit time integration procedure is smaller than the price of solving elliptic sub-problem), while giving the same results as the high-resolution computations. In these computations we use a linear-cost fast Poisson solver, which is indeed one of the fastest elliptic solvers, and the CGP method yields a 3-8 fold reduction in computational cost. As discussed earlier, we highlight that the speed ups would be greater if we used a quadratic-cost sub-optimal elliptic solver.

\subsection{CGP experiments for two-layer QG model}
\label{sec:results2}

The main goal of this section is to test the CGP method in the numerical simulation of the two-layer QG model, which is a standard prototype representing many characteristics of more realistic ocean dynamics including stratification and baroclinic effects. To evaluate the performance of CGP method, we utilize two different parameter sets, corresponding to two physical oceanic settings: (i) Experiment 1 represents a moderate ocean basin with the physical parameters used by \cite{özgökmen1998emergence}, (ii) Experiment 2 represents a large ocean basin with the physical parameters used by \cite{tanaka2010alternating}. In terms of the classification given by \cite{berloff1999large}, both sets of experiments lie under the chaotic regime. The physical parameters and corresponding dimensionless parameters are summarized in Table~\ref{tab:sets-2}.

\begin{table}[!t]
\centering
\caption{Physical parameter sets used in the numerical experiments. }
\label{tab:sets-2}       
\begin{tabular}{lll}
\hline\noalign{\smallskip}
Variable (unit) & Experiment 1 & Experiment 2 \\
\noalign{\smallskip}\hline\noalign{\smallskip}
  $L$ ($km$)  & 2000  & 5000 \\
  $H_1$ ($km$)  & 1.0  & 0.6   \\
  $H_2$ ($km$) & 4.0   & 3.4      \\
  $f_0$ ($s^{-1}$) & $9.35\times10^{-5}$ & $9.35\times10^{-5}$  \\
  $\beta$ ($m^{-1}s^{-1}$) & $1.75\times10^{-11}$ & $1.75\times10^{-11}$  \\
  $\rho_1$ ($kgm^{-3}$)   & 1030 & 1030      \\
  $g'$ ($ms^{-2}$)    & 0.02    & 0.02  \\
  $\tau_0$ ($Nm^{-2}$)    & 0.1    & 0.1  \\
  $\gamma$ ($s^{-1}$)   & $5\times10^{-8}$ & $4\times10^{-7}$     \\
  $\nu$ ($m^{2}s^{-1}$)   & 50 & 100      \\
  $\delta_{M}$ ($km$)    & 14.19 & 17.88     \\
  $\delta_{S}$ ($km$)     & 2.86 & 22.86   \\
  $\delta_{I}$ ($km$)    & 31.56 & 25.77    \\
  $R_{d}$ ($km$)   & 42.79 & 31.16    \\
  $V$ ($ms^{-1}$)    & 0.0174 & 0.0116   \\
  $L/V$ ($year$)   & 3.64  & 13.64   \\
  Ro                    & $2.49\times10^{-4}$  & $2.66\times10^{-5}$   \\
  Fr                   & 0.087   & 0.073   \\
  $\sigma$              & $1.43\times10^{-3}$  & $4.57\times10^{-3}$  \\
  $A$          & $3.57\times10^{-7}$ & $4.57\times10^{-8}$      \\
  $\delta$          & 0.2    & 0.15  \\
  Re            & 697.16  & 580.97  \\
\noalign{\smallskip}\hline
\end{tabular}
\end{table}


\begin{figure*}[!t]
\centering
\mbox{
\subfigure[Experiment 1]{\includegraphics[width=0.5\textwidth]{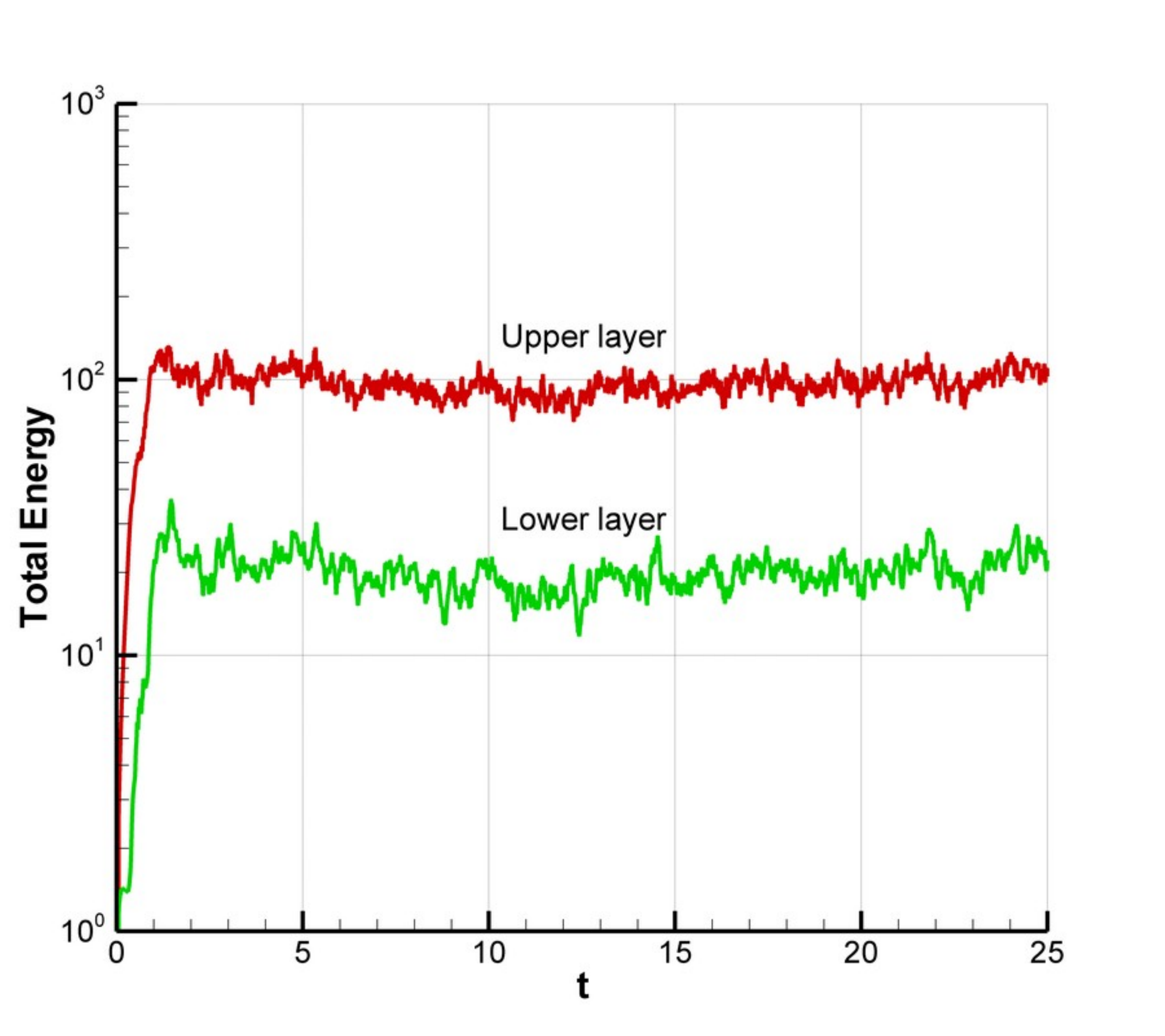}}
\subfigure[Experiment 2]{\includegraphics[width=0.5\textwidth]{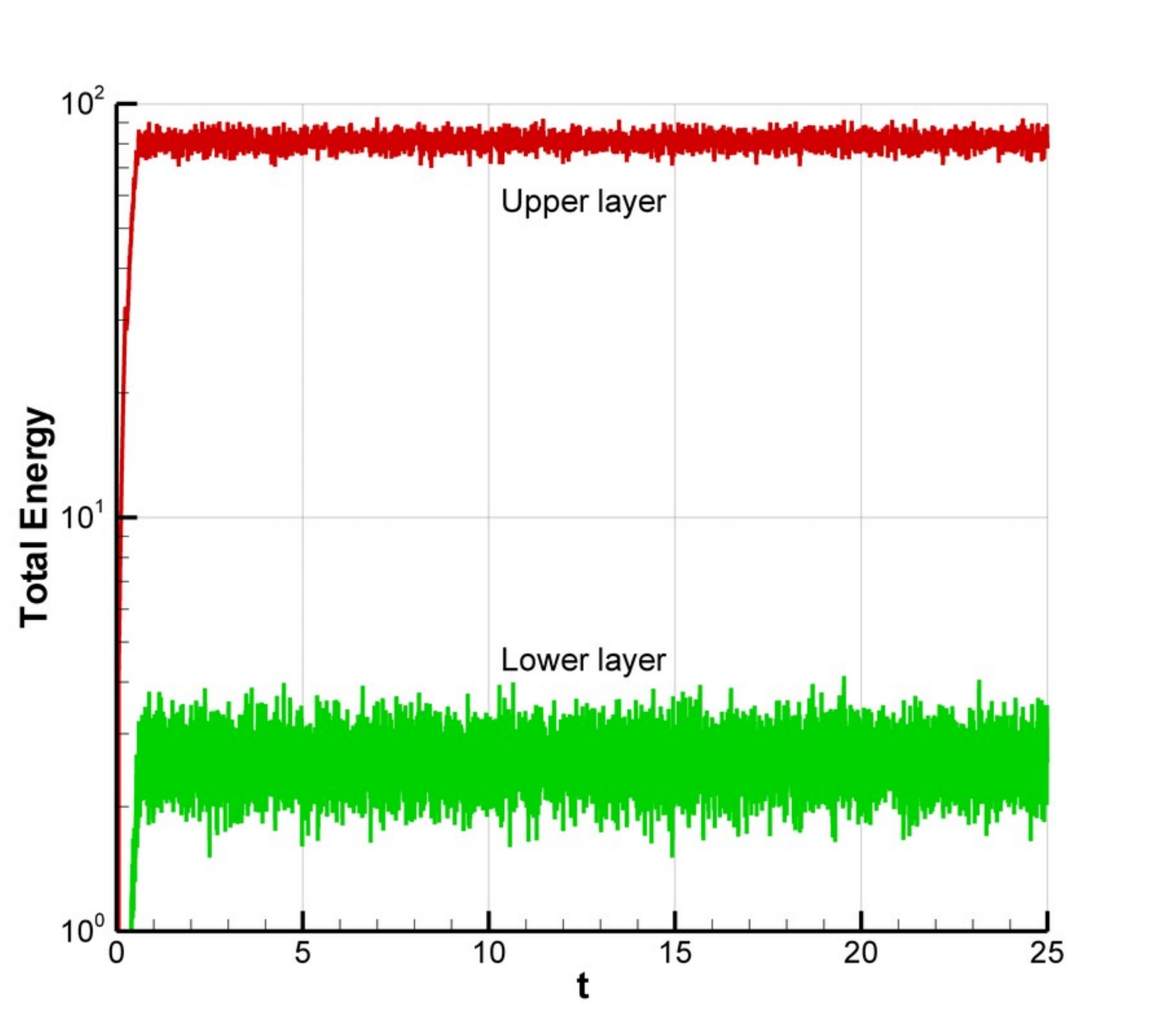}}
}
\caption{
Time histories of basin integrated total kinetic energy for upper and lower layers.
}
\label{fig:hist-2}
\end{figure*}

\begin{figure*}[!t]
\centering
\mbox{
\subfigure{\includegraphics[width=0.4\textwidth]{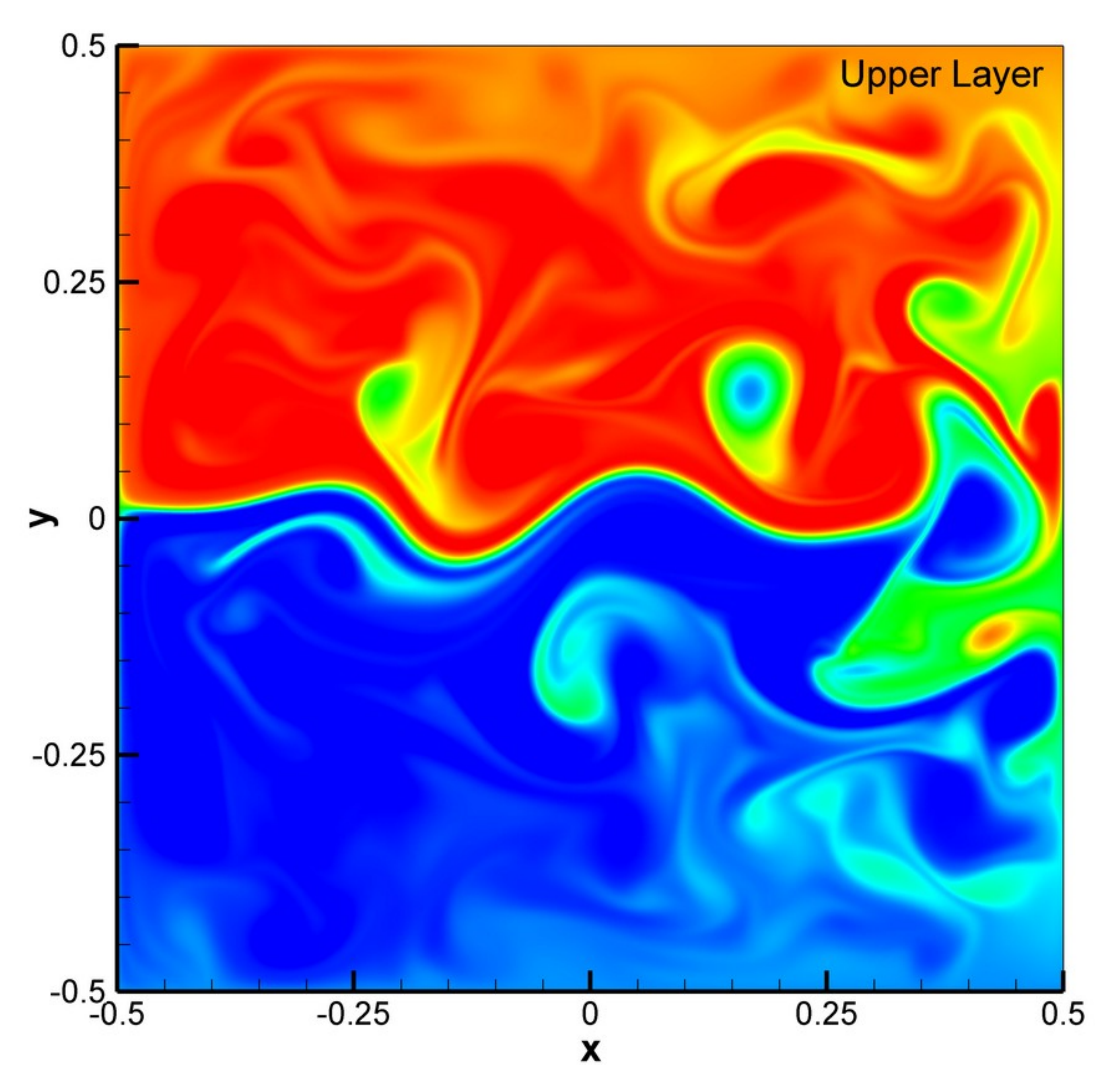}}
\subfigure{\includegraphics[width=0.4\textwidth]{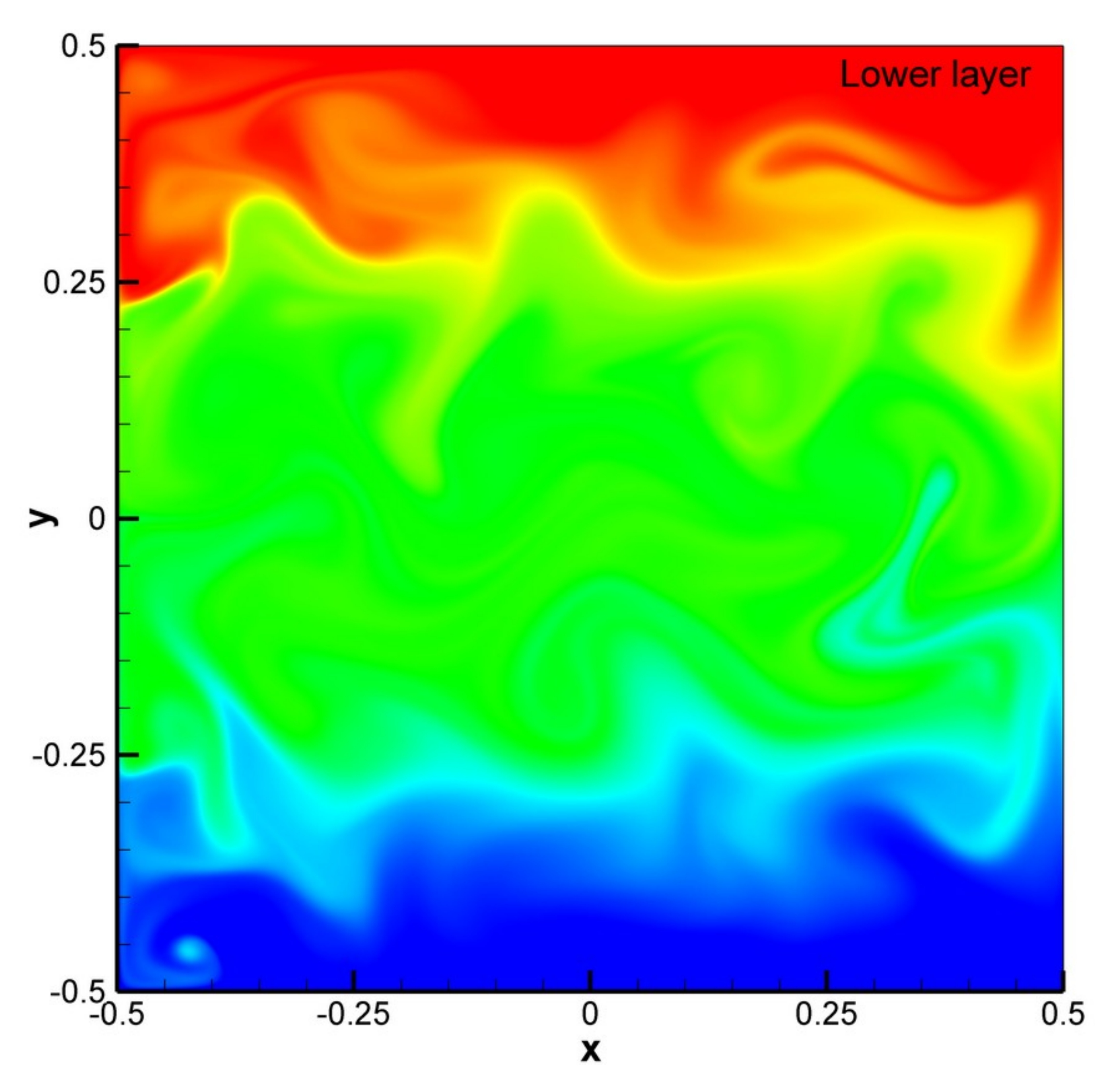}}
}
\caption{
Instantaneous potential vorticity contour plots at time $t=25$ for Experiment 1.
}
\label{fig:inst-2a}
\end{figure*}

\begin{figure*}[!t]
\centering
\mbox{
\subfigure{\includegraphics[width=0.4\textwidth]{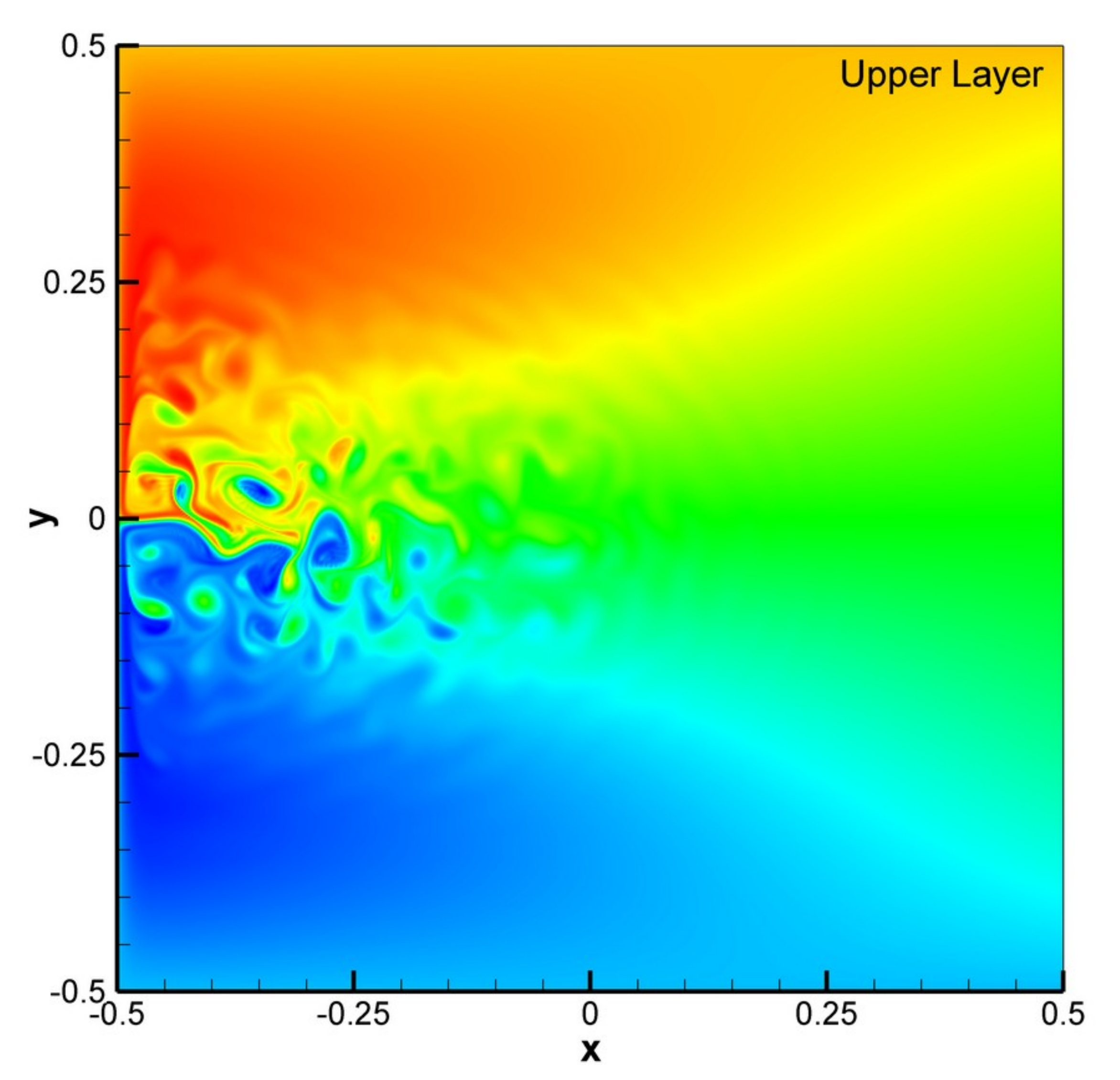}}
\subfigure{\includegraphics[width=0.4\textwidth]{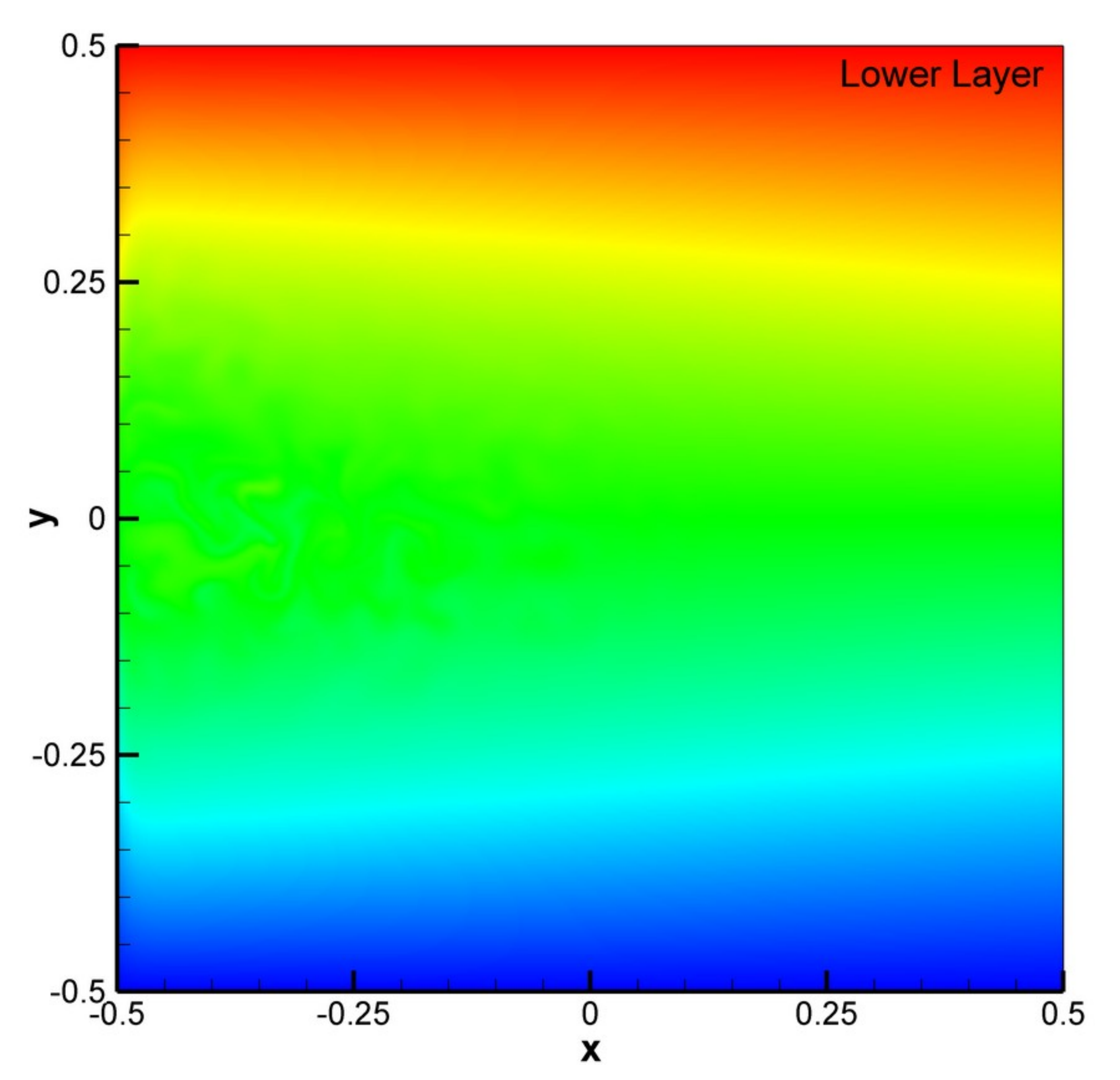}}
}
\caption{
Instantaneous potential vorticity contour plots at time $t=25$ for Experiment 2.
}
\label{fig:inst-2b}
\end{figure*}

\begin{figure}
\centering
\mbox{
\subfigure{\includegraphics[width=0.4\textwidth]{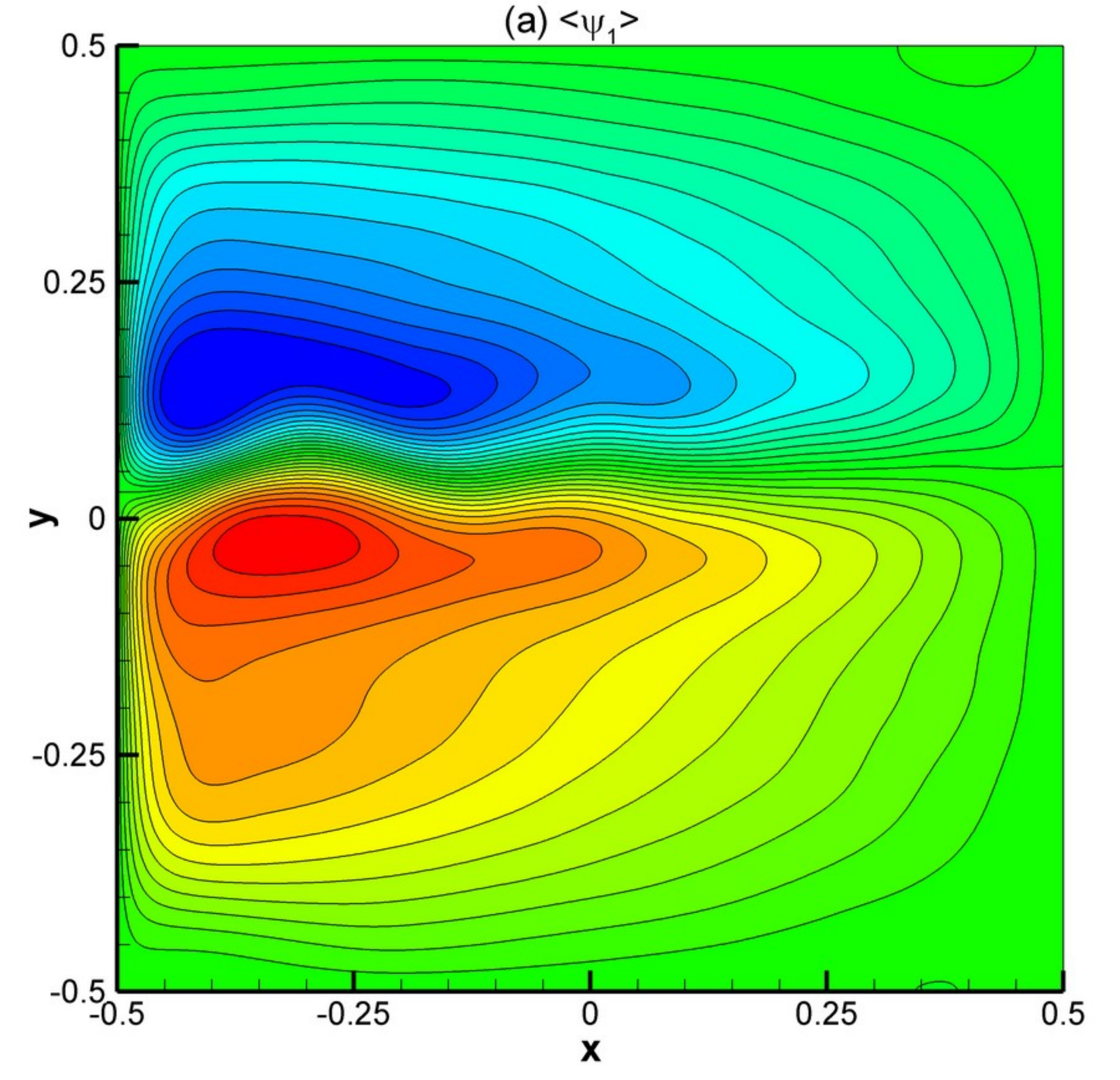}}
\subfigure{\includegraphics[width=0.4\textwidth]{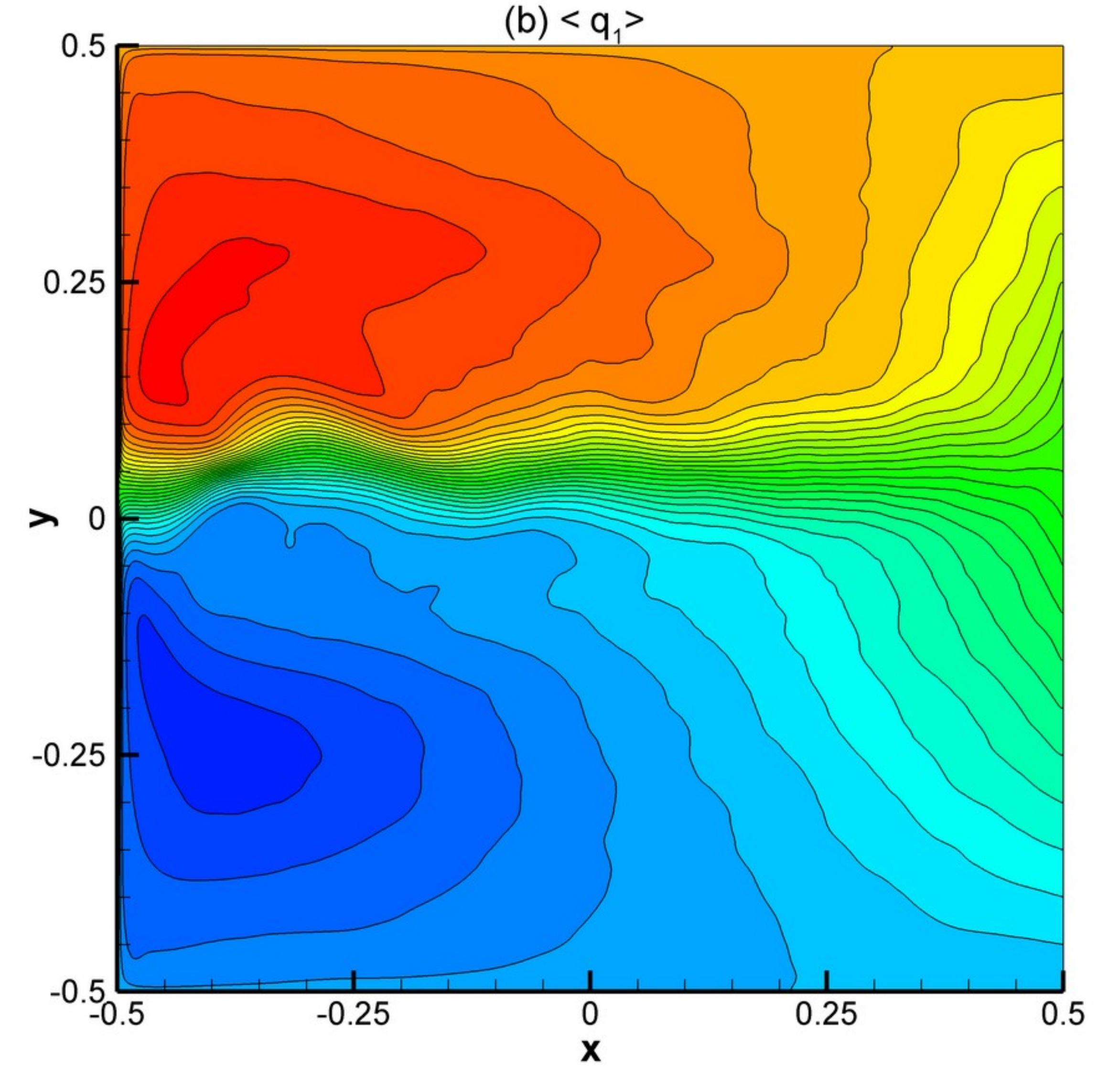}}
}
\mbox{
\subfigure{\includegraphics[width=0.4\textwidth]{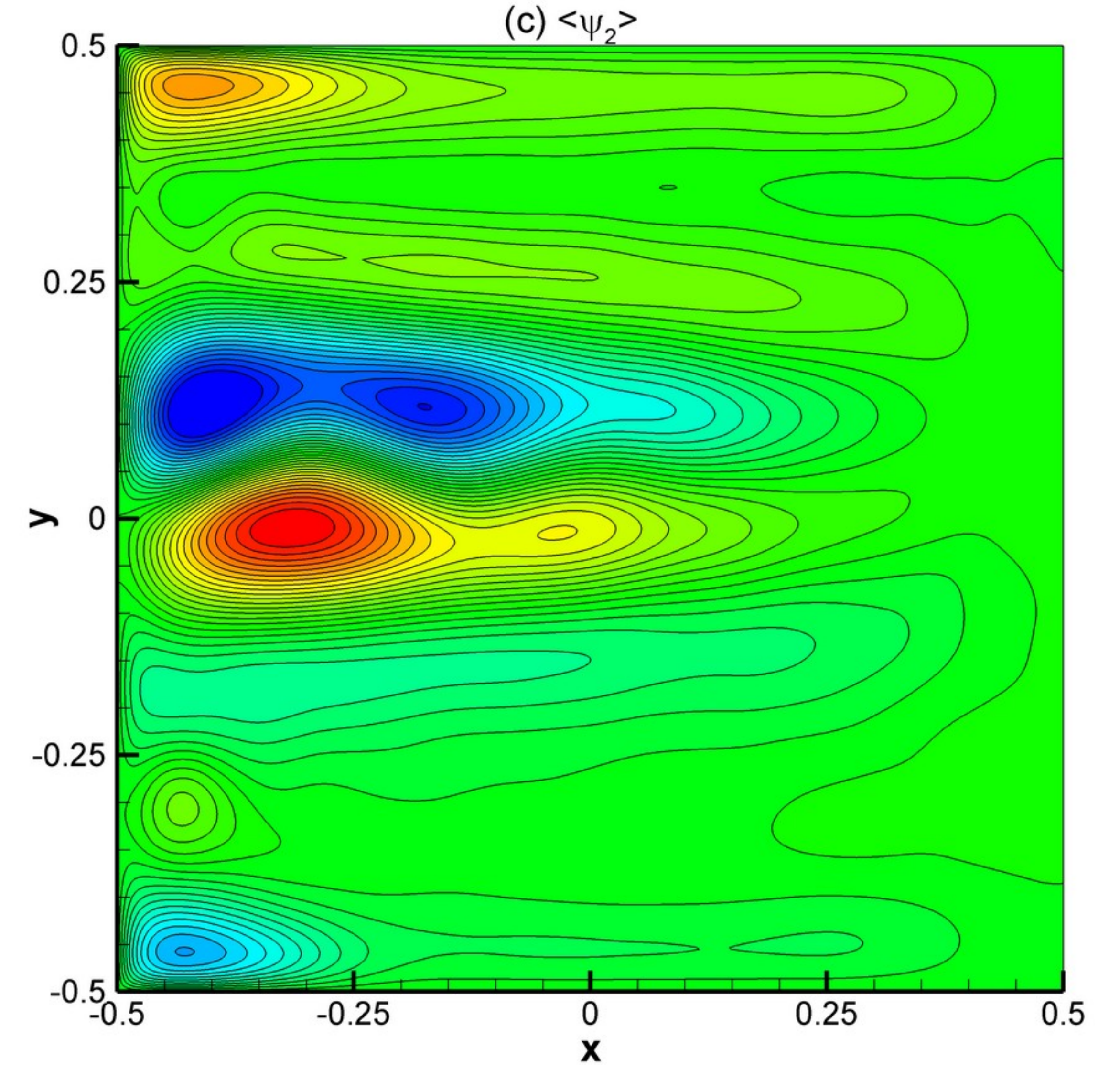}}
\subfigure{\includegraphics[width=0.4\textwidth]{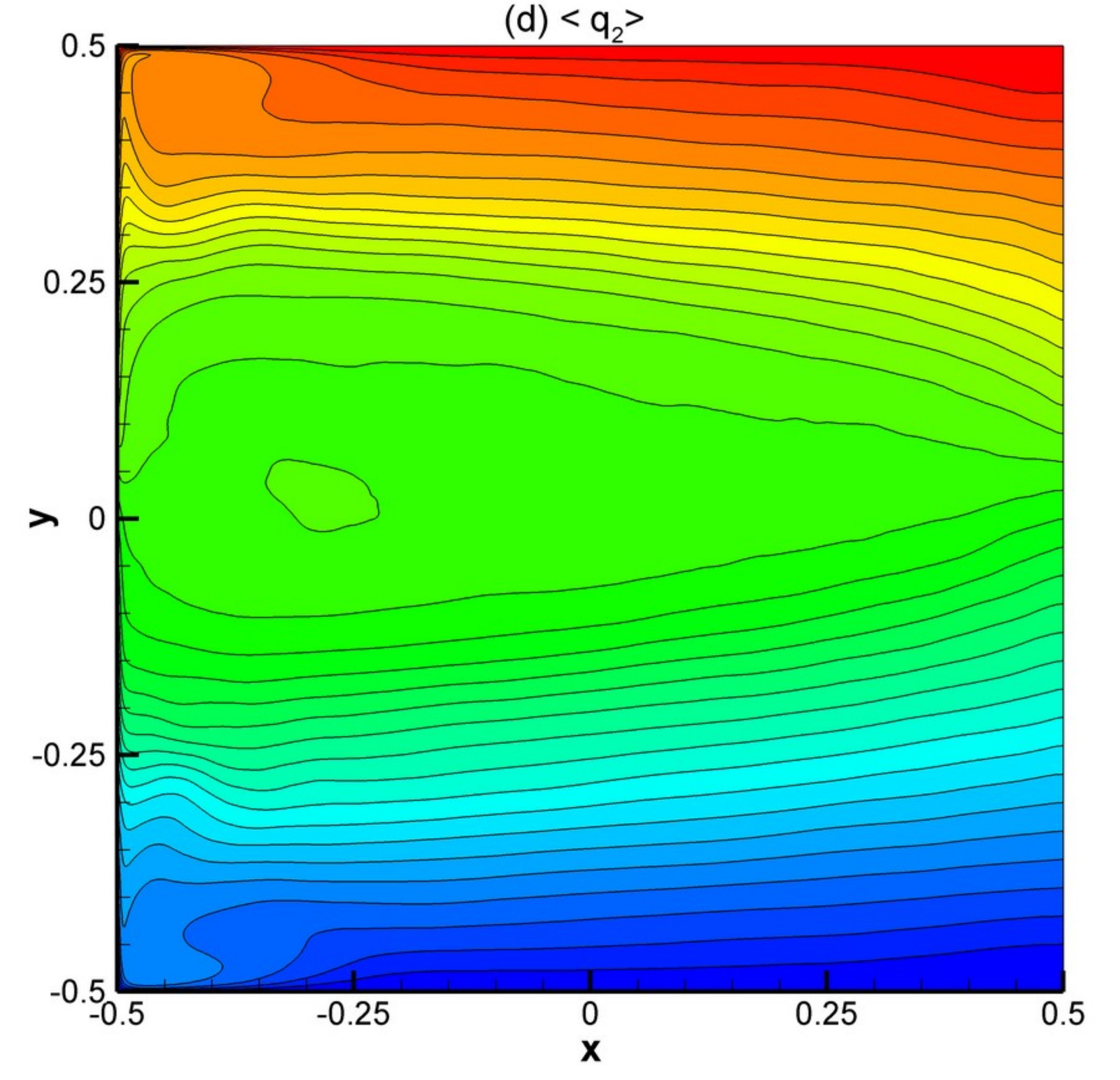}}
}
\caption{
Experiment 1: DNS results for
(a) mean stream function contours for the upper layer,
(b) mean potential vorticity contours for the upper layer,
(c) mean stream function contours for the lower layer, and
(d) mean potential vorticity contours for the lower layer.
}
\label{fig:mean-1}
\end{figure}

\begin{figure}
\centering
\mbox{
\subfigure{\includegraphics[width=0.4\textwidth]{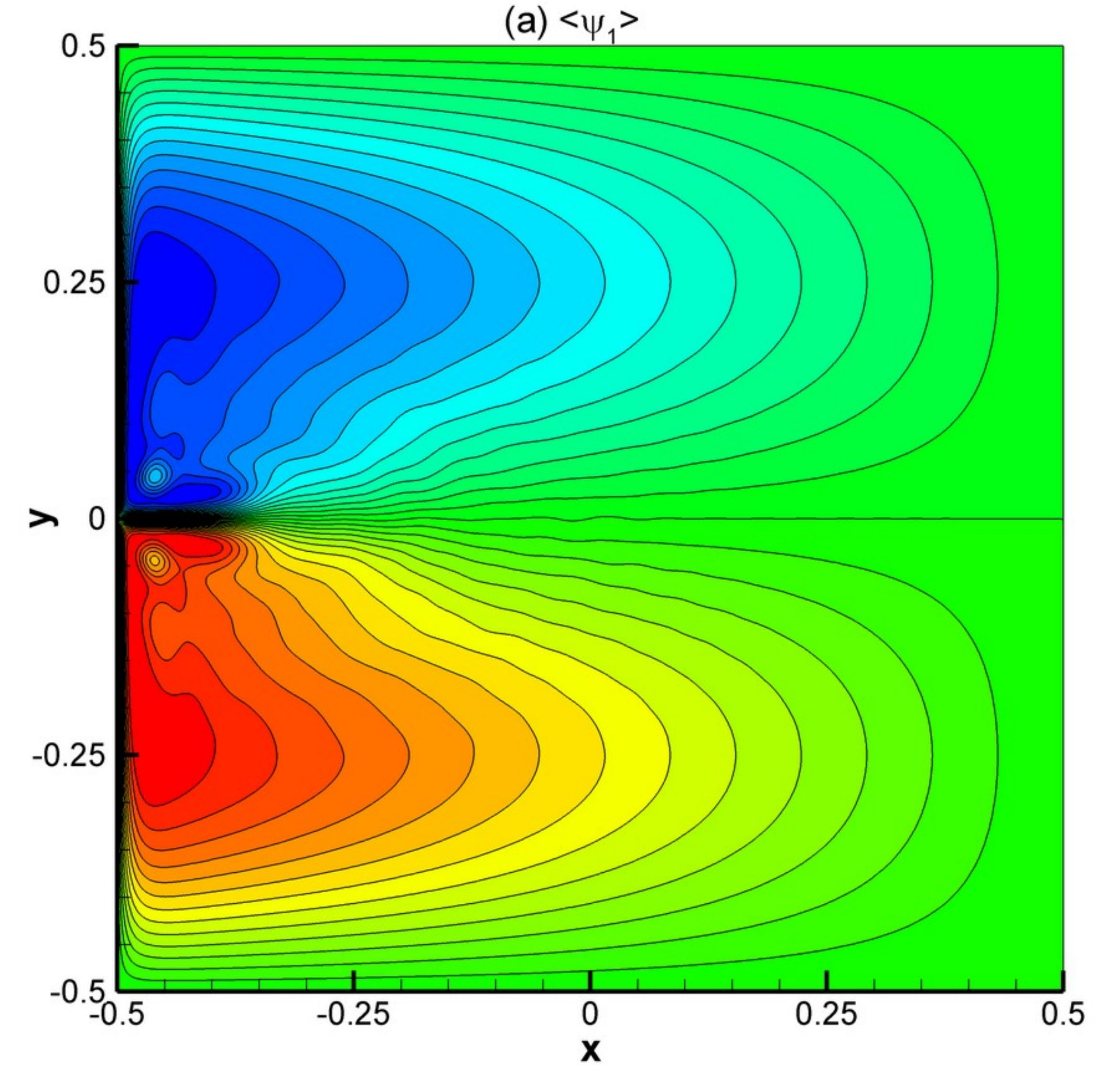}}
\subfigure{\includegraphics[width=0.4\textwidth]{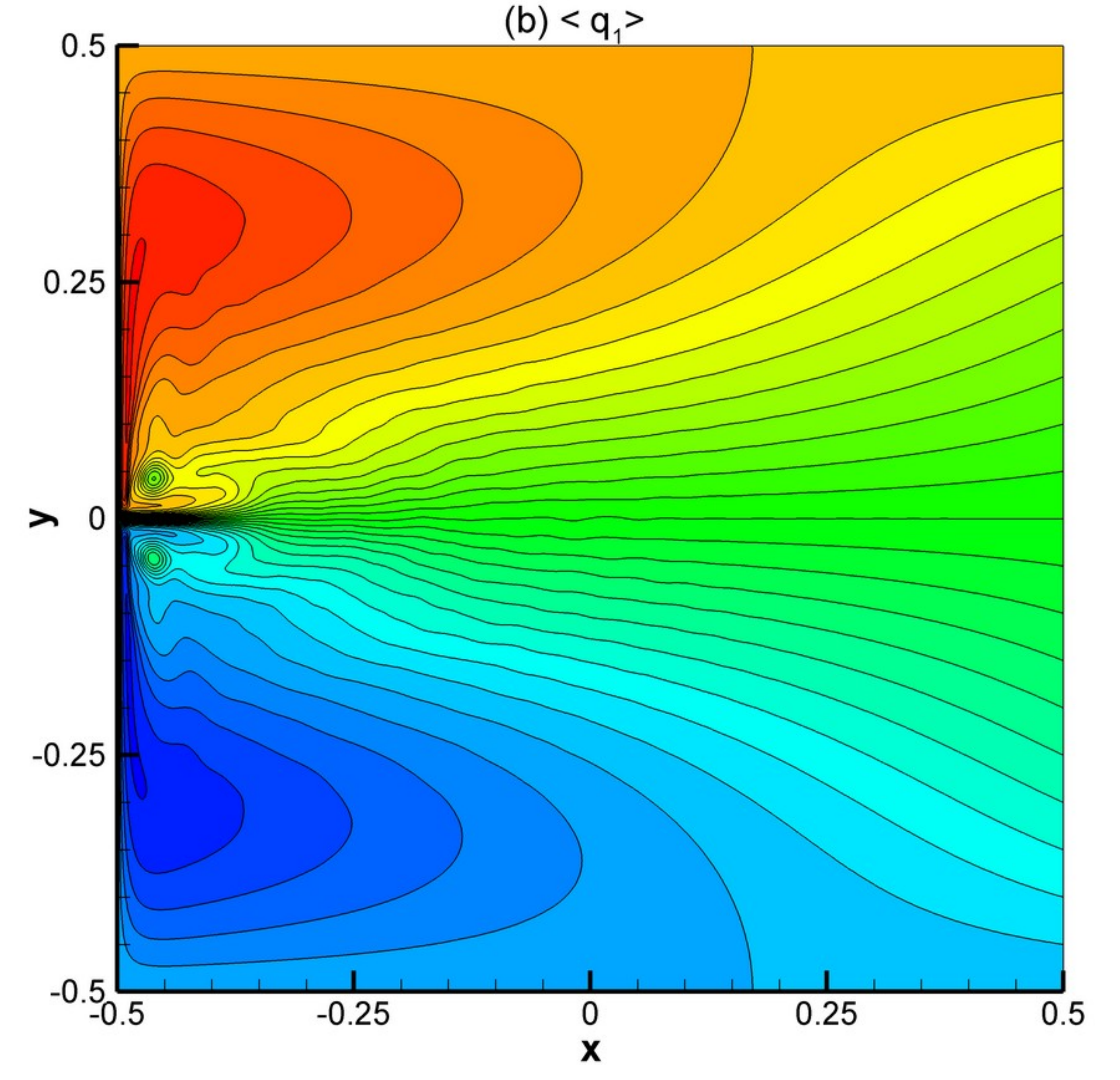}}
}
\mbox{
\subfigure{\includegraphics[width=0.4\textwidth]{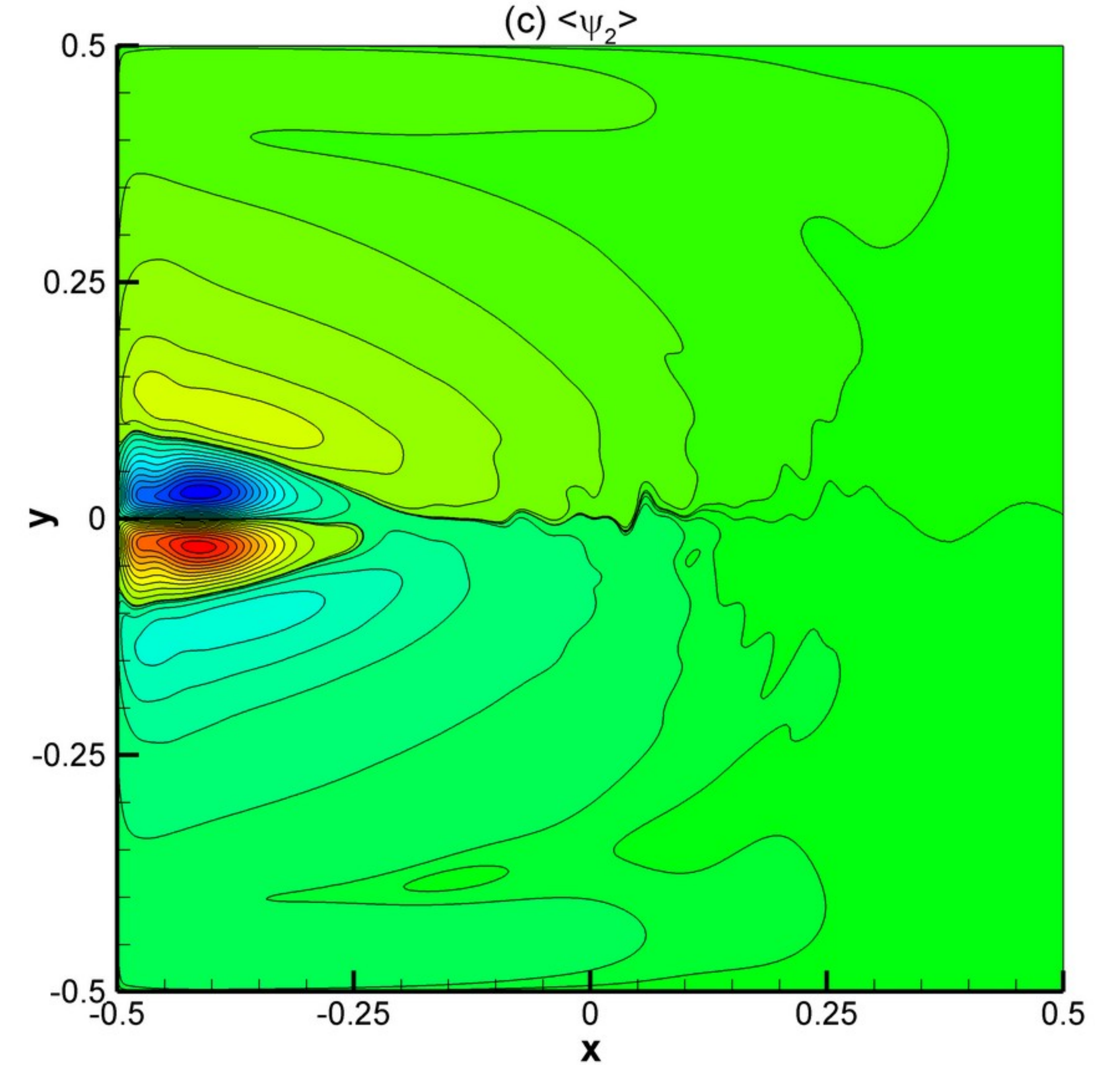}}
\subfigure{\includegraphics[width=0.4\textwidth]{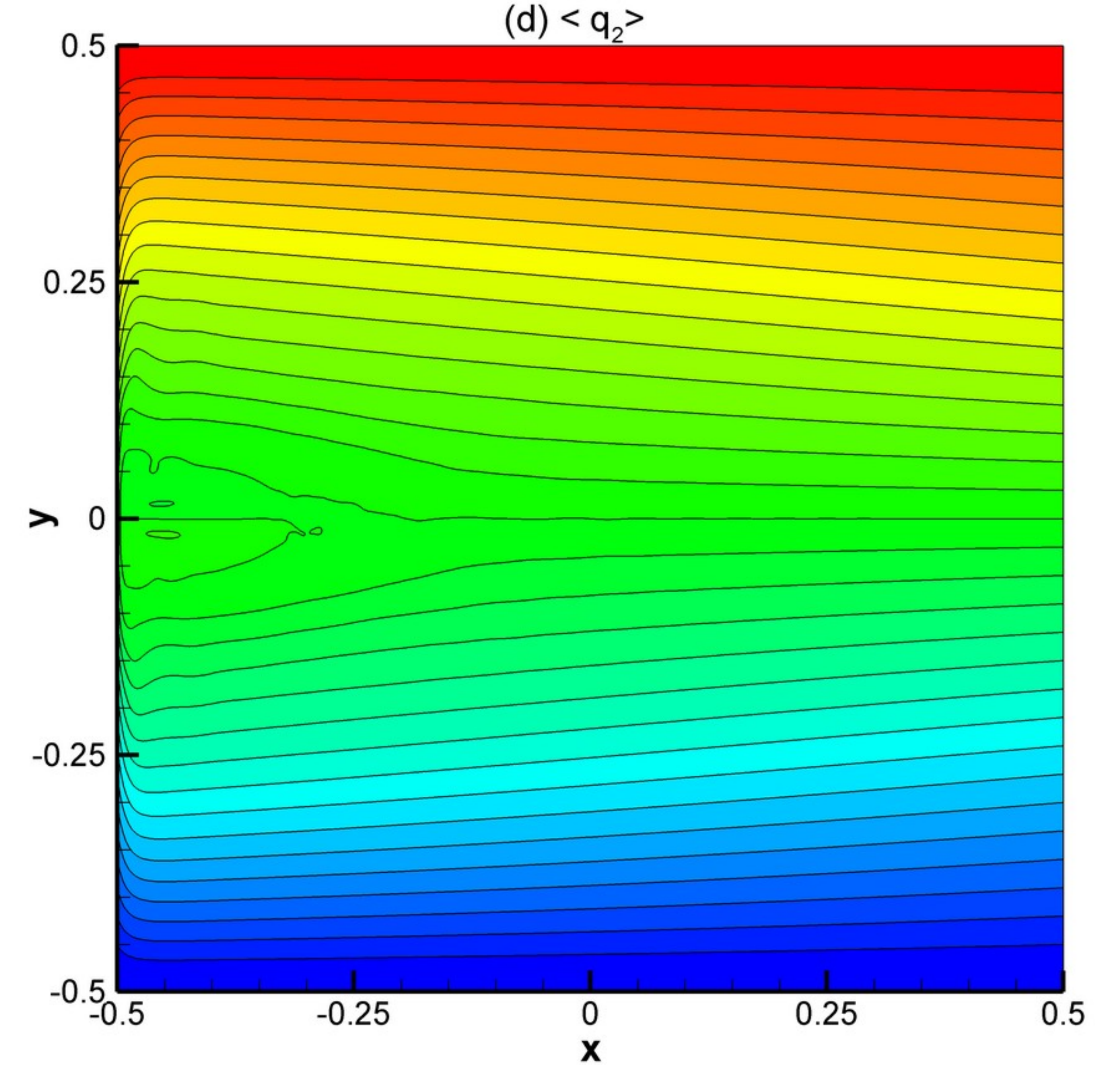}}
}
\caption{
Experiment 2: DNS results for
(a) mean stream function contours for the upper layer,
(b) mean potential vorticity contours for the upper layer,
(c) mean stream function contours for the lower layer, and
(d) mean potential vorticity contours for the lower layer.
}
\label{fig:mean-2}
\end{figure}

\begin{figure*}
\centering
\mbox{
\subfigure{\includegraphics[width=0.33\textwidth]{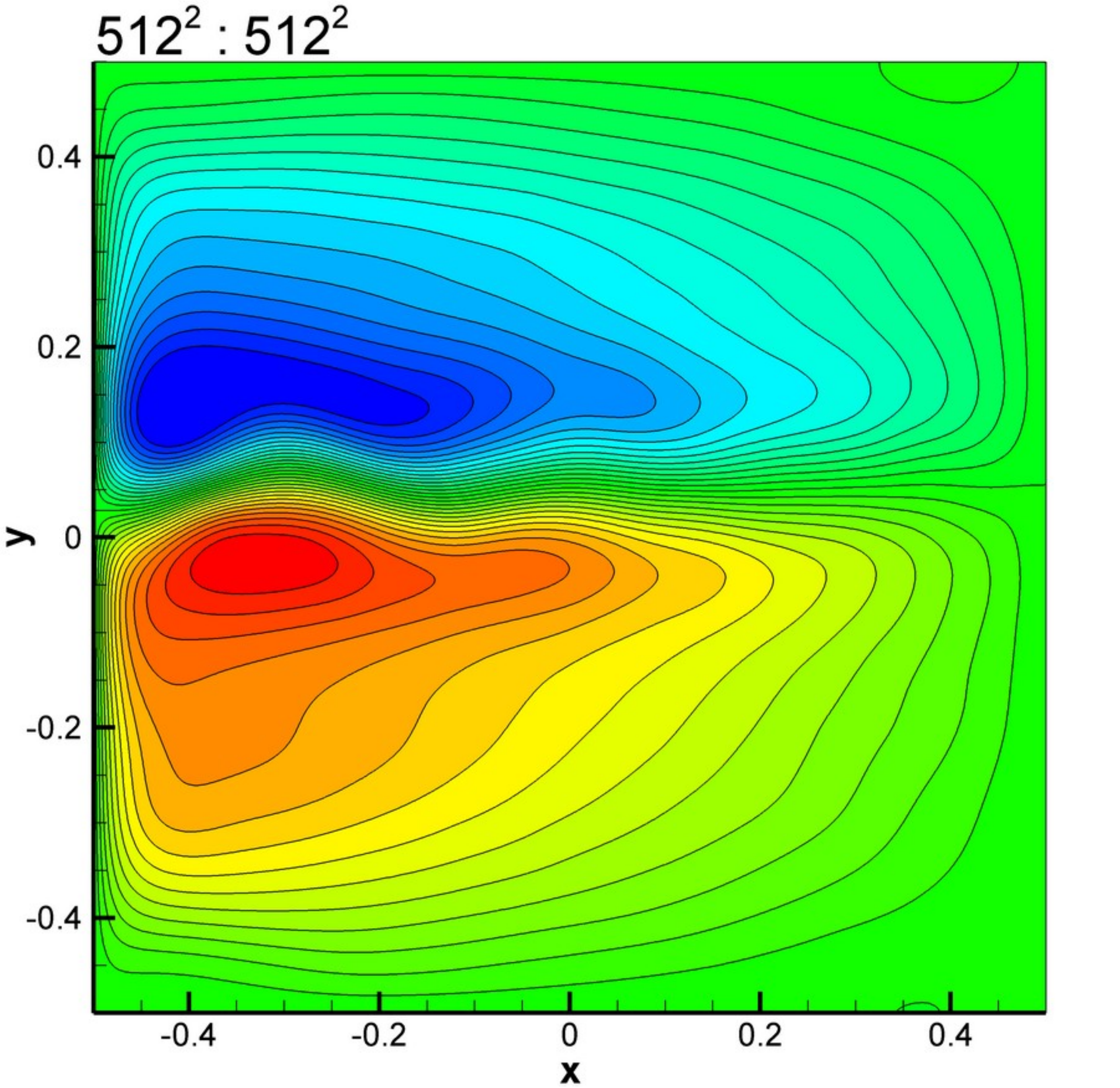}}
\subfigure{\includegraphics[width=0.33\textwidth]{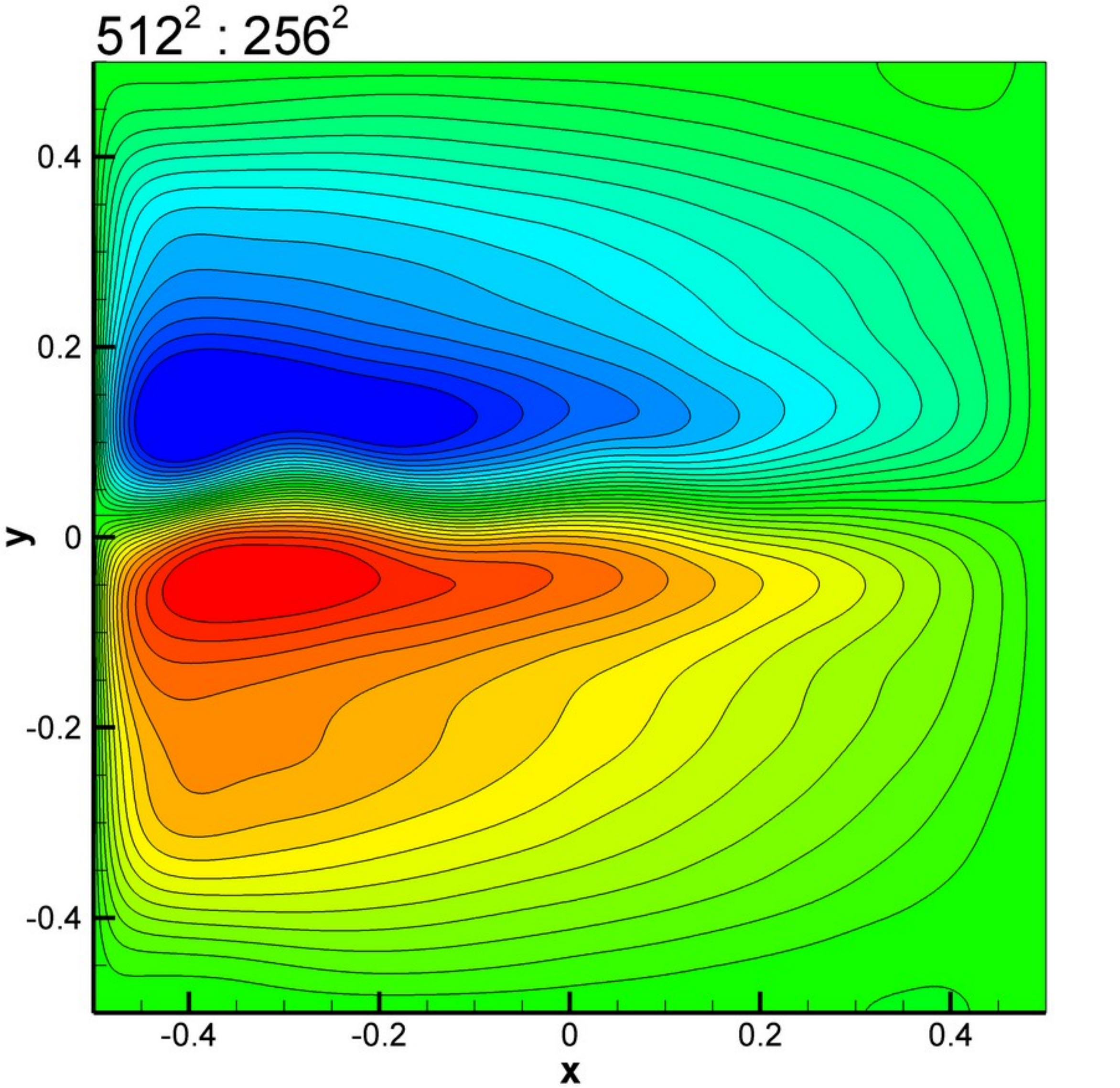}}
\subfigure{\includegraphics[width=0.33\textwidth]{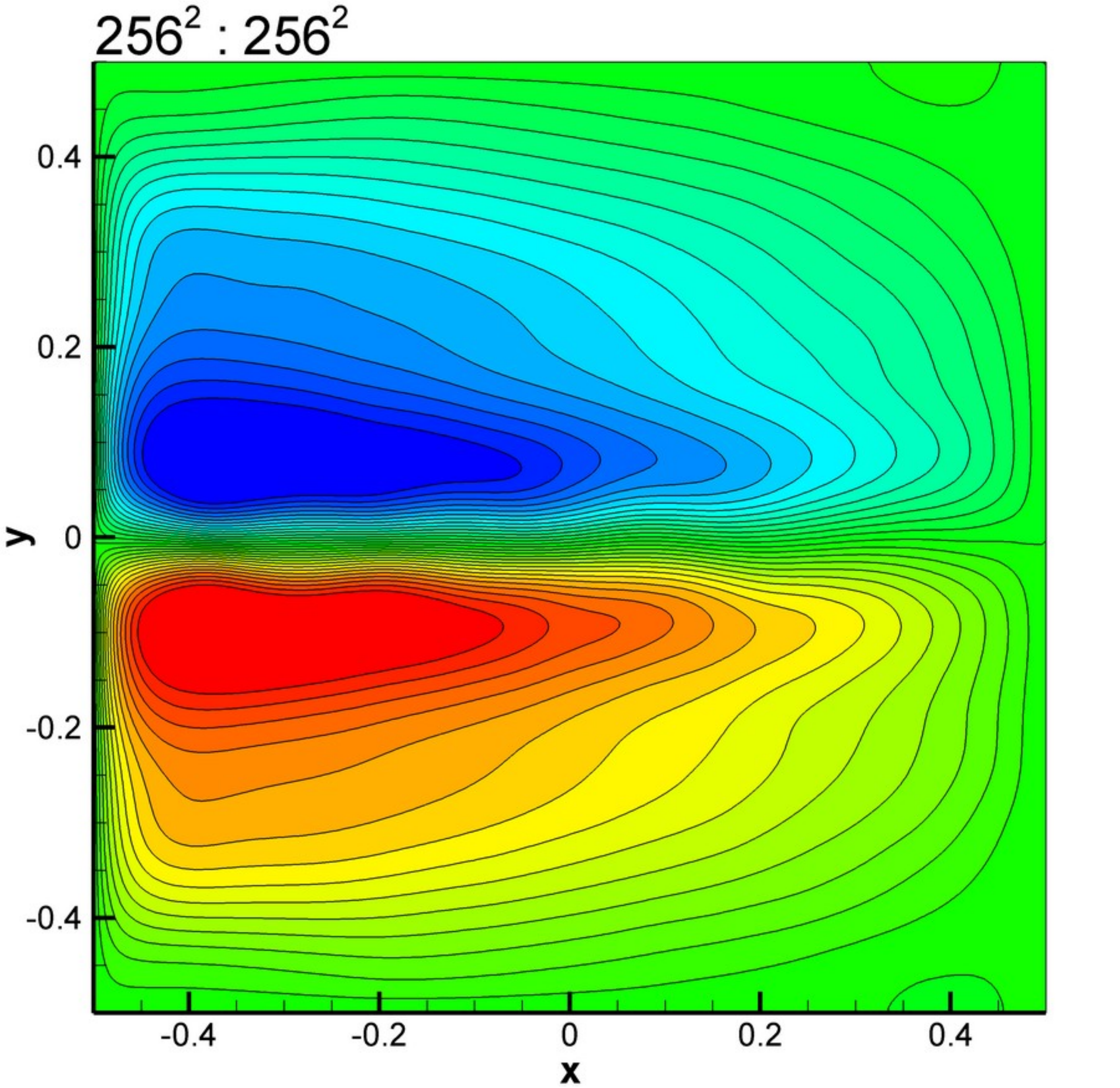}} }
\\
\mbox{
\subfigure{\includegraphics[width=0.33\textwidth]{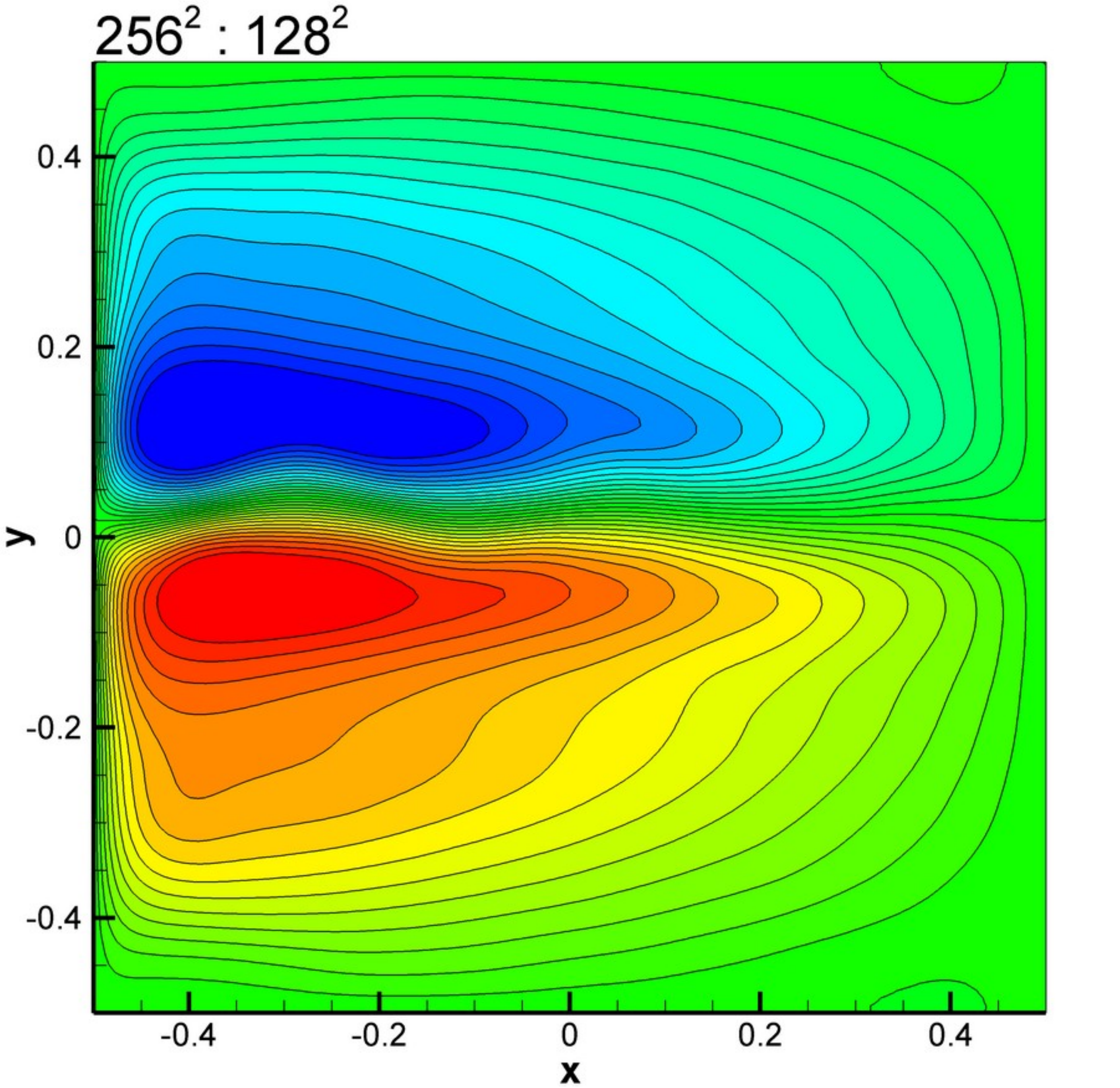}}
\subfigure{\includegraphics[width=0.33\textwidth]{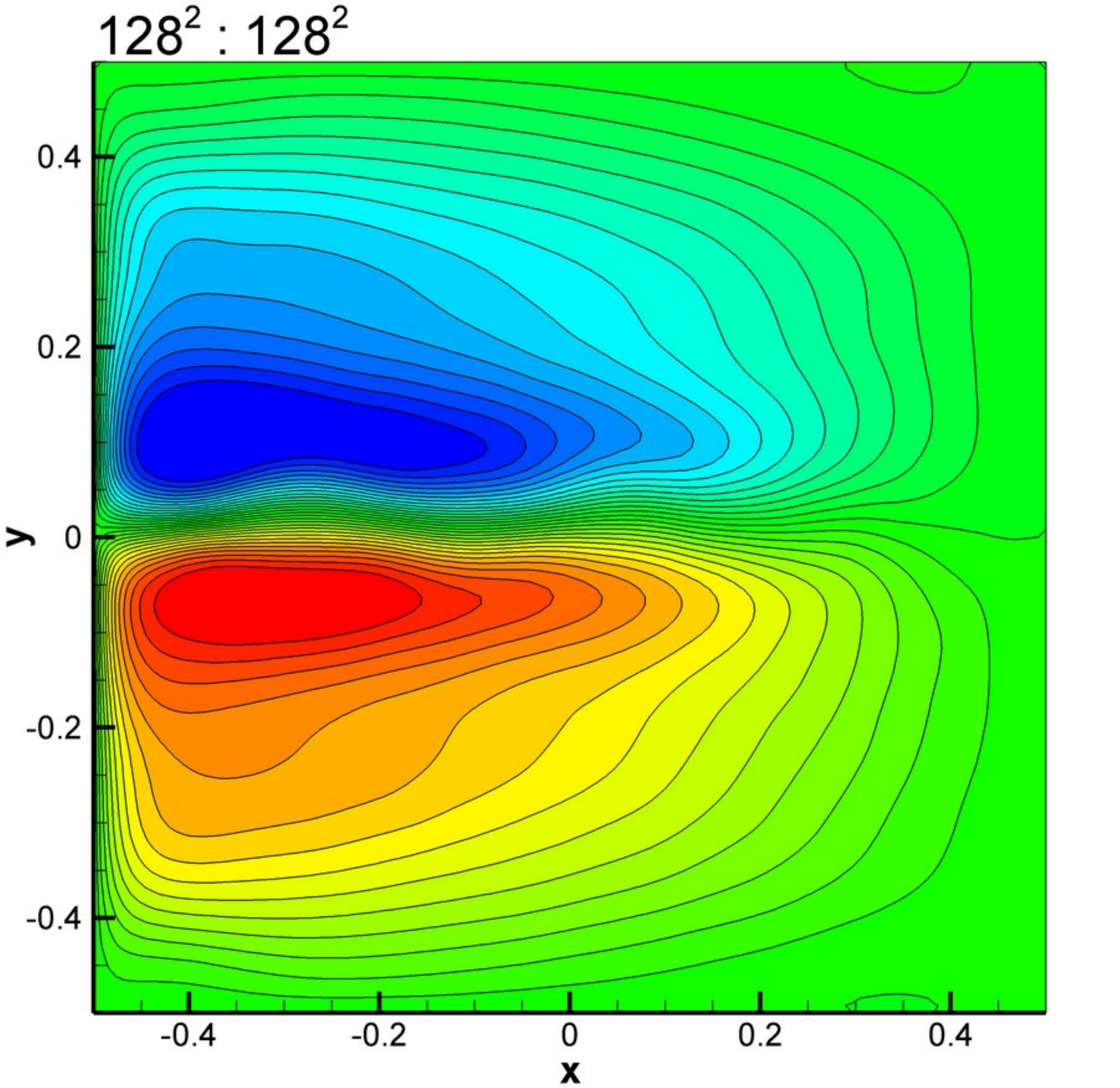}}
\subfigure{\includegraphics[width=0.33\textwidth]{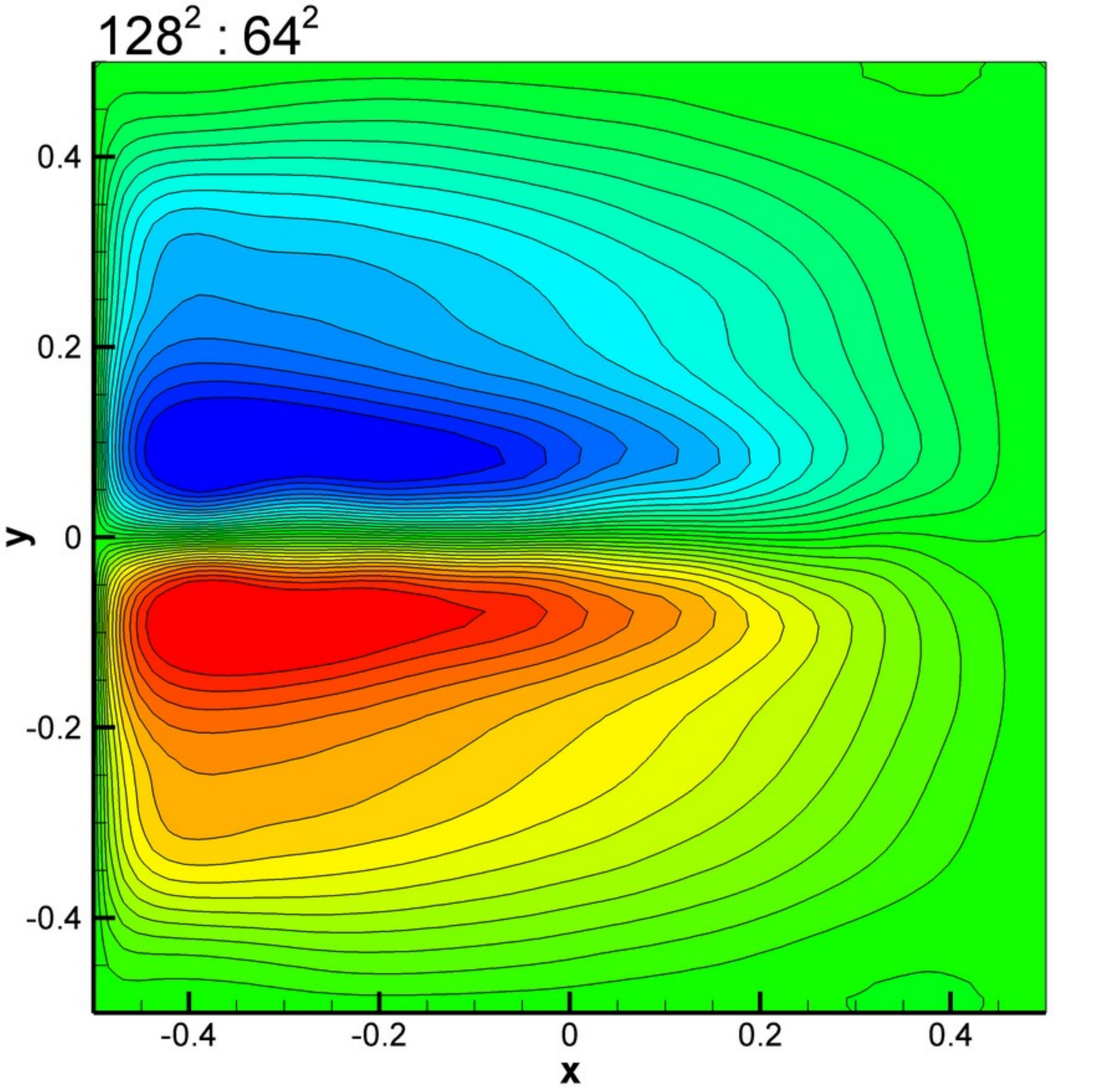}} }
\\
\mbox{
\subfigure{\includegraphics[width=0.33\textwidth]{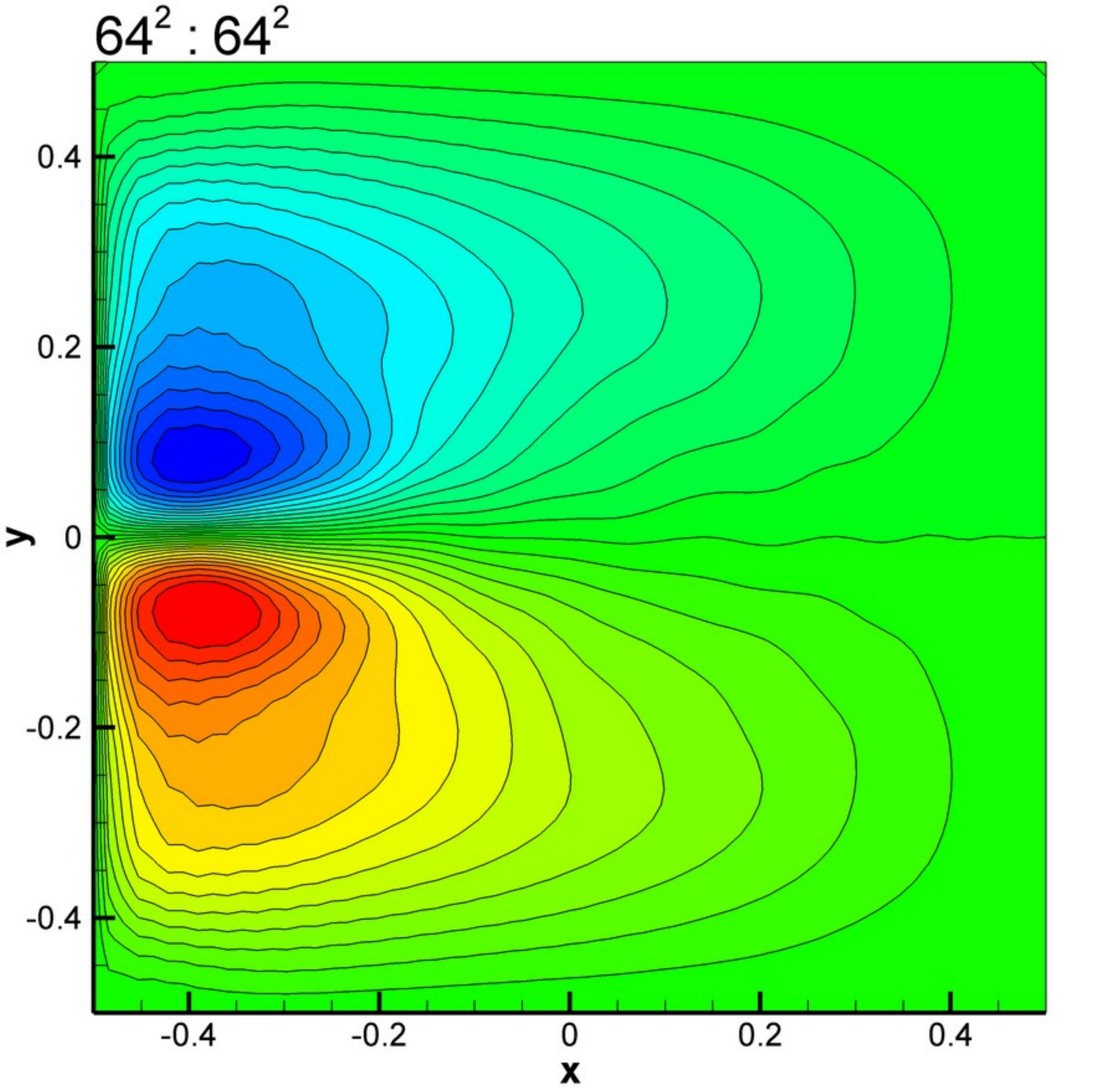}}
\subfigure{\includegraphics[width=0.33\textwidth]{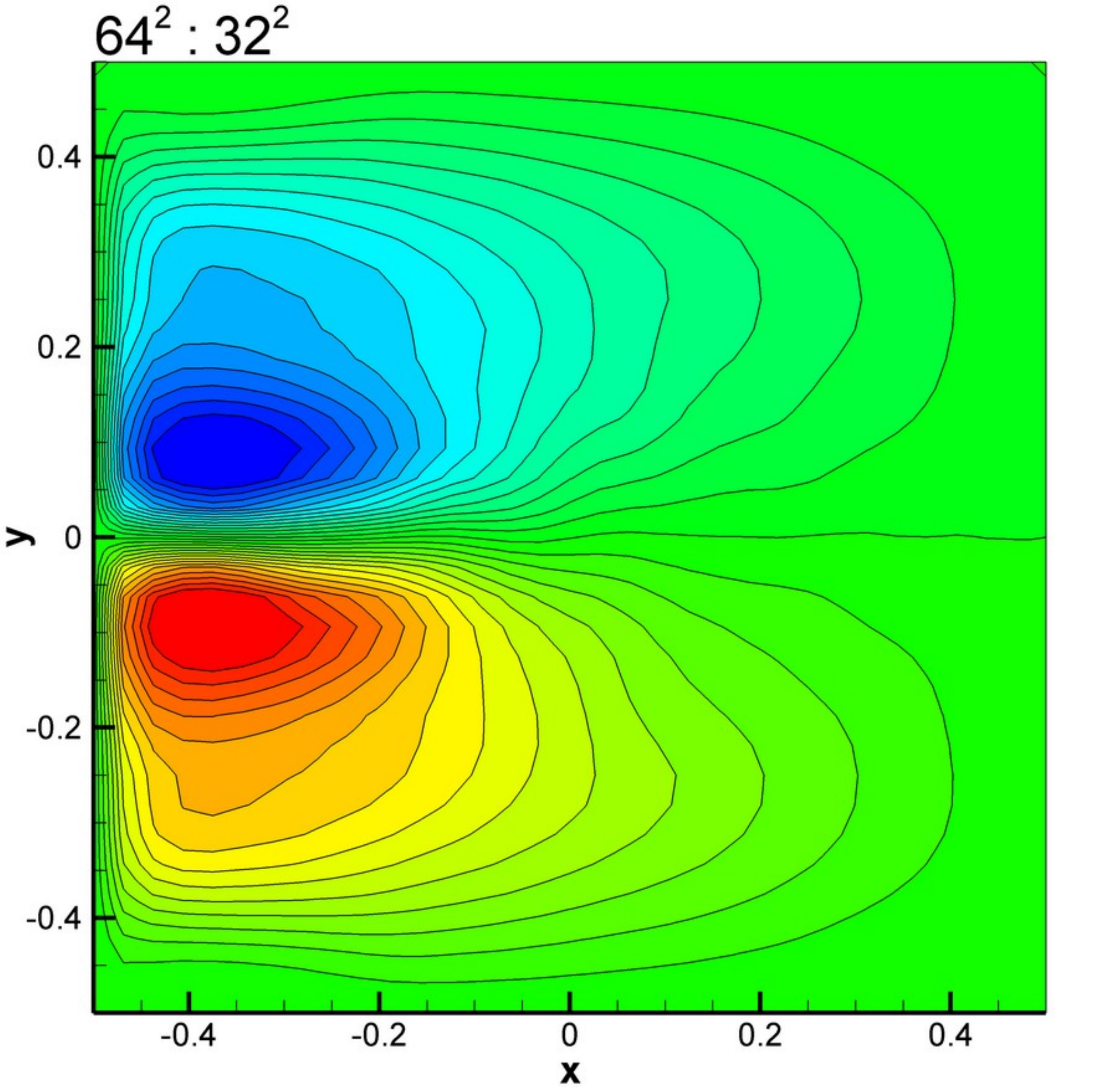}}
\subfigure{\includegraphics[width=0.33\textwidth]{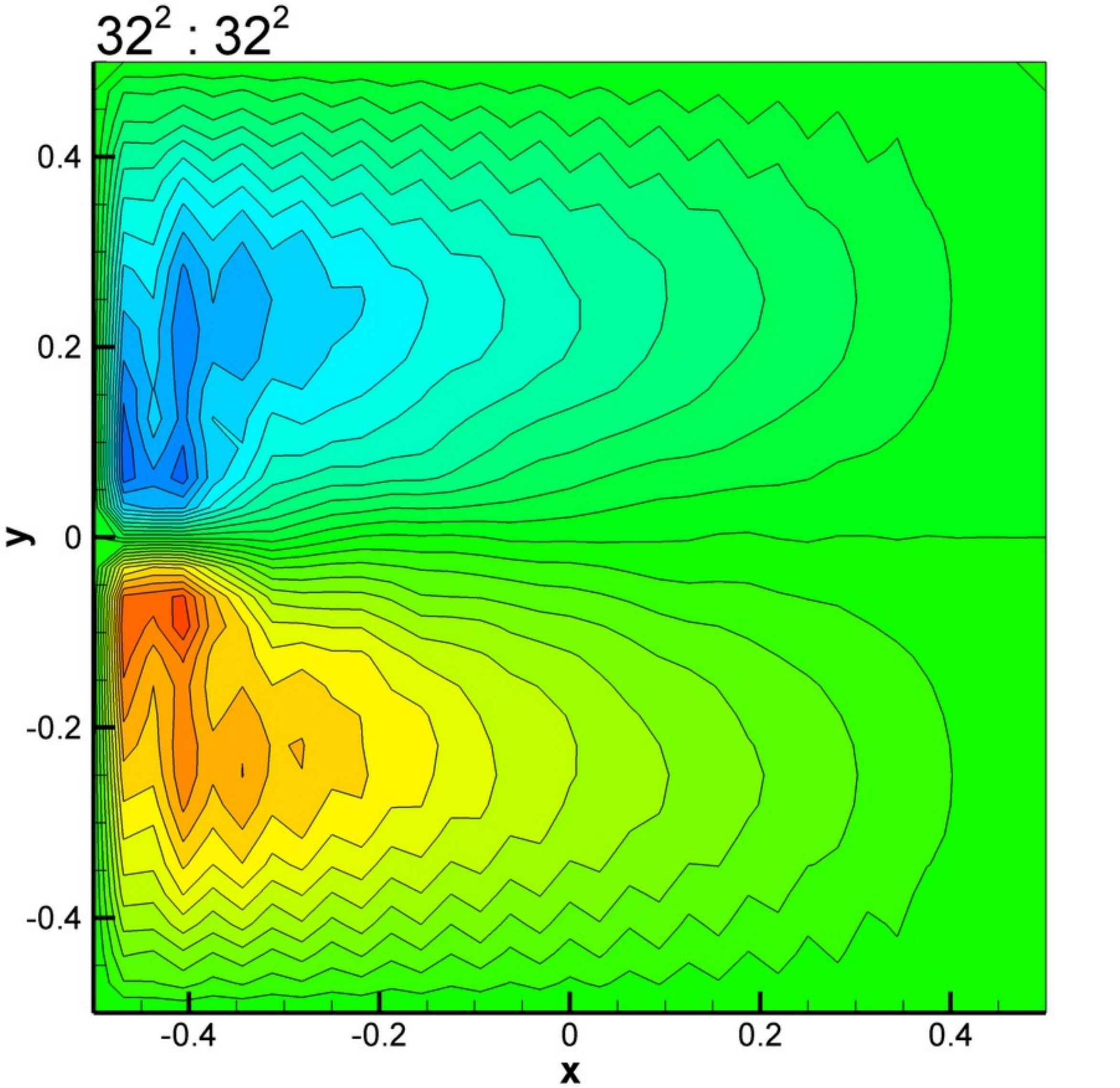}} }
\caption{
Comparison of mean stream function for the upper layer for the Experiment 1. Labels include the resolutions for both parts of the solver in the form $N^2 : M^2$, where $N^2$ is the resolution for the vorticity transport equations, and $M^2$ is the resolution for the elliptic sub-problems.
}
\label{fig:s-O}
\end{figure*}

\begin{figure*}
\centering
\mbox{
\subfigure{\includegraphics[width=0.33\textwidth]{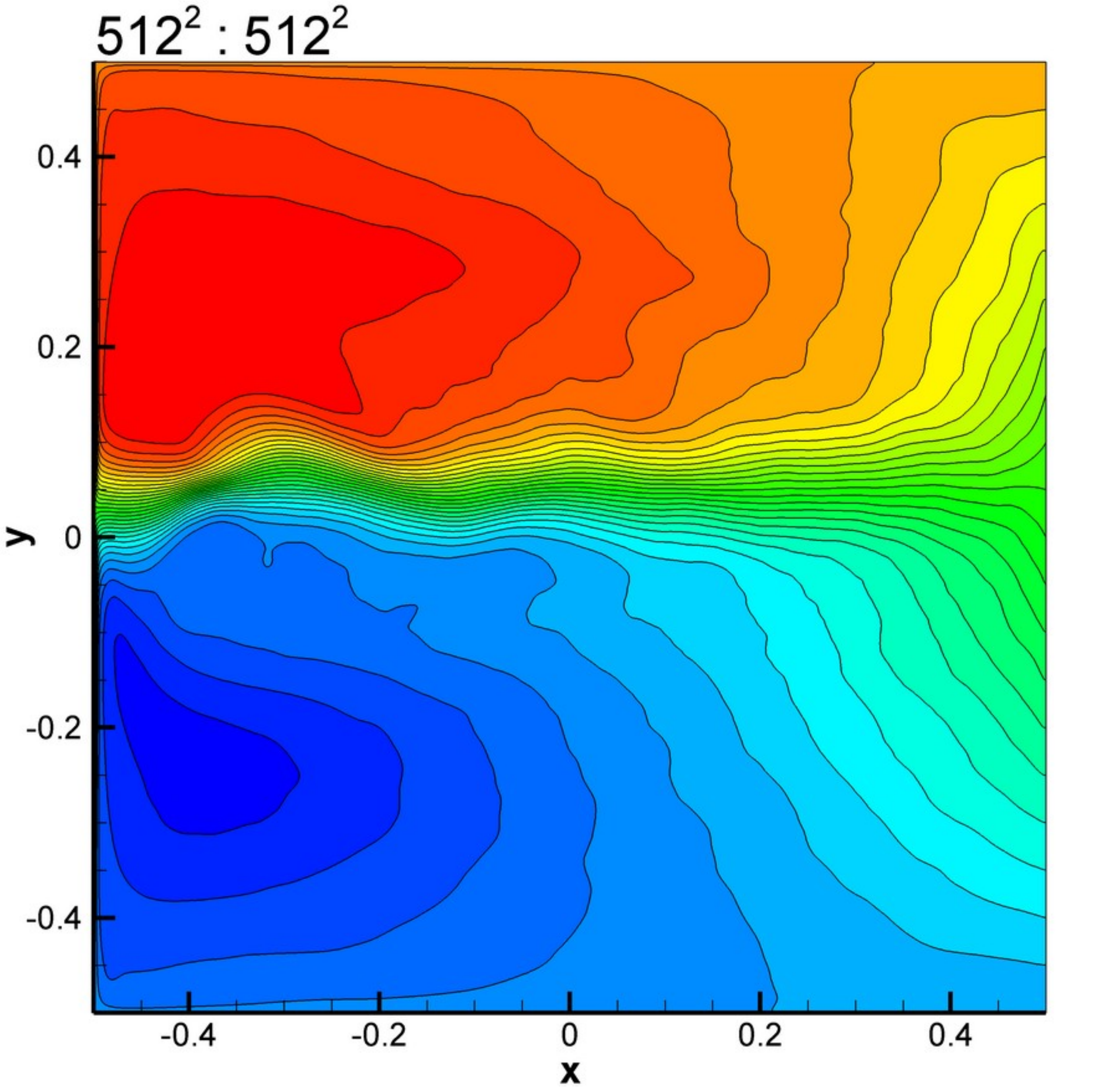}}
\subfigure{\includegraphics[width=0.33\textwidth]{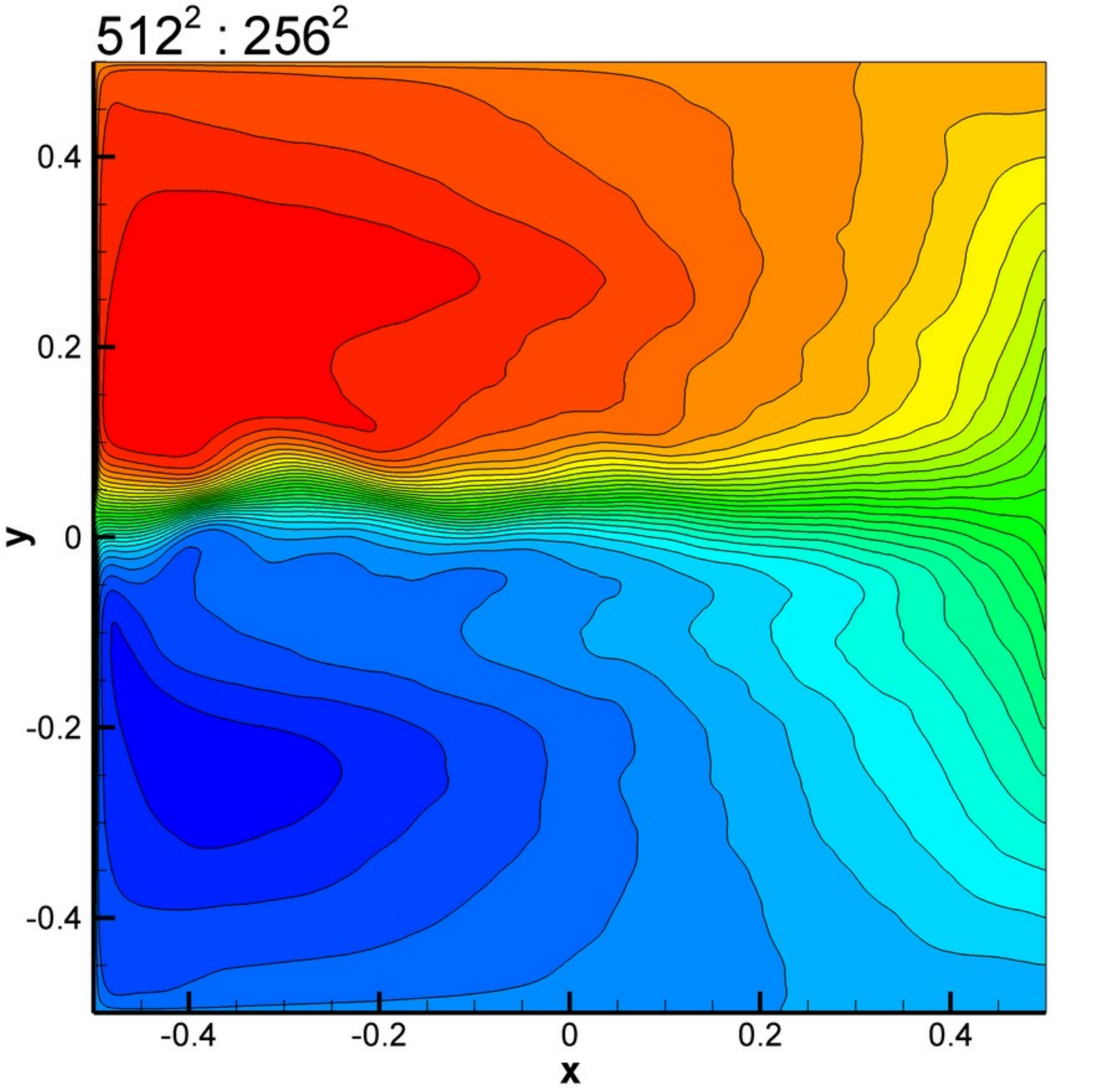}}
\subfigure{\includegraphics[width=0.33\textwidth]{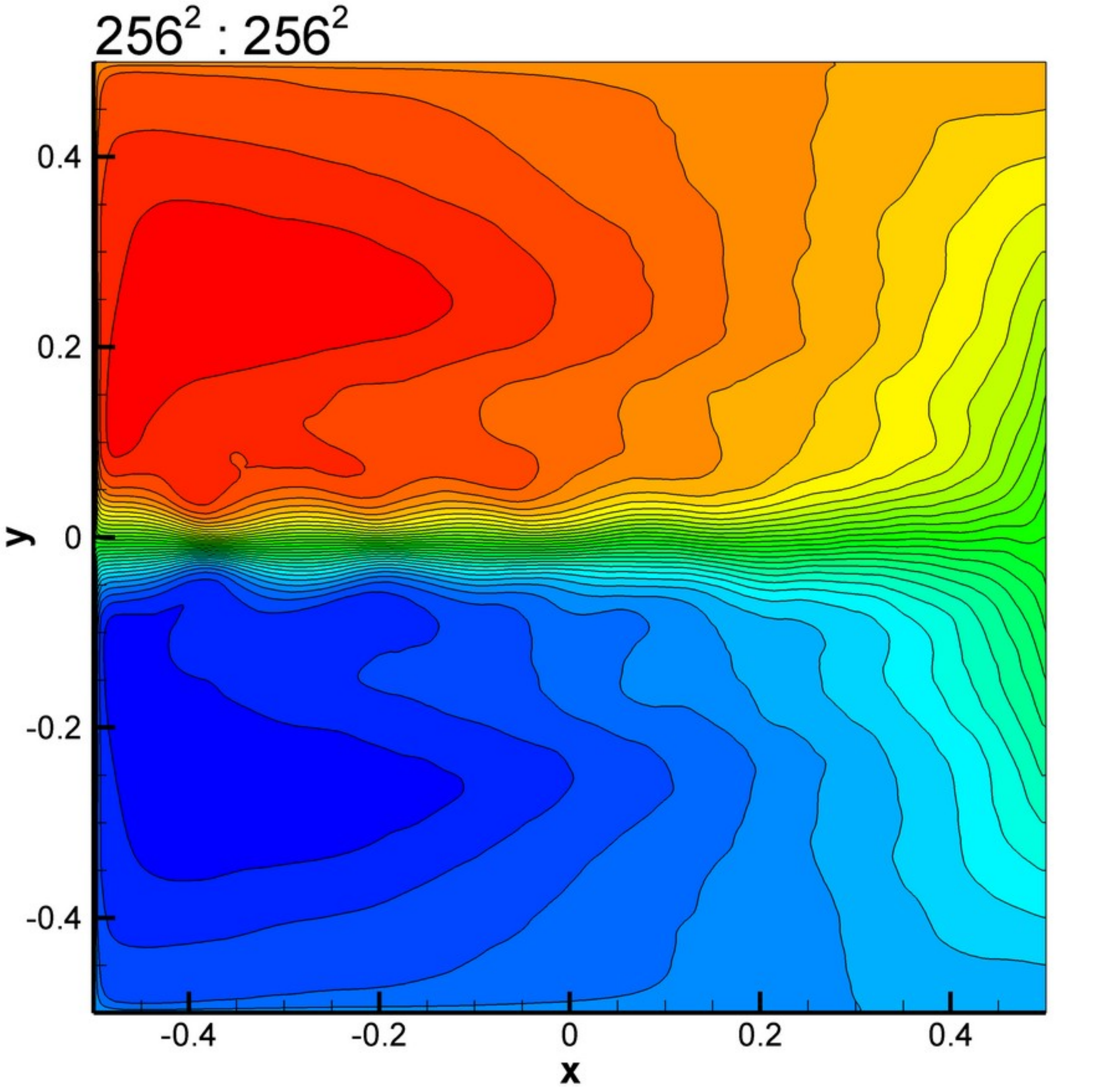}} }
\\
\mbox{
\subfigure{\includegraphics[width=0.33\textwidth]{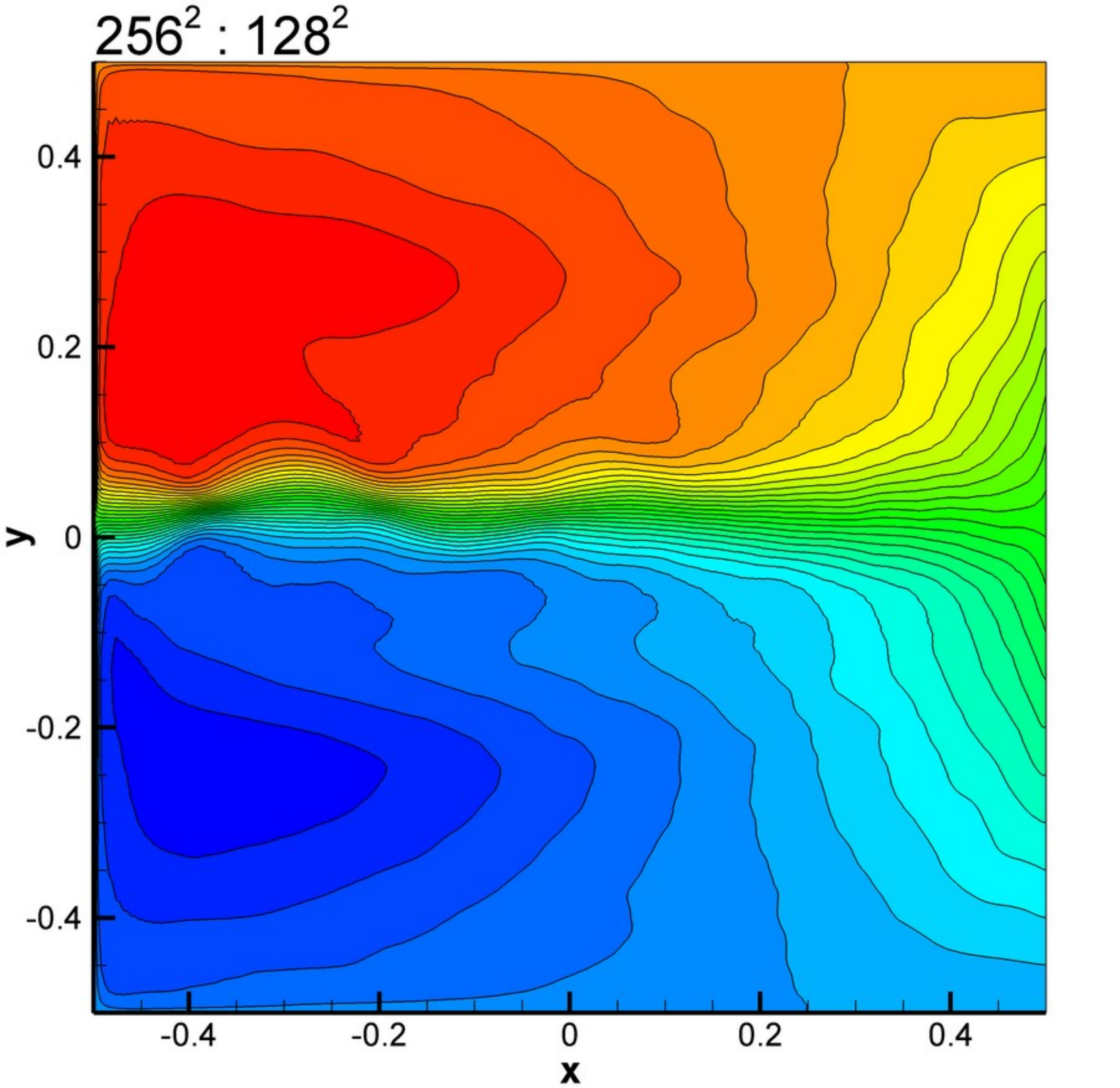}}
\subfigure{\includegraphics[width=0.33\textwidth]{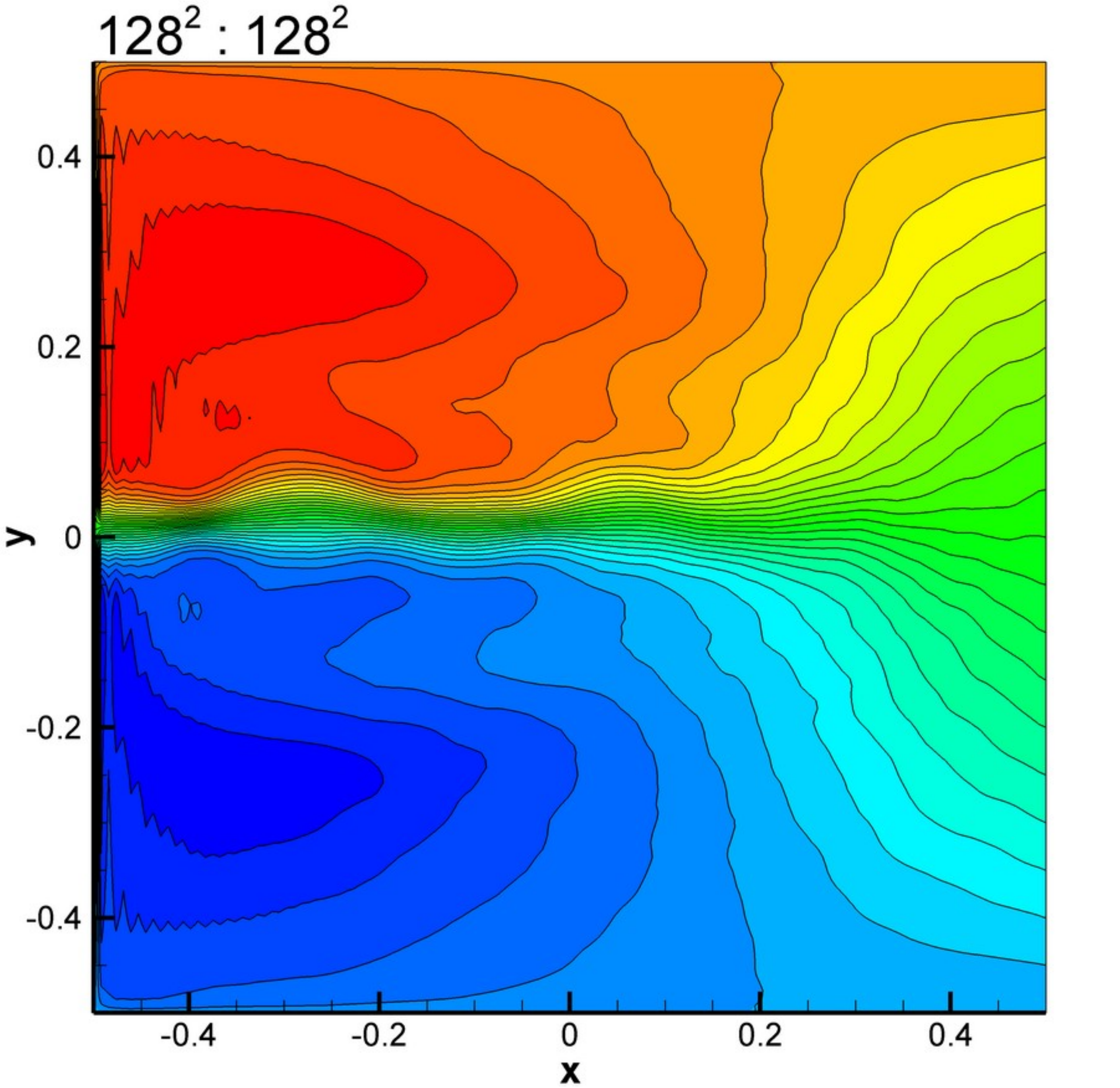}}
\subfigure{\includegraphics[width=0.33\textwidth]{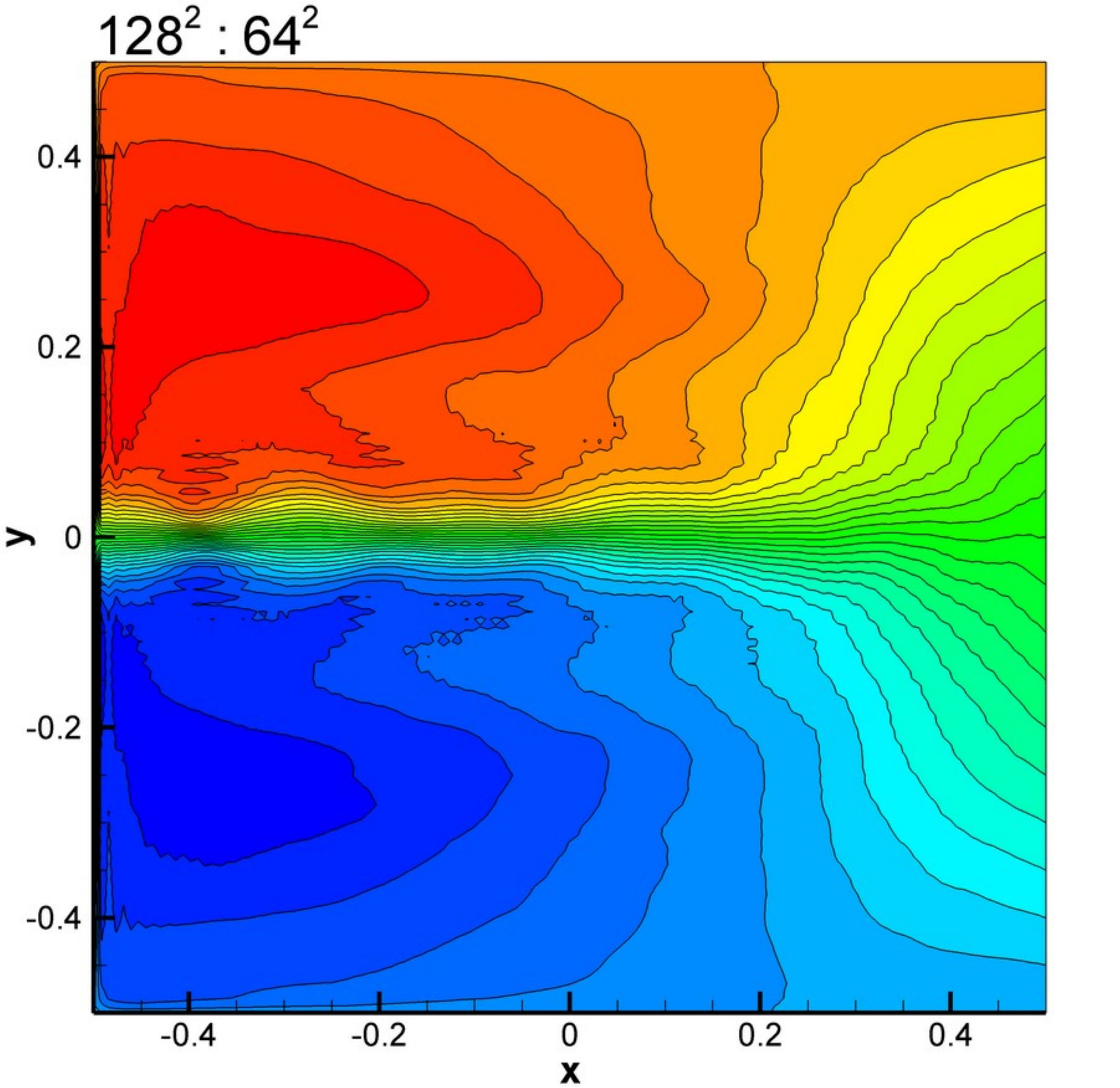}} }
\\
\mbox{
\subfigure{\includegraphics[width=0.33\textwidth]{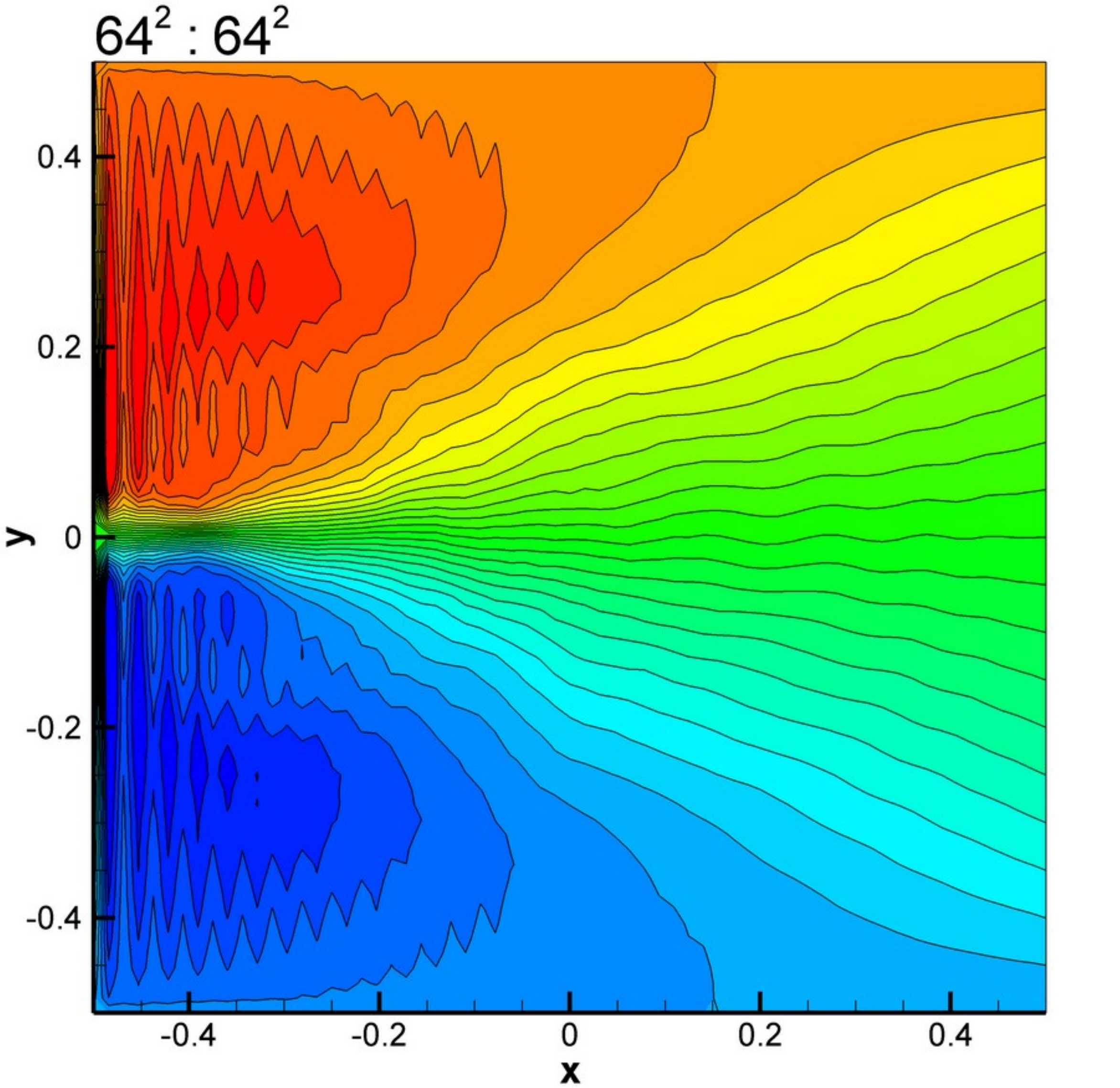}}
\subfigure{\includegraphics[width=0.33\textwidth]{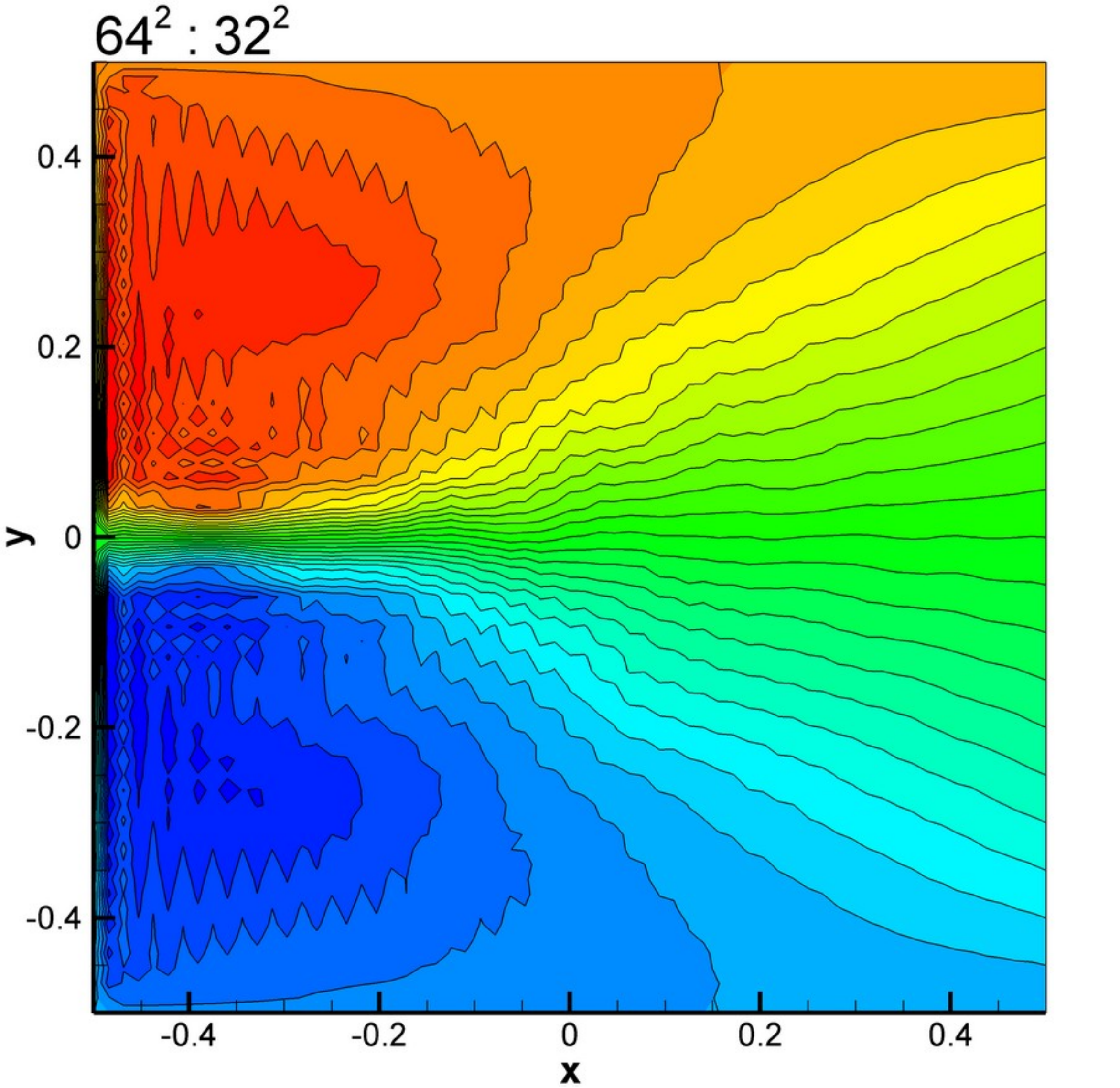}}
\subfigure{\includegraphics[width=0.33\textwidth]{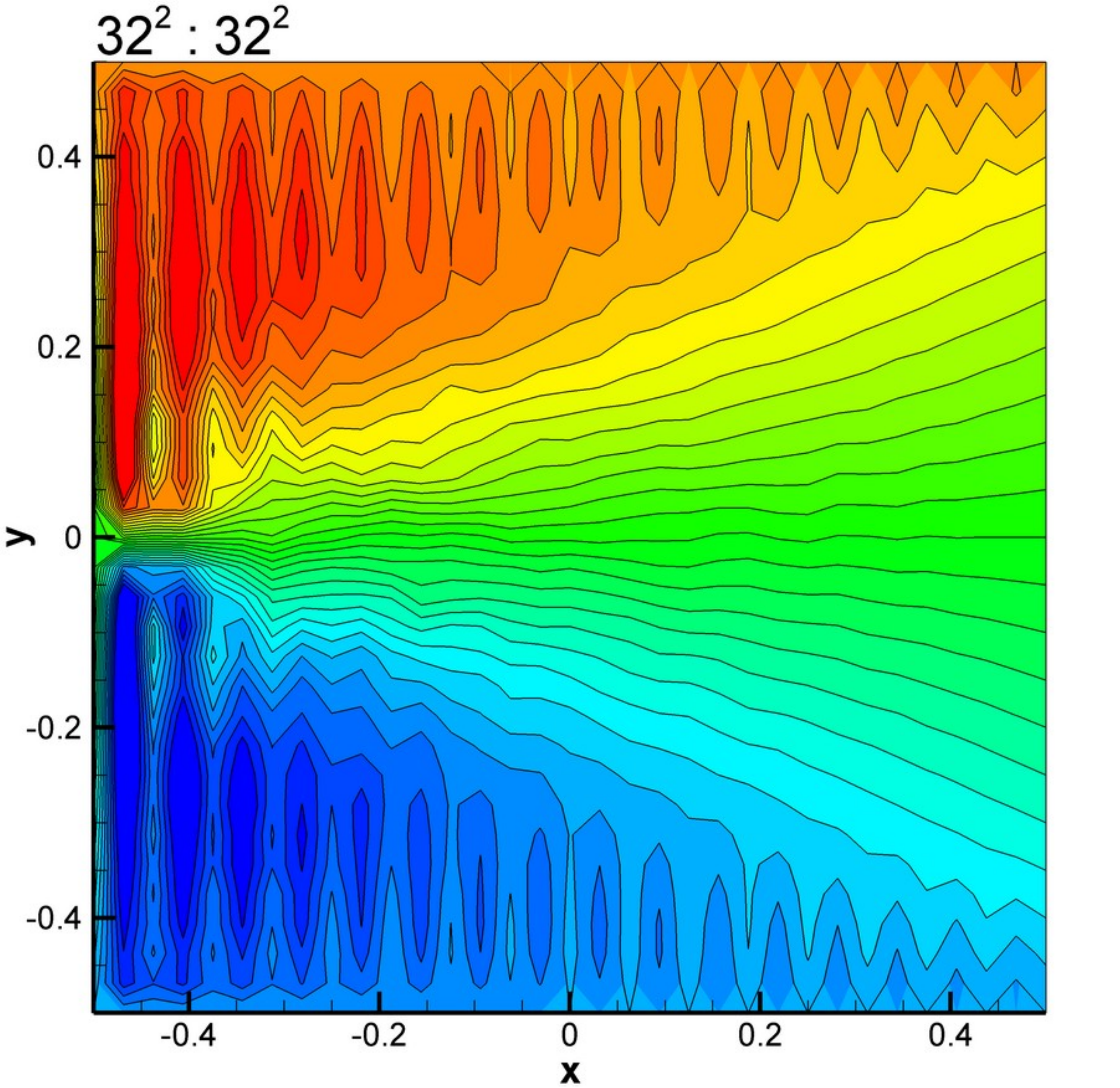}} }
\caption{
Comparison of mean potential vorticity for the upper layer for the Experiment 1. Labels include the resolutions for both parts of the solver in the form $N^2 : M^2$, where $N^2$ is the resolution for the vorticity transport equations, and $M^2$ is the resolution for the elliptic sub-problems.
}
\label{fig:q-O}
\end{figure*}

\begin{figure*}
\centering
\mbox{
\subfigure{\includegraphics[width=0.33\textwidth]{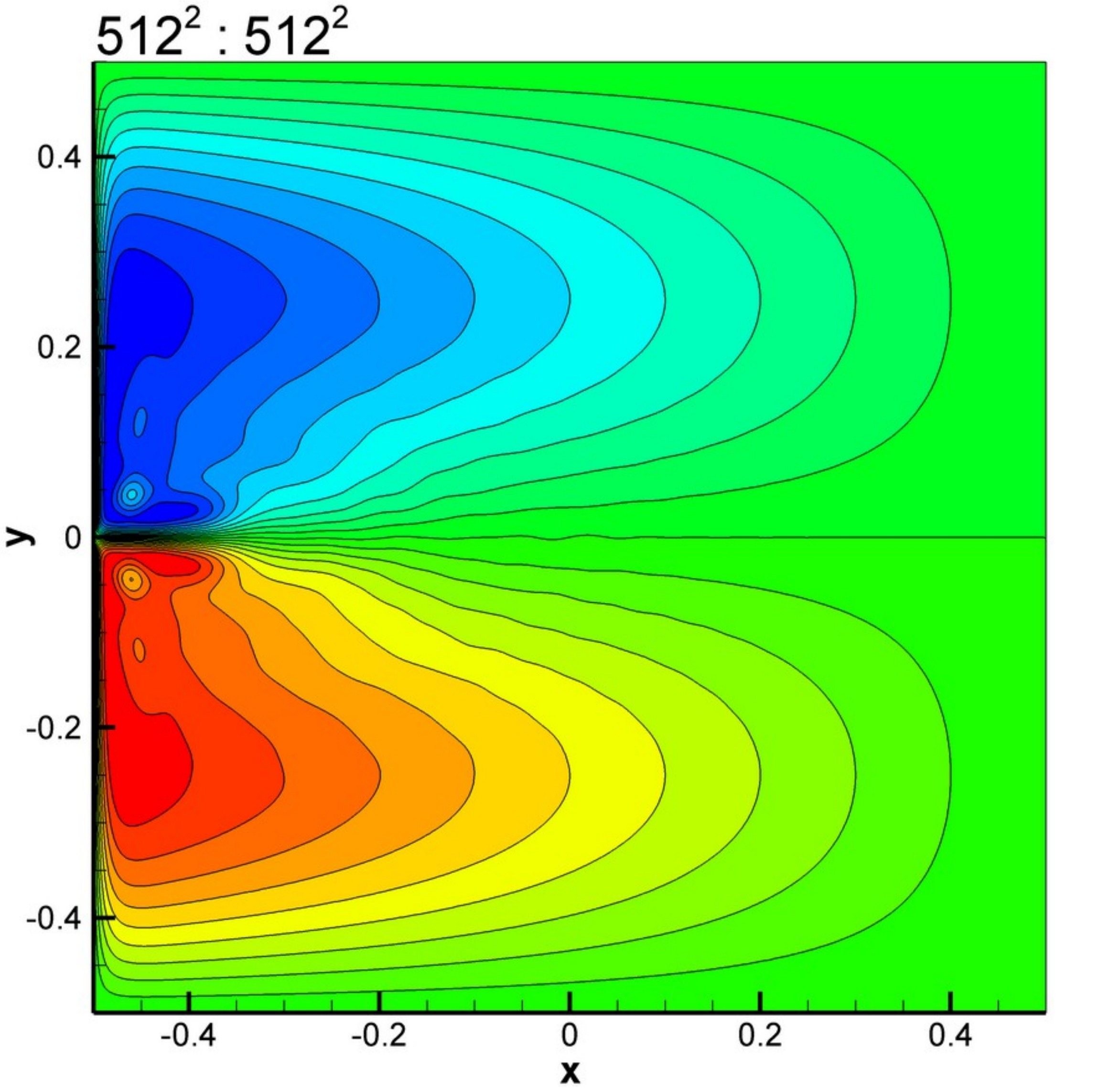}}
\subfigure{\includegraphics[width=0.33\textwidth]{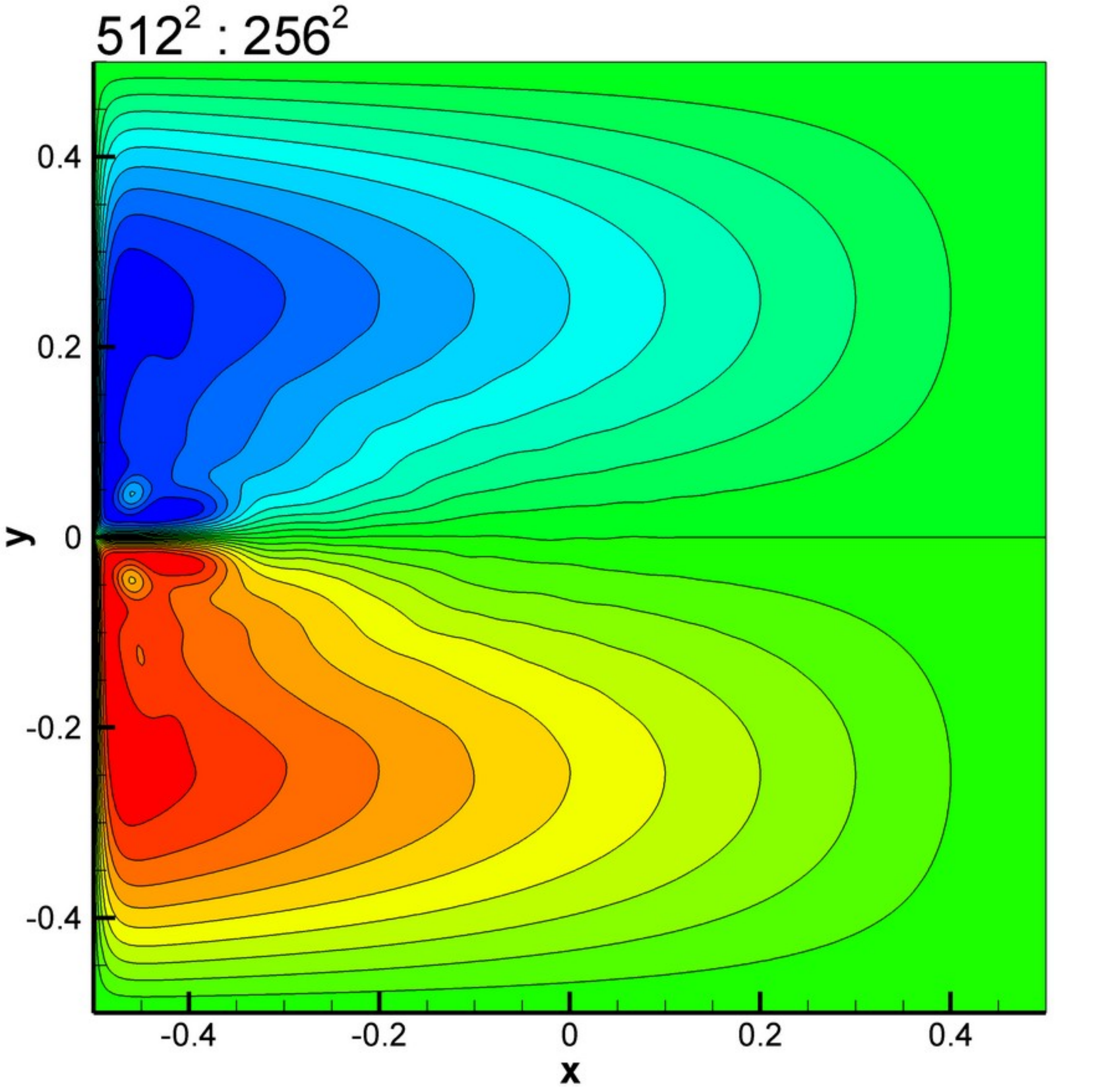}}
\subfigure{\includegraphics[width=0.33\textwidth]{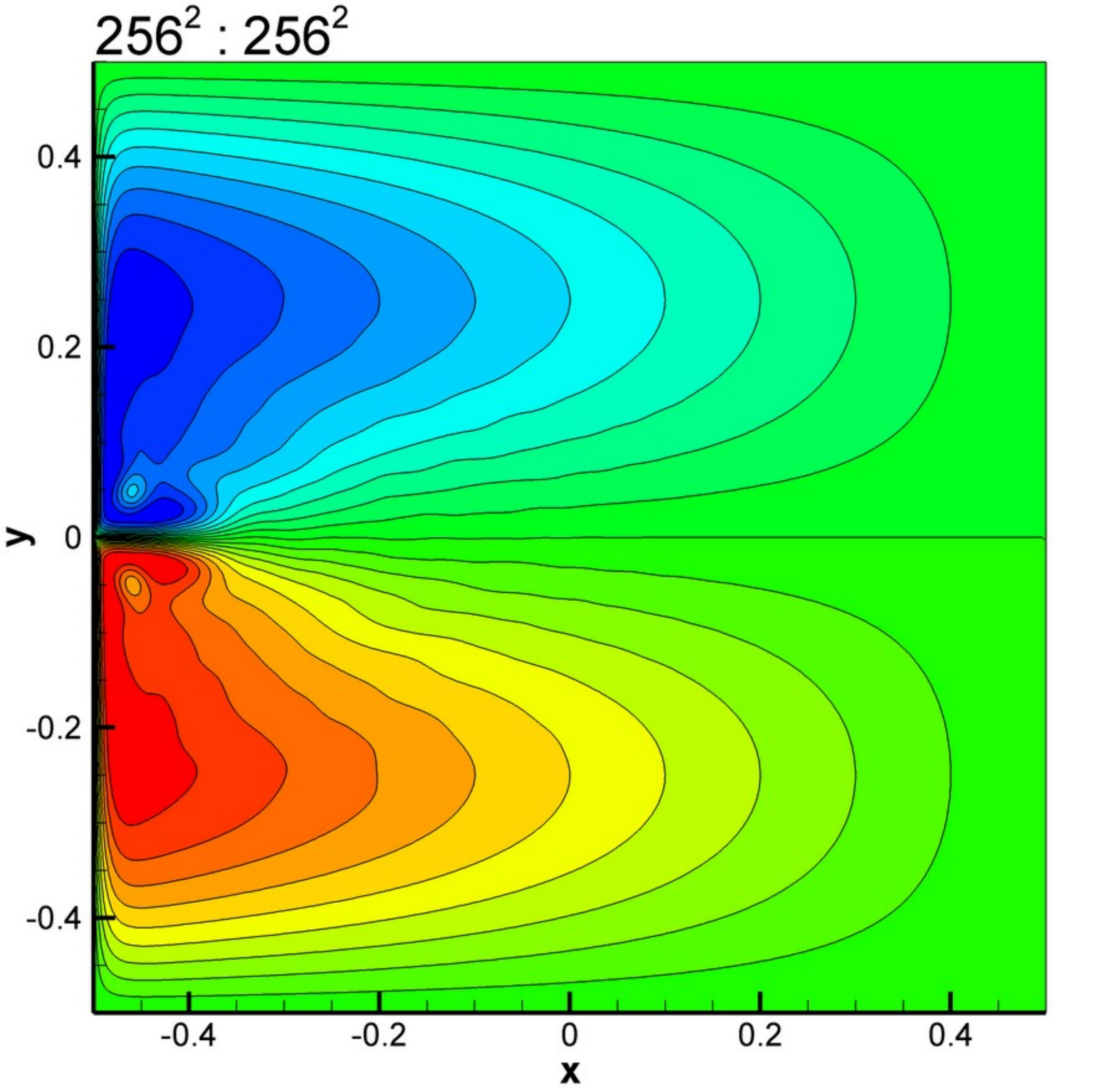}} }
\\
\mbox{
\subfigure{\includegraphics[width=0.33\textwidth]{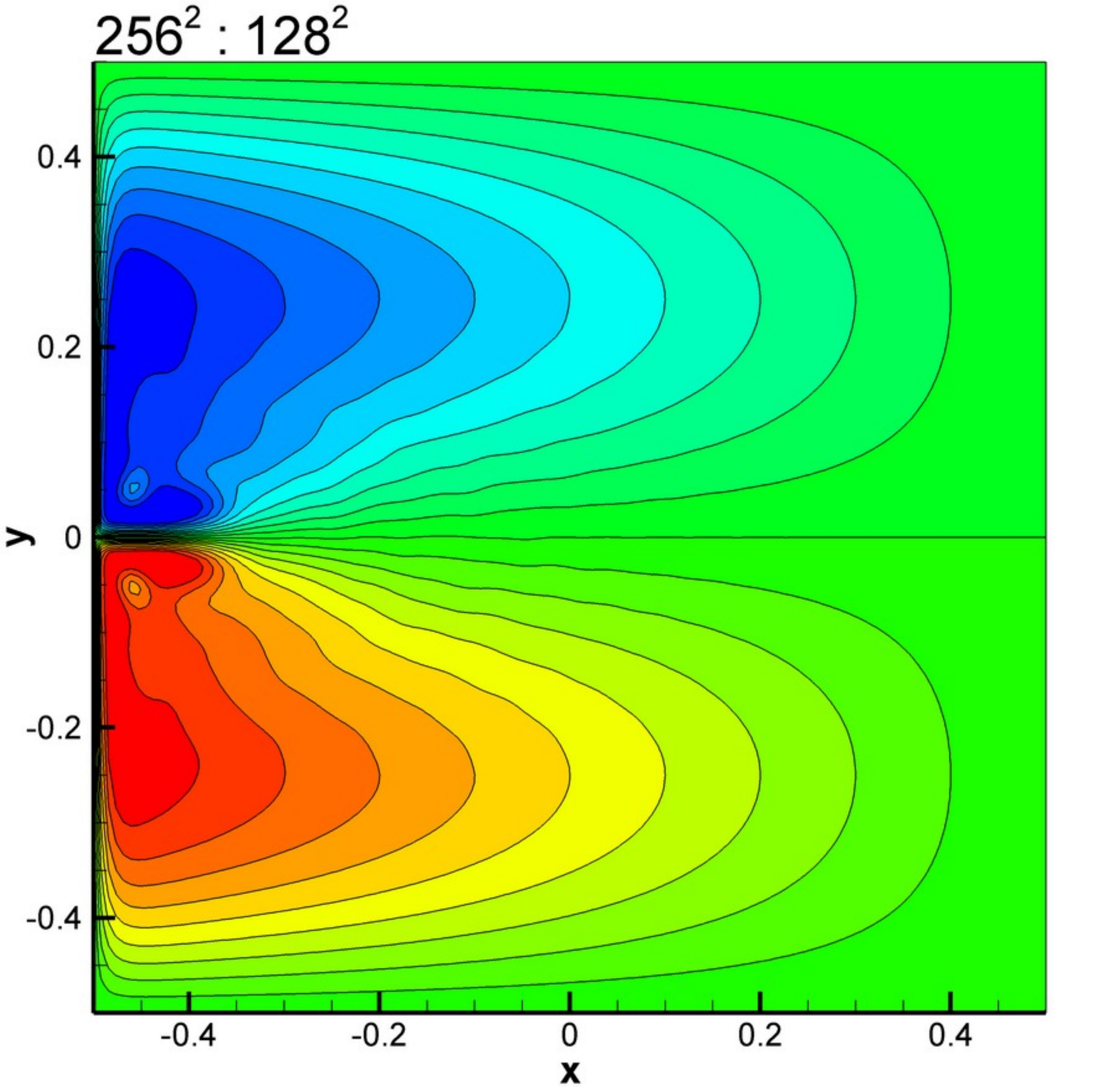}}
\subfigure{\includegraphics[width=0.33\textwidth]{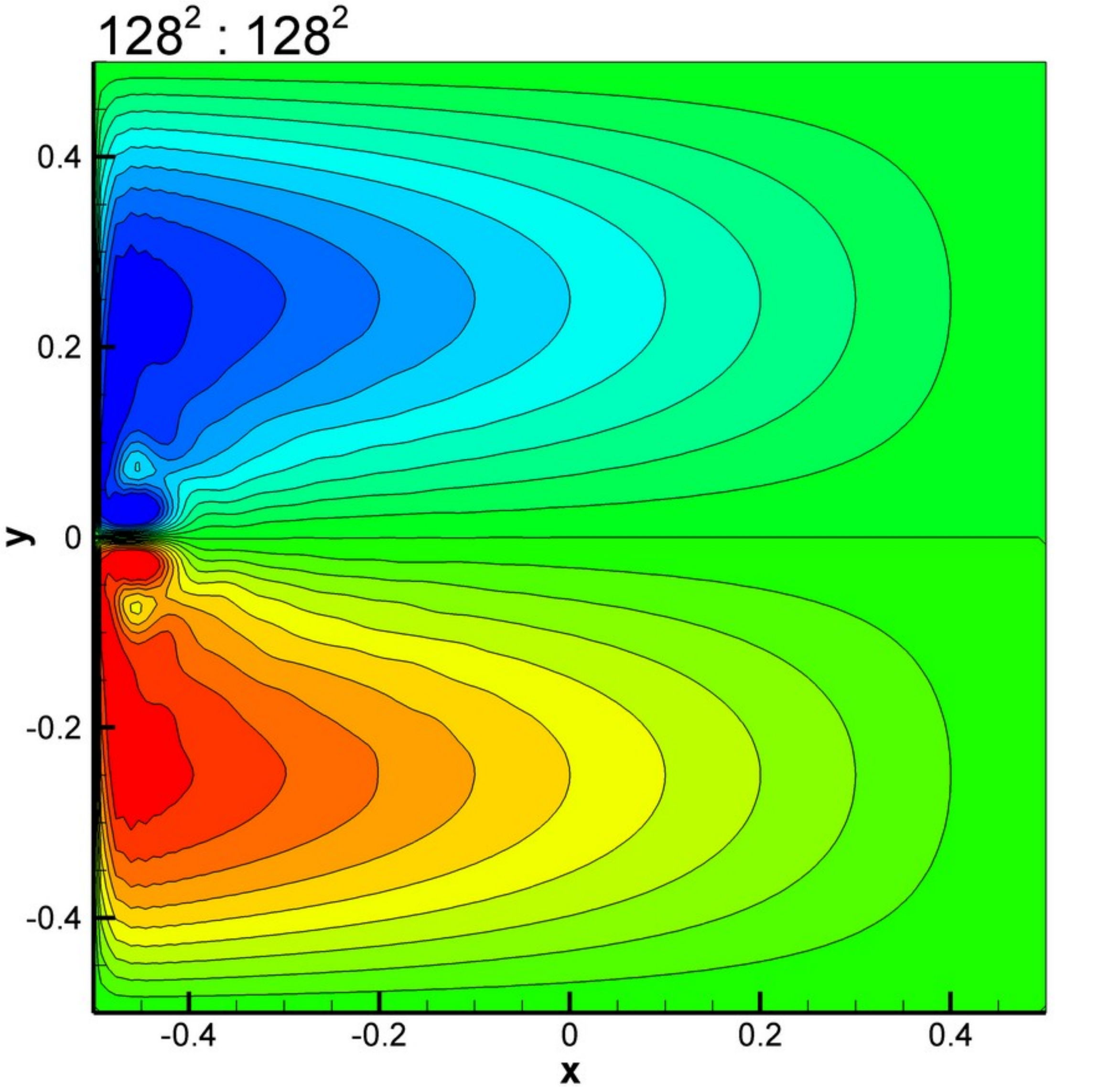}}
\subfigure{\includegraphics[width=0.33\textwidth]{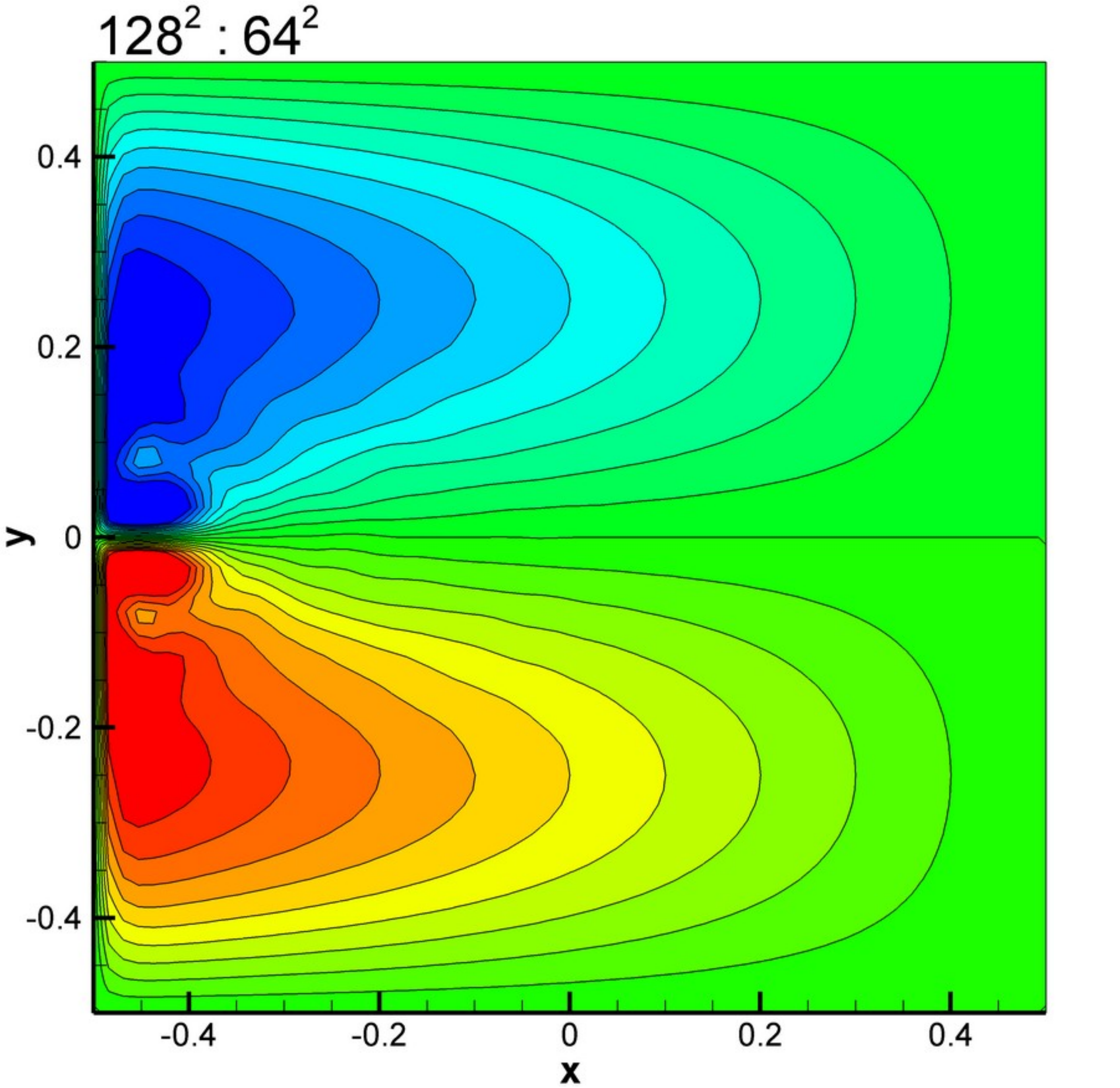}} }
\\
\mbox{
\subfigure{\includegraphics[width=0.33\textwidth]{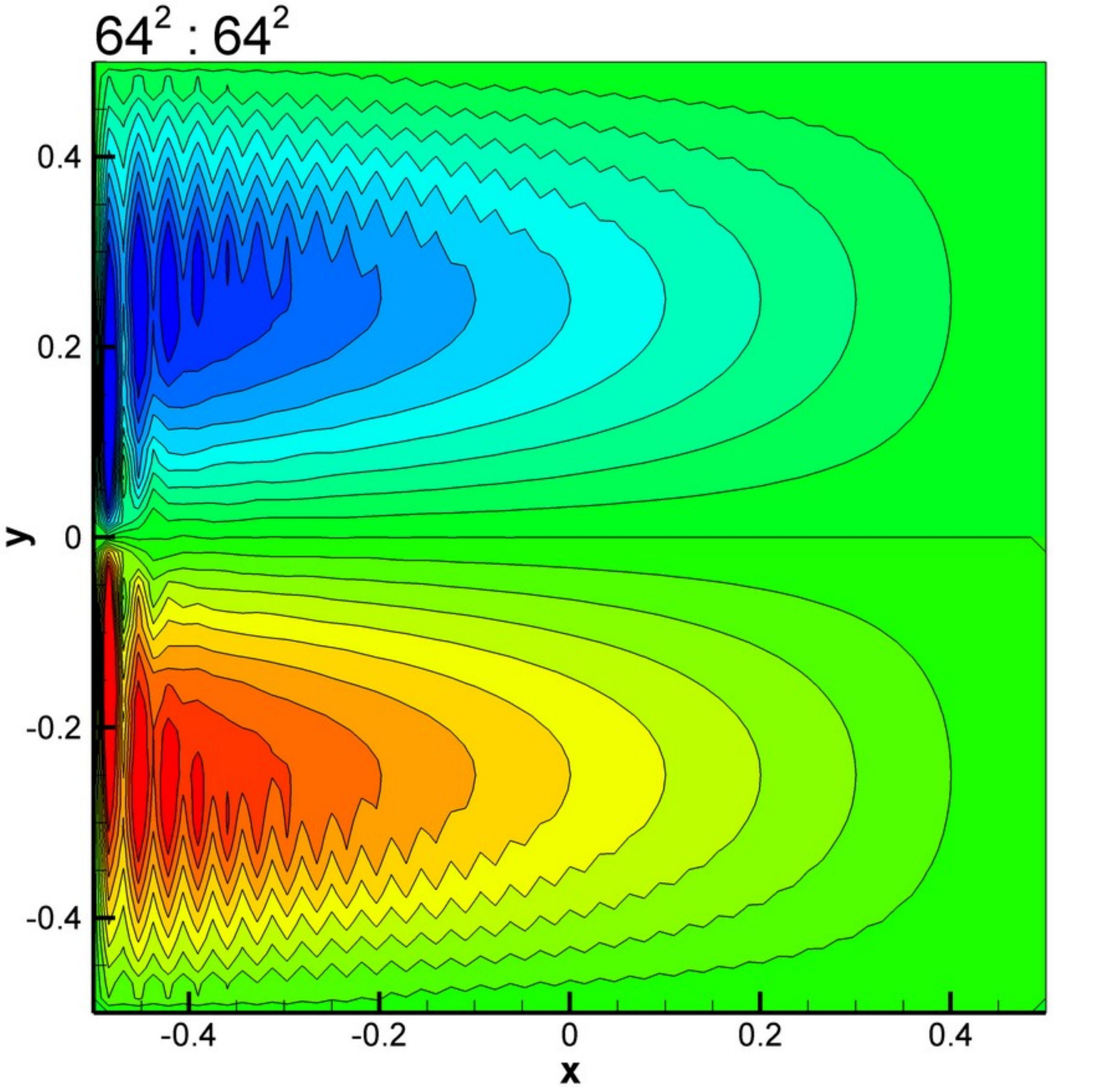}}
\subfigure{\includegraphics[width=0.33\textwidth]{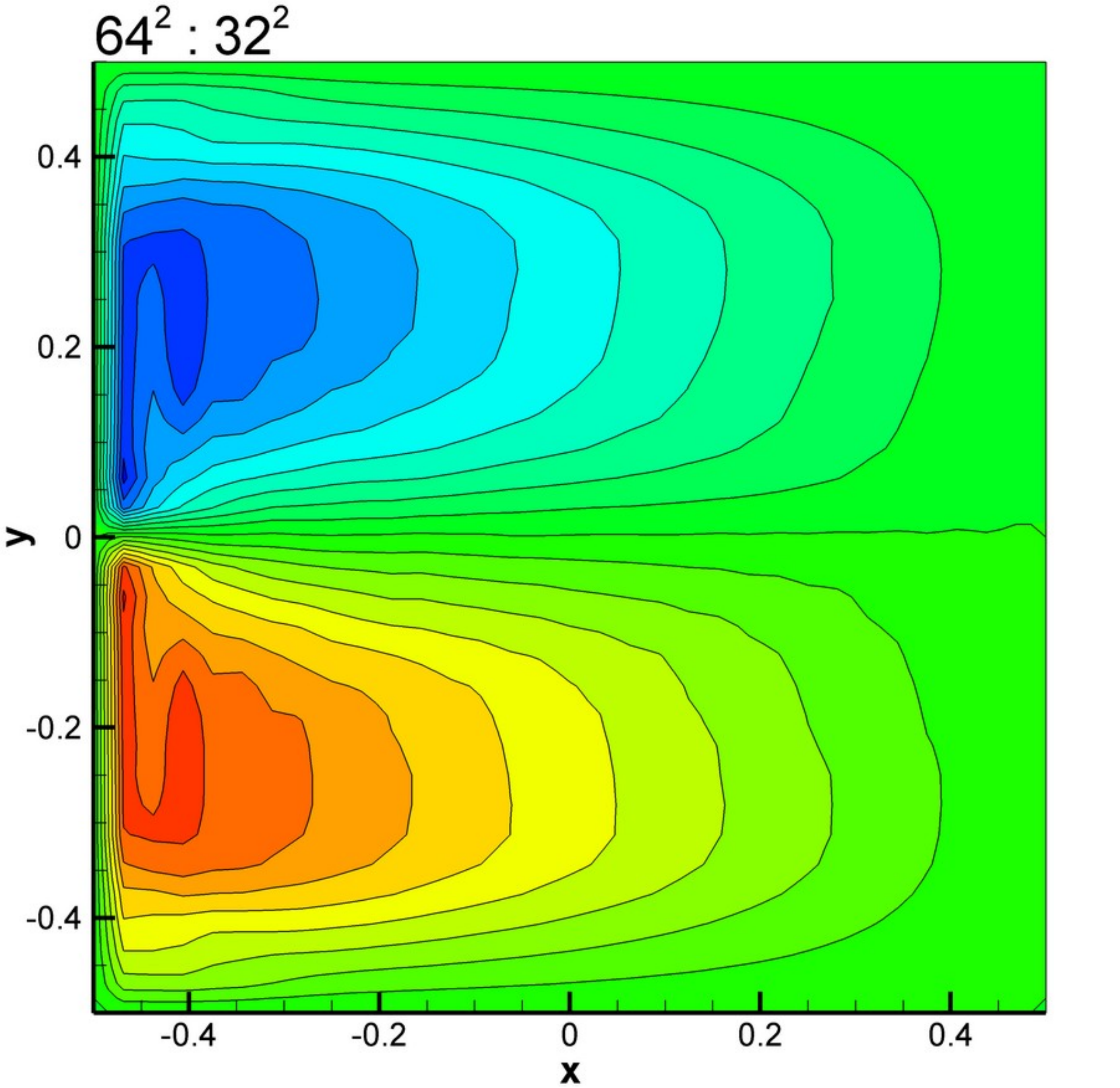}}
\subfigure{\includegraphics[width=0.33\textwidth]{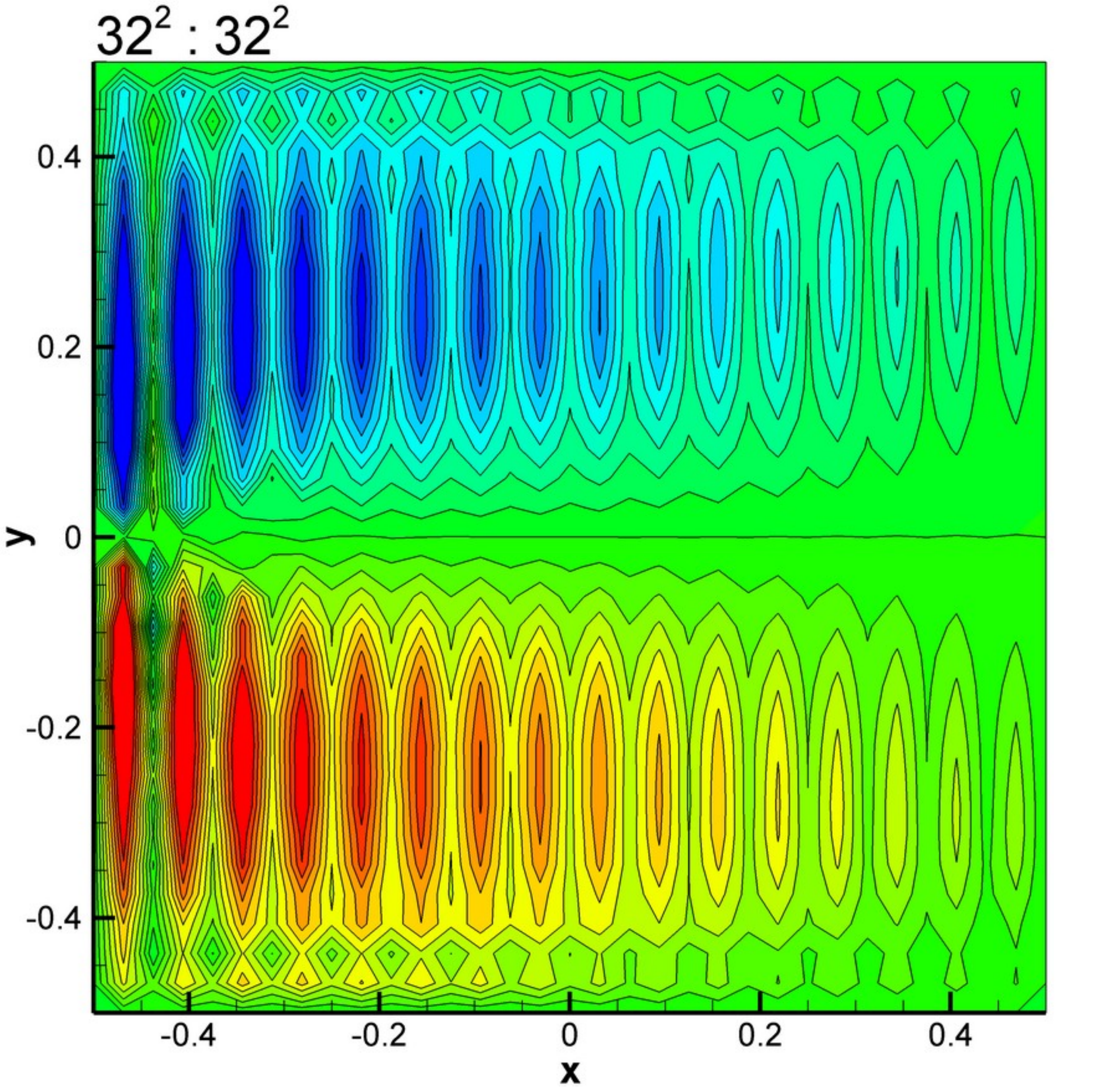}} }
\caption{
Comparison of mean stream function for the upper layer for the Experiment 2. Labels include the resolutions for both parts of the solver in the form $N^2 : M^2$, where $N^2$ is the resolution for the vorticity transport equations, and $M^2$ is the resolution for the elliptic sub-problems.
}
\label{fig:s-T}
\end{figure*}

\begin{figure*}
\centering
\mbox{
\subfigure{\includegraphics[width=0.33\textwidth]{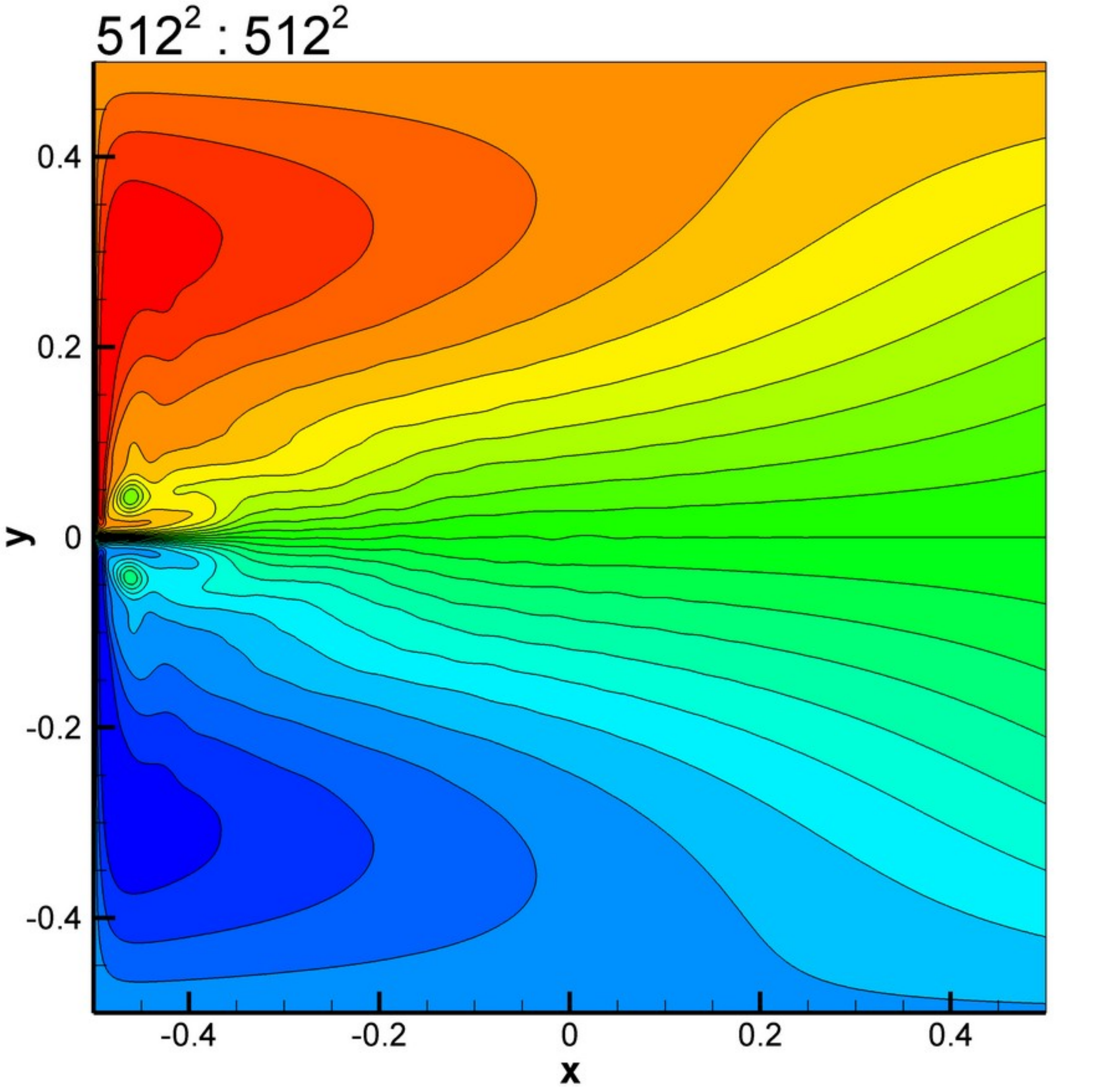}}
\subfigure{\includegraphics[width=0.33\textwidth]{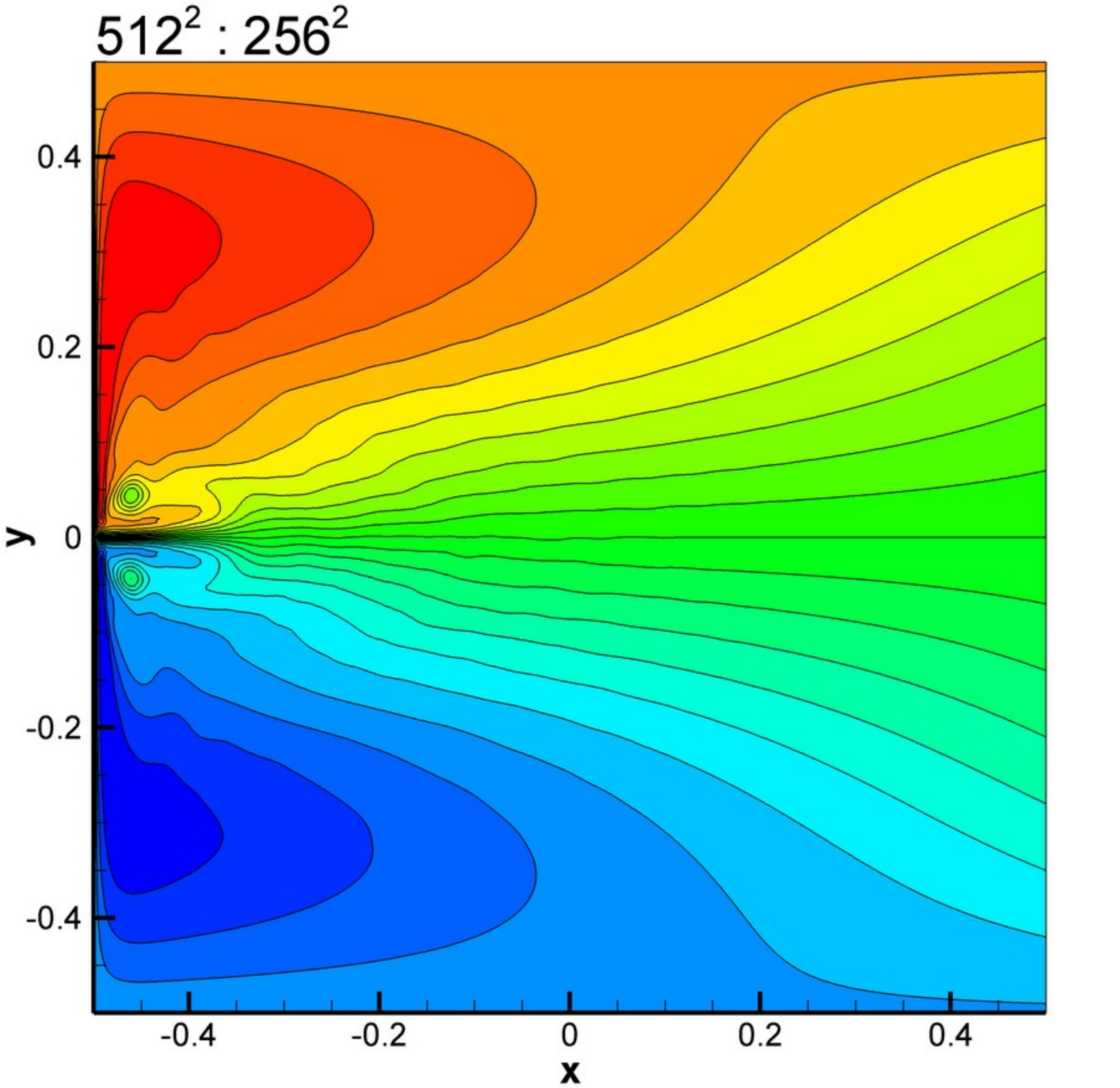}}
\subfigure{\includegraphics[width=0.33\textwidth]{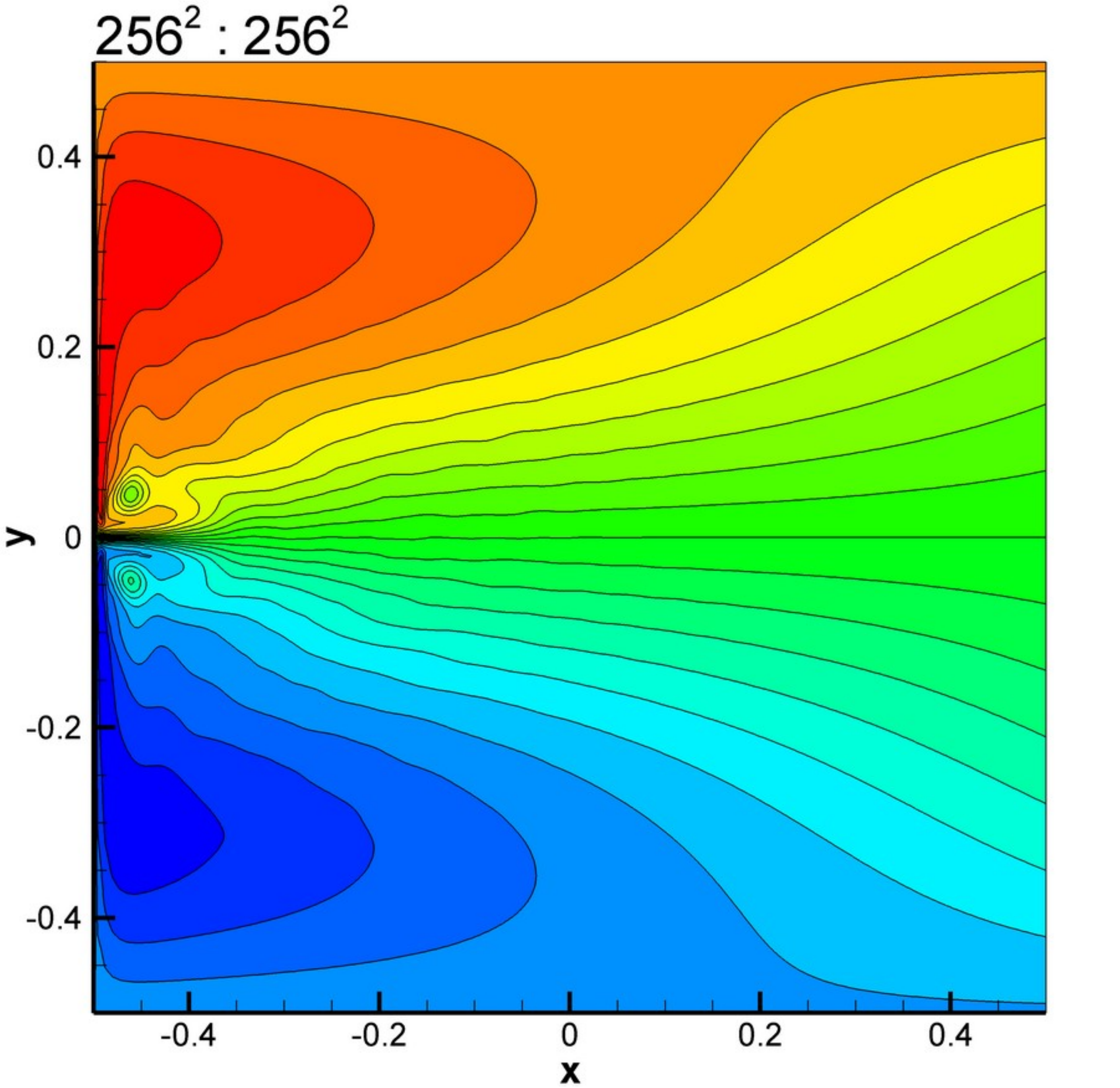}} }
\\
\mbox{
\subfigure{\includegraphics[width=0.33\textwidth]{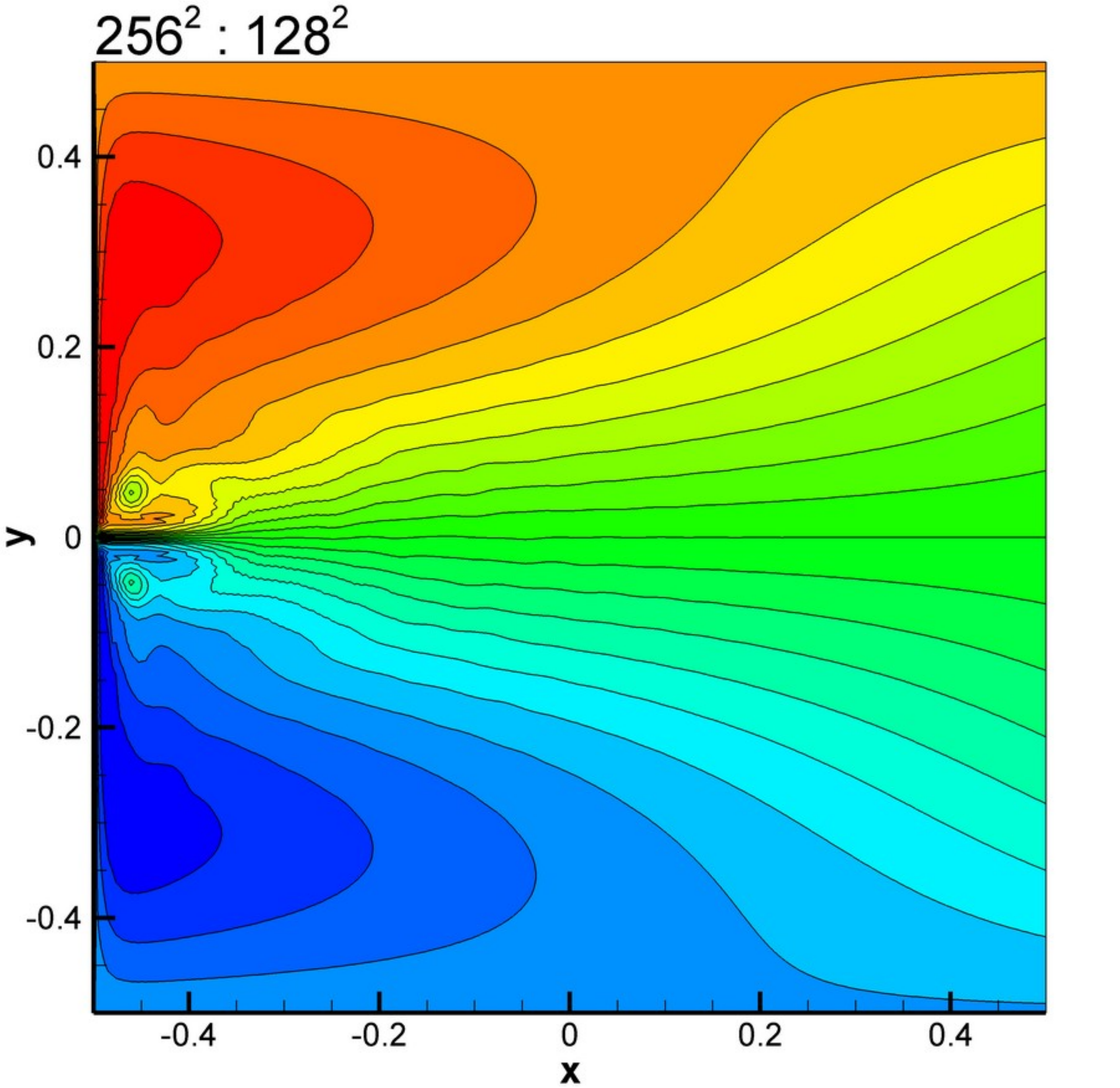}}
\subfigure{\includegraphics[width=0.33\textwidth]{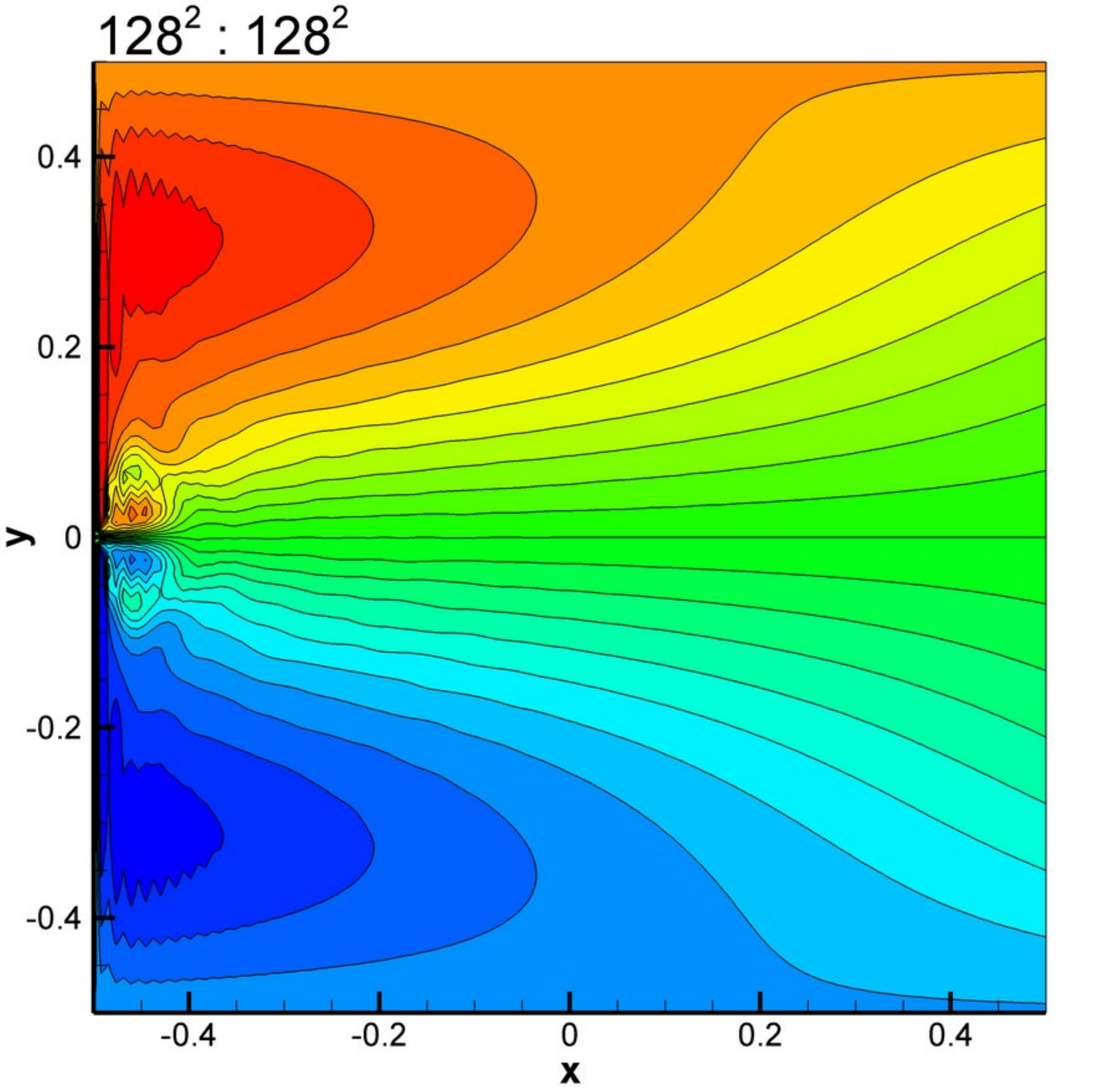}}
\subfigure{\includegraphics[width=0.33\textwidth]{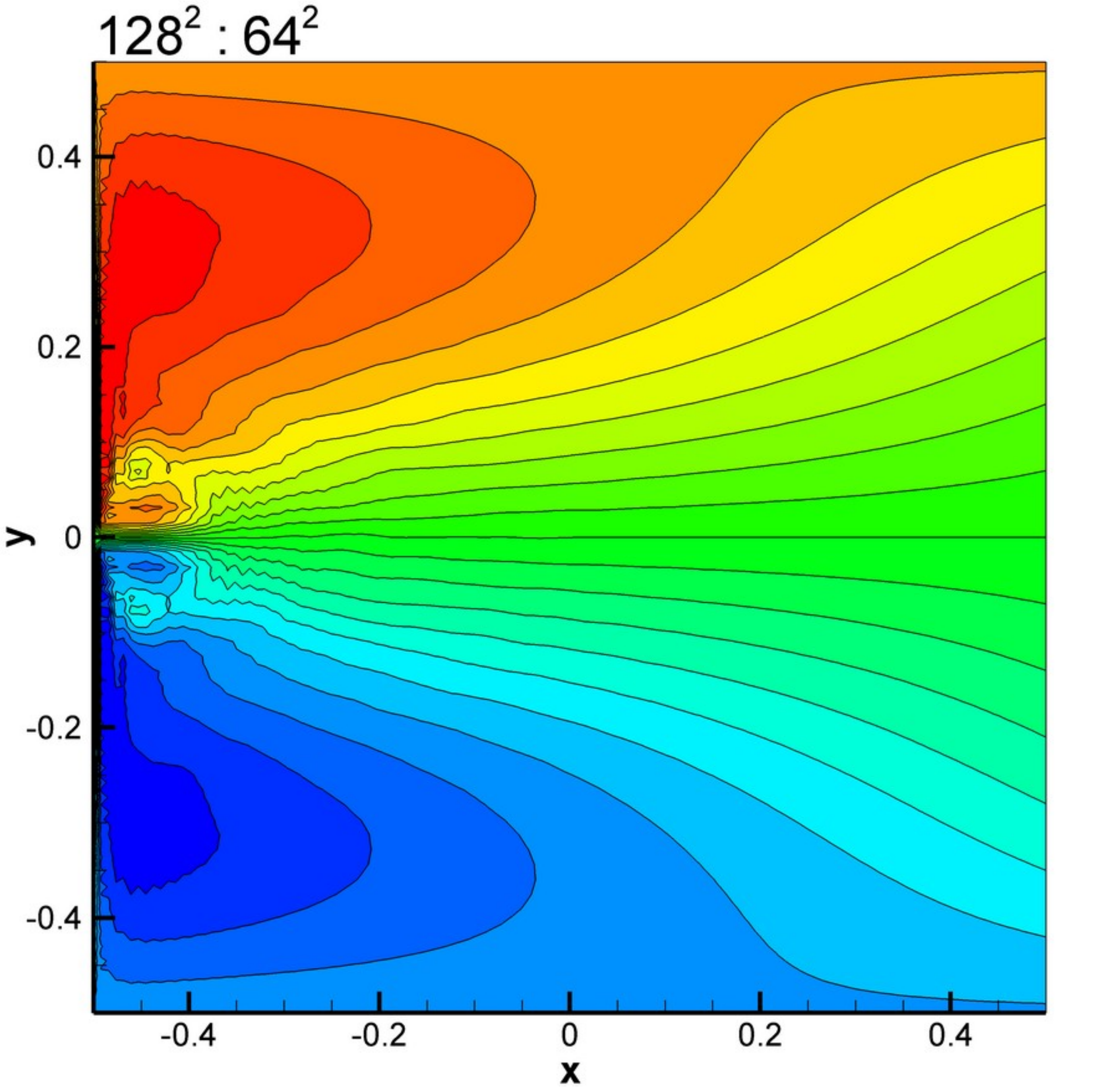}} }
\\
\mbox{
\subfigure{\includegraphics[width=0.33\textwidth]{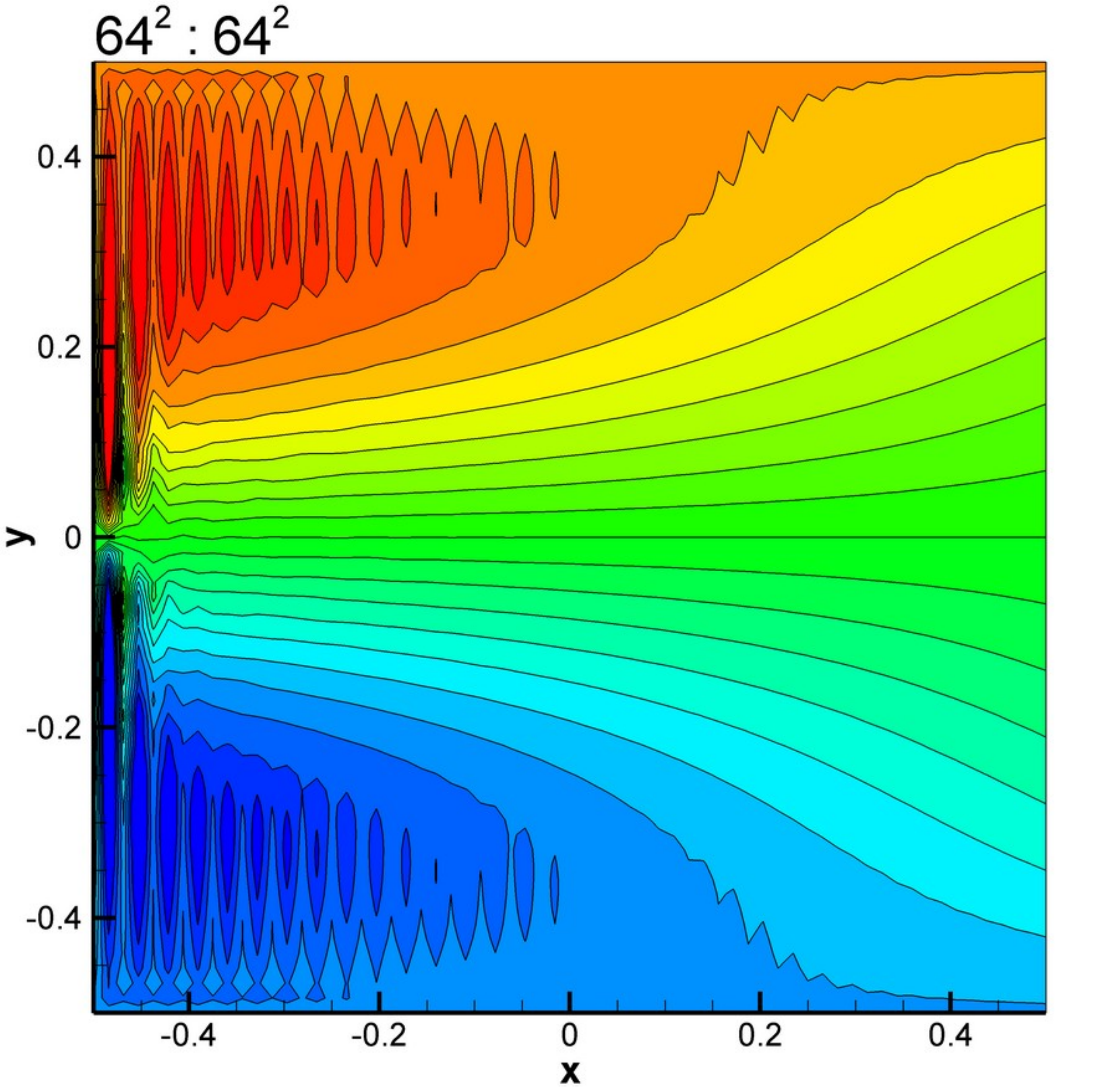}}
\subfigure{\includegraphics[width=0.33\textwidth]{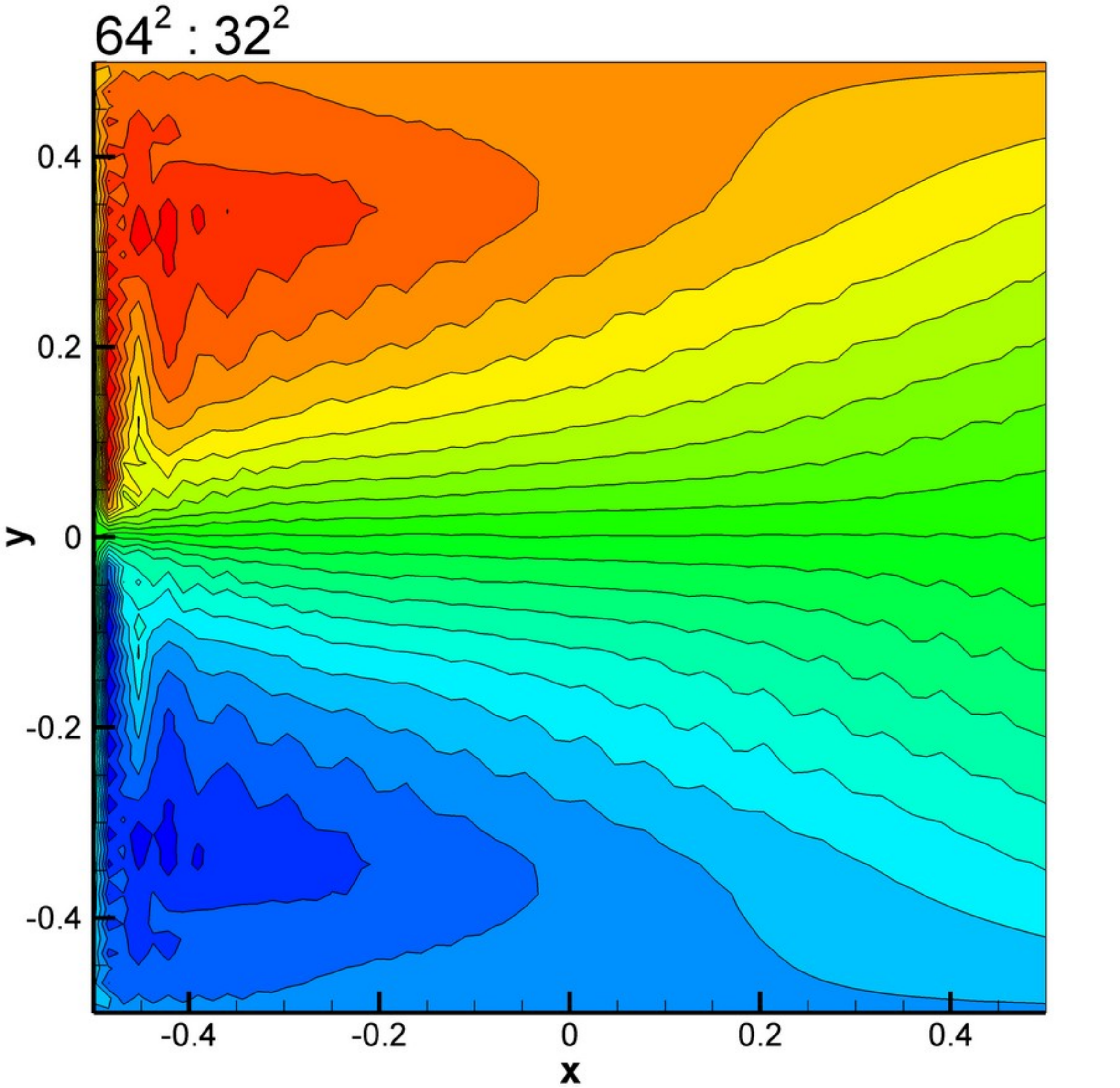}}
\subfigure{\includegraphics[width=0.33\textwidth]{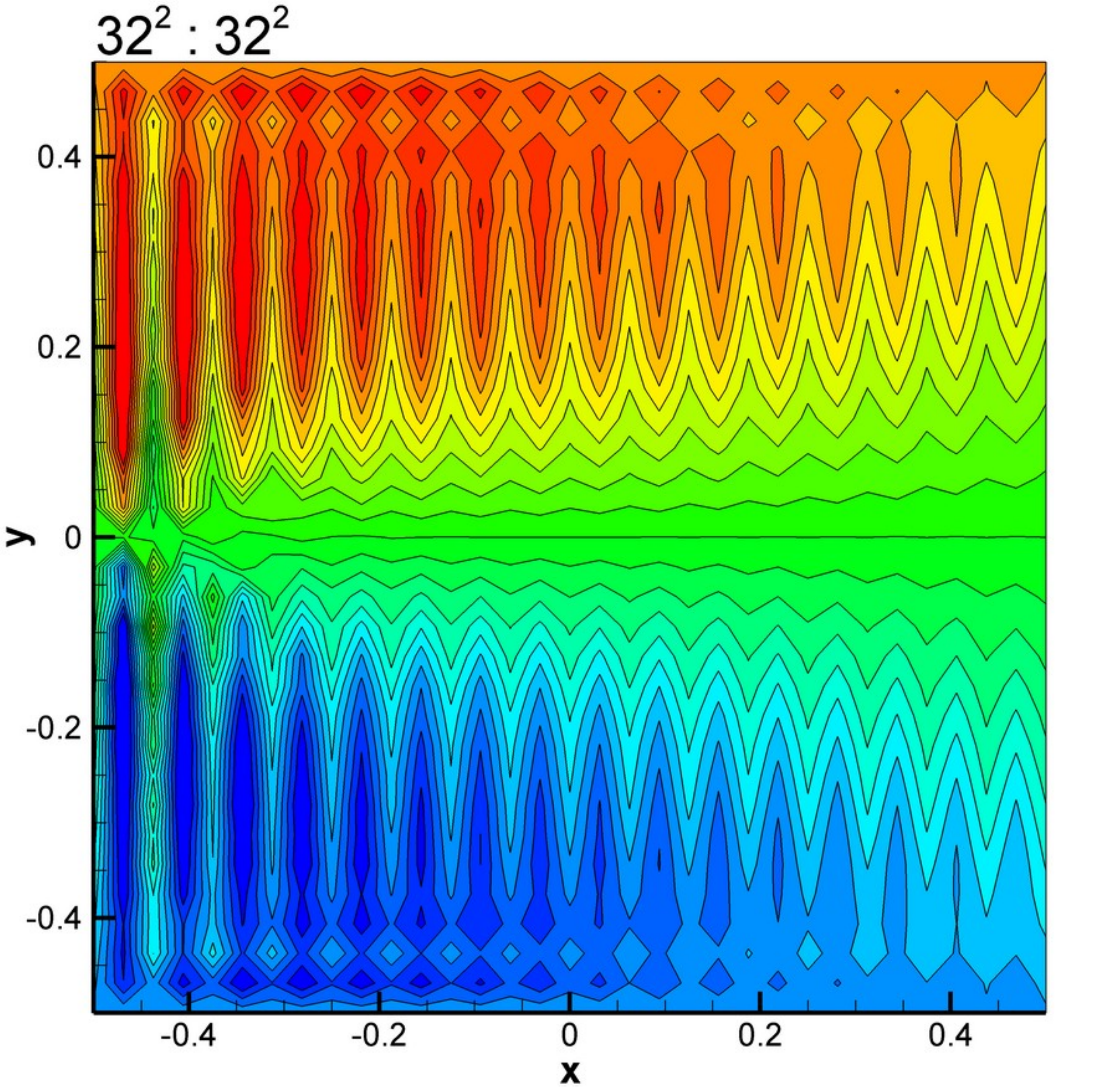}} }
\caption{
Comparison of mean potential vorticity for the upper layer for the Experiment 2. Labels include the resolutions for both parts of the solver in the form $N^2 : M^2$, where $N^2$ is the resolution for the vorticity transport equations, and $M^2$ is the resolution for the elliptic sub-problems.
}
\label{fig:q-T}
\end{figure*}

We start by performing a DNS on a fine mesh of $512^2$ spatial resolution.
Similar to the one-layer analysis, we emphasize that the term DNS in this study is not meant to indicate that a fully detailed solution is being computed on the molecular viscosity scale, but instead refers to resolving the simulation down to the Munk scale via the specified lateral eddy viscosity parameterization. A statistically steady state solution is obtained after an initial transient spin-up process. The basin integrated total kinetic energy is
\begin{equation}
E_{i}(t)
= \frac{1}{2} \iint \left(\frac{\partial \psi_i}{\partial x}\right)^2
+ \left(\frac{\partial \psi_i}{\partial y} \right)^2 dx \, dy ,
\label{eq:26}
\end{equation}
where the subscript $i$ represents the layer index. Figure~\ref{fig:hist-2} shows the time history of the basin integrated kinetic energy for the upper and lower layers for both of the oceanic settings. As shown in this figure, the system reaches the statistically steady state after the dimensionless time $t=1$. As expected, the total energy of the upper layer is much larger than that in the lower layer. It is important to emphasize that, second layer in Experiment 1 is more active than that of Experiment 2. Instantaneous contour plots for the potential vorticities in the upper and lower layers are shown in Figure~\ref{fig:inst-2a} and Figure~\ref{fig:inst-2b} for Experiment 1 and Experiment 2, respectively. The length scales in these two experiments are quite different. For example, the ratio of the basin length scale $L$ to the Rossby deformation radius $R_d$ is $L/R_d=46.74$ for Experiment 1 and $L/R_d=160.5$ for Experiment 2. Therefore, the structure of the eastward jet formation on the western boundary for Experiment 1 is different from that of Experiment 2. This difference becomes more obvious in the mean flow field. The results for time-averaged mean field data obtained from $75$ thousands snapshots between time $t=10$ and $t=25$ in the statistically steady state are given in Figure~\ref{fig:mean-1} and Figure~\ref{fig:mean-2}. The results show strong western boundary currents with cyclonic (counter-clockwise rotating) subpolar gyres and anticyclonic (clockwise rotating) subtropical gyres producing a strong eastward jet in both experiments. However, the produced eastward jet formation in Experiment 1 shows swirling structure and almost reaches the eastern boundary of the basin. Compared to Experiment 2, the bottom layer is more active in Experiment 1. Since in Experiment 1 we used the same parameters and boundary conditions as in \cite{özgökmen1998emergence}, the plot in Figure~\ref{fig:mean-1} is similar to Figure 2 in  \cite{özgökmen1998emergence}. Although in Experiment 2 we have used the same parameters as those used in \cite{tanaka2010alternating}, the boundary conditions we used are different from their boundary conditions: we used the slip boundary conditions, whereas they used the no-slip boundary conditions. Thus, the plot in Figure~\ref{fig:mean-2} is different from the corresponding one in \cite{tanaka2010alternating}.

To test the CGP method, we employ the standard coarsening methodology. The criterion used in assessing the success of the new CGP model is its ability to produce more accurate (i.e., closer to the DNS) results than those for standard method without CGP having the same resolution for elliptic sub-problem, without a significant increase in computational time. For Experiment 1, we plot the mean stream function and potential vorticity contours in Figs.~(\ref{fig:s-O}) and~(\ref{fig:q-O}), respectively. The various resolutions used in the computations are written in the labels of the subfigures. The labels include the resolution for the time dependant part and the resolution for the elliptic part, as described in the captions. As was the case in previous analysis for one-layer model, we compare the CGP computations with standard computations without CGP. For example, the flow field obtained by a standard computation with $128^2:128^2$ agrees well with that obtained by the CGP method with $128^2:64^2$, and better than that of the standard computation with $64^2:64^2$. Similar observations holds for higher and lower resolution computations as well. It can clearly be seen that the CGP approach provides results at the same level of accuracy, but at a reduced computational cost compared to the standard, fine-resolution simulations, by reducing the resolution of the elliptic sub-problem.

Similarly, we plot the mean stream function and potential vorticity contours in Figs.~(\ref{fig:s-T}) and~(\ref{fig:q-T}) for Experiment 2. We note that the proposed CGP model yields again improved results by smoothing out the numerical oscillations present in the under-resolved standard simulations without CGP. The numerical results for both experiments clearly suggest that the the CGP model can provide relatively accurate results for stratified geophysical flows at a low computational cost.

\section{Conclusions}
\label{sec:summary}

A new coarse grid projection (CGP) multiscale method was introduced for large-scale ocean circulation models. The CGP method was tested in the numerical simulation of the wind-driven circulation in both one-layer and two-layer ocean basins, standard prototypes of more realistic ocean dynamics. The first mathematical model employed was the barotropic vorticity equation (BVE) for the one-layer quasigeostropic model, which is driven by a symmetric double-gyre wind forcing that yielded a four-gyre circulation in the time mean. The second mathematical model used was the two-layer quasigeostropic equations for a stratified ocean model accounting for baroclinic effects.

In the CGP methodology the cost of large-scale ocean dynamics computations is reduced by coarsening the number of grid points used for the solution of the elliptic sub-problem in quasigeostropic (QG) ocean models. The CGP approach is general and in fact constitutes a family of methods, since in addition to choosing the coarsening and prolongation operators and the time integration scheme, the elliptic solver used in the approach can vary. In this work, we investigated the performance of a particular CGP method that uses an optimal FFT based elliptic solver, the full weighting operation for the coarsening operator, and bilinear interpolation for the prolongation operator. We used the spatially second-order accurate fully conservative Arakawa scheme along with the third-order TVD Runge-Kutta time integration scheme. The CGP method was tested on meshes that were coarser than those used for the direct numerical simulation (DNS) computations.
The CGP method yielded numerical results that were in close agreement with those of the DNS for both barotropic and stratified ocean circulation models. In particular, we found that the new CGP models with both one-level and multiple-level coarsening yield improved results by smoothing out the numerical oscillations present in the results obtained by under-resolved standard computations without the CGP procedure. This first step in the numerical assessment of the proposed ocean circulation models shows that CGP methodology could represent a viable tool for QG models of more realistic turbulent geophysical flows.



\section*{Acknowledgements}
This research was partially supported by the Institute for Critical Technology and Applied Science (ICTAS) at Virginia Tech via grant number 118709.












\end{document}